\documentclass[amssymb,amsmath,amstext,amsfont,amsthm,12pt]{amsart}
\usepackage{graphicx}
\usepackage{accents}
\usepackage{enumerate}

\newtheorem{theorem}{Theorem}

\DeclareMathOperator{\tr}{tr}

\allowdisplaybreaks[4]

 \numberwithin{equation}{section}

\begin{document}
%\begin{titlepage}
\title[Axisymmetric perturbations]{A Positive-Definite Energy Functional for the Axisymmetric Perturbations of Kerr-Newman Black Holes}
\author{Vincent Moncrief}
%\affiliation{Department of Physics and Department of Mathematics, \\ Yale %University, P.O. Box 208120, New Haven, CT 06520, USA. \\ E-mail address: %vincent.moncrief@yale.edu}
\address{Department of Physics and Department of Mathematics, Yale University, P.O. Box 208120, New Haven, CT 06520, USA.}
\email{vincent.moncrief@yale.edu}
\author{Nishanth Gudapati}
\address{Center of Mathematical Sciences and Applications, Harvard University, 20 Garden Street, Cambridge, MA-02138, USA}
\email{nishanth.gudapati@cmsa.fas.harvard.edu}
%\affiliation{Department of Mathematics, \\ Yeshiva University, 500 West 185th Street, New York, NY 10033, USA. \\ and \\ Department of Mathematics, \\ University of L'Aquila, Via Vetoio, 67010 L'Aquila, AQ ITALY. \\ E-mail address: marini@yu.edu}
%\author{Rachel Maitra}
%\affiliation{Department of Physics, \\ Albion College, 611 E. Porter Street, Albion, MI 49224, USA. \\ E-mail address: rmaitra@albion.edu}
\date{\today}
\begin{abstract}
We consider the \textit{axisymmetric}, linear perturbations of Kerr-Newman black holes, allowing for arbitrarily large (but subextremal) angular momentum and electric charge. By exploiting the famous Carter-Robinson identities, developed previously for the proofs of (stationary) black hole uniqueness results, we construct a positive-definite energy functional for these perturbations and establish its conservation for a class of (coupled, gravitational and electromagnetic) solutions to the linearized field equations. Our analysis utilizes the familiar (Hamiltonian) reduction of the field equations (for axisymmetric geometries) to a system of \textit{wave map} fields coupled to a 2+1-dimensional Lorentzian metric on the relevant quotient 3-manifold. The propagating `dynamical degrees of freedom' of this system are entirely captured by the wave map fields, which take their values in a four dimensional, negatively curved (complex hyperbolic) Riemannian target space whereas the base-space Lorentzian metric is entirely determined, in our setup, by elliptic constraints and gauge conditions.

The associated linearized equations are analyzed with insight derived from the so-called `linearization stability' program for such (generally covariant) systems. In particular this program provides a natural connection between the (conserved, positive-definite) energy defined for first order perturbations and the correction to the ADM mass induced therefrom at second order. A well-known technique allows one to generate, for sufficiently smooth perturbations, a sequence of higher order (conserved, positive-definite) energies that, in turn, bound certain higher order (weighted) Sobolev norms of the linearized solutions. We anticipate that our results may prove useful in analyzing the dynamical stability of (arbitrarily rapidly rotating) Kerr-Newman black holes with respect to axisymmetric perturbations. Establishing such stability at the linearized level is expected to be an essential first step in dealing, ultimately, with the nonlinear problem.
\end{abstract}
%\pacs{02.30.Mv, 02.30.Xx, 03.65.Sq}
%\end{titlepage}
\maketitle

\section{Introduction}
\label{sec:introduction}
Impressive observational and experimental evidence has accumulated for the existence of black holes as dynamically stable entities in the Universe. But are these the black holes predicted by general relativity? To conclude that they are would seem to hinge, in large measure, on the success of ongoing mathematical efforts to prove that the purely theoretical, Einsteinian black holes are, themselves, dynamically stable. A natural first step in this direction would be to establish such stability at the level of linear perturbation theory---a long-standing research program that began with the pioneering work of Regge and Wheeler \cite{Regge-Wheeler_57}, Vishveshwara \cite{Vishveshvara_70} and Zerilli \cite{Zerilli_70} for the case of Schwarzschild perturbations and with the discovery, by Teukolsky \cite{Teukolsky_72,Teukolsky_73}, of a separable wave equation for Kerr perturbations. Subsequently the coupled gravitational and electromagnetic perturbations of (electrically charged but non-rotating) Reissner-Nordstr\"{o}m black holes were analyzed by Zerilli through working in a special gauge \cite{Zerilli_74} and by one of us who developed a gauge-independent, Hamiltonian formalism for the perturbative study of such spherically symmetric `backgrounds' \cite{Moncrief_74_3,Moncrief_74_1,Moncrief_74_2}.

A corresponding treatment of (charged \textit{and} rotating) Kerr-Newman black holes has, up until now, been lacking. Indeed, as recently as 2006, Brandon Carter could write that the coupled system of electromagnetic and gravitational Kerr-Newman perturbations `has so far been found to be entirely intractable' \cite{Carter_06}. Much of the early work on black hole perturbation theory is summarized and extended in interesting ways in the classic monograph by Chandrasekhar \cite{Chandrasekhar_83} which, though it includes an independent derivation of the Reissner-Nordstr\"{o}m results, devotes only a few pages to the unsolved, Kerr-Newman problem.

The earlier, somewhat formal, `mode stability' analysis for Schwarzschild perturbations has recently been upgraded to a genuine proof of linear stability by Dafermos, Holzegel and Rodnianski \cite{HDR_16} and, independently, by Hung, Keller and Wang \cite{HKW_20}. On the other hand, much of the recent work on Kerr stability has focused on analyzing the evolution of various, lower spin `probe' fields propagating in given (Kerr) black hole `backgrounds'. Important results of this type have been obtained for scalar \cite{LB_15_1,DR_11,DRS_16,FKS_05,FKS_08}, electromagnetic \cite{LB_15_2} and wave map \cite{IK_15,JLuk_10} fields. The methods employed in the electromagnetic and wave map cases have required that the background black hole be `slowly rotating' in a suitable sense whereas those ultimately developed for scalar field perturbations allow `arbitrarily rapid' rotation (consistent with the preservation of an event horizon).

For the actual gravitational perturbations of Kerr black holes Hollands and Wald have emphasized a crucial distinction between the analysis of axisymmetric versus fully non-symmetric metric perturbations that arises primarily because of the suppression of `superradiance' in the axisymmetric case \cite{WH_13}. They have argued that the existence of a conserved, positive definite `canonical' energy functional for \textit{axisymmetric}, linear perturbations is in fact a \textit{necessary condition} for Kerr stability. For non-rotating (spherically symmetric) backgrounds, on the other hand, the phenomenon of superradiance (whereby a black hole can absorb \textit{negative} radiative energy) disappears (unless electromagnetically charged fields are considered \cite{BR_16}) and the importance of distinguishing between axisymmetric and non-symmetric perturbations is largely dissolved.

One of the main results of Refs. \cite{Moncrief_74_3,Moncrief_74_1,Moncrief_74_2} was in fact the derivation of a conserved, gauge-invariant, positive definite energy functional for the coupled, \textit{dynamical}, gravitational and electromagnetic perturbations of Reissner-Nordstr\"{o}m black holes. Using totally different (`Hertz potential') methods Wald and Prabhu have recently announced that the conserved, `canonical' energy formula for purely gravitational perturbations given by Hollands and Wald in \cite{WH_13} is indeed positive definite when specialized to a Schwarzschild background and they conjecture that a corresponding result should hold for \textit{axisymmetric} Kerr perturbations \cite{WP_13}.

Even for exclusively axisymmetric perturbations, though, a serious obstacle for the construction of a positive definite energy functional for Kerr (or Kerr-Newman) perturbations is the presence of an `ergo-region' lying outside of any (rotating) black hole's event horizon. This is the region in which the `time-translational' Killing field of the unperturbed (Kerr-Newman) spacetime becomes spacelike and conventional local energy density expressions built from it can lose their definiteness. To a limited extent this shortcoming can be handled by introducing `weighted' energy densities that, by exploiting timelike linear combinations of the `time-translational' and rotational Killing fields of the background, interpolate between positive definite density expressions inside the ergo-region and exterior to it. But this technique does not seem to be capable of treating arbitrarily rapid rotation and, since such energies are not strictly conserved, needs additional, technically intricate, Morawetz type estimates for the extraction of uniform bounds on the fields and their derivatives.

By imposing axial symmetry at the outset Dain and his collaborators applied well-known Kaluza-Klein reduction techniques to re-formulate the (fully nonlinear) vacuum field equations as a 2 + 1---dimensional Einstein---wave map system for which the wave map target space is the hyperbolic plane \cite{DA_14,DA_15}. In this formulation the scalar wave map variables represent the truly dynamical gravitational wave degrees of freedom whereas the 2 + 1---dimensional Lorentzian metric to which they are coupled is fully determined by gauge conditions and elliptic constraints. After using this setup in elegant ways to study Penrose inequalities and black hole thermodynamics in the axisymmetric case, they linearized their system and applied it to the Kerr black hole stability problem. By utilizing an extension \cite{DA_14} of the classic Brill mass formula \cite{BD_59} for axisymmetric, vacuum spacetimes expressed in terms of the wave map variables they computed the first and second variations of this functional about a Kerr background and derived therefrom a conserved, positive definite energy functional for the linearized, purely gravitational perturbations of an \textit{extremal} (i.e., maximally rotating) Kerr black hole.

A key step in the logic of their derivation was the observation that, for fixed \textit{angular momentum} (a strictly conserved quantity for axially symmetric evolutions), the extended Brill mass functional is \textit{minimized}, for Cauchy data containing an apparent horizon, precisely at the initial data for an extremal, Kerr black hole. Through an application of Carter's remarkable identity \cite{Car_71}  (that played a fundamental role in the proof of the uniqueness of the Kerr family among \textit{stationary}, asymptotically flat, vacuum black holes without naked singularities) they showed, by an explicit calculation, that the second variation of the extended Brill mass density functional was, up to a spatial divergence term, positive definite. Upon discarding the boundary integral that resulted from integrating this density over a Cauchy surface for the black hole's \textit{domain of outer communications} (DOC) they arrived at an energy expression for the (axisymmetric) linear perturbations of the \textit{extremal} Kerr spacetime's DOC that was both conserved and positive definite.

On the other hand, even though the concept of extremality applies equally well to the Reissner-Nordstr\"{o}m family (with electrical charge playing a role analogous to that of angular momentum for the Kerr case) no such limitation (to extremal black holes) was needed for the derivation of the earlier results which had been obtained by a somewhat analogous variational calculation. Partly for this reason the authors realized that it should be entirely feasible to remove this limitation in the rotating case and treat \textit{sub-extremal} (as well as electrically charged) black holes. We present the results of our analysis herein by deriving an \textit{explicit, positive definite, conserved energy functional} for the axisymmetric (coupled gravitational and electromagnetic) perturbations of arbitrary sub-extremal Kerr-Newman black holes. While the occurrence of a non-negligible electric charge for a black hole is of doubtful astrophysical significance, sub-extremal holes are certainly more astrophysically significant than extremal ones which, in fact, are thought to be unachievable via realistic natural evolutions, an expectation encoded in the `third law' of black hole mechanics \cite{BCH_73}.

While it may not be strictly necessary for our program, we have found it very illuminating to appeal to a straightforward modification of the mathematical `machinery' developed long ago for the study of the so-called \textit{linearization stability} (LS) problem in general relativity \cite{BD_73,FM_79,FMM_80,Arms_77,AMM_81,AMM_82,Mon_75,Mon_76}. In particular, this technology (which was developed initially for the study of perturbations of spatially compact, `cosmological' spacetimes) provides one with a rather clear understanding of a somewhat mysterious step in the Dain, et al analysis, wherein one multiplies the variations of the Brill energy density by an explicit weight factor that plays, for those authors, its desired role \textit{only} in the extremal case. As we shall see the natural interpretation of this weight factor is that it serves as (a special case of) an element (C, Z) of the kernel of the adjoint operator of the linearized Einstein constraint map wherein C is the normal and Z the tangential projection (at an arbitrary Cauchy hypersurface for the unperturbed spacetime) of the (asymptotically timelike) Killing field of the background \cite{FMM_80,Mon_75,Mon_76}. With this recognition of the significance of such Killing Initial Data Sets (or KIDS as they are now often called) one can remove the limitation to extremality and derive, by methods analogous to those given in \cite{DA_15} combined with Carter's identity for the wave map variables, an energy functional that is both conserved and positive definite. In fact, by exploiting Robinson's renowned generalization of Carter's identity \cite{Rob_74} together with the Kaluza-Klein reduced form of the axisymmetric Einstein-Maxwell equations to a still larger 2 + 1---dimensional Einstein---wave map system (now with complex hyperbolic space as the naturally occurring target), one can extend the aforementioned results to cover the coupled, axisymmetric gravitational and electromagnetic perturbations of fully general Kerr-Newman black holes.

Although we shall not attempt to fully exploit it here, the recognition of the (spacetime covariant) geometrical significance of the kernel (C, Z) of the relevant adjoint operator allows one also to remove any apparent dependence upon the `slicing' employed for the background spacetime and, in particular, to allow for hypersurfaces of the black hole's DOC that could, for example, penetrate its (future) event horizon or intercept (future) null infinity (`Scri') or perhaps both. In the present paper though we shall, for simplicity, only deal with the Boyer-Lindquist type slicings that, in contrast to the above, are `locked down' at the horizon's bifurcation two-sphere and at spacelike infinity. These are actual Cauchy surfaces for the DOC's of interest here and allow for a strictly conserved energy functional whereas energies defined with respect to the more general slicings mentioned above would normally \textit{decay} through the occurrence of outgoing fluxes at the horizon and at Scri \cite{WH_13}.

Another advantage of the use of the LS `technology' is that it shows clearly how to relate the linearized energy expression obtained therefrom to a perturbation of the asymptotically defined ADM mass of the perturbed spacetime which, as we shall see, is necessarily induced at second order from the presence of non-vanishing energy at first order. The absence of such compensating boundary integral expressions in the spatially compact, `cosmological' cases originally considered for the LS problem was what gave rise to the curious phenomenon of linearization \textit{instability} wherein any linear perturbation with a \textit{non-vanishing} Killing conserved quantity was shown to be `spurious' in that it could not, even in principle, be extended to higher order \cite{Arms_77,AMM_81,AMM_82,BD_73,FM_79,FMM_80,Mon_75,Mon_76}. For the spatially non-compact problems of interest here such conserved energy integrals are \textit{not}, of course, forced to vanish but, when non-vanishing and combined with suitable boundary conditions on the perturbations at the black hole's event horizon, coerce a corresponding perturbation in the ADM mass at second order.

Though we shall focus exclusively on the derivation of this fundamental energy expression herein, there is a well-known technique for generating, for sufficiently smooth perturbations, a sequence of higher order energy expressions by successively Lie differentiating the linearized field equations with respect to the (asymptotically timelike) Killing field of the background, essentially `time' differentiating the unknowns sequentially and evaluating their `energies', and then using the linearized equations to `trade' time derivatives for spatial ones in defining the ultimate, higher order, energy expressions. Though we shall not pursue this strategy in detail herein we shall sketch, in the concluding section, it's potential application for extracting (higher order) Sobolev type bounds upon the perturbations from the corresponding energy integrals. The derivation of such bounds would serve, through the application of standard Sobolev inequalities, to establish \textit{uniform boundedness} of the perturbations and their derivatives and will be the subject of a subsequent article. A well-known difficulty in deriving such bounds arises through the natural occurrence of certain `weight factors' in the higher order energies that degenerate at the horizon and thereby force the need for a more subtle analysis for the extraction of the desired Sobolev estimates.

In section \ref{sec:pure-electromagnetic-kerr-spacetimes} we shall begin by focusing on the special case of purely electromagnetic perturbations of a Kerr background spacetime. These have the distinct advantage of allowing a straightforward representation in terms of (electromagnetic) gauge and infinitesimal diffeomorphism-invariant variables that satisfy an elegant system of partial differential equations derived directly from Maxwell's equations in the axisymmetric case. Even for this problem, however, Robinson's identity, specialized to the case at hand, is needed to handle the ergo-region difficulties and demonstrate positivity of the resulting, `regularized' energy expression defined therein. By contrast, the linearized wave map variables for the more general Kerr-Newman problem analyzed in sections~\ref{sec:energy-functional-kerr-newman} and \ref{sec:conservation} are gauge-dependent (since they correspond to the perturbations of non-constant background scalar fields) and accordingly, for the elliptic gauges considered herein, satisfy `non-local' evolution equations incorporating the linearized lapse and shift variables. While one could have employed a non-elliptic gauge of `spacetime harmonic' type (i.e., the analogue of Lorenz gauge for Maxwell's equations) this would have significantly enlarged the system to be analyzed and thus the number of evolving variables to be controlled by energy arguments in an ultimate stability analysis. In our setup, however, only the independent dynamical, linearized wave map variables need to be controlled by the energy (and its higher order generalizations).

Somewhat remarkably most of the elliptic problems involved in our formulation reduce to 2-dimensional flat space Poisson equations for which the relevant fundamental solution (Green's function) is explicitly known. Indeed, this is true for \textit{all} of the elliptic problems in the special case of what we shall call the 2+1---dimensional, \textit{maximal} slicing gauge condition.  For more general gauge conditions (such as 3+1---dimensional maximal slicing) the linear elliptic equation for the perturbed lapse function need not be of this elementary, explicitly solvable type. The elliptic analysis needed for dealing with the linearized constraints and the imposed gauge conditions is developed in Appendices \ref{app:analysis-linearized-constraint-equations}, \ref{app:transforming} and \ref{app:complactly-supported-solutions} while Appendix \ref{app:kerr-newman-spacetimes} presents the Kerr-Newman black hole solutions in the coordinate systems of interest and Appendix~\ref{app:reduced-hamiltonian} reviews the Hamiltonian formalism for the (axial) symmetry-reduced Einstein---wave map system that is the main object of our study. Appendix~\ref{app:global} reviews the global Cauchy problem for the linearized field equations specialized to a `hyperbolic' gauge of Lorenz type whereas Appendix~\ref{app:coveriance} establishes the equivalence between our Hamiltonian formulation of the `twist potential' wave map variables and the more conventional Lagrangian definition of these fields and Appendix~\ref{app:electric-charge} reviews the charge and angular momentum conservation laws in our formalism. Appendix~\ref{app:gauge-conditions} introduces the (Weyl-Papapetrou) gauge condition needed to determine our linearized shift field. Appendix~\ref{app:vanishing} establishes the vanishing of a certain integral invariant the result of which is needed to justify our chosen (Weyl-Papapetrou) gauge condition. Appendix~\ref{app:maximal-slicing} analyzes maximal slicing gauge conditions in both the 2+1 and 3+1 dimensional sense, whereas Appendix~\ref{app:weyl-tensor} lays the foundation for relating our formulation of the linearized field equations to that involving the perturbed Weyl tensor. It has seemed advisable to us to relegate some of these more technical discussions to appendices in order not to unduly interrupt the logical flow of the arguments given in the main body of the article.

In section~\ref{sec:conclusion} we briefly discuss some possible further extensions of our work. In particular, we describe some of the modifications that would be needed for the inclusion of a (positive) cosmological constant and the corresponding derivation of a (conserved, positive definite) energy functional for Kerr-Newman-de Sitter spacetimes. A key point here is that the Robinson identity, which is normally applied to purely electrovacuum problems, only generates, thanks to a favorable sign in one of its terms that vanishes for electrovacuum backgrounds, a new term of positive sign in the presence of a positive cosmological constant. While the remarkable work of Hintz and Vasy has already demonstrated the stability of slowly rotating Kerr-de Sitter black holes with respect to \textit{fully nonlinear} and \textit{non-symmetric} perturbations \cite{HV_16} there may be some potential contribution of our approach to the study, at least at linearized level, of rapidly rotating Kerr-de Sitter solutions as well as to their Kerr-Newman-de Sitter generalizations. We propose to pursue this issue in a future work.

Though our treatment of the U(1)---symmetric Einstein---wave map formalism is herein limited to linearized equations we remark that work by Choquet Bruhat and one of us applied this same setup (in the vacuum case) to establish the (fully nonlinear) stability of a family of (spatially compact) cosmological models in the temporal direction of cosmological expansion \cite{Ycb-Mon_01}. The future stability of a still different set of vacuum cosmological background solutions was proven, for fully non-symmetric perturbations, by Andersson and one of us by using energies of a (generalized) Bel-Robinson type \cite{AM_04,AM_11}. Separately, \textit{large data} global existence for the (nonlinear) equivariant Einstein---wave map system was proven by Andersson, Gudapati and Szeftel \cite{AGS_15} by building on the non-concentration of energy result established by one of us in \cite{diss_13}. An entirely different approach to Kerr mode stability, made possible through Whiting's remarkable transformation of the Teukolsky equation \cite{Whit_89}, has recently been further developed by Andersson, Ma, Paganini and Whiting \cite{AMPW_16}.

We also briefly discuss, in the concluding section, the potential application of our approach to the study of black holes in higher than 4 spacetime dimensions. It has long been realized, for example, that when an n---2 dimensional, commutative, spacelike isometry group is imposed upon the solutions of the Einstein \cite{Maison_79} or Einstein-Maxwell \cite{IU_03} equations in n + 1 dimensions (with \(n > 3\)), these systems can be reduced, \`{a} la Kaluza-Klein, to another wave map system coupled to a Lorentzian 3---metric. In fact \textit{stationary} black holes and more general black objects, at least in the vacuum, analytic case, can be proven to automatically admit such toroidal isometry groups when the associated angular momentum parameters are non-vanishing \cite{MI_08,HIW_07}. Furthermore, generalizations of the Carter and Robinson identities have been systematically derived for the proofs of corresponding black hole uniqueness theorems \cite{HI_12,HY_08,Hollands_11,Chrusciel_12}. Thus all of the needed `machinery' for the extension of our results to such higher dimensional problems seems already to be available. On the other hand, as pointed out by Hollands and Ishibashi, such a high dimensional toroidal isometry group is compatible with asymptotic flatness (in the standard sense for spacetimes with a well-defined `Scri' diffeomorphic to \(S^{n-1} \times \mathbb{R}\)) only in 4 and 5 spacetime dimensions \cite{HI_12}. But the stability of the famous 5-dimensional Myers-Perry rotating black hole solution \cite{MyPe_86} (and its electrovacuum generalization \cite{Hendi_14}) is an important unsolved mathematical problem whereas the \textit{instability} of still higher dimensional, \textit{rotating} black objects has, to a considerable extent, been established  \cite{EmRe_08}. Thus we conjecture that our methods can be applied to shed light on these open questions at least for perturbations preserving the \(T^{n-2}\) `axial' isometry group of the chosen, axi-symmetric background. We propose to investigate this in detail in future work.

The senior author is especially grateful to Abraham Taub for his penetrating insights on the relationship between variational methods and gravitational conservation laws, to Jerrold Marsden for his deep understanding of linearization stability problems and their geometrical significance and to Sergio Dain for his insightful work on black hole perturbations that significantly influenced the present article. This article is dedicated to their memories.  %no equations
\section{Pure Electromagnetic Perturbations of Kerr Spacetimes}%section 1
\label{sec:pure-electromagnetic-kerr-spacetimes}
As is well-known and easily seen, linearization of the Einstein-Maxwell equations about an arbitrary, \textit{vacuum} solution leads to a decoupled system of perturbation equations of which the electromagnetic component consists simply of Maxwell's field equations formulated on the chosen (vacuum) background. The corresponding linearized Einstein component for the metric perturbation is homogeneous in this approximation and thus always compatible with taking the metric perturbation to vanish identically. Specializing the background to be a Kerr, black hole spacetime and demanding, for simplicity, that the metric perturbation be trivial we thus arrive at the important special case of analyzing Maxwell's equations on a given Kerr background.

With this aim in mind it is natural to look for a conserved, positive definite energy functional for Maxwell fields on the domain of outer communications (DOC) of an arbitrary Kerr black hole. As far as we know however, no such energy functional has heretofore been constructed, even for the case of purely axisymmetric perturbations, thanks to the well-known difficulties presented by the \textit{ergo-region} that always surrounds a (rotating) black hole. Thus the solution to this problem that we present here (for the axisymmetric case) may be of interest in its own right as well as providing an example, in a somewhat simpler setting, of the full linearized Kerr-Newman energy functional construction that is the main aim of this paper.

While one could simply specialize our comprehensive, Kerr-Newman construction to the case at hand it will perhaps prove more illuminating to start `from scratch' and derive the pure Maxwell energy functional from first principles, leaving its reconciliation with our general, Kerr-Newman results until later (c.f., the discussion at the end of Appendix \ref{app:analysis-linearized-constraint-equations}). The action for electromagnetic fields on an \textit{arbitrary}, 3+1-dimensional, globally hyperbolic spacetime \(\lbrace\tilde{M} \times \mathbb{R}, {}^{(4)}\!g\rbrace\), with \(\tilde{M}\) a smooth, connected 3-manifold, is given, in Hamiltonian form (c.f. Eqs.~(\ref{eq:a07}--\ref{eq:a10}) by:
\begin{equation}\label{eq:101}
I_\Omega^{\mathrm{Maxwell}} := \int_\Omega d^4x\; \lbrace A'_i\; \mathcal{E}^{i'}_{,t} - N\mathcal{H}^{\mathrm{Max}} - N^i\mathcal{H}_i^{\mathrm{Max}} - A'_0\; \mathcal{E}^{i'}_{,i}\rbrace
\end{equation}
where
\begin{equation}\label{eq:102}
\mathcal{H}^{\mathrm{Max}} := \frac{1}{2}\; \frac{g_{ij}}{\mu_{{}^{(3)}\!g}}\; (\mathcal{E}^{i'}\mathcal{E}^{j'} + \mathcal{B}^{i'}\mathcal{B}^{j'})
\end{equation}
and
\begin{equation}\label{eq:103}
\mathcal{H}^{\mathrm{Max}}_i := -\epsilon_{ijk}\mathcal{E}^{j'}\mathcal{B}^{k'}.
\end{equation}
Here the Lorentzian metric \({}^{(4)}\!g\) has been expressed in ADM (Arnowitt, Deser, Misner) form (c.f., Eq.~(\ref{eq:a03})) and, to ensure convergence, the integral has been restricted to an arbitrary compact domain, \(\Omega \subset \tilde{M} \times \mathbb{R}\), having a piecewise smooth boundary. The `primes' attached to the Maxwell fields, superfluous for the moment, are intended to signify that we regard these as linear perturbations of an identically vanishing background.

We now specialize \(\lbrace \tilde{M} \times \mathbb{R}, {}^{(4)}\!g\rbrace\) to be the domain of outer communications of an arbitrary, rotating Kerr black hole and constrain the (perturbative) Maxwell fields under consideration to be \textit{axisymmetric} and thus, relative to the coordinate systems discussed in Appendix~\ref{app:kerr-newman-spacetimes}, to satisfy
\begin{equation}\label{eq:104}
\frac{\partial}{\partial\varphi} A'_\mu = \frac{\partial}{\partial\varphi}\mathcal{E}^{i'} = \frac{\partial}{\partial\varphi}\mathcal{B}^{i'} = 0.
\end{equation}
Here \(\psi = \frac{\partial}{\partial\varphi}\) together with \(\zeta = \frac{\partial}{\partial t}\) are the axial and time translational Killing fields of the general Kerr solution and, as elaborated in Appendix~\ref{app:reduced-hamiltonian}, it is natural to pass to the quotient space for the circle action generated by \(\psi\) and to formulate the Maxwell equations on the base manifold (with boundary)
\begin{equation}\label{eq:105}
V/U(1) = \mathbb{R} \times M_b
\end{equation}
defined therein.

Variation of \(I_\Omega^{\mathrm{Maxwell}}\) with respect to \(A'_0\) leads immediately to the (Gauss law) constraint equation \(\mathcal{E}^{i'}_{,i} = 0\) which, under our axial symmetry assumption, simplifies to \(\mathcal{E}^{a'}_{,a} = 0\) with \(\lbrace x^a\rbrace = \lbrace x^1,x^2\rbrace\) while \(x^3 = \varphi\). On the simply connected space \(M_b\) one can always solve this constraint, without loss of generality, by introducing a potential function \(\eta'\) and setting
\begin{equation}\label{eq:106}
\mathcal{E}^{a'} = \epsilon^{ab}\eta'_{,b}.
\end{equation}
This follows from applying the Poincar\'{e} lemma to the dual, closed one-form \(\epsilon_{ac}\mathcal{E}^{a'} dx^c\) and expressing it as the exact form \(\eta'_{,c} dx^c\). Writing \(\lambda'\) for the azimuthal component, \(A'_3\), of the `linearized' vector potential one arrives at
\begin{equation}\label{eq:107}
\mathcal{B}^{a'} = \epsilon^{ab}\lambda'_{,b}
\end{equation}
for the corresponding magnetic field components.

Linearizing the defining equations for the electromagnetic momentum variables \(\lbrace\tilde{u},\tilde{v}\rbrace\) (defined through Eqs.~(\ref{eq:a28}),(\ref{eq:a31}), (\ref{eq:a18}) and (\ref{eq:a19})) and recalling that the metric one-form \(\beta_a dx^a\) vanishes on the Kerr background (compare Eqs.~(\ref{eq:b11}) and (\ref{eq:a13})), one finds that
\begin{equation}\label{eq:108}
\tilde{u}' = \mathcal{B}^{3'},\> \tilde{v}' = -\mathcal{E}^{3'}.
\end{equation}
Taking \(\Omega\) to be invariant with respect to the circle action generated by \(\psi = \frac{\partial}{\partial x^3} = \frac{\partial}{\partial\varphi}\) and assuming that it projects to a domain in the quotient space of the form \(\mathcal{D}\!\times [t_0,t_1]\), with \(\mathcal{D}\) compact in \(M_b\), one can reexpress the action integral as \begin{equation}\label{eq:109}
\begin{split}
\tilde{I}_\Omega^{\mathrm{Maxwell}} &= 2\pi\; \int_{t_0}^{t_1} dt\; \int_{\mathcal{D}} d^2x\; \left\lbrace\tilde{u}'\eta'_{,t} + \tilde{v}'\lambda'_{,t} - \left\lbrack\frac{1}{2}\; \frac{\tilde{N}}{\mu_{{}^{(2)}\!\tilde{g}}} e^{2\gamma}\left( (\tilde{u}')^2 + (\tilde{v}')^2\right)\right.\right.\\
 &\left.\left. \vphantom{\left\lbrace\left\lbrack\frac{\tilde{N}}{\mu_{{}^{(2)}\!\tilde{g}}}\right\rbrack\right\rbrace} + \frac{1}{2} \tilde{N}\mu_{{}^{(2)}\!\tilde{g}}\; \tilde{g}^{ab} e^{-2\gamma} (\eta'_{,a}\eta'_{,b} + \lambda'_{,a}\lambda'_{,b}) - \beta_0\epsilon^{ab}\eta'_{,a}\lambda'_{,b}\right\rbrack\right\rbrace\\
 & {} + 2\pi\; \int_{t_0}^{t_1} dt\; \int_{\mathcal{D}} d^2x\; \left\lbrace -(\lambda'\tilde{v}')_{,t} + (A'_a\epsilon^{ab}\eta'_{,t})_{,b}\right\rbrace
\end{split}
\end{equation}
where we have now exploited the parametrization introduced via Eq.~(\ref{eq:a13}) to denote the background metric components in `quotient space' format. The Kerr values for these metric components (in Weyl-Papapetrou coordinates) can be read off from Eq.~(\ref{eq:b11}) (upon taking the background charge Q to vanish).

Since the second integral in (\ref{eq:109}) equates to a pure boundary term and thus makes no contribution to the field equations one may discard it and define, accordingly, the `reduced Maxwell action functional'
\begin{equation}\label{eq:110}
\begin{split}
\tilde{J}_\Omega^{\mathrm{Maxwell}} &:= \int_{t_0}^{t_1} dt\; \int_{\mathcal{D}} d^2x\; \left\lbrace \vphantom{\left\lbrack\frac{1}{2}\right\rbrack}\tilde{u}'\eta'_{,t} + \tilde{v}'\lambda'_{,t}\right.\\
 &- \left\lbrack\frac{1}{2}\; \frac{\tilde{N}}{\mu_{{}^{(2)}\!\tilde{g}}} e^{2\gamma} \left((\tilde{u}')^2 + (\tilde{v}')^2\right) + \frac{1}{2}\; \tilde{N}\mu_{{}^{(2)}\!\tilde{g}}\; \tilde{g}^{ab} e^{-2\gamma} (\eta'_{,a}\eta'_{,b} + \lambda'_{,a}\lambda'_{,b})\right.\\
 &\left.\left.\vphantom{\left\lbrack\frac{1}{2}\right\rbrack}- \beta_0\epsilon^{ab} \eta'_{,a}\lambda'_{,b}\right\rbrack\right\rbrace = (\tilde{I}_\Omega^{\mathrm{Maxwell}}/2\pi) - (\text{boundary term}).
\end{split}
\end{equation}
It may be helpful to note here that the metric functions employed above, namely \(\lbrace\gamma, \tilde{g}_{ab},\beta_a,\beta_0,\tilde{N}^a,\tilde{N}\rbrace\) are related to the conventional ADM metric functions \(\lbrace g_{ij},N^i,N\rbrace\) through
\begin{align}
e^{2\gamma} &= g_{\varphi\varphi},\; \beta_a = e^{-2\gamma}g_{a\varphi},\label{eq:111}\\
\tilde{g}_{ab} &= e^{2\gamma} g_{ab} - e^{4\gamma}\beta_a\beta_b = e^{2\gamma} g_{ab} - g_{a\varphi}g_{b\varphi},\label{eq:112}\\
\tilde{N}^a &= N^a,\; \beta_0 = N^{\varphi} + N^a\beta_a = N^{\varphi} + N^ae^{-2\gamma} g_{a\varphi},\label{eq:113}\\
\tilde{N} &= e^\gamma N\label{eq:114}
\end{align}
and that we write \(\mu_{{}^{(2)}\!\tilde{g}}\) for the 2-dimensional `volume' element \(\sqrt{\det{|\tilde{g}_{ab}|}}\). Furthermore, in the standard coordinate systems discussed in Appendix~\ref{app:kerr-newman-spacetimes}, \(\beta_a = 0\) and \(\tilde{N}^a = N^a = 0\) for a metric of the Kerr type whereas the shift vector \(N^i \frac{\partial}{\partial x^i}\) reduces to \(N^\varphi \frac{\partial}{\partial\varphi} \longrightarrow \beta_0 \frac{\partial}{\partial\varphi}\).

To this point no actual field equations have needed to be imposed on the background metric --- the axisymmetric Maxwell equations for an arbitrary such background may thus be derived by variation of \(\tilde{J}_\Omega^{\mathrm{Maxwell}}\) with respect to the (unconstrained) canonical variables \(\lbrace(\eta',\tilde{u}'),(\lambda',\tilde{v}')\rbrace\). For most of the arguments to follow, however, satisfaction of the vacuum field equations (specifically by the Kerr metric) will play an \textit{essential} role. In terms of the twist potential \(\omega\), defined via Eqs.~(\ref{eq:a24}) and (\ref{eq:a18}), the field equations satisfied by the Kerr metric are given, after setting \(\eta = \lambda = 0\), by Eqs.~(\ref{eq:a64})--(\ref{eq:a71}). Of these, the most immediately relevant are
\begin{align}
4(\tilde{N}\mu_{{}^{(2)}\!\tilde{g}}\; \tilde{g}^{ab}\gamma_{,a})_{,b} + 2\tilde{N}\mu_{{}^{(2)}\!\tilde{g}}\; \tilde{g}^{ab}e^{-4\gamma} \omega_{,a}\omega_{,b} =& 0,\label{eq:115}\\
(\tilde{N}\mu_{{}^{(2)}\!\tilde{g}}\; \tilde{g}^{ab} e^{-4\gamma}\omega_{,a})_{,b} =& 0\label{eq:116}
\end{align}
and
\begin{equation}\label{eq:117}
\tilde{N}^{|c}_{\hphantom{|c}|c} = \frac{1}{\mu_{{}^{(2)}\!\tilde{g}}} \left(\mu_{{}^{(2)}\!\tilde{g}}\; \tilde{g}^{ab} \tilde{N}_{,a}\right)_{,b} = 0.
\end{equation}
In addition Eq.~(\ref{eq:a37}), specialized to the (stationary, vacuum) case at hand, reduces to
\begin{equation}\label{eq:118}
\beta_{0,a} + \tilde{N} e^{-4\gamma} \epsilon_{ab}\; \mu_{{}^{(2)}\!\tilde{g}}\; \tilde{g}^{bc} \omega_{,c} = 0.
\end{equation}
Explicit formulas for the relevant quantities appearing herein are given, in Boyer-Lindquist coordinates, by Eqs.~(\ref{eq:a39})--(\ref{eq:a46}), upon setting the charge Q to 0. Henceforth, in this section, we restrict the background metric to be specifically that of a Kerr black hole.

In a suitable function space setting the Hamiltonian
\begin{equation}\label{eq:119}
\tilde{H}^{\mathrm{Maxwell}} := \int_{M_b} d^2x\; \lbrace N\mathcal{H}^{\mathrm{Max}} + N^\varphi\mathcal{H}^{\mathrm{Max}}_\varphi\rbrace,
\end{equation}
where now
\begin{equation}\label{eq:120}
N\mathcal{H}^{\mathrm{Max}} + N^\varphi\mathcal{H}^{\mathrm{Max}}_\varphi = \left\lbrace\frac{1}{2}\; \frac{\tilde{N} e^{2\gamma}}{\mu_{{}^{(2)}\!\tilde{g}}} \left((\tilde{u}')^2 + (\tilde{v}')^2\right) + \frac{1}{2} \tilde{N} \mu_{{}^{(2)}\!\tilde{g}}\; \tilde{g}^{ab} e^{-2\gamma} (\eta'_{,a}\eta'_{,b} + \lambda'_{,a}\lambda'_{,b}) - \beta_0 \epsilon^{ab} \eta'_{,a}\lambda'_{,b}\right\rbrace,
\end{equation}
will be well-defined and yield the symmetry-reduced Maxwell equations in the form of Hamilton's equations for the canonical pairs \(\lbrace(\eta',\tilde{u}'),(\lambda',\tilde{v}')\rbrace\). These latter are readily found to be:
\begin{align}
\eta'_{,t} &= \frac{\delta\tilde{H}^{\mathrm{Maxwell}}}{\delta\tilde{u}'} = \frac{\tilde{N} e^{2\gamma}}{\mu_{{}^{(2)}\!g}} \tilde{u}',\label{eq:121}\\
\lambda'_{,t} &= \frac{\delta\tilde{H}^{\mathrm{Maxwell}}}{\delta\tilde{v}'} = \frac{\tilde{N} e^{2\gamma}}{\mu_{{}^{(2)}\!\tilde{g}}} \tilde{v}',\label{eq:122}\\
\tilde{u}'_{,t} &= {} - \frac{\delta\tilde{H}^{\mathrm{Maxwell}}}{\delta\eta'} = (\tilde{N}\mu_{{}^{(2)}\!\tilde{g}}\; \tilde{g}^{ab} e^{-2\gamma} \eta'_{,a})_{,b} + \tilde{N}\mu_{{}^{(2)}\!\tilde{g}}\; \tilde{g}^{ab} e^{-4\gamma} \omega_{,a}\lambda'_{,b},\label{eq:123}\\
\tilde{v}'_{,t} &= {} - \frac{\delta\tilde{H}^{\mathrm{Maxwell}}}{\delta\lambda'} = (\tilde{N}\mu_{{}^{(2)}\!\tilde{g}}\; \tilde{g}^{ab} e^{-2\gamma} \lambda'_{,a})_{,b} - \tilde{N}\mu_{{}^{(2)}\!\tilde{g}}\; \tilde{g}^{ab} e^{-4\gamma} \omega_{,a} \eta'_{,b}\label{eq:124}
\end{align}
wherein we have exploited the background equation (\ref{eq:118}) to reexpress the resulting forms, eliminating \(\beta_0\) in favor of the twist potential, \(\omega\).

Though \(\tilde{H}^{\mathrm{Maxwell}}\) would seem to be a natural candidate for the energy functional we are seeking to construct, its density (\ref{eq:120}) can be shown to attain negative values inside the Kerr ergo-region leaving positivity of the total energy in doubt. To see this explicitly assume for definiteness that the Kerr rotation parameter a is positive and evaluate the Hamiltonian density (\ref{eq:120}) in Weyl-Papapetrou coordinates \(\lbrace\rho,z\rbrace\), for which
\begin{equation}\label{eq:125}
\mu_{{}^{(2)}\!\tilde{g}}\; \tilde{g}^{ab} \frac{\partial}{\partial x^a} \otimes \frac{\partial}{\partial x^b} \longrightarrow \frac{\partial}{\partial\rho} \otimes \frac{\partial}{\partial\rho} + \frac{\partial}{\partial z} \otimes \frac{\partial}{\partial z},
\end{equation}
taking (locally defined) Cauchy data of the form
\begin{equation}\label{eq:126}
\tilde{u}' = \tilde{v}' = 0
\end{equation}
with \(\eta'\) and \(\lambda'\) satisfying the Cauchy Riemann equations,
\begin{equation}\label{eq:127}
\lambda'_{,\rho} = \eta'_{,z}\> \text{ and }\> \lambda'_{,z} = -\eta'_{,\rho},
\end{equation}
within some open subset of the ergo-region. With these substitutions the Hamiltonian density reduces to
\begin{equation}\label{eq:128}
N\mathcal{H}^{\mathrm{Max}} + N^\varphi\mathcal{H}^{\mathrm{Max}}_\varphi \longrightarrow (\tilde{N} e^{-2\gamma} + \beta_0) \left((\lambda'_{,\rho})^2 + (\lambda'_{,z})^2\right)
\end{equation}
and one has \(\beta_0 + \tilde{N} e^{-2\gamma} < 0\) inside the ergo-region. To treat the case \(a < 0\) one need only reverse the roles of \(\eta'\) and \(\lambda'\) to generate a similar, negative result. Thus whereas for a \textit{single}, axisymmetric scalar field the troublesome term in the shift vector, \(\beta_0 \frac{\partial}{\partial\varphi}\), drops out of the corresponding Hamiltonian density this is not true for the pair of electromagnetic scalars \(\eta'\) and \(\lambda'\) for which the shift induces the natural, Maxwellian coupling between them.

Note that, by virtue of the background field equation (\ref{eq:118}) one can write
\begin{equation}\label{eq:129}
\begin{split}
-\beta_0 \epsilon^{ab} \eta'_{,a}\lambda'_{,b} &= \beta_0 \epsilon^{ab} \eta'_{,b}\lambda'_{,a}\\
 &= (\beta_0 \epsilon^{ab} \eta'_{,b}\lambda')_{,a} + \tilde{N} e^{-4\gamma} \mu_{{}^{(2)}\!\tilde{g}}\; \tilde{g}^{ab} \omega_{,a}\eta'_{,b}\lambda'
\end{split}
\end{equation}
and use this identity to replace the `shift term' in the Hamiltonian density (\ref{eq:120}) by a term involving the background gravitational `twist potential' \(\omega\) together with  a spatial divergence. Since the latter integrates to a pure boundary expression that will not contribute to the equations of motion we may discard it and define an alternative Hamiltonian, \(H^{\mathrm{Alt}}\), given by
\begin{equation}\label{eq:130}
H^{\mathrm{Alt}} := \int_{M_b} d^2x\; \lbrace\mathcal{H}^{\mathrm{Alt}}\rbrace
\end{equation}
where
\begin{equation}\label{eq:131}
\mathcal{H}^{\mathrm{Alt}} = \left\lbrace\frac{1}{2}\; \frac{\tilde{N} e^{2\gamma}}{\mu_{{}^{(2)}\!\tilde{g}}} \left((\tilde{u}')^2 + (\tilde{v}')^2\right) + \frac{1}{2} \tilde{N} \mu_{{}^{(2)}\!\tilde{g}}\; \tilde{g}^{ab} e^{-2\gamma} (\eta'_{,a}\eta'_{,b} + \lambda'_{,a}\lambda'_{,b}) + \tilde{N} e^{-4\gamma} \mu_{{}^{(2)}\!\tilde{g}}\; \tilde{g}^{ab} \omega_{,a}\eta'_{,b}\lambda'\right\rbrace
\end{equation}

As we shall see later this arises as a special case of the general Kerr-Newman perturbational Hamiltonian that we shall derive below in Appendix \ref{app:analysis-linearized-constraint-equations}. At first sight though it appears to amount to a step backwards since, if we exploit the freedom to shift the (undifferentiated) \(\lambda'\) by an additive constant, we could make \(\mathcal{H}^{\mathrm{Alt}}\) locally negative \textit{even outside the ergo-region}!

Now, however, we are in the fortuitous position of being able to exploit Robinson's identity which, specialized to the case of a vacuum background and purely electromagnetic perturbations and reexpressed in our notation, reads:
\begin{equation}\label{eq:132}
\begin{split}
&\tilde{N} \mu_{{}^{(2)}\!\tilde{g}}\; \tilde{g}^{ab} e^{-2\gamma} (\eta'_{,a}\eta'_{,b} + \lambda'_{,a}\lambda'_{,b}) + 2\tilde{N} \mu_{{}^{(2)}\!\tilde{g}}\; \tilde{g}^{ab} e^{-4\gamma} \omega_{,a}\eta'_{,b}\lambda' + L_1 \left\lbrack\frac{1}{2} (\eta')^2 + \frac{1}{2} (\lambda')^2\right\rbrack \\
& {} - \lambda'\eta' L_2 + \frac{1}{2} \frac{\partial}{\partial x^b} \left\lbrace -2\tilde{N} \mu_{{}^{(2)}\!\tilde{g}}\; \tilde{g}^{ab} e^{-4\gamma} \omega_{,a}\eta'\lambda' + \tilde{N} (\mu_{{}^{(2)}\!\tilde{g}}\; \tilde{g}^{ab})(e^{-2\gamma})_{,a} \left((\eta')^2 + (\lambda')^2\right)\right\rbrace\\
& {} = \frac{1}{2} \tilde{N} e^{2\gamma} \mu_{{}^{(2)}\!\tilde{g}}\; \tilde{g}^{ab} \left\lbrace\partial_a (e^{-2\gamma}\lambda') \partial_b (e^{-2\gamma}\lambda') + \partial_a (e^{-2\gamma}\eta') \partial_b (e^{-2\gamma}\eta')\right\rbrace\\
& {} + \frac{1}{2} \tilde{N} e^{-2\gamma} \mu_{{}^{(2)}\!\tilde{g}}\; \tilde{g}^{ab} \left\lbrace(\eta'_{,a} + \lambda' e^{-2\gamma} \omega_{,a})(\eta'_{,b} + \lambda' e^{-2\gamma} \omega_{,b}) + (\lambda'_{,a} - \eta' e^{-2\gamma} \omega_{,a})(\lambda'_{,b} - \eta' e^{-2\gamma} \omega_{,b})\right\rbrace\\
\end{split}
\end{equation}
where
\begin{equation}\label{eq:133}
L_1 := \frac{e^{-2\gamma}}{2} \left\lbrace 4(\tilde{N} \mu_{{}^{(2)}\!\tilde{g}}\; \tilde{g}^{ab} \gamma_{,a})_{,b} + 2\tilde{N} e^{-4\gamma} \mu_{{}^{(2)}\!\tilde{g}}\; \tilde{g}^{ab} \omega_{,a}\omega_{,b}\right\rbrace
\end{equation}
and
\begin{equation}\label{eq:134}
L_2 := -(\tilde{N} \mu_{{}^{(2)}\!\tilde{g}}\; \tilde{g}^{ab} e^{-4\gamma} \omega_{,a})_{,b}
\end{equation}
Note that \(L_1\) and \(L_2\) both vanish when the background field equations (\ref{eq:115})--(\ref{eq:116}) are enforced. Thus for a vacuum background Robinson's identity lets us replace the indefinite potential energy density in \(\mathcal{H}^{\mathrm{Alt}}\) (i.e., the first two terms appearing in (\ref{eq:132})) with a spatial divergence and \textit{a sum of positive terms}.

Again discarding the boundary term resulting from the integrated divergence, we define our ultimate, regulated, Maxwell Hamiltonian as the integral over \(M_b\) of the density, \(\mathcal{H}^{\mathrm{Reg}}\), thus constructed, setting
\begin{equation}\label{eq:135}
H^{\mathrm{Reg}} = \int_{M_b} d^2x\; \lbrace\mathcal{H}^{\mathrm{Reg}}\rbrace,
\end{equation}
with
\begin{equation}\label{eq:136}
\begin{split}
\mathcal{H}^{\mathrm{Reg}} &:= \frac{1}{2} \frac{\tilde{N}}{\mu_{{}^{(2)}\!\tilde{g}}} e^{2\gamma} \left((\tilde{u}')^2 + (\tilde{v}')^2\right)\\
 & {} + \frac{1}{2} \tilde{N} \mu_{{}^{(2)}\!\tilde{g}}\; \tilde{g}^{ab} \left\lbrace\left\lbrack\partial_a(e^{-\gamma}\lambda') - \frac{1}{2} (e^{-\gamma}\eta') e^{-2\gamma} \omega_{,a}\right\rbrack\left\lbrack\partial_b (e^{-\gamma}\lambda') - \frac{1}{2} (e^{-\gamma} \eta') e^{-2\gamma}\omega_{,b}\right\rbrack\right.\\
 & {}+ \left\lbrack\partial_a (e^{-\gamma}\eta') + \frac{1}{2} (e^{-\gamma} \lambda') e^{-2\gamma} \omega_{,a}\right\rbrack\left\lbrack\partial_b (e^{-\gamma}\eta') + \frac{1}{2} (e^{-\gamma} \lambda') e^{-2\gamma} \omega_{,b}\right\rbrack\\
 & {}\left.\vphantom{\left\lbrack\frac{1}{2}\right\rbrack}+ \frac{1}{2} \left(2\gamma_{,a}\gamma_{,b} + \frac{1}{2} e^{-4\gamma} \omega_{,a}\omega_{,b}\right)\left\lbrack(e^{-\gamma}\lambda')^2 + (e^{-\gamma}\eta')^2\right\rbrack\right\rbrace
\end{split}
\end{equation}
Note that \(H^{\mathrm{Reg}}\) can be more compactly expressed in terms of the `rescaled' canonical pairs \(\left\lbrace(\underline{\eta}',\underline{\tilde{u}}'), (\underline{\lambda}',\underline{\tilde{v}}')\right\rbrace\) defined by
\begin{equation}\label{eq:137}
\underline{\eta}' := \frac{\eta'}{e^\gamma},\; \underline{\tilde{u}}' := e^\gamma\tilde{u}',\; \underline{\lambda}' := \frac{\lambda'}{e^\gamma},\; \underline{\tilde{v}}' = e^\gamma\tilde{v}'
\end{equation}
for which the regulated density becomes simply
\begin{equation}\label{eq:138}
\begin{split}
\mathcal{H}^{\mathrm{Reg}} &= \frac{1}{2} \frac{\tilde{N}}{\mu_{{}^{(2)}\!\tilde{g}}} \left((\underline{\tilde{u}}')^2 + (\underline{\tilde{v}}')^2\right)\\
 & {}+ \frac{1}{2} \tilde{N} \mu_{{}^{(2)}\!\tilde{g}}\; \tilde{g}^{ab} \left\lbrace\left(\partial_a \underline{\lambda}' - \frac{1}{2} \underline{\eta}' e^{-2\gamma} \omega_{,a}\right)\left(\partial_b \underline{\lambda}' - \frac{1}{2} \underline{\eta}' e^{-2\gamma} \omega_{,b}\right)\right.\\
 & {}+ \left(\partial_a \underline{\eta}' + \frac{1}{2} \underline{\lambda}' e^{-2\gamma} \omega_{,a}\right)\left(\partial_b \underline{\eta}' + \frac{1}{2} \underline{\lambda}' e^{-2\gamma} \omega_{,b}\right)\\
 & {}\left.\vphantom{\left\lbrack\frac{1}{2}\right\rbrack}+ \frac{1}{2} \left(2\gamma_{,a}\gamma_{,b} + \frac{1}{2} e^{-4\gamma}\omega_{,a}\omega_{,b}\right)\left((\underline{\lambda}')^2 + (\underline{\eta}')^2\right)\right\rbrace
\end{split}
\end{equation}

Hamilton's equations, which now take the form
\begin{align}
\underline{\eta}'_{,t} &= \frac{\delta H^{\mathrm{Reg}}}{\delta\underline{\tilde{u}}'},\; \underline{\lambda}'_{,t} = \frac{\delta H^{\mathrm{Reg}}}{\delta\underline{\tilde{v}}'},\label{eq:139}\\
\underline{\tilde{u}}'_{,t} &= {} - \frac{\delta H^{\mathrm{Reg}}}{\delta\underline{\eta}'},\; \underline{\tilde{v}}'_{,t} = {} - \frac{\delta H^{\mathrm{Reg}}}{\delta\underline{\lambda}'},\label{eq:140}
\end{align}
regenerate the Maxwell equations (\ref{eq:121})--(\ref{eq:124}) given previously but now with a positive definite Hamiltonian.

Using these equations to compute the time derivative of \(\mathcal{H}^{\mathrm{Reg}}\) one arrives at
\begin{equation}\label{eq:141}
\mathcal{H}^{\mathrm{Reg}}_{,t} = \frac{\partial}{\partial x^b} \left\lbrace\tilde{N}^2\tilde{g}^{ab}\underline{\tilde{u}}' \left(\underline{\eta}'_{,a} + \frac{1}{2} \underline{\lambda}' e^{-2\gamma} \omega_{,a}\right) + \tilde{N}^2\tilde{g}^{ab}\underline{\tilde{v}}' \left(\underline{\lambda}'_{,a} - \frac{1}{2}\underline{\eta}' e^{-2\gamma} \omega_{,a}\right)\right\rbrace
\end{equation}
which leads one to define the divergence-free vector density current, \(\text{\boldmath
$\mathcal{J}$}\!_{\mathrm{Reg}}\), via
\begin{align}
\mathcal{J}_{\mathrm{Reg}}^0 &:= \mathcal{H}^{\mathrm{Reg}}\label{eq:142}\\
\mathcal{J}_{\mathrm{Reg}}^b &:= -\tilde{N}^2\tilde{g}^{ab} \left\lbrack\underline{\tilde{u}}' \left(\underline{\eta}'_{,a} + \frac{1}{2}\underline{\lambda}' e^{-2\gamma}\omega_{,a}\right) + \underline{\tilde{v}}' \left(\underline{\lambda}'_{,a} - \frac{1}{2}\underline{\eta}' e^{-2\gamma}\omega_{,a}\right)\right\rbrack\label{eq:143}
\end{align}
with
\begin{equation}\label{eq:144}
\text{\boldmath
$\mathcal{J}$}\!_{\mathrm{Reg}} = \mathcal{J}_{\mathrm{Reg}}^0 \frac{\partial}{\partial t} + \mathcal{J}_{\mathrm{Reg}}^b \frac{\partial}{\partial x^b}
\end{equation}
satisfying, by construction,
\begin{equation}\label{eq:145}
\frac{\partial}{\partial t}\mathcal{J}_{\mathrm{Reg}}^0 + \frac{\partial}{\partial x^a}\mathcal{J}_{\mathrm{Reg}}^a = 0.
\end{equation}
The regularity conditions at the axis satisfied by the (rescaled) canonical variables \(\left\lbrace(\underline{\eta}',\underline{\tilde{u}}'), (\underline{\lambda}',\underline{\tilde{v}}')\right\rbrace\) together with their asymptotic behaviors at the Kerr event horizon and at infinity are discussed in detail in Appendix~\ref{app:global}. Appealing to these results it is straightforward to apply the continuity equation (\ref{eq:145}) to show that the total electromagnetic `energy', defined by \(H^{\mathrm{Reg}}\), is strictly conserved. This energy could only differ in value from those defined by \(H^{\mathrm{Alt}}\) and \(H^{\mathrm{Maxwell}}\) by possible boundary contributions at (spacelike) infinity or at the bifurcation 2-sphere lying in the black hole's horizon. It has, however, the significant analytical advantage over these latter quantities of being manifestly \textit{positive definite}.

Combining Eqs.~(\ref{eq:121}), (\ref{eq:122}), and (\ref{eq:106})--(\ref{eq:108}) one finds that
\begin{align}
\eta'_{,t} &= \frac{\tilde{N} e^{2\gamma}\mathcal{B}^{3'}}{\mu_{{}^{(2)}\!\tilde{g}}},\; \eta'_{,b} = \epsilon_{ab}\mathcal{E}^{a'},\label{eq:146}\\
\lambda'_{,t} &= {} - \frac{\tilde{N} e^{2\gamma}\mathcal{E}^{3'}}{\mu_{{}^{(2)}\!\tilde{g}}},\; \lambda'_{,b} = \epsilon_{ab}\mathcal{B}^{a'},\label{eq:147}
\end{align}
and thus that, in any open connected domain of \(\mathbb{R} \times M_b\) in which the (projected) electromagnetic field components (\(\mathcal{E}^{i'}, \mathcal{B}^{i'}\)) all vanish, the potentials \(\eta'\) and \(\lambda'\) must both be spacetime constants. Since the Maxwell fields \(\mathcal{E}^{i'}\) and \(\mathcal{B}^{i'}\) propagate causally on the domain of outer communications of a Kerr black hole it is easily verified that these fields, projected to the quotient space \(\mathbb{R} \times M_b\), propagate causally with respect to the induced, 2+1-dimensional Lorentz metric \({}^{(3)}\!k\) defined by
\begin{equation}\label{eq:148}
{}^{(3)}\!k := -\tilde{N}^2dt \otimes dt + \tilde{g}_{ab} (dx^a + \tilde{N}^a dt) \otimes (dx^b + \tilde{N}^b dt)
\end{equation}
(c.f. Eq.~(\ref{eq:a13})). It follows that any non-constant disturbance in the potentials \(\eta'\) and \(\lambda'\) must propagate causally on (\(\mathbb{R} \times M_b, {}^{(3)}\!k\)).

Another way of verifying the causal propagation of energy in this quotient space is to calculate the flux density of the current \(\text{\boldmath
$\mathcal{J}$}\!_{\mathrm{Reg}}\) across an arbitrary null hypersurface in (\(\mathbb{R} \times M_b, {}^{(3)}\!k\)) with (future directed) null normal \(\ell^\mu \frac{\partial}{\partial x^\mu}\), i.e., to evaluate \(-\ell_\mu \mathcal{J}_{\mathrm{Reg}}^\mu = -{}^{(3)}\!k_{\mu\nu} \ell^\mu \mathcal{J}_{\mathrm{Reg}}^\nu\) for an arbitrary tangent field \(\ell^\mu \frac{\partial}{\partial x^\mu}\) satisfying
\begin{equation}\label{eq:153}
\ell^0 > 0, \qquad {}^{(3)}\!k_{\mu\nu} \ell^\mu \ell^\nu = 0.
\end{equation}
Appealing to the defining equations (\ref{eq:138}), (\ref{eq:142}) and (\ref{eq:143}) and recalling that \(\tilde{N}^a = 0\) for the metrics of interest herein it is straightforward to verify directly that this flux density, \(-\ell_\mu \mathcal{J}_{\mathrm{Reg}}^\mu\), is always non-negative and thus that the corresponding energy can only flow causally through such null hypersurfaces.

For the coordinate systems discussed in Appendix~\ref{app:kerr-newman-spacetimes} it is well-known that null geodesics originating in the Kerr black hole's DOC cannot reach infinity or the event horizon in finite (coordinate) time but only in the limit as \(t \longrightarrow \pm\infty\). Projected to the quotient space this result implies, in particular, that Cauchy data \(\left\lbrace(\eta',\tilde{u}'), (\lambda',\tilde{v}')\right\rbrace\) specified at \(t = t_0\) and having \textit{compact support} in \(M_b\) at this `initial' instant will evolve in such a way as to preserve this property \(\forall\> t\). In other words the support of these fields, evaluated at any time \textit{t}, will remain bounded and disjoint from the horizon. For these solutions in particular it is evident that the various `energies', \(\tilde{H}^{\mathrm{Maxwell}}\), \(H^{\mathrm{Alt}}\) and \(H^{\mathrm{Reg}}\) that we have defined will all coincide.

Note however that whereas the densities \(N\mathcal{H}^{\mathrm{Max}} + N^\varphi\mathcal{H}^{\mathrm{Max}}_\varphi\) and \(\mathcal{H}^{\mathrm{Alt}}\) both vanish at any point at which \(\mathcal{E}^i = \mathcal{B}^i = 0\), this need not be true of \(\mathcal{H}^{\mathrm{Reg}}\). Indeed, as we have seen, the fields \(\mathcal{E}^i \frac{\partial}{\partial x^i}\) and \(\mathcal{B}^i \frac{\partial}{\partial x^i}\) could both vanish throughout an entire open subset of \(\mathbb{R} \times M_b\) implying only that \(\eta'\) and \(\lambda'\) be spacetime constants in this domain. Unless both these constants also vanish the density \(\mathcal{H}^{\mathrm{Reg}}\) will be strictly positive (though time independent) throughout this region. In other words \(\mathcal{H}^{\mathrm{Reg}}\) is actually non-local in the fundamental fields \(\mathcal{E}^i \frac{\partial}{\partial x^i}\) and \(\mathcal{B}^i \frac{\partial}{\partial x^i}\). One could nevertheless express it explicitly in terms of these fields by applying the methods of Appendix~\ref{app:transforming} to solve the two Poisson equations,
\begin{equation}\label{eq:149}
(\mu_{{}^{(2)}\!\tilde{g}}\; \tilde{g}^{ab}\eta'_{,a})_{,b} = (\mu_{{}^{(2)}\!\tilde{g}} \tilde{g}^{cd}\epsilon_{ac}\; \mathcal{E}^{a'})_{,d}
\end{equation}
and
\begin{equation}\label{eq:150}
(\mu_{{}^{(2)}\!\tilde{g}}\; \tilde{g}^{ab}\lambda'_{,a})_{,b} = (\mu_{{}^{(2)}\!\tilde{g}} \tilde{g}^{cd}\epsilon_{ac}\; \mathcal{B}^{a'})_{,d}
\end{equation}
that follow from the defining formulas (\ref{eq:106}) and (\ref{eq:107}). Alternatively one could appeal to the exactness of the one-forms \(\epsilon_{ac}\; \mathcal{E}^{a'}\; dx^c\) and \(\epsilon_{ac}\; \mathcal{B}^{a'}\; dx^c\) and integrate the defining equations
\begin{equation}\label{eq:151}
\eta'_{,c} = \epsilon_{ac}\; \mathcal{E}^{a'},
\end{equation}
and
\begin{equation}\label{eq:152}
\lambda'_{,c} = \epsilon_{ac}\; \mathcal{B}^{a'}
\end{equation}
along conveniently chosen paths from (say) points on the axis where these potentials both vanish (c.f., the discussion in Appendix~\ref{app:electric-charge}).

To summarize the results of this section, we have proven the following:
%\begin{quote}
\begin{theorem}\label{thm:01}
Maxwell's equations for the axisymmetric, purely electromagnetic perturbations of an arbitrary Kerr black hole are derivable from a Hamiltonian (\(H^{\mathrm{Reg}}\) defined by Eqs.~(\ref{eq:135})--(\ref{eq:137})) that is positive definite and strictly conserved but non-local when expressed in terms of the electric and magnetic fields.
\end{theorem}
%\end{quote}

Since this work was completed one of us (N.G.) has shown how to derive a corresponding (positive definite energy) result, in the presence of a positive cosmological constant, for the axisymmetric, Maxwellian perturbations of Kerr-de Sitter black holes \cite{Gudapati_18}. Independently, Wald and Prabhu have derived positive definite energy functionals for the axisymmetric, electromagnetic perturbations of Kerr black holes that are not only compatible with ours in 4-dimensions but which apply as well to sufficiently symmetric electromagnetic perturbations of higher dimensional black holes \cite{PW_18}.  %equations starting 101
\section{An Energy Functional for Axisymmetric Kerr-Newman Perturbations}
\label{sec:energy-functional-kerr-newman}
In this section we construct a conserved, positive definite energy functional for linear, axisymmetric perturbations of arbitrary Kerr-Newman black holes. While we focus, for technical reasons, on the most astro-physically relevant, \textit{sub-extremal} cases our main calculational results are equally applicable to extremal black holes for which they generalize, to the electro-vacuum framework, those given in Ref.~\cite{DA_14}.

To set the stage for our derivation we first recall how similar ideas were developed, long ago, for the special case of (charged but non-rotating) Reissner-Nordstr\"{o}m black holes.

\subsection{Background on Reissner-Nordstr\"{o}m Perturbations}
\label{subsec:background-reissner-nordstrom}
The derivation of a conserved, positive definite energy functional for the coupled (electromagnetic and gravitational) dynamical perturbations of Reissner-Nordstr\"{o}m (RN) black hole spacetimes was given by one of us in Refs.~\cite{Moncrief_74_3,Moncrief_74_1,Moncrief_74_2}. It followed from computing the second variation of the Einstein-Maxwell action functional about a Reissner-Nordstr\"{o}m black hole background and restricting the resulting expression to the latter's \textit{domain of outer communications} (DOC). It has been realized from the time of Jacobi that such a 2nd variation functional serves, in turn, as an action for the corresponding linearized equations --- in the present context for those of the linear perturbations of an arbitrary RN background.

By exploiting the spherical symmetry of the RN geometry and expanding the perturbations in Regge-Wheeler tensor harmonics one was able to carry out an explicit canonical transformation to a new set of variables wherein a certain (unconstrained, gauge invariant) subset of canonical pairs was found to represent the radiative, dynamical degrees of freedom. This was complemented by a further subset comprised of the (equally gauge-invariant) linearized constraints and their (gauge-variant but unconstrained) canonically conjugate partners. The linearized lapse, shift and electromagnetic `scalar potential' served as Lagrange multipliers in the perturbative action functional where, paired with the (gauge group generating) linearized constraints, they could be adjusted to fix the evolution of gauge-dependent canonical variables.

The Hamiltonian emerging naturally from this formulation of the linearized field equations (for the radiative, dynamical degrees of freedom unconstrained by the Birkhoffian rigidity of the complementary, purely spherically symmetric perturbations) was found, by explicit calculation, to be positive definite, conserved and to bound a naturally associated Sobolev norm of the gauge-invariant dynamical variables. That a positive definite energy functional emerged rather straightforwardly from this analysis was however due, in no small measure, to the absence of an ergo-region in a Reissner-Nordstr\"{o}m black hole's DOC. An interesting feature of the resulting field equations, found independently by Zerilli who derived them by working in a special gauge \cite{Zerilli_74}, was that certain specific linear combinations of the perturbative, gravitational and electomagnetic variables decoupled from one another and satisfied single component wave equations of Regge-Wheeler-Zerilli type.

On the other hand the heavy reliance on the use of tensor spherical harmonics seemed to limit the application of the aforementioned methods to spherically symmetric backgrounds. While there is no particular difficulty involved in computing the 2nd variation of the Einstein-Maxwell action about the more general, Kerr-Newman backgrounds of principal interest herein, a demonstration that the resulting `canonical energy' expression is, at least for purely axisymmetric perturbations, actually positive definite has not heretofore been realized.

By exploiting the reducibility of the axisymmetric field equations to a 2+1-dimensional Einstein---wave map system (c.f., Appendix~\ref{app:reduced-hamiltonian}), computing the 1st and 2nd variations of the corresponding field equations (c.f., Appendix~\ref{app:analysis-linearized-constraint-equations}) about a Kerr-Newman black hole background (c.f., Appendix~\ref{app:kerr-newman-spacetimes}) and applying Robinson's identity to the resulting energy expression, we shall derive below an energy functional with the desired properties. This result will subsume that for purely electromagnetic perturbations of Kerr backgrounds, given in section~\ref{sec:pure-electromagnetic-kerr-spacetimes}, as a special case and now incorporate the coupled gravitational and electromagnetic perturbations of general Kerr-Newman black hole spacetimes in linear approximation. Since our strategy for deriving the desired energy expression will entail an extension of certain mathematical methods developed for the study of the so-called \textit{linearization stability} problem for Einstein's equations we briefly review some of the central ideas of that analysis in the following subsection. These will not only show the way for deriving the desired energy formula but also that for interpreting its geometrical significance.

\subsection{Brief Review of the Linearization Stability Problem}
\label{subsec:review-linearization-stability}
At around the same time that this early work on Reissner-Nordstr\"{o}m perturbations was being carried out some seemingly unrelated technology was being developed for the rigorous analysis of what came to be called the `linearization stability problem' for Einstein \cite{AMM_82,FM_79,FMM_80,Mon_75,Mon_76}, Einstein-Maxwell \cite{Arms_77} and, still more generally, Einstein-Yang-Mills \cite{AMM_82} spacetimes having compact Cauchy hypersurfaces. These studies dealt with the fact that, for spatially compact solutions of the relevant field equations that admitted one or more \textit{globally defined} Killing vector fields, one could show that the associated linearized equations admitted so-called `spurious solutions' that were not tangent to any differentiable curve of exact solutions to the corresponding, nonlinear field equations.

Such spurious solutions could be identified and excluded precisely by demanding that the Noether-like conserved quantities for the perturbation problem --- one for each Killing field of the background --- vanish. This additional condition supplemented the linearized field equations themselves with a (non-vacuous) set of conserved, gauge invariant quadratic integral restrictions upon the first order perturbations that was eventually shown to have a natural geometric interpretation \cite{AMM_81,AMM_82,FMM_80}. The geometrical meaning of this result was that it showed that the manifold structure of the solution space of the (nonlinear) constraint equations broke down at any point corresponding to Cauchy data for an exact solution admitting nontrivial, global Killing symmetries (or, in the case of the Einstein-Yang-Mills equations, generalized gauge symmetries \cite{AMM_82}). Roughly speaking the space of solutions to the nonlinear constraints was shown to exhibit a conical singularity at any such point and the role of the supplementary quadratic conditions on the linear perturbations was to restrict them to actually be tangent to this conical structure \cite{AMM_82,FM_79,FMM_80,Mon_75,Mon_76}.

This geometrically nontrivial conclusion did not carry over to the case of non-compact Cauchy surfaces since, roughly stated, certain boundary integrals linear in the second order perturbations now arose to `take up the slack' and allow the otherwise spurious first order perturbations to tangentially approximate curves of exact solutions while forcing the boundary contributions at second order (which were absent in the compact case) to take on certain specific values. More precisely the conserved quadratic integral expressions in the first order perturbations --- one for each global Killing field of the background exact solution --- were no longer coerced to vanish, since their actual values could always now be compensated by those of the (2nd order) boundary integral expressions. One could forcibly recover the conical singularity structure in the solution space for the nonlinear constraints only by \textit{artificially} restricting certain asymptotically defined, \textit{conserved} quantities for the nonlinear problem (e.g., the ADM mass for asymptotically flat solutions) to have fixed values.

A key step in the development of this linearization stability technology was the proof (originally for the vacuum case) that the kernel of the adjoint of the linearized constraint operator was precisely the space of Cauchy data for the Killing fields of the corresponding, vacuum spacetime \cite{AMM_82,FM_79,FMM_80,Mon_75,Mon_76}. Each such Killing initial data set (or KID as they are nowadays known) consists of a function \textit{C} and a spatial vector field \textit{Z} such that the pair \((C, Z)\) provides the normal and tangential components (at the chosen Cauchy hypersurface) of a Killing vector field on the resulting Einstein spacetime. For the special case of a timelike Killing field \(\zeta\), formulated in coordinates for which \(\zeta = \frac{\partial}{\partial t}\), the pair \((C, Z)\) is nothing other than the lapse and shift of the spacetime metric expressed in those coordinates.

While the published literature does not seem to include precisely the analogous (non-vacuum) case of interest to us here we shall simply verify the needed result, for the problem at hand, by a direct calculation rather than appeal to general theory. In this regard a key role will be played by the fact that the \textit{reduced} lapse function, \(\tilde{N}\), for Kerr-Newman backgrounds (c.f., Eqs.~(\ref{eq:a13}) and (\ref{eq:a42})) is \textit{harmonic} (c.f., Eq.~(\ref{eq:a71})). Since the corresponding \textit{reduced} shift field vanishes, the pair \((\tilde{N},O)\) will prove to provide precisely the needed kernel of the adjoint to the linearized constraint operator. It is straightforward to show that, in the \(\lbrace t, \varphi, \bar{\rho}, \bar{z}\rbrace\) coordinate system defined in Appendix~\ref{app:kerr-newman-spacetimes}, this kernel takes the simple form \((\bar{\rho}, O)\).

\subsection{An `Alternative' Energy Functional and its Regularization}
\label{subsec:alternative-energy-functional}
While a fully general, `canonical' energy density for non-symmetric perturbations of Kerr-Newman black holes could readily be constructed by reinstating the terms that were omitted in the static background (RN) limit analyzed in Ref.~\cite{Moncrief_74_3}, this would have little hope of directly yielding a positive definite total energy. By constraining the study to axisymmetric perturbations, however, and transforming the field equations to the 2+1-dimensional Einstein---wave map form reviewed in Appendix~\ref{app:reduced-hamiltonian} we are led to introduce the `alternative' energy density, `\(\mathcal{E}^{\mathrm{Alt}}\)', defined in Eqs.~(\ref{eq:c19}) and (\ref{eq:c22}) of Appendix~\ref{app:analysis-linearized-constraint-equations}.

This expression is itself indefinite though since in particular it reduces to \(\mathcal{H}^{\mathrm{Alt}}\) for the purely electromagnetic perturbations of Kerr backgrounds. The single negative term in the kinetic energy density, \(-\frac{\tilde{N}}{2} e^{2\nu} \sqrt{{}^{(2)}\!h} (\tau^\prime)^2\), can, however, be disposed of by imposing a suitable time-coordinate gauge condition. The simplest choice, \(\tau^\prime = 0\), corresponds to enforcing 2+1-dimensional, \textit{maximal} slicing\footnote{Here \(\tau'\) designates the first variation of the 2+1-dimensional mean curvature, \(\tau\), defined by Eq.~(\ref{eq:j01}). Its 3+1-dimensional analogue is given by Eq.~(\ref{eq:j02})\label{note06}} whereas setting
\begin{equation}
\tau^\prime = \mbox{} - \frac{\tilde{p}^\prime e^{-2\nu}}{4\sqrt{{}^{(2)}\!h}}
\label{eq:201}
\end{equation}
implies the imposition of maximal slicing in the `lifted', 3+1-dimensional sense. Though the latter choice leaves a negative term in the kinetic density it is easily seen to be dominated by the complementary, positive terms leaving a net positive definite kinetic energy density.

Let us abbreviate by \(\frac{\tilde{N}}{2} D^2 \mathcal{V}(\mathbf{q}, {}^{(2)}\!h) \cdot (\mathbf{q}^\prime, \mathbf{q}^\prime)\) the potential energy density given explicitly by
\begin{align}
\frac{\tilde{N}}{2} D^2 \mathcal{V}(\mathbf{q}, {}^{(2)}\!h) \cdot (\mathbf{q}^\prime, \mathbf{q}^\prime) &:= \left\lbrace\frac{\tilde{N}}{2} \sqrt{{}^{(2)}\!h}\; h^{ab}\left\lbrack\vphantom{\frac{1}{2}} 4\gamma^\prime_{,a} \gamma^\prime_{,b} + 2e^{-2\gamma} (\gamma^\prime)^2 (\eta_{,a}\eta_{,b} + \lambda_{,a}\lambda_{,b})\right.\right.\nonumber\\
& \mbox{} - 4e^{-2\gamma} \gamma^\prime (\eta_{,a}\eta^\prime_{,b} + \lambda_{,a}\lambda^\prime_{,b}) + e^{-2\gamma} (\eta^\prime_{,a}\eta^\prime_{,b} + \lambda^\prime_{,a}\lambda^\prime_{,b})\nonumber\\
& \mbox{} + 8e^{-4\gamma} (\gamma^\prime)^2 (\omega_{,a} + \lambda\eta_{,a})(\omega_{,b} + \lambda\eta_{,b})\nonumber\\
& \mbox{} - 8e^{-4\gamma} \gamma^\prime (\omega_{,a} + \lambda\eta_{,a})(\omega^\prime_{,b} + \lambda\eta^\prime_{,b} + \lambda^\prime\eta_{,b})\nonumber\\
& \mbox{} + e^{-4\gamma} (\omega^\prime_{,a} + \lambda\eta^\prime_{,a} + \lambda^\prime\eta_{,a})(\omega^\prime_{,b} + \lambda\eta^\prime_{,b} + \lambda^\prime\eta_{,b})\nonumber\\
& \left.\left.\mbox{} + e^{-4\gamma} (\omega_{,a} + \lambda\eta_{,a})(2\lambda^\prime\eta^\prime_{,b})\vphantom{\frac{1}{2}}\right\rbrack\right\rbrace
\label{eq:202}
\end{align}
and by
\begin{align}
K_1 &:= \frac{4}{\sqrt{{}^{(2)}\!h}} (\tilde{N} \sqrt{{}^{(2)}\!h}\; h^{ab} \gamma_{,a})_{,b} + \tilde{N} e^{-2\gamma}\; h^{ab} (\eta_{,a}\eta_{,b} + \lambda_{,a}\lambda_{,b})\nonumber\\
& \mbox{} + 2\tilde{N} e^{-4\gamma}\; h^{ab} (\omega_{,a} + \lambda\eta_{,a})(\omega_{,b} + \lambda\eta_{,b}),\label{eq:203}\\
K_2 &:= \frac{1}{\sqrt{{}^{(2)}\!h}} \left(\tilde{N} \sqrt{{}^{(2)}\!h}\; h^{ab} e^{-4\gamma} (\omega_{,a} + \lambda\eta_{,a})\right)_{,b}, \label{eq:204}\\
K_3 &:= \frac{1}{\sqrt{{}^{(2)}\!h}} (\tilde{N} \sqrt{{}^{(2)}\!h}\; h^{ab} e^{-2\gamma} \eta_{,a})_{,b} + \tilde{N} h^{ab} e^{-4\gamma} \lambda_{,b} (\omega_{,a} + \lambda\eta_{,a}), \label{eq:205}\\
K_4 &:= \frac{1}{\sqrt{{}^{(2)}\!h}} (\tilde{N} \sqrt{{}^{(2)}\!h}\; h^{ab} e^{-2\gamma} \lambda_{,a})_{,b} - \tilde{N} h^{ab} e^{-4\gamma} \eta_{,b} (\omega_{,a} + \lambda\eta_{,a})\label{eq:206}
\end{align}
the wavemap expressions which vanish for any Kerr-Newman background (c.f., Eqs.~(\ref{eq:a65}--\ref{eq:a68})). Finally, denote by \(\lbrace{}^{(1)}\!V, \ldots, {}^{(8)}\!V\rbrace\) the 1-forms defined explicitly by
\begin{align}
{}^{(1)}\!V_a &:= 2\gamma^\prime_{,a} + e^{-4\gamma} (\omega^\prime + \lambda\eta^\prime)(\omega_{,a} + \lambda\eta_{,a}) + \frac{1}{2} e^{-2\gamma} (\eta^\prime\eta_{,a} + \lambda^\prime\lambda_{,a}),\label{eq:207}\\
&\nonumber\\
{}^{(2)}\!V_a &:= -\left(e^{-2\gamma} (\omega^\prime + \lambda\eta^\prime)\right)_{,a} + \frac{1}{2} e^{-2\gamma} (\eta^\prime\lambda_{,a} - \lambda^\prime\eta_{,a}) + 2e^{-2\gamma} \gamma^\prime (\omega_{,a} + \lambda\eta_{,a}), \label{eq:208}\\
&\nonumber\\
{}^{(3)}\!V_a &:= \frac{1}{2} e^{2\gamma} (\eta^\prime_{,a}) - e^{2\gamma} \gamma^\prime (\eta_{,a}) + \left(\frac{\lambda^\prime}{2}\right) \left(\omega_{,a} + \lambda (\eta_{,a})\right),\label{eq:209}\\
&\nonumber\\
{}^{(4)}\!V_a &:= \frac{1}{2} e^{2\gamma} (\lambda^\prime_{,a}) - e^{2\gamma} \gamma^\prime (\lambda_{,a}) - \left(\frac{\eta^\prime}{2}\right) \left(\omega_{,a} + \lambda (\eta_{,a})\right),\label{eq:210}\\
&\nonumber\\
{}^{(5)}\!V_a &:= \frac{1}{4} (\eta^\prime\lambda_{,a} - \lambda^\prime\eta_{,a}),\label{eq:211}\\
&\nonumber\\
{}^{(6)}\!V_a &:= \frac{1}{2} (\eta^\prime_{,a}) - \eta^\prime \gamma_{,a} + \frac{1}{2} e^{-2\gamma} \lambda_{,a} (\omega^\prime + \lambda\eta^\prime),\label{eq:212}\\
&\nonumber\\
{}^{(7)}\!V_a &:= \frac{1}{2} (\lambda^\prime_{,a}) - \lambda^\prime \gamma_{,a} - \frac{1}{2} e^{-2\gamma} \eta_{,a} (\omega^\prime + \lambda\eta^\prime),\label{eq:213}\\
&\nonumber\\
{}^{(8)}\!V_a &:= 2\gamma_{,a} (\omega^\prime + \lambda\eta^\prime) - \frac{1}{2} (\eta^\prime\lambda_{,a} - \lambda^\prime\eta_{,a}) - 2\gamma^\prime (\omega_{,a} + \lambda\eta_{,a}).\label{eq:214}
\end{align}

In this notation Robinson's identity \cite{Rob_74} takes the form:
\begin{equation}
\begin{split}
& \tilde{N}D^2 \mathcal{V}(\mathbf{q}, {}^{(2)}\!h) \cdot (\mathbf{q}^\prime, \mathbf{q}^\prime) + \frac{\partial}{\partial x^b}\left\lbrace\vphantom{\frac{1}{2}}\tilde{N} \sqrt{{}^{(2)}\!h}\; h^{ab} \left\lbrack\vphantom{\frac{1}{2}} -2e^{-4\gamma} \gamma_{,a} (\omega^\prime + \lambda\eta^\prime)^2\right.\right.\\
& \mbox{} - e^{-2\gamma} \gamma_{,a} \left((\eta^\prime)^2 + (\lambda^\prime)^2\right) + e^{-4\gamma} (\omega^\prime + \lambda\eta^\prime)(\lambda_{,a}\eta^\prime - \lambda^\prime\eta_{,a})\\
& \mbox{} - e^{-4\gamma} (\omega_{,a} + \lambda\eta_{,a}) \eta^\prime\lambda^\prime + 4e^{-4\gamma} \gamma^\prime (\omega_{,a} + \lambda\eta_{,a})(\omega^\prime + \lambda\eta^\prime)\\
& \mbox{}  + \left.\vphantom{\frac{1}{2}}\left.\vphantom{\frac{1}{2}}2e^{-2\gamma} \gamma^\prime\eta^\prime\eta_{,a} + 2e^{-2\gamma} \gamma^\prime\lambda^\prime\lambda_{,a}\right\rbrack\right\rbrace\\
& \mbox{} + \frac{\sqrt{{}^{(2)}\!h}}{2e^{2\gamma}} K_1\; \left\lbrack e^{-2\gamma} (\omega^\prime + \lambda\eta^\prime)^2 + \frac{1}{2} \left((\eta^\prime)^2 + (\lambda^\prime)^2\right)\right\rbrack\\
& \mbox{} + \sqrt{{}^{(2)}\!h}\; K_2\; \left\lbrack\eta^\prime\lambda^\prime - 4\gamma^\prime (\omega^\prime + \lambda\eta^\prime)\right\rbrack\\
& \mbox{} + \sqrt{{}^{(2)}\!h}\; K_3\; \left\lbrack\lambda^\prime e^{-2\gamma} (\omega^\prime + \lambda\eta^\prime) - 2\gamma^\prime\eta^\prime\right\rbrack\\
& \mbox{} \sqrt{{}^{(2)}\!h}\; K_4\; \left\lbrack\eta^\prime e^{-2\gamma} (-\omega^\prime - \lambda\eta^\prime) - 2\gamma^\prime\lambda^\prime\right\rbrack\\
& \mbox{} \equiv \tilde{N} \sqrt{{}^{(2)}\!h}\; h^{ab} \left\lbrace{}^{(1)}\!V_a {}^{(1)}\!V_b + {}^{(2)}\!V_a {}^{(2)}\!V_b\right.\\
& \mbox{} + 2e^{-6\gamma}\; {}^{(3)}\!V_a {}^{(3)}\!V_b + 2e^{-6\gamma}\; {}^{(4)}\!V_a {}^{(4)}\!V_b\\
& \mbox{} + 12e^{-4\gamma}\; {}^{(5)}\!V_a {}^{(5)}\!V_b + 2e^{-2\gamma}\; {}^{(6)}\!V_a {}^{(6)}\!V_b\\
& \left.\mbox{} + 2e^{-2\gamma}\; {}^{(7)}\!V_a {}^{(7)}\!V_b + e^{-4\gamma}\; {}^{(8)}\!V_a {}^{(8)}\!V_b\right\rbrace
\label{eq:215}
\end{split}
\end{equation}
It follows immediately that, if we drop the terms that vanish by virtue of the background field equations (i.e., set \(K_1 = K_2 = K_3 = K_4 = 0\)), then we can express the potential energy density occurring in \(\mathcal{E}^{\mathrm{Alt}}\) as the sum of a spatial divergence and a \textit{positive definite} quadratic expression in the one-forms \({}^{(1)}\!V, \ldots, {}^{(8)}\!V\).

 Since the integrated divergence will only contribute a boundary term in the total energy expression we set it aside here and define our regulated energy density, \(\mathcal{E}^{\mathrm{Reg}}\), by
 \begin{equation}
 \begin{split}
 \mathcal{E}^{\mathrm{Reg}} &:= \left\lbrace\frac{\tilde{N}}{\sqrt{{}^{(2)}\!h}} e^{-2\nu}\; \left\lbrack{\tilde{r}^\prime}_a^b {\tilde{r}^\prime}_b^a + \frac{1}{8} (\tilde{p}^\prime)^2 + \frac{1}{2} e^{4\gamma} (\tilde{r}^\prime)^2\right.\right.\\
 & \mbox{} + \left.\left.\frac{1}{2} e^{2\gamma} \left((\tilde{v}^\prime)^2 + (\tilde{u}^\prime - \lambda\tilde{r}^\prime)^2\right)\right\rbrack - \frac{\tilde{N}}{2} e^{2\nu} \sqrt{{}^{(2)}\!h} (\tau^\prime)^2\right\rbrace\\
 & \mbox{} + \tilde{N} \sqrt{{}^{(2)}\!h}\; h^{ab} \left\lbrace\frac{1}{2}\; {}^{(1)}\!V_a {}^{(1)}\!V_b + \frac{1}{2}\; {}^{(2)}\!V_a {}^{(2)}\!V_b + e^{-6\gamma}\; {}^{(3)}\!V_a {}^{(3)}\!V_b\right.\\
 & \mbox{} + e^{-6\gamma}\; {}^{(4)}\!V_a {}^{(4)}\!V_b + 6e^{-4\gamma}\; {}^{(5)}\!V_a {}^{(5)}\!V_b + e^{-2\gamma}\; {}^{(6)}\!V_a {}^{(6)}\!V_b\\
 & \mbox{} + e^{-2\gamma}\; {}^{(7)}\!V_a {}^{(7)}\!V_b + \frac{1}{2} e^{-4\gamma}\; {}^{(8)}\!V_a {}^{(8)}\!V_b\left.\vphantom{\frac{1}{2}}\right\rbrace
 \label{eq:216}
 \end{split}
 \end{equation}
 and corresponding total regulated energy by
 \begin{equation}
 E^{\mathrm{Reg}} := \int_{M_b} d^2x\; \lbrace\mathcal{E}^{\mathrm{Reg}}\rbrace.
 \label{eq:217}
 \end{equation}
 Evaluated with respect to either of the two maximal slicing gauges discussed above \(\mathcal{E}^{\mathrm{Reg}}\) and thus \(E^{\mathrm{Reg}}\) are manifestly positive definite.
 %equations starting 201
\section{Conservation of the Total Energy}
\label{sec:conservation}
The `alternative' energy, \(E^{\mathrm{Alt}}\), defined by Eq.~(\ref{eq:c23}) has its density, \(\mathcal{E}^{\mathrm{Alt}}\), given explicitly via Eqs.~(\ref{eq:c19}) and (\ref{eq:c22}). While \(E^{\mathrm{Alt}}\) potentially differs by a boundary integral from its `regulated' counterpart \(E^{\mathrm{Reg}}\) (c.f., Eqs.~(\ref{eq:216})--(\ref{eq:217})), we shall see below that this difference actually vanishes for the class of (asymptotically-pure-gauge) perturbations that we consider here. Thus for our present purposes it will suffice to show that \(E^{\mathrm{Alt}}\) is conserved since this will imply the corresponding result for \(E^{\mathrm{Reg}}\).

As discussed more completely in Appendices~\ref{app:gauge-conditions} and \ref{app:transforming} we work in a partially gauge-fixed setting wherein the flat (2-dimensional) `conformal' desensitized spatial metric,
\begin{equation}\label{eq:404}
{}^{(2)}\tilde{h} := \frac{h_{ab}}{\sqrt{{}^{(2)}\!h}}\; dx^a \otimes dx^b,
\end{equation}
is held fixed during the evolution through a judicious choice of the linearized shift field \(X' := \tilde{N}^{c'} \frac{\partial}{\partial x^c}\). By contrast, at least for now, we leave the perturbative time gauge unspecified by allowing the linearized lapse function, \(\tilde{N}'\), to remain arbitrary. That our \textit{total} energy will be found to be conserved independently of the interior behavior of the linearized time gauge chosen corresponds to its essential gauge invariance. Otherwise one could modify its evolution by merely making a change of gauge.

Computing \(\frac{\partial}{\partial t} \mathcal{E}^{\mathrm{Alt}}\) directly by means of the linearized field equations one gets:
\begin{equation}\label{eq:401}
\begin{split}
\frac{\partial}{\partial t} \mathcal{E}^{\mathrm{Alt}} &= \frac{\partial}{\partial x^b} \left\lbrace\frac{\tilde{N}^2}{\sqrt{{}^{(2)}\tilde{g}}} \left\lbrack\tilde{p}' \left(\sqrt{{}^{(2)}h} h^{ab} \gamma_{,a}\right)'\right.\right.\\
 & {} + e^{4\gamma} \tilde{r}' \left(e^{-4\gamma} \sqrt{{}^{(2)}h} h^{ab} (\omega_{,a} + \lambda\eta_{,a})\right)' + e^{2\gamma}\tilde{v}' \left(\sqrt{{}^{(2)}h} h^{ab} e^{-2\gamma} \lambda_{,a}\right)'\\
 & {} + e^{2\gamma} (\tilde{u}' - \lambda\tilde{r}') \left(\sqrt{{}^{(2)}h} h^{ab} e^{-2\gamma} \eta_{,a}\right)'\\
 & {} + \left.\vphantom{\frac{1}{\sqrt{2}}} e^{2\gamma} (\tilde{u}' - \lambda\tilde{r}') \lambda' \sqrt{{}^{(2)}h} h^{ab} e^{-4\gamma} (\omega_{,a} + \lambda\eta_{,a})\right\rbrack\\
 & {} + \gamma' \mathcal{L}_{X'} \left( 4 \tilde{N} \sqrt{{}^{(2)}h} h^{ab} \gamma_{,a}\right) + \omega' \mathcal{L}_{X'} \left(\tilde{N} \sqrt{{}^{(2)}h} h^{ab} e^{-4\gamma} (\omega_{,a} + \lambda\eta_{,a})\right)\\
 & {} + \lambda' \mathcal{L}_{X'} \left(\tilde{N} e^{-2\gamma} \sqrt{{}^{(2)}h} h^{ab} \lambda_{,a}\right) + \eta' \mathcal{L}_{X'} \left(\tilde{N} e^{-2\gamma} \sqrt{{}^{(2)}h} h^{ab} \eta_{,a}\right.\\
 & \quad {} + \left.\vphantom{\sqrt{2}} \tilde{N} e^{-4\gamma} \sqrt{{}^{(2)}h} h^{ab} \lambda (\omega_{,a} + \lambda\eta_{,a})\right)\\
 & {} + 2 (\mathcal{L}_{X'} \tilde{N}) \left(\sqrt{{}^{(2)}h} h^{ba} \nu'_{,a}\right) + 2 (\mathcal{L}_{X'} \nu') \sqrt{{}^{(2)}h} h^{ab} \tilde{N}_{,a}\\
 & {} - 2 X^{b'} \left(\sqrt{{}^{(2)}h} h^{ac} \nu'_{,a} \tilde{N}_{,c}\right)\\
 & {} + 2\tilde{N} h^{bc} (\tilde{r}_c^a)' e^{-2\nu} \tilde{N}'_{,a} + (\tilde{N} \tilde{N}'_{,a} - \tilde{N}'\tilde{N}_{,a}) \tau' \sqrt{{}^{(2)}h} h^{ab}\\
 & {} \left.\vphantom{\frac{1}{\sqrt{2}}} - 2\tilde{N}' h^{bc} (\tilde{r}_c^a)' e^{-2\nu} \tilde{N}_{,a}\right\rbrace\\
 & {} - \left\lbrace\tilde{\mathcal{H}}' \mathcal{L}_{X'} \tilde{N} + (\tilde{N}'\tilde{N}_{,a} - \tilde{N}\tilde{N}'_{,a}) e^{-2\nu} h^{ab} \tilde{\mathcal{H}}'_b\right\rbrace
\end{split}
\end{equation}
where, since the variations of \(\sqrt{{}^{(2)}\!h}\; h^{ab}\; \frac{\partial}{\partial x^a} \otimes \frac{\partial}{\partial x^b}\) are taken to vanish (c.f., Appendices~\ref{app:gauge-conditions} and \ref{app:transforming}), one has
\begin{align}
\left(\sqrt{{}^{(2)}h} h^{ab} \gamma_{,a}\right)' &= \sqrt{{}^{(2)}h} h^{ab} \gamma'_{,a},\label{eq:402}\\
\begin{split}\label{eq:403}
\left(\sqrt{{}^{(2)}h} h^{ab} e^{-2\gamma} \lambda_{,a}\right)' &= \sqrt{{}^{(2)}h} h^{ab} e^{-2\gamma} \lambda'_{,a}\\
 & {} - 2 \sqrt{{}^{(2)}h} h^{ab} e^{-2\gamma} \gamma' \lambda_{,a},\quad \hbox{etc.}
 \end{split}
\end{align}
Note that the terms in the last \(\lbrace~\rbrace\) bracket in Eq.~(\ref{eq:401}) vanish by virtue of the linearized constraints leaving only a total spatial divergence whose integral over \(M_b\) will result in pure boundary expressions. As we shall show below these boundary integrals vanish for the class of (asymptotically-pure-gauge) perturbations that we consider herein.

\subsection{Evaluating the `Dynamical' Boundary Flux Terms}
\label{subsec:evaluating_dynamical}

Consider first the boundary flux contributions from Eq.~(\ref{eq:401}) that are each linear in the perturbed momenta \(\left\lbrace\tilde{p}', \tilde{r}', \tilde{v}', \tilde{u}' - \lambda\tilde{r}'\right\rbrace\) --- we shall refer to these as `dynamical' flux terms. From Eqs.~(\ref{eq:f72})--(\ref{eq:f75}) we see that the momentum factors occurring in each of these terms can, in the \textit{asymptotic regions} near \(R = R_+\) and \(R = \infty\) defined therein, be expressed as a pairing of the vector field (c.f., \ref{eq:f77})
\begin{equation}\label{eq:405}
{}^{(2)}\!\mathcal{D}^a \frac{\partial}{\partial x^a} := \tilde{N}^2 \tilde{g}^{ab} {}^{(4)}\!Y^0_{,b} \frac{\partial}{\partial x^a}
\end{equation}
with a corresponding one-form taken from the list\hfill\break
 \(\left\lbrace 4\gamma_{,a} dx^a, \lambda_{,a} dx^a, \eta_{,a} dx^a, (\omega_{,a} + \lambda\eta_{,a}) dx^a\right\rbrace\). Here \({}^{(4)}\!Y^0\) is the time component of the gauge transforming vector field \({}^{(4)}\!Y = {}^{(4)}\!Y^\mu \frac{\partial}{\partial x^\mu}\) introduced in this appendix. From Eq.~(\ref{eq:f23}) we see that (in the asymptotic regions defined in Appendix \ref{app:transforming}  where the compactly supported `hyperbolic gauge perturbation' \({}^{(4)}\!k_{\alpha\beta}\) vanishes) this component is determined from integrating
\begin{equation}\label{eq:406}
{}^{(4)}\!Y_{,0}^0 \rightarrow \frac{\tilde{N}'}{\tilde{N}} - {}^{(2)}\!Y^a \frac{\tilde{N}_{,a}}{\tilde{N}}
\end{equation}
with \({}^{(2)}\!Y = {}^{(2)}\!Y^a \frac{\partial}{\partial x^a} = {}^{(4)}\!Y^a \frac{\partial}{\partial x^a}\) given in turn by (c.f., Eqs.~(\ref{eq:f34})--(\ref{eq:f35}), (\ref{eq:f44}), (\ref{eq:f50})--(\ref{eq:f56}) and recall the vanishing of \(a_0 (R_+)\) for the perturbations of interest herein established in Appendix~\ref{app:vanishing}):
\begin{align}
\frac{{}^{(2)}\!Y^R}{R} &= \sum_{n=1}^\infty \alpha_n^{(+)}(t) R_+^n \left(\frac{R^n}{R_+^n} - \frac{R_+^n}{R^n}\right) \cos{(n\theta)},\label{eq:407}\\
{}^{(2)}\!Y^\theta &= \sum_{n=1}^\infty \alpha_n^{(+)} (t) R_+^n \left(\frac{R^n}{R_+^n} + \frac{R_+^n}{R^n}\right) \sin{(n\theta)}\label{eq:408}\\
\intertext{near \(R = R_+\), and by}
\frac{{}^{(2)}\!Y^R}{R} &= {} - \sum_{n=1}^\infty \beta_n^{(-)}(t) R^{-n} \cos{(n\theta)},\label{eq:409}\\
{}^{(2)}\!Y^\theta &= \sum_{n=1}^\infty \beta_n^{(-)}(t) R^{-n} \sin{(n\theta)}\label{eq:410}
\end{align}
near \(R = \infty\). Recall that the \textit{t}-dependent coefficients \(\left\lbrace\alpha_n^{(+)}, \beta_n^{(-)}\right\rbrace\) are computable in terms of specified `sources' (determined by \({}^{(4)}\!k_{\alpha\beta}\)) via Eqs.~(\ref{eq:f57}), (\ref{eq:f59})--(\ref{eq:f80}).

Defining
\begin{equation}\label{eq:411}
{}^{(2)}\!\mathcal{Y}^a(t, R, \theta) := \int_{t_0}^t dt' {}^{(2)}\!Y^a (t', R, \theta)
\end{equation}
we see that
\begin{equation}\label{eq:412}
{}^{(4)}\!Y^0 (t, R, \theta) = \left.{}^{(4)}\!Y^0\right|_{t=t_0} (R, \theta) + \int_{t_0}^t dt' \left(\frac{\tilde{N}'}{\tilde{N}} (t', R, \theta)\right) - {}^{(2)}\!\mathcal{Y}^a (t , R, \theta) \frac{\tilde{N}_{,a}}{\tilde{N}} (R, \theta)
\end{equation}
where \(\left.{}^{(4)}\!Y^0\right|_{t=t_0} (R, \theta)\) is initial data chosen for \({}^{(4)}\!Y^0\). Defining
\begin{align}
\tilde{\alpha}_n^{(+)}(t) &:= \int_{t_0}^t dt'\; \alpha_n^{(+)}(t'),\label{eq:413}\\
\tilde{\beta}_n^{(-)}(t) &:= \int_{t_0}^t dt'\; \beta_n^{(-)}(t'),\label{eq:414}\\
\tilde{c}_0(R, t) &:= \int_{t_0}^t dt'\; c_0(R, t')\label{eq:415}
\end{align}
one arrives at explicit (Fourier representation) formulas for the \({}^{(2)}\!\mathcal{Y}^a\) by making the replacements \(\alpha_n^{(+)} \rightarrow \tilde{\alpha}_{n}^{(+)}\) and \(\beta_n^{(-)} \rightarrow \tilde{\beta}_n^{(-)}\) in Eqs.~(\ref{eq:407})--(\ref{eq:410}).

For definiteness we shall herein eventually impose the \(2+1\)-dimensional maximal slicing gauge condition which, taken together with suitable (homogeneous) boundary conditions for the linearized lapse function, \(\delta\tilde{N} := \tilde{N}'\), will imply that \(\tilde{N}' = 0\) (c.f., the discussion given in Appendix~\ref{app:maximal-slicing}). For now however we shall retain the contributions of (a non-vanishing) \(\tilde{N}'\) to \({}^{(4)}\!Y^0\) so that they can be easily reinstated if alternative gauge conditions (e.g., \(3+1\)-dimensional, maximal slicing, as discussed in Appendix~(\ref{app:maximal-slicing})) are desired in the future.

One is of course free to impose essentially arbitrary boundary conditions upon the initial data \(\left.{}^{(4)}\!Y^0\right|_{t=t_0}\). We shall assume in the following that these have been chosen so that, together with the boundary behavior of the linearized lapse function, one has
\begin{align}
\mathcal{K}_R &:= \left.{}^{(4)}\!Y^0_{,R}\right|_{t=t_0} + \int_{t_0}^t dt'\; \left(\frac{\tilde{N}'}{\tilde{N}}\right)_{,R} \sim O \left(\frac{1}{R^3}\right),\label{eq:416}\\
\mathcal{K}_\theta &:= \left.{}^{(4)}\!Y^0_{,\theta}\right|_{t=t_0} + \int_{t_0}^t dt'\; \left(\frac{\tilde{N}'}{\tilde{N}}\right)_{,\theta} \sim O \left(\frac{1}{R^2}\right)\label{eq:417}
\end{align}
as \(R \rightarrow \infty\) with \(\mathcal{K}_R\) and \(\mathcal{K}_\theta\) both bounded as \(R \rightarrow R_+\) and with \(\mathcal{K}_R\) bounded and \(\mathcal{K}_\theta\) vanishing like
\begin{equation}\label{eq:418}
\mathcal{K}_\theta \sim \sin{(\theta)} \times \hbox{(regular factor)}
\end{equation}
at the axes corresponding to \(\theta = 0, \pi\).

For vanishing \(\tilde{N}'\) these are trivial to ensure by choice of \(\left.{}^{(4)}\!Y^0\right|_{t=t_0}\) but would need to be verified on a case-by-case basis for alternative time gauges. On the other hand these conditions are only \textit{sufficient} for the arguments to follow and could be somewhat relaxed without disturbing the main results.

The components of the vector field \({}^{(2)}\!\mathcal{D}^a \frac{\partial}{\partial x^a}\) are given explicitly by
\begin{align}
\begin{split}
{}^{(2)}\!\mathcal{D}^R &= \frac{R^4}{\left\lbrack (r^2 + a^2)^2 - a^2 \Delta\sin^2{(\theta)}\right\rbrack} \left\lbrace\left( 1 - \frac{R_+^2}{R^2}\right)^2 \mathcal{K}_R\right.\\
& \left.{} - \left\lbrack\left( 1 - \frac{R_+^4}{R^4}\right) \left(\frac{{}^{(2)}\!\mathcal{Y}^R}{R}\right)_{,R} - \frac{4R_+^2}{R^3} \left(\frac{{}^{(2)}\!\mathcal{Y}^R}{R}\right) + \frac{\cos{\theta}}{\sin{\theta}} \left( 1 - \frac{R_+^2}{R^2}\right)^2 {}^{(2)}\!\mathcal{Y}^\theta_{,R}\right\rbrack\right\rbrace,
\end{split}\label{eq:419}\\
\intertext{and}
\begin{split}
{}^{(2)}\!\mathcal{D}^\theta &= \frac{R^2}{\left\lbrack (r^2 + a^2)^2 - a^2 \Delta\sin^2{(\theta)}\right\rbrack} \left\lbrace\left( 1 - \frac{R_+^2}{R^2}\right)^2 \mathcal{K}_\theta\right.\\
& \left.{} - \left\lbrack\left(\frac{{}^{(2)}\!\mathcal{Y}^R}{R}\right)_{,\theta} \left( 1 - \frac{R^4_+}{R^4}\right) + \left(\frac{{}^{(2)}\!\mathcal{Y}^\theta \cos{(\theta)}}{\sin{(\theta)}}\right)_{,\theta} \left( 1 - \frac{R_+^2}{R^2}\right)^2\right\rbrack\right\rbrace
\end{split}\label{eq:420}
\end{align}
where \(\Delta := r^2 - 2Mr + a^2 + Q^2\) (c.f., Eq.~(\ref{eq:b04})). Evaluated in the asymptotic region at large \textit{R} via Eqs.~(\ref{eq:409})--(\ref{eq:411}), (\ref{eq:414}) these become
\begin{align}
\begin{split}
{}^{(2)}\!\mathcal{D}^R &= \frac{R^4}{\left\lbrack (r^2 + a^2)^2 - a^2 \Delta\sin^2{(\theta)}\right\rbrack} \left\lbrace\left( 1 - \frac{R_+^2}{R^2}\right)^2 \mathcal{K}_R\right.\\
& \hphantom{=}{} - \left\lbrack\left( 1 - \frac{R_+^4}{R^4}\right) \sum_{n=2}^\infty \tilde{\beta}_n^{(-)} nR^{-n-1} \cos{(n\theta)}\right.\\
& \hphantom{=}{}  - 4 \frac{R_+^2}{R^3} \left( - \sum_{n=2}^\infty \tilde{\beta}_n^{(-)} R^{-n} \cos{(n\theta)}\right)\\
& \hphantom{=}{}  + \left( 1 - \frac{R_+^2}{R^2}\right)^2 \left(-\sum_{n=2}^\infty \tilde{\beta}_n^{(-)} nR^{-n-1} \frac{\cos{(\theta)}}{\sin{(\theta)}} \sin{(n\theta)}\right)\\
& \left.\left.{} + 2 \frac{R_+^2}{R^2} \left( 3 - \frac{R_+^2}{R^2}\right) \frac{\tilde{\beta}_1^{(-)}}{R^2} \cos{(\theta)}\right\rbrack\right\rbrace
\end{split}\label{eq:421}
\intertext{and}
\begin{split}
{}^{(2)}\!\mathcal{D}^\theta &= \frac{R^2}{\left\lbrack (r^2 + a^2)^2 - a^2 \Delta\sin^2{(\theta)}\right\rbrack} \left\lbrace\left( 1 - \frac{R_+^2}{R^2}\right)^2 \mathcal{K}_\theta\right.\\
& \hphantom{=}{}  - \left\lbrack\frac{2R_+^2}{R^3} \tilde{\beta}_1^{(-)} \sin{(\theta)} \left( 1 - \frac{R_+^2}{R^2}\right)\right.\\
& \hphantom{=}{}  + \left( 1 - \frac{R_+^4}{R^4}\right) \sum_{n=2}^\infty \tilde{\beta}_n^{(-)} nR^{-n} \sin{(n\theta)}\\
& \left.\left.{} + \left( 1 - \frac{R_+^2}{R^2}\right)^2 \sum_{n=2}^\infty \tilde{\beta}_n^{(-)} R^{-n} \left(\frac{\cos{(\theta)}}{\sin{(\theta)}} \sin{(n\theta)}\right)_{,\theta}\right\rbrace\right\rbrack
\end{split}\label{eq:422}
\end{align}
whereas for \textit{R} near \(R_+\) they take the forms (c.f., Eqs.~(\ref{eq:407})--(\ref{eq:408}), (\ref{eq:411}), (\ref{eq:413}))
\begin{align}
\begin{split}
{}^{(2)}\!\mathcal{D}^R &= \frac{R^4}{\left\lbrack (r^2 + a^2)^2 - a^2 \Delta\sin^2{(\theta)}\right\rbrack} \left\lbrace\left( 1 - \frac{R_+^2}{R^2}\right)^2 \mathcal{K}_R\right.\\
& \hphantom{=}{}  - \left\lbrack\left( 1 - \frac{R_+^2}{R^2}\right) \left\lbrace\left( 1 + \frac{R_+^2}{R^2}\right) \sum_{n=1}^\infty \tilde{\alpha}_n^{(+)} \frac{R_+^n}{R} n \left(\frac{R^n}{R_+^n} + \frac{R_+^n}{R^n}\right) \cos{(n\theta)}\right.\right\rbrace\\
& {} - \frac{4R_+^2}{R^3} \sum_{n=1}^\infty \tilde{\alpha}_n^{(+)} R_+^n \left(\frac{R^n}{R_+^n} - \frac{R_+^n}{R^n}\right) \cos{(n\theta)}\\
& \left.\left. {} + \left( 1 - \frac{R_+^2}{R^2}\right)^2 \sum_{n=1}^\infty \left(\frac{\cos{(\theta)}}{\sin{(\theta)}} \sin{(n\theta)}\right) \frac{n\tilde{\alpha}_n^{(+)}}{R} R_+^n \left(\frac{R^n}{R_+^n} - \frac{R_+^n}{R^n}\right)\right\rbrack\right\rbrace
\end{split}\label{eq:423}
\intertext{and}
\begin{split}
{}^{(2)}\!\mathcal{D}^\theta &= \frac{R^2}{\left\lbrack (r^2 + a^2)^2 - a^2 \Delta\sin^2{(\theta)}\right\rbrack} \left\lbrace\left( 1 - \frac{R_+^2}{R^2}\right)^2 \mathcal{K}_\theta\right.\\
& \hphantom{=}{}  - \left( 1 - \frac{R_+^2}{R^2}\right) \left\lbrace\sum_{n=1}^\infty \tilde{\alpha}_n^{(+)} \left( R^n + \frac{R_+^{2n}}{R^n}\right) \left(\frac{\cos{(\theta)}}{\sin{(\theta)}} \sin{(n\theta)}\right)_{,\theta} \cdot \left( 1 - \frac{R_+^2}{R^2}\right)\right.\\
& \left.\left.{} - \left( 1 + \frac{R_+^2}{R^2}\right) \sum_{n=1}^\infty \tilde{\alpha}_n^{(+)} R_+^{n} \left(\frac{R^n}{R_+^n} - \frac{R_+^n}{R^n}\right) n\sin{(n\theta)}\right\rbrace\right\rbrace
\end{split}\label{eq:424}
\end{align}
A standard trigonometric identity shows that \(\frac{\cos{(\theta)}}{\sin{(\theta)}} \sin{(n\theta)}\) is expressible as a polynomial of degree \textit{n} in \(\cos{(\theta)}\) and thus is regular at the axes \(\theta = 0,\pi\). From this same result it follows that \(\left(\frac{\cos{(\theta)}}{\sin{(\theta)}} \sin{(n\theta)}\right)_{,\theta}\) vanishes like \(\sin{(\theta)}\) at these axes. In addition it is straightforward to verify that \(\frac{\left(\frac{R^n}{R_+^n} - \frac{R_+^n}{R^n}\right)}{\left(\frac{R}{R_+} - \frac{R_+}{R}\right)}\) is smooth \(\forall\> R\) and, in particular, has the limits
\begin{align}
\lim_{R \rightarrow R_+} \frac{\left(\frac{R^n}{R_+^n} - \frac{R_+^n}{R^n}\right)}{\left(\frac{R}{R_+} - \frac{R_+}{R}\right)} &= n\label{eq:425}\\
\intertext{and}
\lim_{R \rightarrow R_+} \left\lbrack\frac{\frac{R^n}{R_+^n} - \frac{R_+^n}{R^n}}{\left(\frac{R}{R_+} - \frac{R_+}{R}\right)}\right\rbrack_{,R} &= 0\label{eq:426}
\end{align}
\(\forall\> n = 1, 2, \ldots\).

From the explicit formulas for \(\left\lbrace\gamma, \lambda, \eta, \omega\right\rbrace\) one finds that each of the quantities \(\left\lbrace 4\gamma_{,R}, \lambda_{,R}, \eta_{,R}, (\omega_{,R} + \lambda\eta_{,R})\right\rbrace\) vanishes like \(\left( 1 - \frac{R_+^2}{R^2}\right)\) as \(R \rightarrow R_+\) whereas \(\left\lbrace 4\gamma_{,\theta}, \lambda_{,\theta}, \eta_{,\theta}, (\omega_{,\theta} + \lambda\eta_{,\theta})\right\rbrace\) are all regular in this limit. Assuming that \(\left.{}^{(4)}\!Y\right|_{t=t_0}\) and \(\frac{\tilde{N}'}{\tilde{N}}\) have been chosen so that both \(\mathcal{K}_R\) and \(\mathcal{K}_\theta\) (c.f., Eqs.~(\ref{eq:416})--(\ref{eq:417})) are regular at \(R_+\) we see that each of the quantities
\begin{equation}\label{eq:427}
{}^{(2)}\!\mathcal{D}^a \chi_a \sim \left( 1 - \frac{R_+^2}{R^2}\right) \times \hbox{(regular factor)}
\end{equation}
as \(R \rightarrow R_+\) where \(\chi_a = 4\gamma_{,a}, \lambda_{,a}, \eta_{,a}\) and \((\omega_{,a} + \lambda\eta_{,a})\). Thus each of the (linearized momentum) factors \(\left\lbrace\frac{\tilde{N}\tilde{p}'}{\sqrt{{}^{(2)}\!\tilde{g}}}, \frac{\tilde{N} e^{2\gamma}\tilde{v}'}{\sqrt{{}^{(2)}\!\tilde{g}}}, \frac{\tilde{N} e^{2\gamma}}{\sqrt{{}^{(2)}\!\tilde{g}}} (\tilde{u}' - \lambda\tilde{r}'), \frac{\tilde{N} e^{4\gamma}\tilde{r}'}{\sqrt{{}^{(2)}\!\tilde{g}}}\right\rbrace\) (c.f., Eqs.~(\ref{eq:f72})--(\ref{eq:f75})) vanishes like \(\left( 1 - \frac{R_+^2}{R^2}\right) \times \hbox{(regular factor)}\) as \(R \rightarrow R_+\). Note that in the corresponding flux terms these are each multiplied by an `additional' factor of  \(\tilde{N} = R\sin{(\theta)}\left( 1 - \frac{R_+^2}{R^2}\right)\) and paired (respectively) with factors of the form \(\left(\sqrt{{}^{(2)}\!h}\; h^{ab} \gamma_{,a}\right)'\), \(\left(\sqrt{{}^{(2)}\!h}\; h^{ab} e^{-2\gamma} \lambda_{,a}\right)'\), etc. to yield their ultimate contributions to the flux integrals at the various boundaries.

Recalling that our gauge conditions enforce the constraint that
\begin{equation}\label{eq:428}
\left(\sqrt{{}^{(2)}\!h}\; h^{ab}\right)' = 0
\end{equation}
we see that (c.f., Eqs.~(\ref{eq:402}) and (\ref{eq:403}))
\begin{align}
\left(\sqrt{{}^{(2)}\!h}\; h^{ab} \gamma_{,a}\right)' &= \sqrt{{}^{(2)}\!h}\; h^{ab} \gamma'_{,a},\label{eq:429}\\
\begin{split}
\left(\sqrt{{}^{(2)}\!h}\; h^{ab} e^{-2\gamma} \lambda_{,a}\right)' &= \sqrt{{}^{(2)}\!h}\; h^{ab} e^{-2\gamma} \lambda'_{,a}\\
& \hphantom{=}{}  -2 \sqrt{{}^{(2)}\!h}\; h^{ab} e^{-2\gamma} \gamma'\lambda_{,a},
\end{split}\label{eq:430}\\
\intertext{etc.}\nonumber
\end{align}
From the explicit formulas for the (background) metric functions (\ref{eq:a40}), (\ref{eq:a43}), (\ref{eq:a45}) and (\ref{eq:a46}) the asymptotic forms for the (pure gauge) perturbations, (\ref{eq:f17})--(\ref{eq:f18}), and the boundary conditions for \({}^{(2)}\!Y^a \frac{\partial}{\partial x^a}\) which yield
\begin{equation}\label{eq:431}
{}^{(2)}\!Y^R (R_+, \theta) = 0, \qquad {}^{(2)}\!Y^\theta_{\hphantom{\theta},R} (R_+, \theta) = 0
\end{equation}
we find that each of \(\left\lbrace\gamma', \lambda', \eta', \omega'\right\rbrace\) has a smooth limit as \(R \rightarrow R_+\) whereas their first radial derivatives \(\left\lbrace\gamma'_{,R}, \lambda'_{,R}, \eta'_{,R}, \omega'_{,R}\right\rbrace\) are all vanishing in this limit.\footnote{Note that these results allow for a symmetrical extension of the perturbations through the background spacetime's bifurcation 2-sphere.\label{note01}}
One also finds that, in the asymptotic regions  near \(R = R_+\) and \(R \rightarrow \infty\), one has \(\left\lbrace\lambda', \lambda'_{,\theta}, \eta', \eta'_{,\theta}, \omega', \omega'_{,\theta}\right\rbrace\) vanishing at the axes \(\theta \rightarrow 0,\pi\) whereas \(\gamma'\) is regular in this limit with \(\gamma'_{,\theta} \rightarrow 0\) at \(\theta = 0,\pi\).

Given these results it is straightforward to verify that each of the `dynamical' flux terms vanishes, pointwise, at the horizon boundary corresponding to \(R \rightarrow R_+\). One might still wonder whether the factors of \(\frac{1}{\sin^2{\theta}}\) and \(\frac{1}{\sin^4{\theta}}\) occurring (respectively) in the coefficients \(e^{-2\gamma}\) and \(e^{-4\gamma}\) induce some irregularity at the axes but it is not difficult to verify that such potential singularities are in fact cancelled by the rapidly vanishing angular dependences of \(\left\lbrace\lambda, \lambda', \eta, \eta', \omega, \omega'\right\rbrace\) as \(\theta \rightarrow 0, \pi\).

Turning to the behavior at the outer boundary one finds that the linearized momentum factors decay, in the asymptotic region as \(R \rightarrow \infty\), according to
\begin{align}
\frac{\tilde{N}\tilde{p}'}{\sqrt{{}^{(2)}\!\tilde{g}}} &\longrightarrow O \left(\frac{1}{R^4}\right),\label{eq:432}\\
\frac{\tilde{N}e^{2\gamma}\tilde{v}'}{\sqrt{{}^{(2)}\!\tilde{g}}} &\longrightarrow O \left(\frac{1}{R^5}\right) \sin^2{\theta},\label{eq:433}\\
\frac{\tilde{N}e^{2\gamma}(\tilde{u}' - \lambda\tilde{r}')}{\sqrt{{}^{(2)}\!\tilde{g}}} &\longrightarrow O \left(\frac{1}{R^4}\right) \sin^2{\theta},\label{eq:434}\\
\frac{\tilde{N}e^{4\gamma}\tilde{r}'}{\sqrt{{}^{(2)}\!\tilde{g}}} &\longrightarrow O \left(\frac{1}{R^4}\right) \sin^4{\theta}.\label{eq:435}
\end{align}
As noted earlier these are each multiplied by an `additional' factor of \(\tilde{N} = R \sin{\theta} \left(1 - \frac{R^2_+}{R^2}\right)\) but then paired with (the radial components) of terms of the form (\ref{eq:429})--(\ref{eq:430}), etc. to determine the radial flux integrands as \(R \rightarrow \infty\).

The (pure gauge) metric and wavemap perturbations, together with their needed radial derivatives, behave asymptotically as
\begin{align}
\gamma' &\longrightarrow \frac{b^{(-)}_2 + Mb^{(-)}_1 \cos{(\theta)}}{R^2} + O \left(\frac{1}{R^3}\right) \label{eq:436}\\
\gamma'_{,R} &\longrightarrow \frac{-2 \left( b^{(-)}_2 + Mb^{(-)}_1 \cos{(\theta)}\right)}{R^3} + O \left(\frac{1}{R^4}\right)\label{eq:437}\\
\lambda' &\longrightarrow \sin^2{(\theta)} \left\lbrace\frac{6Qa b^{(-)}_1 \cos{(\theta)}}{R^2} + O \left(\frac{1}{R^3}\right)\right\rbrace\label{eq:438}\\
\lambda'_{,R} &\longrightarrow \sin^2{(\theta)} \left\lbrace\frac{-12 Qab^{(-)}_1 \cos{(\theta)}}{R^3} + O \left(\frac{1}{R^4}\right)\right\rbrace\label{eq:439}\\
\eta' &\longrightarrow \sin^2{(\theta)} \left\lbrace\frac{2Qb^{(-)}_1}{R} + O \left(\frac{1}{R^2}\right)\right\rbrace\label{eq:440}\\
\eta'_{,R} &\longrightarrow \sin^2{(\theta)} \left\lbrace\frac{-2Qb^{(-)}_1}{R^2} + O \left(\frac{1}{R^3}\right)\right\rbrace\label{eq:441}\\
(\omega' + \lambda\eta') &\longrightarrow \sin^4{(\theta)} \left\lbrace\frac{-6Ma b^{(-)}_1}{R} + O \left(\frac{1}{R^2}\right)\right\rbrace\label{eq:442}\\
\omega'_{,R} &\longrightarrow \sin^4{(\theta)} \left\lbrace\frac{6Ma b^{(-)}_1}{R^2} +  O \left(\frac{1}{R^3}\right)\right\rbrace\label{eq:443}
\end{align}

Combining these results one finds that the `dynamical' boundary flux terms have the asymptotic decay properties
\begin{align}
\frac{\tilde{N}^2}{\sqrt{{}^{(2)}\!\tilde{g}}} \tilde{p}' \left(\sqrt{{}^{(2)}\!h}\; h^{Ra} \gamma_{,a}\right)' &\longrightarrow O \left(\frac{1}{R^5}\right) \sin{\theta},\label{eq:444}\\
\frac{\tilde{N}^2}{\sqrt{{}^{(2)}\!\tilde{g}}} e^{2\gamma} (\tilde{u}' - \lambda\tilde{r}') \left(\sqrt{{}^{(2)}\!h}\; h^{Ra} e^{-2\gamma} \eta_{,a}\right)' &\longrightarrow O \left(\frac{1}{R^6}\right) \sin^3{\theta},\label{eq:445}\\
\frac{\tilde{N}^2 e^{2\gamma}}{\sqrt{{}^{(2)}\!\tilde{g}}} \tilde{v}' \left(\sqrt{{}^{(2)}\!h}\; h^{Ra} e^{-2\gamma} \lambda_{,a}\right)' &\longrightarrow O \left(\frac{1}{R^8}\right) \sin^3{\theta} \cos{\theta},\label{eq:446}\\
\frac{\tilde{N}^2 e^{4\gamma} \tilde{r}'}{\sqrt{{}^{(2)}\!\tilde{g}}} \left(e^{-4\gamma} \sqrt{{}^{(2)}\!h}\; h^{Ra} (\omega_{,a} + \lambda\eta_{,a})\right)' &\longrightarrow O \left(\frac{1}{R^8}\right) \sin^5{\theta},\label{eq:447}\\
\frac{\tilde{N}^2 e^{2\gamma} (\tilde{u}' - \lambda\tilde{r}')}{\sqrt{{}^{(2)}\!\tilde{g}}} \lambda' \left(\sqrt{{}^{(2)}\!h}\; h^{Ra} e^{-4\gamma} (\omega_{,a} + \lambda\eta_{,a})\right) &\longrightarrow O \left(\frac{1}{R^{11}}\right) \sin^5{\theta} \cos^2{\theta}.\label{eq:448}
\end{align}
These rapid rates of decay, which clearly yield pointwise vanishing flux contributions at the outer boundary, reflect the fact that the perturbations are \textit{pure gauge} in the asymptotic region \(R \rightarrow \infty\).

Utilizing the pointwise decay rates of the perturbative quantities \(\lbrace\gamma', \eta', \lambda', \omega' + \lambda\eta', \omega'\rbrace\) at the boundaries \(R \searrow R_+\) and \(R \nearrow \infty\) uncovered in this section it is straightforward to verify that all of the corresponding boundary integrals potentially distinguishing \(E^{\mathrm{Alt}}\) from \(E^{\mathrm{Reg}}\) actually \textit{vanish} for the class of (asymptotically pure gauge) perturbations considered here. This result follows from a detailed evaluation of the asymptotic behavior (indeed rapid decay) of the coefficients of the perturbative expressions \(\left\lbrace (\eta')^2 + (\lambda')^2, (\omega' + \lambda\eta')^2, \cdots\right\rbrace\) occurring in the associated flux integrals.

Taking into account the regularity at the axes (corresponding to \(\theta \rightarrow 0, \pi\)) of the perturbative quantities \(\lbrace\gamma', \eta', \lambda', \omega' + \lambda\eta', \omega'\rbrace\) developed in detail below in Section~(\ref{subsec:axis-regularity}) and evaluating the corresponding behavior of their coefficients in the flux expressions that potentially distinguish \(E^{\mathrm{Alt}}\) from \(E^{\mathrm{Reg}}\) it is straightforward to verify that these integrals also vanish for the class of perturbations considered herein. Thus, for the class of (asymptotically pure gauge) perturbations that we consider \(E^{\mathrm{Alt}} = E^{\mathrm{Reg}}\). 
\subsection{Evaluating the `Kinematical' Boundary Flux Terms}
\label{subsec:kinematical}
Consider next the boundary flux contributions from Eq.~(\ref{eq:401}) that are each linear in the Lie derivative of a vector density \(\tilde{\mathcal{V}} = \mathcal{V}^a \frac{\partial}{\partial x^a}\) taken from the list
\begin{align}
\tilde{\mathcal{V}}_I &:= 4\tilde{N} \sqrt{{}^{(2)}\!h} h^{ac} \gamma_{,c} \frac{\partial}{\partial x^a},\label{eq:449}\\
\tilde{\mathcal{V}}_{II} &:= \tilde{N} \sqrt{{}^{(2)}\!h} h^{ac} e^{-4\gamma} (\omega_{,c} + \lambda\eta_{,c}) \frac{\partial}{\partial x^a},\label{eq:450}\\
\tilde{\mathcal{V}}_{III} &:= \tilde{N} \sqrt{{}^{(2)}\!h} h^{ac} e^{-2\gamma} \lambda_{,c} \frac{\partial}{\partial x^a},\label{eq:451}\\
\tilde{\mathcal{V}}_{IV} &:= \tilde{N} \sqrt{{}^{(2)}\!h} h^{ac} e^{-2\gamma} \eta_{,c} \frac{\partial}{\partial x^a},\label{eq:452}\\
\begin{split}
\tilde{\mathcal{V}}_V &:= \lambda\tilde{\mathcal{V}}_{II}\\
 &= \frac{2Qra\sin^2{\theta}}{(r^2 + a^2 \cos^2{\theta})} \tilde{\mathcal{V}}_{II},
\end{split}\label{eq:453}
\end{align}
with respect to the linearized shift vector field \(X' := \tilde{N}^{c'} \frac{\partial}{\partial x^c}\) given, in the chosen gauge, by Eq.~(\ref{eq:f24}). In the asymptotic regions where \({}^{(4)}\!k_{\alpha\beta}\) vanishes this expression reduces to
\begin{equation}\label{eq:454}
\begin{split}
X' &= \tilde{N}^{a'} \frac{\partial}{\partial x^a} = \left({}^{(4)}\!Y_{,0}^a - \tilde{N}^2 \tilde{g}^{ab} {}^{(4)}\!Y_{,b}^0\right) \frac{\partial}{\partial x^a}\\
&= \left({}^{(2)}\!Y_{,0}^a - {}^{(2)}\!\mathcal{D}^a\right) \frac{\partial}{\partial x^a}
\end{split}
\end{equation}
where the explicit formulas for \({}^{(4)}\!Y^a = {}^{(2)}\!Y^a\) and \({}^{(2)}\!\mathcal{D}^a\) are given by Eqs.~(\ref{eq:407}--\ref{eq:410}) and (\ref{eq:419}--\ref{eq:424}).

Recalling that, for a vector \textit{density} \(\tilde{\mathcal{V}} = \mathcal{V}^a \frac{\partial}{\partial x^a}\), one has
\begin{equation}\label{eq:455}
(\mathcal{L}_{X'} \tilde{\mathcal{V}})^a = (X'^c \mathcal{V}^a)_{,c} - X'^a_{\hphantom{a},c} \mathcal{V}^c
\end{equation}
it is straightforward to compute \(\mathcal{L}_{X'} \tilde{\mathcal{V}}_I, \cdots \mathcal{L}_{X'} \tilde{\mathcal{V}}_V\) and to pair these with their associated factors taken from the list \(\lbrace\gamma', \omega', \lambda', \eta'\rbrace\). From the regularity of these latter quantities at the horizon and the readily verified vanishing of each of the radial components \(\left\lbrace (\mathcal{L}_{X'} \tilde{\mathcal{V}}_I)^R, \cdots (\mathcal{L}_{X'} \tilde{\mathcal{V}}_V)^R\right\rbrace\) at this inner boundary it follows that the flux contributions of these Lie derivative terms are each (pointwise) vanishing at the horizon.\footnote{\label{note02}We are assuming that \({}^{(4)}\!Y^0|_{t=t_0}\) and \(\delta\tilde{N} := \tilde{N}'\) have been chosen so that the one-form components \({}^{(4)}\!Y_{\hphantom{0},a}^0|_{t=t_0} + \int_{t_0}^t dt' \left(\frac{\delta\tilde{N}}{\tilde{N}}\right)_{,a}\) and their \(\theta\) derivatives are each non-singular at the horizon.}

Recalling Eqs.~(\ref{eq:436}), (\ref{eq:438}), (\ref{eq:440}) and (\ref{eq:442}) we see that each of \(\lbrace\gamma', \omega', \lambda', \eta'\rbrace\) decays of order \(O(1/R)\) or faster as \(R \rightarrow \infty\). It follows that the corresponding (Lie derivative) flux integral expressions will make no contributions at the outer boundary provided that the associated radial components \(\left\lbrace (\mathcal{L}_{X'} \tilde{\mathcal{V}}_I)^R , (\mathcal{L}_{X'} \tilde{\mathcal{V}}_{II})^R\right.\), \((\mathcal{L}_{X'} \tilde{\mathcal{V}}_{III})^R\), \((\mathcal{L}_{X'} \tilde{\mathcal{V}}_{IV})^R\), \(\left.(\mathcal{L}_{X'} \tilde{\mathcal{V}}_V)^R\right\rbrace\) are all bounded as \(R \rightarrow \infty\). A straightforward computation of these quantities shows that this is indeed the case provided that \({}^{(4)}\!Y^0\) and \(\delta\tilde{N}\) are chosen so that the one-form components \({}^{(4)}\!Y^0_{,a}|_{t=t_0} + \int_{t_0}^t dt' \left(\frac{\delta\tilde{N}}{\tilde{N}}\right)_{,a}\) and their \(\theta\)-derivatives are sufficiently regular in this limit. Since we shall eventually take \(\delta\tilde{N} = 0\) and since \({}^{(4)}\!Y^0|_{t=t_0}\) is at our discretion, this latter condition is trivial to arrange. 
\subsection{Evaluating the `Conformal' Boundary Flux Terms}
\label{subsec:evaluation-conformal}
Consider next the boundary flux contributions from Eq.~(\ref{eq:401}) that are each linear in the perturbed conformal factor \(\nu'\). These arise from the integrated divergence of the vector density
\begin{equation}\label{eq:456}
\begin{split}
Q^b \frac{\partial}{\partial x^b} &:= \left\lbrace 2 (\mathcal{L}_{X'} \tilde{N}) (\sqrt{{}^{(2)}h}\; h^{ba}\nu'_{,a})\right.\\
& \left.{} + 2 (\mathcal{L}_{X'} \nu') \sqrt{{}^{(2)}h}\; h^{ba} \tilde{N}_{,a} - 2X^{\prime b} (\sqrt{{}^{(2)}h}\; h^{ac} \nu'_{,a} \tilde{N}_{,c})\right\rbrace \frac{\partial}{\partial x^b}.
\end{split}
\end{equation}
One might first think to identify \(\nu'\), in the asymptotic regions where the `background perturbations' \({}^{(4)}k_{\alpha\beta}\) and \({}^{(4)}\ell_\alpha\) both vanish, with the `pure gauge perturbation', \({}^{(2)}Y^a \nu_{,a}\), of the unperturbed conformal factor \(\nu\). Indeed one can, not surprisingly, verify directly that this choice combined with the complementary pure gauge perturbations\hfill\newline \(\left\lbrace {}^{(2)}Y^a \gamma_{,a}, {}^{(2)}Y^a \omega_{,a}, \cdots\right\rbrace\) satisfies all of the linearized constraint equations. As we shall see though this choice would leave an uncancelled flux contribution at the horizon boundary corresponding to \(R \searrow R_+\) and even to uncancelled, regularity violating flux terms at the symmetry axes corresponding to \(\theta = 0,\pi\).

The subtlety here is that the supposedly defining equation for the conformal factor \(\nu\) (at the fully nonlinear level) was the decomposition of the Riemannian metric \({}^{(2)}\tilde{g}\), via \(\tilde{g}_{ab} = e^{2\nu}h_{ab}\), into a conformal factor \(e^{2\nu}\) and a `conformal metric' \(h_{ab}\) that was required to be \textit{flat} (c.f., the discussion in Appendix~\ref{app:gauge-conditions}). But, thanks to a well-known conformal identity, valid in 2-dimensions, any metric of the form \(h_{ab}^\lambda = e^{2\lambda}h_{ab}\) conformal to a flat metric \(h_{ab}\) will still be flat if and only if the function \(\lambda\) is \textit{harmonic} (with respect to \(h_{ab}\) or, equivalently, to any metric conformal thereto). In other words the decomposition recalled above does not \textit{uniquely} determine \(\nu\) (and therefore also \(h_{ab}\)) at the nonlinear level and, of course, therefore also at the corresponding linearized level. Indeed, as one can easily see from the explicit form of the linearized Hamiltonian constraint (\ref{eq:c07}), \(\nu'\) is only determined, without further information, up to a harmonic function of the metric \(\tilde{g}_{ab}\) (or, equivalently, of any metric conformal thereto).

As a special case of the above recall that a \textit{pure gauge} transformation of the flat-metric \(h_{ab}\) generated by an \textit{analytic} change of coordinates (i.e., coordinates satisfying the Cauchy Riemann equations) automatically preserves \(h_{ab}\) up to a conformal factor of the above type (i.e., a factor \(e^{2\lambda}\) with \(\lambda\) in fact harmonic). But such a conformal transformation to \(h_{ab}\) can, \textit{by convention}, be absorbed unambiguously into an inhomogeneous transformation of the logarithm \(\nu\) occurring in the `unified' expression for the `physical' 2-metric \(\tilde{g}_{ab} = e^{2\nu} h_{ab}\) wherein \(h_{ab}\), by conventional fiat, remains fixed while \(\nu\) picks up an (additive) non-tensorial complement.

At the linearized level, wherein a pure gauge perturbation of \(\sqrt{{}^{(2)}\tilde{g}}\) has the (unambiguous) form \(\mathcal{L}_{{}^{(2)}Y} \sqrt{{}^{(2)}\tilde{g}}\), with
\begin{equation}\label{eq:497}
\begin{split}
\mathcal{L}_{{}^{(2)}Y} \sqrt{{}^{(2)}\tilde{g}} &= \left({}^{(2)}Y^a \sqrt{{}^{(2)}\tilde{g}}\right)_{,a} = \left({}^{(2)}Y^a e^{2\nu} \sqrt{{}^{(2)}h}\right)_{,a}\\
& = e^{2\nu} \left({}^{(2)}Y^a\; \sqrt{{}^{(2)}h}\right)_{,a} {} + 2e^{2\nu} {}^{(2)}Y^a \nu_{,a} \sqrt{{}^{(2)}h},
\end{split}
\end{equation}
but where (by the convention of holding \(\sqrt{{}^{(2)}h}\) fixed) we regard this as a (pure gauge) variation of \(\sqrt{{}^{(2)}\tilde{g}}\) of the form
\begin{equation}\label{eq:498}
\mathcal{L}_{{}^{(2)}Y} \sqrt{{}^{(2)}\tilde{g}} = \left.\delta \sqrt{{}^{(2)}\tilde{g}} = 2e^{2\nu} \sqrt{{}^{(2)}h}\; \delta\nu\right|_{\mathrm{pure~ gauge}}
\end{equation}
it follows that this pure gauge perturbation, \(\left.\nu'\right|_{\mathrm{gauge}}\), of \(\nu\) takes the form
\begin{equation}\label{eq:499}
\left.\nu'\right|_{\mathrm{gauge}} = \left.\delta\nu\right|_{\mathrm{pure~ gauge}} = {}^{(2)}Y^a \nu_{,a} + \frac{1}{2} \frac{1}{\sqrt{{}^{(2)}h}} \partial_a \left(\sqrt{{}^{(2)}h} {}^{(2)}Y^a\right).
\end{equation}
In the asymptotic regions near \(R_+\) and \(\infty\) where (as discussed in Appendix~\ref{app:transforming}) \({}^{(2)}Y\) is a conformal Killing field of the flat-metric \(h_{ab}\) it is straightforward to verify that the supplementary, `correction' term
\begin{equation}\label{eq:500}
\frac{1}{2} \frac{1}{\sqrt{{}^{(2)}h}} \partial_a \left(\sqrt{{}^{(2)}h} {}^{(2)}Y^a\right) = \frac{1}{2} {}^{(2)}\nabla_a (h) {}^{(2)}Y^a
\end{equation}
is indeed harmonic with respect to \(h_{ab}\), i.e., to check that
\begin{equation}\label{eq:501}
{}^{(2)}\nabla_c (h) {}^{(2)}\nabla^c (h) \left\lbrack\frac{1}{2} {}^{(2)}\nabla_a (h) {}^{(2)}Y^a\right\rbrack = 0.
\end{equation}
It follows therefrom that satisfaction of the linearized field equations, in particular the (linearized) Hamiltonian constraint, is not disturbed by the inclusion of this correction to \(\nu'\).

The integrals with respect to \(\theta\) of the radial component of \(Q^b \frac{\partial}{\partial x^b}\), namely
\begin{equation}\label{eq:502}
\begin{split}
Q^R &= 2\nu'_{,\theta} \left\lbrack X^{\prime\theta} R \sin{\theta} \left( 1 + \frac{R_+^2}{R^2}\right)\right.\\
 &\hphantom{=} \left. {} - X^{\prime R} \cos{\theta} \left( 1 - \frac{R_+^2}{R^2}\right)\right\rbrack\\
 & {} + 2 \nu'_{,R} \left\lbrack X^{\prime R} \sin{\theta}\; R \left( 1 + \frac{R_+^2}{R^2}\right)\right.\\
 &\hphantom{=} \left. {} + X^{\prime\theta} \cos{\theta}\; R^2 \left( 1 - \frac{R_+^2}{R^2}\right)\right\rbrack,
\end{split}
\end{equation}
evaluated in the limits \(R\nearrow\infty\) and \(R\searrow R_+\) yield the potential flux contributions at these boundaries for the given, specific (`corrected', pure gauge) choice for \(\nu'\)
\begin{equation}\label{eq:503}
\nu' \rightarrow {}^{(2)}Y^a \nu_{,a} + \frac{1}{2} {}^{(2)}\nabla_a (h) {}^{(2)}Y^a.
\end{equation}

By using only the basic Green's function asymptotics for the vector fields \({}^{(2)}Y\) and \({}^{(2)}\mathcal{Y}\) characterized in Appendix~\ref{app:transforming} one finds that \(Q^R\) vanishes (pointwise) at least of order \(O\left(\frac{1}{R}\right)\) as \(R\nearrow\infty\). Using however the more immediately detailed Fourier expansion formulas for these quantities (c.f., Eqs.~(\ref{eq:407})--(\ref{eq:411})) one finds that, thanks to a leading order cancellation of some terms involving the Fourier coefficient \(\beta_1^{(-)}\), the actual rate of decay is \(O\left(\frac{1}{R^2}\right)\). By either reckoning the corrected, pure gauge perturbation,
\(\nu'  = {}^{(2)}Y^a \nu_{,a} + \frac{1}{2} {}^{(2)}\nabla_a (h) {}^{(2)}Y^a\), `contributes' pointwise vanishing flux at spatial infinity.

At the other limit, on the other hand, the boundary flux angular integrand reduces to
\begin{equation}\label{eq:504}
\begin{split}
Q^R \xrightarrow[R \searrow R_+]{} & \left\lbrace 4 R_+ \sin{\theta}\; \nu'_{,R}\right.\\
 & \left.\left. {} \times \left({}^{(2)}Y^R_{\hphantom{R},0} + \frac{R_+^4}{(r^2 + a^2)^2} \frac{4}{R_+} \left(\frac{{}^{(2)}\mathcal{Y}^R}{R_+}\right)\right)\right\rbrace\right|_{R \searrow R_+}.
 \end{split}
\end{equation}
Exploiting Eq.~(\ref{eq:f30}) one can show that
\begin{equation}\label{eq:505}
\begin{split}
\frac{\partial}{\partial R} & \left({}^{(2)}Y^a \nu_{,a} + \frac{1}{2} {}^{(2)}\nabla_a (h) {}^{(2)}Y^a\right) \xrightarrow[R\searrow R_+]{} \left\lbrace\frac{\partial}{\partial R}\left\lbrace {} - \frac{{}^{(2)}Y^R}{R} + {}^{(2)}Y^R \left( 1 - \frac{R_+^2}{R^2}\right)\right.\right.\\
&\hphantom{\longrightarrow} \times\left\lbrack\frac{2r(r^2+a^2) - a^2\sin^2{\theta} (r-M)}{\left\lbrack (r^2+a^2)^2 - a^2\Delta\sin^2{\theta}\right\rbrack}\right\rbrack\\
&\left.\left.\left.\hphantom{\longrightarrow} + \frac{{}^{(2)}Y^\theta\cos{\theta}}{\sin{\theta}} \frac{\left\lbrack (r^2+a^2)^2 - 2a^2\Delta\sin^2{\theta}\right\rbrack}{\left\lbrack (r^2+a^2)^2 - a^2\Delta\sin^2{\theta}\right\rbrack} + {}^{(2)}Y^R_{,R}\right\rbrace\right\rbrace\right|_{R\searrow R_+}.
\end{split}
\end{equation}
Recalling that \(\frac{\partial r}{\partial R} = 1 - \frac{R_+^2}{R^2}\) and employing Eq.~(\ref{eq:f31}) to express (in the asymptotic regions) \({}^{(2)}Y_{,R}^\theta\) via
\begin{equation}\label{eq:506}
{}^{(2)}Y_{,R}^\theta \frac{\cos{\theta}}{\sin{\theta}} \longrightarrow {} - \frac{1}{R} \left(\frac{\cos{\theta}}{\sin{\theta}} \left(\frac{{}^{(2)}Y^R}{R}\right)_{,\theta}\right)
\end{equation}
we can exploit the chosen boundary condition (c.f., Appendix~\ref{app:transforming}),
\begin{equation}\label{eq:507}
\left.{}^{(2)}Y^R\right|_{R\searrow R_+} \longrightarrow 0,
\end{equation}
to deduce that
\begin{equation}\label{eq:508}
\begin{split}
\left.\nu'_{,R}\right|_{R\searrow R_+} &= \left.\left({}^{(2)}Y^a \nu_{,a} + \frac{1}{2} {}^{(2)}\nabla_a (h) {}^{(2)}Y^a\right)_{,R}\right|_{R \searrow R_+}\\
&= \left.\frac{\partial}{\partial R} \left\lbrace R\left(\frac{{}^{(2)}Y^R}{R}\right)_{,R}\right\rbrace\right|_{R\searrow R_+}\\
&= \left.\left({}^{(2)}Y_{,\theta R}^\theta\right)\right|_{R\searrow R_+} = \left.\left(\frac{\partial}{\partial\theta} \left({}^{(2)}Y_{,R}^\theta\right)\right)\right|_{R\searrow R_+}\\
&= \left.\left( {} - \frac{1}{R} \left(\frac{{}^{(2)}Y^R}{R}\right)_{,\theta\theta}\right)\right|_{R\searrow R_+} = 0
\end{split}
\end{equation}
where we have again exploited Eqs.~(\ref{eq:f30}) and (\ref{eq:f31}) to reexpress derivatives of \({}^{(2)}Y^R\) in terms of those of \({}^{(2)}Y^\theta\) and vice versa and appealed to the boundary condition (\ref{eq:507}) in the final step. Recalling Eq.~(\ref{eq:504}) we thus see that the corresponding `conformal' boundary flux integrand \(\left. Q^R\right|_{R\searrow R_+}\) vanishes at the horizon boundary.

To give the result (\ref{eq:508}) a precise geometrical interpretation recall that, in our notation (c.f. Eqs.~(\ref{eq:111})--(\ref{eq:112}) and (\ref{eq:a13}), the `spatial' Riemannian metric induced on a \(t = \mathrm{const.}\) hypersurface is given by
\begin{equation}\label{eq:509}
\begin{split}
g_{ij}\; dx^i \otimes dx^j &= e^{-2\gamma} \tilde{g}_{ab}\; dx^a \otimes dx^b\\
 &\hphantom{=} + e^{2\gamma} (d\varphi + \beta_a dx^a) \otimes (d\varphi + \beta_b dx^b)
\end{split}
\end{equation}
where, in Weyl-Papapetrou coordinates \(\lbrace x^a\rbrace = \lbrace R,\theta\rbrace\),
\begin{equation}\label{eq:510}
\tilde{g}_{ab}\; dx^a \otimes dx^b = e^{2\nu} (dR \otimes dR + R^2 d\theta \otimes d\theta).
\end{equation}
It is straightforward to evaluate the first fundamental form, \(\mu_{AB}\; dx^A \otimes dx^B\), and second fundamental form, \(\lambda_{AB}\; dx^A \otimes dx^B\), induced thereby upon a (topologically spherical) surface \(R = R_0 = \mathrm{const.}\) and to calculate the \textit{mean curvature}, \(\tr_\mu \lambda := \mu^{AB}\lambda_{AB}\) of the latter (where \(\lbrace x^A\rbrace = \lbrace\theta, \varphi\rbrace\)). The result is
\begin{equation}\label{eq:511}
\tr_\mu \lambda = {} - e^{\gamma -\nu} \frac{1}{Re^\nu} \frac{\partial}{\partial R} (Re^\nu)
\end{equation}
which of course vanishes at the event horizon, \(R\searrow R_+\), of a Kerr-Newman black hole, the latter being a \textit{minimal surface}. Linearizing (\ref{eq:511}) about this background one finds that
\begin{equation}\label{eq:512}
(\tr_\mu \lambda)' \xrightarrow[R\searrow R_+]{} - \left.\left( e^{\gamma -\nu} \frac{\partial}{\partial R} \nu'\right)\right|_{R=R_+}.
\end{equation}
Thus the boundary condition \(\left.\nu'_{,R}\right|_{R=R_+} = 0\) corresponds precisely to preserving minimality of the surface \(R = R_+\) at the linearized level.
\subsection{Axis Regularity and Evaluation of Flux Terms at the `Artificial Boundaries'}
\label{subsec:axis-regularity}
The sections above have dealt with the evaluation of (potential) energy flux contributions at the \textit{actual} boundaries corresponding to \(R \nearrow \infty\) and \(R \searrow R_+\) and established the pointwise vanishing of these flux expressions for the boundary conditions chosen. But the full flux integral formula (\ref{eq:401}) also includes a potential contribution from the `artificial boundaries' provided by the axes of symmetry corresponding to \(\theta = 0,\pi\). Needless to say the evaluation of these potentially energy violating flux contributions hinges upon the \textit{regularity} of the various fields involved at these axes of symmetry. Since, by assumption, we begin with a globally regular solution to the linearized field equations expressed in (say) a `hyperbolic gauge' (c.f., Appendix~\ref{app:global}) and transform this solution to the desired gauge with an everywhere smooth gauge transformation (c.f.,Appendix~\ref{app:transforming}) the regularity of the resulting perturbations (as smooth tensorial fields on the spacetime manifold) is not in question.

But is this smoothness and its implicit axial regularity sufficient to ensure the vanishing of the potential flux contributions? In this section we shall verify that this is indeed the case.

An especially useful resource in this regard is the article \cite{Rinne_05} by Rinne and Stewart which derives the natural regularity conditions satisfied (at an axis of symmetry) by various smooth tensor fields (including scalar fields, vector fields, one forms and symmetric, second rank tensor fields) on a smooth, axi-symmetric spacetime under the assumption that the various `perturbations' are themselves axi-symmetric. Exploiting their  results in conjunction with our linearized field equations it is straightforward to evaluate the various `dynamical' boundary flux terms and establish their (pointwise) vanishing at the symmetry axes to the following orders:
\begin{align}
\tilde{N} \left(\frac{\tilde{N}\tilde{p}'}{\mu_{{}^{(2)}\tilde{g}}}\right) \left(\sqrt{{}^{(2)}h} h^{\theta a} \gamma_{,a}\right)' &\rightarrow O(\sin^2{\theta})\label{eq:484}\\
\tilde{N} \left(\frac{\tilde{N}e^{4\gamma}\tilde{r}'}{\mu_{{}^{(2)}\tilde{g}}}\right) \left( e^{-4\gamma} \sqrt{{}^{(2)}h} h^{\theta a} (\omega_{,a} + \lambda\eta_{,a})\right)' &\rightarrow O(\sin^2{\theta})\label{eq:485}\\
\tilde{N} \left(\frac{\tilde{N} e^{2\gamma}\tilde{v}'}{\mu_{{}^{(2)}\tilde{g}}}\right) \left(\sqrt{{}^{(2)}h} h^{\theta a} e^{-2\gamma} \lambda_{,a}\right)' &\rightarrow O(\sin^2{\theta})\label{eq:488}\\
\tilde{N} \left(\frac{\tilde{N} e^{2\gamma} (\tilde{u}' - \lambda\tilde{r}')}{\mu_{{}^{(2)}\tilde{g}}}\right) \left(\sqrt{{}^{(2)}h} h^{\theta a} e^{-2\gamma} \eta_{,a}\right)' &\rightarrow O(\sin^2{\theta})\label{eq:489}\\
\tilde{N} \left(\frac{\tilde{N} e^{2\gamma}(\tilde{u}' - \lambda\tilde{r}')}{\mu_{{}^{(2)}\tilde{g}}}\right) \lambda' \sqrt{{}^{(2)}h} h^{\theta a} e^{-4\gamma} (\omega_{,a} + \lambda\eta_{,a}) &\rightarrow O(\sin^4{\theta})\label{eq:490}
\end{align}
Thus these versions `dynamical' flux terms provide \textit{no} (energy violating) contributions at the axes of symmetry, \(\theta = 0,\pi\).

Turning to the `kinematical' flux terms involving the Lie derivatives of the vector densities (\ref{eq:449})--(\ref{eq:453}) one finds, in the analogous way, that each of the factors \(\mathcal{L}_{X'}\tilde{\mathcal{V}}_I, \cdots , \mathcal{L}_{X'}\tilde{\mathcal{V}}_V\) has a regular (but, in general non-vanishing) limit at the axes of symmetry corresponding to \(\theta = 0,\pi\). On the other hand, as discussed fully in Appendix~\ref{app:electric-charge}, each of the multiplicative factors \(\lbrace\omega', \lambda', \eta'\rbrace\) is, for the class of perturbations considered herein, required to vanish on these axes of symmetry. In fact, as smooth scalar fields, they must vanish at least of order \(O(\sin^2{\theta})\) as \(\theta\rightarrow 0, \pi\). Thus the corresponding energy flux terms vanish (pointwise) at these artificial boundaries.

The remaining factor, \(\gamma'\), however has a smooth (but in general, non-vanishing) limit as \(\theta\rightarrow 0,\pi\). The corresponding flux term has the limiting values
\begin{equation}\label{eq:491}
\left.\left(\gamma'\mathcal{L}_{X'}, \tilde{\mathcal{V}}_I\right)_{\theta\rightarrow 0,\pi}^\theta \rightarrow \left\lbrace 4\gamma'\left( X'^R \cos{\theta} \left( 1 - \frac{R_+^2}{R^2}\right)_{,R}\right)\right\rbrace\right|_{\theta = 0,\pi}
\end{equation}
at the respective axes \(\theta = 0,\pi\) over which it is to be integrated from \(R_+\) to \(\infty\).

Fortunately, however, this (in general non-vanishing) net flux contribution combines naturally with the remaining `conformal' flux contribution comprised of the integrals (over the two axes) of (c.f.~Eq.~(\ref{eq:456}))
\begin{equation}\label{eq:492}
\begin{split}
Q_{\theta\rightarrow 0,\pi}^\theta &\rightarrow \left.\left\lbrace 2\nu'_{,R} X'^R \cos{\theta} \left( 1 - \frac{R_+^2}{R^2}\right)\right\rbrace\right|_{\theta = 0,\pi}\\
 &= \left\lbrace\left( 2\nu' X'^R \cos{\theta} \left( 1 - \frac{R_+^2}{R^2}\right)\right)_{,R}\right.\\
 &\hphantom{=} \left.\left.{} - 2\nu' \left( X'^R \cos{\theta} \left( 1 - \frac{R_+^2}{R^2}\right)\right)_{,R}\right\rbrace\right|_{\theta = 0,\pi}.
\end{split}
\end{equation}
The integrals with respect to \textit{R} of the total derivative terms (evaluated at \(\theta = 0,\pi\)) from \(R_+\) to \(\infty\) are readily shown to vanish by virtue of the limiting behavior of the factor \(\left( 2\nu' X'^R\right)\) along the axes, namely
\begin{equation}
\left.\left( 2\nu' X'^R\right)\right|_{\theta = 0,\pi} \xrightarrow[R\nearrow\infty]{} O\left(\frac{1}{R^2}\right)
\end{equation}
with this same quantity vanishing in the limit as \(R \searrow R_+\). In deriving this result one needs to exploit Eq.~(\ref{eq:f30}) together with L'Hospital's  rule to show that, in the asymptotic regions along the axes, one has
\begin{align}
\nu' &\xrightarrow[\hphantom{\theta\to 0,\pi}]{} {}^{(2)}Y^a \nu_{,a} + \frac{1}{2} {}^{(2)}\nabla_a (h) {}^{(2)}Y^a\label{eq:493}\\
&\xrightarrow[\theta\to 0,\pi]{} \left\lbrace 2 {}^{(2)}Y^R_{\hphantom{R},R} - \frac{2}{R} {}^{(2)}Y^R\right.\nonumber\\
&\hphantom{\rightarrow}\left.\left.{} + {}^{(2)}Y^R \left( 1 - \frac{R_+^2}{R^2}\right) \left\lbrack\frac{2r}{r^2 + a^2}\right\rbrack\right\rbrace\right|_{\theta = 0,\pi}.\label{eq:494}
\end{align}
The vanishing at \(R_+\) follows from the vanishing of \(\left(\left. X'^R\right|_{\theta = 0,\pi}\right)\) there together with the regularity of \(\nu'\) in this limit. Note also the additional factor of \(\left( 1 - \frac{R_+^2}{R^2}\right)\) in the resulting  `end point' expression.

It follows from the above that the full \textit{integrated} flux expression will vanish provided that
\begin{equation}\label{eq:495}
\nu' = 2\gamma'
\end{equation}
\textit{along the axes of symmetry}. One can, however, again appeal to the Rinne/Stewart results \cite{Rinne_05} to verify that (expressed in our notation)
\begin{equation}\label{eq:496}
\left.\nu'\right|_{\theta = 0,\pi} = \left. 2\gamma'\right|_{\theta = 0,\pi}
\end{equation}
for any regular (axi-symmetric) metric perturbation. This equivalence can also be checked explicitly in the asymptotic regions (along the axes) where the perturbations are \textit{pure gauge}.

It then follows that we have proven:
\begin{theorem}\label{thm:02}
For the class of axisymmetric, asymptotically-pure-gauge Kerr-Newman perturbations considered herein (c.f., Appendices \ref{app:transforming} and \ref{app:complactly-supported-solutions}) the positive-definite energy functional defined by Eqs.~(\ref{eq:216})--(\ref{eq:217}) is strictly conserved when the Weyl-Papapetrou and 2+1-dimensional maximal slicing gauge conditions are imposed.
\end{theorem}

Remarks: It is somewhat curious to note that the ultimate vanishing of the net (potentially energy conservation violating) flux terms along the artificial boundaries provided by the axes of symmetry is obtained only \textit{after} the `integration by-parts' procedure outlined above is carried out. Note also that the additional boundary flux terms from Eq.~(\ref{eq:401}) that are linear in the perturbed lapse function \(\tilde{N}'\), vanish identically in our (2+1 --dimensional maximal slicing) gauge for which \(\tilde{N}' \equiv 0\). Such terms would need to be considered, however, in alternative gauges for which \(\tilde{N}'\) is non-vanishing. Perhaps the most interesting such choice is the \textit{3+1 --dimensional maximal slicing} gauge which is discussed in some detail in Appendix~\ref{app:maximal-slicing}. In particular we show therein that these additional (potentially energy conservation violating) boundary flux terms do indeed vanish as desired. While we do not actually prove the existence of this gauge for our problem (the necessary elliptic theory being rather technically involved) we thus nevertheless show that, if this gauge does indeed exist (as is most plausible), then our energy functional continues to be conserved upon employing it.

It is clear from the form of Eq.~(\ref{eq:401}) though that conservation of the energy depends only upon \(\tilde{N}'\) through the vanishing of its (potential) boundary flux contributions and not upon the behavior of this quantity in the DOC's `interior'. Thus \textit{any choice} of linearized time guage which secures the vanishing of these boundary flux terms would yield a corresponding conservation result. This is the essential \textit{gauge invariance} of our energy expression alluded to previously.  %equations starting 401
\section{Summary, Concluding Remarks, and Outlook}
\label{sec:conclusion}
The mathematical problem of stability of black hole spacetimes is the subject of a long standing research program that dates back to the 1960s. Historically, an essential first step was to study the stability of such spacetimes with respect to linear scalar wave, Maxwell and linearized Einstein perturbations. To establish the stability of such black hole `backgrounds' it is necessary to verify the boundedness and decay of the perturbations.

Arguably, the most important obstacle to controlling the perturbations of \textit{rotating} black holes is the fact that the energy of even linear waves propagating in such a spacetime is not necessarily positive-definite due to the ergo-region that always surrounds a black hole with non-vanishing angular momentum. This issue, which has both technical and physical ramifications, limits the immediate use of standard techniques for proving the decay of waves.

From a mathematical perspective one should recall that the energy of the waves being not necessarily positive-definite is a consequence of the fact that the Killing vector \(\partial_t\) is not globally timelike throughout a (rotating) black hole's DOC, becoming instead spacelike within its ergo-region. An example of this phenomenon can be seen in the linear, scalar wave perturbations of Kerr black holes which fail to admit a conserved and positive-definite energy. In the special case of axially symmetric scalar wave perturbations, however, this problem evaporates since the troublesome, indefinite term in the energy density actually vanishes but the problem reappears for both axially symmetric Maxwell and linearized Einstein waves.\footnote{Indeed, as shown in Section \ref{sec:pure-electromagnetic-kerr-spacetimes}, the conventional local energy density for axisymmetric Maxwell fields can be negative inside the ergo-region.\label{note07}}

Indeed, the lack of a positive-definite energy and the related so-called `super-radiance effect' could in principle allow the perturbations to blow up exponentially, even in the axisymmetric case \cite{WP_13}. A common technique to exclude this possibility is to introduce a `blended' vector field \(T_\chi\) such as
\begin{equation}\label{601}
T_\chi = \partial_t + \chi\partial_\varphi
\end{equation}
where \(\chi\) is a suitable `cutoff function' chosen so that \(T_\chi\) is globally timelike and the corresponding energy is positive-definite. However, since this energy is not in general conserved, suitable Morawetz-type spacetime integral estimates are needed to establish its boundedness and for the cases of Maxwell and linearized Einstein waves propagating on Kerr backgrounds these techniques seem to be currently limited to the treatment of \textit{small}, subextremal angular momentum, \(|a| \ll M\) and little is known about the stability of Kerr or Kerr-Newman spacetimes with respect to Maxwell and linearized Einstein perturbations in the case of \textit{arbitrary} (subextremal) angular momentum, \(|a| < M\).

In this work, using Hamiltonian methods, we establish the existence of a \textit{conserved} and \textit{positive-definite} total energy for the fully coupled, axially symmetric Einstein-Maxwell perturbations of Kerr-Newman spacetimes for the entire subextremal range (\(|a|, |Q| < M, a^2 + Q^2 < M^2\)). Our proof of energy conservation has necessitated a demonstration that a plethora of (potentially conservation violating) boundary flux terms actually all vanish. This argument was quite intricate in view of the elliptic nature of our chosen (Weyl-Papapetrou) gauge conditions which, in turn, were needed for the employment of the famous Carter-Robinson identities in their traditional form. These identities were needed to transform our energy expression into its desired positive-definite form.

Our use of the Carter-Robinson identities exploits, of course, the wave map structure resulting, in a well-known way, from the dimensional reduction of the Einstein-Maxwell equations with one rotational isometry. The general methods developed herein can be used to study the stability of a variety of black hole spacetimes which exhibit analogous wave map structure. In \cite{NG_2018} for example, it was shown that Lorentzian Einstein manifolds (i.e., those satisfying the Einstein equations with a non-vanishing cosmological constant) with one rotational isometry admit a Lagrangian dimensional reduction to the (2+1-dimensional) Einstein equations coupled to a `modified' wave map system wherein the traditional wave map is \textit{shifted} by a term in the cosmological constant. A crucial observation in this work was that the cosmological constant effectively decouples in such a way that it acts as a `source term' for the wave map without destroying its essential geometric structure.

Another application of the ideas developed herein is that one can use them to derive, for the axisymmetric, purely Maxwellian perturbations of a Kerr spacetime, a conserved, positive-definite energy functional expressible, albeit nonlocally, in terms of the Newman-Penrose scalars for the Maxwell field \cite{NG_2019}. By contrast the conventional energy expression for these quantities, while local, fails to have the corresponding positivity. A first step towards extending this result to deal with the \textit{gravitational} perturbations of Kerr black holes is presented in Appendix \ref{app:weyl-tensor} wherein the Weyl tensor for vacuum axisymmetric spacetimes is expressed in terms of the wave map and 2+1-dimensional metric variables. For this special case of (axisymmetric, gravitational) Kerr perturbations one of us (N.G.) has shown how to correlate positive-definiteness of the perturbative energy to the negativity of the curvature of the corresponding wave map target space (hyperbolic 2-space) \cite{NG_19a}. This argument is naturally covariant with respect to the target space geometry.

As is well-known, for sufficiently smooth but non-stationary solutions to the linearized equations for a stationary background, one can derive a sequence of new solutions to the same equations by sequentially Lie differentiating a given solution with respect to the (asymptotically timelike) Killing field of the background. In standard coordinates adapted to the stationarity of the background, wherein the relevant Killing field, \(\zeta\), takes the form \(\zeta = \partial_t\), this simply amounts to time differentiating the chosen, linearized solution as many times as its smoothness allows. At each stage of this procedure one can apply the linearized field equations themselves to reexpress time derivatives in terms of spatial ones, thus generating a family of solutions to the linearized equations built from sequentially higher order spatial derivatives of the initial one.

For the Kerr-Newman problem in particular one can thus derive a sequence of higher order (conserved, positive definite) energy expressions which, combined with standard Sobolev inequalities, could, in principle, be exploited to derive corresponding uniform bounds on the perturbations.

A well-known complication in this procedure, however, is the sequential occurrence, in each of these higher order energy expressions, of certain `weight factors' arising from the background spacetime's (2+1-dimensional) lapse function, \(\tilde{N}\),
\begin{equation}\label{eq:602}
\tilde{N} = R \sin{\theta} \left( 1 - \frac{R_+^2}{R^2}\right)
\end{equation}
which vanishes at the black hole's horizon (\(R \searrow R_+\)) and at the axes of symmetry (\(\theta \rightarrow 0,\pi\)) and which blows up (linearly) at spatial infinity (\(R \nearrow \infty\)). One can see this phenomenon occurring already at the lowest order wherein the formula (\ref{eq:216}) for \(\mathcal{E}^{\mathrm{Reg}}\) has an overall, multiplicative factor of \(\tilde{N}\). New such factors arise from each sequential time differentiation of the chosen perturbation when one applies the linearized field equations to replace time derivatives with spatial ones.

Fortunately, however, the associated, so-called \textit{redshift effect} arising from the vanishing of \(\tilde{N}\) at the black hole's horizon is a familiar one and has been analyzed in other, `model' stability problems. Even so the use of Sobolev inequalities for the extraction of optimal uniform bounds on the perturbations from the higher order energy expressions is a technically intricate problem which we shall not pursue here. It is worth remarking though that, since the particular class of perturbations that we consider is, by construction, \textit{pure gauge} in the asymptotic regions near the horizon and `near' infinity, not to mention constructively regular at the axes of symmetry, the behavior of these perturbations in these asymptotic regions (and at the axes of symmetry) is not expected to be problematic. On the other hand the natural longer range aim of applicability of our (higher order) energies would encompass the treatment of more general classes of perturbative solutions and thus necessitate a more detailed analysis of this redshift effect in the asymptotic region near the horizon as well as one of the behavior at infinity and near the axes of symmetry. An interesting first step in this direction would be to carry out the corresponding analysis for the purely Maxwellian perturbations of the Kerr backgrounds considered in Section \ref{sec:pure-electromagnetic-kerr-spacetimes}.

Another potentially interesting application of our approach would be to the perturbations of (arbitrarily rapidly rotating) Kerr-Newman-de Sitter black holes arising through the inclusion of a \textit{positive} cosmological constant \(\Lambda\) in the Einstein-Maxwell equations. As we have already mentioned in the Introduction, a fortuitous feature of the Carter-Robinson identity that plays such a crucial role in our program but which is normally applied to purely electrovacuum problems (i.e., those having \(\Lambda = 0\)), is that it only generates, thanks to a favorable sign in one of its terms that vanishes for electrovacuum backgrounds, a new term of positive sign in the presence of a positive cosmological constant. This feature (of the Carter-Robinson identity) has already been exploited by one of us (N.G.) to extend the arguments of Section \ref{sec:pure-electromagnetic-kerr-spacetimes} above to the treatment of the purely Maxwellian perturbations of Kerr-de Sitter black hole backgrounds \cite{Gudapati_18}.

As we have also alluded to in the Introduction there is the interesting potential of applying our approach to the analysis of stability of black holes in higher than 4 spacetime dimensions. The most significant open question in this regard would seem to be the stability of the famous 5-dimensional Myers-Perry rotating black hole solution in \cite{MyPe_86} and its (still not explicitly known) electrovacuum generalization \cite{Hendi_14}. For perturbations preserving the \(T^2\), `axial' isometry of such an axially symmetric background (c.f., \cite{MI_08} and \cite{HIW_07}), one can apply well-known Kaluza-Klein reduction techniques to reduce the field equations to those of a wave map coupled to a 2+1-dimensional Lorentzian metric that closely resembles the system we have already treated \cite{IU_03,Maison_79}. Furthermore the needed Carter-Robinson type identities for these (higher dimensional, reduced) field equations have already been derived and systematically applied to the development of corresponding black hole uniqueness theorems \cite{HI_12,Hollands_11,HY_08,Chrusciel_12}. In addition, the relevant linearization stability (LS) `technology' can be readily extended to the higher dimensional setting of interest so that one should be able to generalize the arguments given herein to the treatment of such higher dimensional black holes.

An attractive feature of the LS `machinery' alluded to above  is that, being essentially spacetime covariant in nature, it lends itself to the treatment of alternative slicings of the background such as those foliated by hypersurfaces that either intersect the future horizon or future null infinity, \(\mathrm{Scri}^+\), or both instead of being `locked down' at the bifurcation 2-sphere and at spacelike infinity, \(i_0\), as ours were required to do. Such alternative slices are not true Cauchy surfaces for the full DOC of a Kerr-Newman black hole but perturbative data given on them does uniquely control the evolution of such data to their causal futures. Furthermore the corresponding energy fluxes at the future horizon and at \(\mathrm{Scri}^+\) are expected to have good signs (for the axisymmetric perturbations to which our formalism naturally applies) and thus to yield \textit{decaying} (as opposed to strictly conserved)  energy expressions (and their higher order generalizations).  %equations starting 601

\section*{Acknowlegements}

This article is the outcome of several years of joint work and we would like to take this opportunity to express our gratitude to colleagues and institutions that provided conducive conditions to complete this work. We thank IHES for the gracious hospitality extended to both of us during the fall of 2016.

N. Gudapati  gratefully acknowledges the support  of Deutsche Forschungsgemeinschaft (DFG) Fellowships GU 1513/1-1 and GU 1513/2-1, hosted by the Department of Mathematics, Yale University and  the Albert Einstein Institute (AEI, Golm) respectively.  N. Gudapati also acknowledges the support from the Gordon and Betty Moore Foundation and the John Templeton Foundation through the Black Hole Initiative of Harvard University, during his postdoctoral stay at the Center of Mathematical Sciences and Applications (CMSA). Finally, N.Gudapati expresses special thanks to  Igor Frenkel for his postdoctoral stay at Yale University in the academic year 2017-2018 and Hermann Nicolai for hosting him at Albert Einstein Institute in the summer of 2016 and the fall of 2019.

%\end{acknowlegments}

\appendix
\section{Explicit Representations of Kerr-Newman Spacetimes}
\label{app:kerr-newman-spacetimes}
Several different coordinate systems for the Kerr-Newman, black hole spacetimes are employed in the present paper. We give these coordinate expressions here together with the explicit transformations connecting them. Except for the elementary degeneracies of the familiar angular coordinates for topological 2-spheres, each of these covers the domain of outer communications of the corresponding black hole in a non-singular way. They each, however, break down at the black hole's event horizon which would necessitate a further transformation to be properly covered. We shall work throughout in `geometrical' units for which Newton's constant \textit{G} and the speed of light \textit{c} are both set to unity.

Each Kerr-Newman black hole is characterized by three parameters, \((M,a,Q)\), where \textit{M} designates the mass, \textit{Q} its electric charge and where \textit{a} determines its angular momentum \(\mathcal{S}\), along its axis of rotational symmetry, through \(\mathcal{S} = aM\). These are subject to the inequalities \(M > 0\) and \(M^2 \geq a^2 + Q^2\) with \(M^2 = a^2 + Q^2\) corresponding to the \textit{extremal} case. Solutions violating either of these do not correspond to black holes.

In Boyer-Lindquist coordinates the line element and vector potential are given by
\begin{align}
ds^2 &= -\left(\frac{\Delta - a^2\sin^2{\theta}}{\Sigma}\right) dt^2 - \frac{2a \sin^2{\theta} (r^2 + a^2 - \Delta)}{\Sigma} dt d\varphi\nonumber\\
& + \left\lbrack\frac{(r^2 + a^2)^2 - \Delta a^2 \sin^2{\theta}}{\Sigma}\right\rbrack \sin^{2}{\theta} d\varphi^2 + \frac{\Sigma}{\Delta} dr^2 + \Sigma d\theta^2
\label{eq:b01}\\
A &= \frac{-Qr}{\Sigma} \lbrack dt - a \sin^2{\theta} d\varphi\rbrack
\label{eq:b02}
\end{align}
where
\begin{equation}
\Sigma := r^2 + a^2 \cos^2{\theta}
\label{eq:b03}
\end{equation}
and
\begin{equation}
\Delta := r^2 - 2Mr + a^2 + Q^2
\label{eq:b04}
\end{equation}
The domain of the outer communications (or black hole `exterior') is the region for which \(t \in \mathbb{R}\),
\begin{equation}
r > r_+ := M + \sqrt{M^2 - (a^2 + Q^2)}
\label{eq:b05}
\end{equation}
and where the angles \(\lbrace\theta,\varphi\rbrace\), with \(\theta \in [0,\pi]\) and \(\varphi \in [0,2\pi)\), label the points of topological 2-spheres having \(t = \text{ constant}\) and \(r = \text{ constant}\). The black hole's event horizon (not properly covered by these coordinates) lies at the limiting coordinate radius \(r = r_+\).

When \(a \neq 0\) the spacetime has precisely two independent Killing fields,
\begin{equation}
\zeta = \frac{\partial}{\partial t}\qquad \text{ and }\qquad \psi = \frac{\partial}{\partial\varphi},
\label{eq:b06}
\end{equation}
which correspond to its stationarity and axial symmetry whereas the special cases (\(a = 0, Q \neq 0\)) and (\(a = 0, Q = 0\)) yield the Reissner-Nordstr\"{o}m and Schwarzschild solutions (respectively) which, each being spherically symmetric, admit two additional, rotational Killing fields. When \(a \neq 0\) the Killing field \(\zeta\), which is timelike at large radius, becomes spacelike inside the so-called `ergo-region' characterized by
\begin{equation}
r > r_+,\qquad \Delta - a^2 \sin^2{\theta} < 0.
\label{eq:b07}
\end{equation}
The presence of this region in these rotating cases causes serious difficulties for the task of finding positive energy expressions for the gravitational and electromagnetic perturbations. The main aim of this article is to construct such an energy for axisymmetric perturbations and to analyze its implications for the black hole stability problem in linear approximation.

A transformation of the radial coordinate given by
\begin{equation}
R = \frac{1}{2} \left(r - M + \sqrt{r^2 - 2Mr + (a^2 + Q^2)}\right)
\label{eq:b08}
\end{equation}
with inverse
\begin{equation}
r = R + M + \frac{(M^2 - a^2 - Q^2)}{4R}
\label{eq:b09}
\end{equation}
combined with the introduction of `isothermal' coordinates defined via
\begin{equation}
\rho = R\sin{\theta},\qquad z = R\cos{\theta},
\label{eq:b10}
\end{equation}
puts the line element into Weyl-Papapetrou form
\begin{equation}
\begin{split}
ds^2 &= \left(\frac{\Sigma}{\left\lbrack(r^2 + a^2)^2 - a^2 \Delta\sin^2{\theta}\right\rbrack}\right) \left\lbrace-\Delta dt^2 + \left\lbrack\frac{(r^2 + a^2)^2 - a^2\Delta\sin^{2}{\theta}}{R^2}\right\rbrack (d\rho^2 + dz^2)\right\rbrace\\
&+ \frac{\sin^2{\theta}}{\Sigma} \left\lbrack(r^2 + a^2)^2 - a^2\Delta\sin^2{\theta}\right\rbrack \left\lbrack d\varphi - \left(\frac{a (2Mr - Q^2) dt}{\left\lbrack(r^2 + a^2)^2 - a^2\Delta\sin^2{\theta}\right\rbrack}\right)\right\rbrack^2
\label{eq:b11}
\end{split}
\end{equation}
where
\begin{equation}
R = \sqrt{\rho^2 + z^2},\: \sin{\theta} = \frac{\rho}{\sqrt{\rho^2 + z^2}},\: \cos{\theta} = \frac{z}{\sqrt{\rho^2 + z^2}} \label{eq:b12}
\end{equation}
and
\begin{equation}
r = \sqrt{\rho^2 + z^2} + M + \frac{(M^2 - a^2 - Q^2)}{4\sqrt{\rho^2 + z^2}} \label{eq:b13}
\end{equation}
In these coordinates the domain of outer communications corresponds to
\begin{equation}
\begin{split}
R &= \sqrt{\rho^2 + z^2} > \frac{1}{2} (r_+ - M) = \frac{1}{2} \sqrt{M^2 - (a^2 + Q^2)}\\
&:= R_+ \geq 0. \label{eq:b14}
\end{split}
\end{equation}
Note that \(R_+ = 0\) only in the extremal case.

The Carter \cite{Car_71} and Robinson \cite{Rob_74} identities, which play a crucial role in the present paper, are traditionally expressed in alternative variations of Weyl-Papapetrou coordinates in which the event horizon at \(r = r_+\) is mapped to an interval (or `cut') along the symmetry axis. Recalling that solutions of the Cauchy-Riemann equations preserve the `isothermal' form of the Riemannian 2-metric \(d\rho^2 + dz^2\) one easily shows that the transformation defined by (conjugate harmonic functions)
\begin{equation}
\bar{\rho} = \rho - \frac{(M^2 - a^2 - Q^2)\rho}{4(\rho^2 + z^2)} \label{eq:b15}
\end{equation}
and
\begin{equation}
\bar{z} = z + \frac{(M^2 - a^2 - Q^2)z}{4(\rho^2 + z^2)} \label{eq:b16}
\end{equation}
induces the conformal mapping
\begin{equation}
d\bar{\rho}^2 + d\bar{z}^2 = \left(1 + \frac{C(\rho^2 - z^2)}{2(\rho^2 + z^2)^2} + \frac{C^2}{16(\rho^2 + z^2)^2}\right) \times (d\rho^2 + dz^2) \label{eq:b17}
\end{equation}
where \(C = M^2 - Q^2 - a^2\). The inverse transformation can be readily derived by exploiting the identity
\begin{equation}
(\rho^2 + z^2) + \frac{C^2}{16(\rho^2 + z^2)} = \frac{1}{2} \left\lbrace(\bar{\rho}^2 + \bar{z}^2) + \sqrt{(\bar{\rho}^2 + \bar{z}^2)^2 + 2C\left\lbrack(\bar{\rho}^2 - \bar{z}^2) + \frac{C}{2}\right\rbrack}\right\rbrace \label{eq:b18}
\end{equation}
to solve a quadratic equation for \(\rho^2 + z^2\) in terms of \(\bar{\rho}\) and \(\bar{z}\).

It is easily verified (for the non-degenerate cases having \(C > 0\)) that the horizon `semi-circle' defined by
\begin{equation}
\rho^2 + z^2 = R_+^2 = \frac{1}{4} \left(M^2 - (a^2 + Q^2)\right) > 0
\label{eq:b19}
\end{equation}
gets mapped to a `cut' on the \(\bar{z}\) axis given by
\begin{equation}
\bar{\rho} = 0,\: \bar{z} \in \left\lbrack-\sqrt{M^2 - (a^2 + Q^2)},\: \sqrt{M^2 - (a^2 + Q^2)}\right\rbrack. \label{eq:b20}
\end{equation}
For the degenerate cases (having \(C = 0\)) transformation (\ref{eq:b15}--\ref{eq:b16}) reduces to the identity and the horizon, in these coordinates, `collapses' to a point.

Finally, setting \(c := \sqrt{M^2 - (a^2 + Q^2)}\), consider the transformation defined by
\begin{eqnarray}
\bar{\rho} &=& (\lambda^2 - c^2)^{1/2} (1 - \mu^2)^{1/2}\nonumber\\
\bar{z} &=& \mu\lambda \label{eq:b21}
\end{eqnarray}
where \(c < \lambda < \infty\), \(-1 \leq \mu \leq 1\). It is readily verified that
\begin{equation}
d\bar{\rho}^2 + d\bar{z}^2 = (\lambda^2 - c^2\mu^2) \left\lbrack\frac{d\lambda^2}{\lambda^2 - c^2} + \frac{d\mu^2}{1 - \mu^2}\right\rbrack \label{eq:b22}
\end{equation}
In these coordinates the two symmetry axis components correspond to \(\mu = \pm 1\) whereas the horizon occurs at \(\lambda \searrow c\). The transformation in (\ref{eq:b21}) is readily inverted through the use of the identity
\begin{equation}
\lambda^2 + \frac{c^2\bar{z}^2}{\lambda^2} = c^2 + \bar{\rho}^2 + \bar{z}^2
\label{eq:b23}
\end{equation}
and the \(\lbrace\lambda,\mu\rbrace\) coordinates play a key role in the Robinson identity presented in \cite{Rob_74}.
 %A; orginally appendix B
\section{The Global Cauchy Problem for the Linearized Einstein-Maxwell Equations}
\label{app:global}
The Einstein-Maxwell equations, in the absence of a charged current source, are expressible, in their most general 4-dimensional form, as
\begin{align}
\begin{split}\label{eq:h01}
\left\lbrack{}^{(4)}Ein ({}^{(4)}g)\right\rbrack_{\alpha\beta} &:= \left\lbrack{}^{(4)}Ric ({}^{(4)}g) - \frac{1}{2} {}^{(4)}g\; {}^{(4)}R ({}^{(4)}g)\right\rbrack_{\alpha\beta}\\
 &= 8\pi\; \left\lbrack{}^{(4)}T ({}^{(4)}g, {}^{(4)}F)\right\rbrack_{\alpha\beta}\\
 &= 2 \left\lbrace{}^{(4)}F_{\alpha\mu}\; {}^{(4)}F_{\beta\nu}\; {}^{(4)}g^{\mu\nu} - \frac{1}{4} {}^{(4)}g_{\alpha\beta}\; {}^{(4)}F_{\mu\nu}\; {}^{(4)}F^{\mu\nu}\right\rbrace,
\end{split}\\
\left\lbrack\delta_{{}^{(4)}g} \cdot {}^{(4)}F\right\rbrack^\alpha &:= {}^{(4)}\nabla_\beta\; {}^{(4)}F^{\alpha\beta} = 0,\label{eq:h02}\\
\begin{split}\label{eq:h03}
\left\lbrack d\; {}^{(4)}F\right\rbrack_{\alpha\beta\gamma} &:= {}^{(4)}F_{\alpha\beta,\gamma} + {}^{(4)}F_{\beta\gamma,\alpha} + {}^{(4)}F_{\gamma\alpha,\beta}\\
 &= 0
\end{split}
\end{align}
where \({}^{(4)}g = {}^{(4)}g_{\mu\nu}\; dx^\mu \otimes dx^\nu\) is the spacetime-metric, \({}^{(4)}Ric({}^{(4)}g)\) and \({}^{(4)}R({}^{(4)}g)\) are its associated Ricci tensor and scalar curvature, \({}^{(4)}F = {}^{(4)}F_{\mu\nu}\; dx^\mu \otimes dx^\nu\) is the electromagnetic 2-form field and where \({}^{(4)}\nabla_\alpha\) (or, equivalently \(;\alpha\)) designates covariant differentiation with respect to \({}^{(4)}g\). In the above and throughout we have set Newton's constant of gravitation, \textit{G}, and the speed of light, \textit{c}, equal to unity by choice of units.

We shall assume here and throughout that the field tensor \({}^{(4)}F\) is derived from a `vector potential' \({}^{(4)}\!A = {}^{(4)}\!A_\mu\; dx^\mu\) such that \({}^{(4)}F = d\; {}^{(4)}\!A\) or, in coordinates,
\begin{equation}\label{eq:h04}
\begin{split}
{}^{(4)}F_{\mu\nu} &= \lbrack d\; {}^{(4)}\!A\rbrack_{\mu\nu}\\
 &= {}^{(4)}\!A_{\nu,\mu} - {}^{(4)}\!A_{\mu,\nu}
\end{split}
\end{equation}
so that Eq.~(\ref{eq:h03}) is satisfied identically. Henceforth we regard \({}^{(4)}F\) as expressed, as above, in terms of \({}^{(4)}\!A\) and regard the pair \(\lbrace{}^{(4)}g, {}^{(4)}\!A\rbrace\) as the `fundamental fields' upon which the field equations are imposed.

Designating the first variations \((\delta\; {}^{(4)}g, \delta\; {}^{(4)}\!A)\) of \(({}^{(4)}g, {}^{(4)}\!A)\) by
\begin{equation}\label{eq:h05}
({}^{(4)}h, {}^{(4)}\!A') = ({}^{(4)}h_{\mu\nu}\; dx^\mu \otimes dx^\nu,\quad {}^{(4)}A'_\mu\; dx^\mu)
\end{equation}
we can express the corresponding linearized equations as
\begin{equation}\label{eq:h06}
D\; {}^{(4)}Ein({}^{(4)}g) \cdot {}^{(4)}h = 8\pi D\; {}^{(4)}T ({}^{(4)}g, {}^{(4)}\!A) \cdot ({}^{(4)}h, {}^{(4)}\!A'),
\end{equation}
and
\begin{equation}\label{eq:h07}
D(\delta_{{}^{(4)}g} \cdot {}^{(4)}F) \cdot ({}^{(4)}h, {}^{(4)}\!A') = 0
\end{equation}
where
\begin{equation}\label{eq:h08}
\begin{split}
&\left(D\; Ein ({}^{(4)}g) \cdot {}^{(4)}h\right)_{\alpha\beta}\\
&\quad = \frac{1}{2} \left\lbrace\vphantom{\left\lbrack{}^{(4)}Ric ({}^{(4)}g)\right\rbrack^{\mu\nu}}{}^{(4)}\bar{h}_{\alpha\mu;\beta}^{\hphantom{\alpha\mu;\beta};\mu} + {}^{(4)}\bar{h}_{\beta\mu;\alpha}^{\hphantom{\beta\mu;\alpha};\mu} - {}^{(4)}\bar{h}_{\alpha\beta;\mu}^{\hphantom{\alpha\beta;\mu};\mu}\right.\\
&\quad\quad{} - {}^{(4)}g_{\alpha\beta}\; {}^{(4)}\bar{h}_{\mu\nu}^{\hphantom{\mu\nu};\mu\nu} - {}^{(4)}R({}^{(4)}g)\; {}^{(4)}\bar{h}_{\alpha\beta}\\
&\quad\quad\left.{} + {}^{(4)}g_{\alpha\beta} \left\lbrack{}^{(4)}Ric ({}^{(4)}g)\right\rbrack^{\mu\nu} {}^{(4)}\bar{h}_{\mu\nu}\right\rbrace
\end{split}
\end{equation}
and
\begin{equation}\label{eq:h09}
\begin{split}
&\left\lbrack D (\delta_{{}^{(4)}g} \cdot {}^{(4)}F) \cdot ({}^{(4)}h, {}^{(4)}\!A')\right\rbrack_\alpha\\
&\quad = {}- {}^{(4)}h_{\mu\nu}\; {}^{(4)}F_\alpha^{\hphantom{\alpha}\mu;\nu} - \frac{1}{2} {}^{(4)}F^{\beta\gamma} ({}^{(4)}h_{\alpha\beta;\gamma} - {}^{(4)}h_{\alpha\gamma;\beta})\\
&\quad\quad{}- {}^{(4)}F_\alpha^{\hphantom{\alpha}\beta} \left({}^{(4)}h_{\nu\beta} - \frac{1}{2} {}^{(4)}g_{\nu\beta}\; {}^{(4)}h_\gamma^\gamma\right)^{;\nu}\\
&\quad\quad{}- ({}^{(4)}\!A'_\alpha)_{;\mu}^{\hphantom{;\mu};\mu} + ({}^{(4)}\!{A'}_\nu^{\hphantom{\nu};\nu})_{;\alpha} + \left({}^{(4)}Ric ({}^{(4)}g)\right)_\alpha^{\hphantom{\alpha}\nu}\; {}^{(4)}\!A'_\nu
\end{split}
\end{equation}
with \(D\; {}^{(4)}T ({}^{(4)}g, {}^{(4)}\!A) \cdot ({}^{(4)}h, {}^{(4)}\!A')\) readily computable algebraically in terms of \({}^{(4)}h\) and
\begin{equation}\label{eq:h10}
\begin{split}
&\left\lbrack D\; {}^{(4)}F ({}^{(4)}\!A) \cdot {}^{(4)}\!A'\right\rbrack_{\mu\nu}\\
&\quad = \partial_\mu\; {}^{(4)}\!A'_\nu - \partial_\nu\; {}^{(4)}\!A'_\mu.
\end{split}
\end{equation}
In the above \({}^{(4)}\bar{h} = {}^{(4)}\bar{h}_{\mu\nu}\; dx^\mu \otimes dx^\nu\) designates the `trace-reversed' metric perturbation defined by
\begin{equation}\label{eq:h11}
\begin{split}
{}^{(4)}\bar{h}_{\mu\nu} &:= {}^{(4)}h_{\mu\nu} - \frac{1}{2} {}^{(4)}g_{\mu\nu}\; {}^{(4)}h_{\alpha\beta}\; {}^{(4)}g^{\alpha\beta}\\
 &= {}^{(4)}h_{\mu\nu} - \frac{1}{2} {}^{(4)}g_{\mu\nu}\; {}^{(4)}h_\gamma^{\hphantom{\gamma}\gamma}
\end{split}
\end{equation}
which is readily inverted to yield
\begin{equation}\label{eq:h12}
{}^{(4)}h_{\mu\nu} = {}^{(4)}\bar{h}_{\mu\nu} - \frac{1}{2} {}^{(4)}g_{\mu\nu}\; {}^{(4)}\bar{h}_\gamma^{\hphantom{\gamma}\gamma}
\end{equation}
with \({}^{(4)}h_\gamma^{\hphantom{\gamma}\gamma} := {}^{(4)}h_{\mu\nu}\; {}^{(4)}g^{\mu\nu} = -{}^{(4)}\bar{h}_\gamma^{\hphantom{\gamma}\gamma} := -{}^{(4)}\bar{h}_{\mu\nu}\; {}^{(4)}g^{\mu\nu}\).

As is well known, when the background field equations (\ref{eq:h01}--\ref{eq:h03}) are satisfied the corresponding linearized equations (\ref{eq:h06}--\ref{eq:h07}) are invariant with respect to an Abelian group of gauge transformations generated by pairs of the form \(\lbrace{}^{(4)}\Lambda, {}^{(4)}Y\rbrace\) where \({}^{(4)}\Lambda\) is a scalar field and \({}^{(4)}Y = {}^{(4)}Y^\mu \frac{\partial}{\partial x^\mu}\) a vector field on the given background spacetime. The fundamental linearized variables \(({}^{(4)}h, {}^{(4)}\!A')\) undergo the gauge transformations
\begin{align}
{}^{(4)}\!A'_\mu &\rightarrow {}^{(4)}\!A'_\mu + \partial_\mu\; {}^{(4)}\Lambda + (\mathcal{L}_{{}^{(4)}\!Y}\; {}^{(4)}\!A)_{\mu},\label{eq:h13}\\
{}^{(4)}F'_{\mu\nu} &:= \partial_\mu {}^{(4)}\!A'_\nu - \partial_\nu {}^{(4)}\!A'_\mu \rightarrow {}^{(4)}F'_{\mu\nu} + (\mathcal{L}_{{}^{(4)}\!Y}\; {}^{(4)}\!F)_{\mu\nu},\label{eq:h14}\\
\begin{split}\label{eq:h15}
{}^{(4)}h_{\mu\nu} &\rightarrow {}^{(4)}h_{\mu\nu} + {}^{(4)}\nabla_\mu\; {}^{(4)}Y_\nu + {}^{(4)}\nabla_\nu\; {}^{(4)}Y_\mu\\
 &= {}^{(4)}h_{\mu\nu} + (\mathcal{L}_{{}^{(4)}Y}\; {}^{(4)}g)_{\mu\nu}
\end{split}
\end{align}
where \(\mathcal{L}_{{}^{(4)}Y}\) designates Lie differentiation with respect to \({}^{(4)}Y\) and where \({}^{(4)}Y_\mu := {}^{(4)}g_{\mu\nu}\; {}^{(4)}Y^\nu\) is the latter's covariant form. One can exploit this gauge invariance to impose the electromagnetic `Lorenz' and gravitational `harmonic' (or de Donder) gauge conditions given (respectively) by
\begin{align}
{}^{(4)}\bar{h}_{\mu\nu}^{\hphantom{\mu\nu};\nu} &= 0\label{eq:h16}\\
\intertext{and}
{}^{(4)}\!{A'}_\nu^{\hphantom{\nu};\nu} &= 0.\label{eq:h17}
\end{align}
This is accomplished by solving the inhomogeneous wave equations
\begin{align}
{}^{(4)}Y_{\mu;\nu}^{\hphantom{\mu;\nu};\nu} + \left\lbrack{}^{(4)}Ric({}^{(4)}g)\right\rbrack_\mu^{\hphantom{\mu}\nu}\; {}^{(4)}Y_\nu &= -{}^{(4)}\bar{h}_{\mu\nu}^{\hphantom{\mu\nu};\nu}\label{eq:h18}\\
\intertext{and}
{}^{(4)}\Lambda_{;\mu}^{\hphantom{;\mu};\mu} &= -{}^{(4)}\!{A'}_\mu^{\hphantom{\mu};\mu} - \left\lbrack(\mathcal{L}_{{}^{(4)}\!Y}\; {}^{(4)}\!A)_{\mu}\right\rbrack^{;\mu}\label{eq:h19}
\end{align}
for \({}^{(4)}Y_\mu\; dx^\mu\) and \({}^{(4)}\Lambda\) respectively. Theorems guaranteeing the \textit{global} existence and uniqueness of solutions to the corresponding \textit{linear} Cauchy problems, formulated on a globally hyperbolic spacetime, are proven in \cite{Ringstrom_09}. The solutions to (\ref{eq:h18}) and (\ref{eq:h19}) are, of course, only unique up to the addition of arbitrary solutions to the corresponding homogeneous equations.

When the foregoing gauge conditions are imposed, the linearized field equations (\ref{eq:h06}) and (\ref{eq:h07}) reduce to the manifestly hyperbolic, coupled system
\begin{align}
\begin{split}\label{eq:h20}
&\frac{1}{2} \left\lbrace -{}^{(4)}\bar{h}_{\alpha\beta;\mu}^{\hphantom{\alpha\beta;\mu};\mu} + \left\lbrack{}^{(4)}Riem ({}^{(4)}g)\right\rbrack_{\alpha\hphantom{\rho\mu}\beta}^{\hphantom{\alpha}\rho\mu}\; {}^{(4)}\bar{h}_{\rho\mu}\right.\\
&\quad {}+ \left\lbrack{}^{(4)}Ric ({}^{(4)}g)\right\rbrack^\rho_{\hphantom{\rho}\beta}\; {}^{(4)}\bar{h}_{\alpha\rho} + \left\lbrack{}^{(4)}Riem ({}^{(4)}g)\right\rbrack_{\beta\hphantom{\rho\mu}\alpha}^{\hphantom{\beta}\rho\mu}\; {}^{(4)}\bar{h}_{\rho\mu}\\
&\quad {}+ \left.\left\lbrack{}^{(4)}Ric ({}^{(4)}g)\right\rbrack^\rho_{\hphantom{\rho}\alpha}\; {}^{(4)}\bar{h}_{\beta\rho} - {}^{(4)}R ({}^{(4)}g)\; {}^{(4)}\bar{h}_{\alpha\beta} + {}^{(4)}g_{\alpha\beta}\; \left\lbrack{}^{(4)}Ric ({}^{(4)}g)\right\rbrack^{\mu\nu}\; {}^{(4)}\bar{h}_{\mu\nu}\right\rbrace\\
&\quad = 8\pi\; \left\lbrack D\; {}^{(4)}T ({}^{(4)}g, {}^{(4)}\!A) \cdot ({}^{(4)}h, {}^{(4)}\!A')\right\rbrack_{\alpha\beta},
\end{split}
\intertext{and}
\begin{split}\label{eq:h21}
&\quad {}- {}^{(4)}\!{A'}_{\mu;\nu}^{\hphantom{\mu;\nu};\nu} + \left\lbrack{}^{(4)}Ric ({}^{(4)}g)\right\rbrack_\mu^{\hphantom{\mu}\nu}\; {}^{(4)}\!A'_\nu\\
&\quad {}- {}^{(4)}h_{\alpha\beta}\; {}^{(4)}F_\mu^{\hphantom{\mu}\alpha;\beta} - {}^{(4)}F_\mu^{\hphantom{\mu}\beta}\; \left({}^{(4)}h_{\nu\beta} - \frac{1}{2} {}^{(4)}g_{\nu\beta}\; {}^{(4)}h_{\hphantom{\gamma}\gamma}^\gamma\right)^{;\nu}\\
&\quad {}- \frac{1}{2}\; {}^{(4)}F^{\beta\gamma}\; \left({}^{(4)}h_{\mu\beta;\gamma} - {}^{(4)}h_{\mu\gamma;\beta}\right)\\
&\quad = 0
\end{split}
\end{align}
where \(\lbrack{}^{(4)}Riem ({}^{(4)}g)\rbrack^\alpha_{\hphantom{\alpha}\beta\gamma\delta}\; \frac{\partial}{\partial x^\alpha} \otimes dx^\beta \otimes dx^\gamma \otimes dx^\delta\) is the Riemann curvature tensor of \({}^{(4)}g\).

To ensure satisfaction of the gauge conditions however one must restrict the choice of allowed Cauchy data for the above system accordingly. If \(\Sigma\) is a Cauchy hypersurface of the background spacetime (assumed here to be globally hyperbolic and time orientable) then one must impose
\begin{align}
\left.{}^{(4)}\!{A'}_\nu^{\hphantom{\nu};\nu}\right|_{\Sigma} = 0,&\qquad n^\alpha \left.\left({}^{(4)}\!{A'}_\nu^{\hphantom{\nu};\nu}\right)_{;\alpha}\right|_{\Sigma} = 0\label{eq:h22}\\
\left.{}^{(4)}\bar{h}_{\mu\nu}^{\hphantom{\mu\nu};\nu}\right|_\Sigma = 0,&\qquad n^\alpha \left.\left({}^{(4)}\bar{h}_{\mu\nu}^{\hphantom{\mu\nu};\nu}\right)_{;\alpha}\right|_\Sigma = 0\label{eq:h23}
\end{align}
where \(n^\alpha \frac{\partial}{\partial x^\alpha}\) is the unit, future pointing normal field to \(\Sigma\).

To show that Eqs.~(\ref{eq:h22}--\ref{eq:h23}) are both necessary and sufficient for the implementation and preservation of the gauge conditions we first derive wave equations satisfied by the quantities \({}^{(4)}\!{A'}_\nu^{\hphantom{\nu};\nu}\) and \({}^{(4)}\bar{h}_{\mu\nu}^{\hphantom{\mu\nu};\nu}\). These are most easily obtained by computing the first variations of the identities
\begin{equation}\label{eq:h24}
\left({}^{(4)}F_{\mu\nu}^{\hphantom{\mu\nu};\nu}\right)^{;\mu} \equiv 0,\qquad {}^{(4)}Ein ({}^{(4)}g)_{\mu\nu}^{\hphantom{\mu\nu};\nu} \equiv 0.
\end{equation}
Reducing the resultant variational identities through imposition of the gauge fixed field equations (\ref{eq:h20}) and (\ref{eq:h21}) leads directly to
\begin{align}
\begin{split}\label{eq:h25}
\left({}^{(4)}\!{A'}_\nu^{\hphantom{\nu};\nu}\right)_{;\mu}^{\hphantom{;\mu};\mu} &= {}^{(4)}h_{\alpha\beta}\; {}^{(4)}K^{\alpha;\beta}\\
&\quad {}+ {}^{(4)}K^\beta\; \left({}^{(4)}h_{\nu\beta} - \frac{1}{2}\; {}^{(4)}g_{\nu\beta}\; {}^{(4)}h_\gamma^{\hphantom{\gamma}\gamma}\right)^{;\nu},
\end{split}
\intertext{and}
\left({}^{(4)}\bar{h}_{\alpha\mu}^{\hphantom{\alpha\mu};\mu}\right)_{;\beta}^{\hphantom{;\beta};\beta} &+ \left\lbrack{}^{(4)}Ric ({}^{(4)}g)\right\rbrack_\alpha^{\hphantom{\alpha}\beta}\; {}^{(4)}\bar{h}_{\beta\mu}^{\hphantom{\beta\mu};\mu} = 0\label{eq:h26}\\
\intertext{where}
{}^{(4)}K_\mu &:= {}^{(4)}F_{\mu\nu}^{\hphantom{\mu\nu};\nu}\label{eq:h27}
\end{align}
which of course vanishes when the background (Maxwell) field equations are satisfied. In deriving Eq.~(\ref{eq:h26}) we have exploited the fact that
\begin{equation}\label{eq:h28}
D\; {}^{(4)}T_{\mu\nu}^{\hphantom{\mu\nu};\nu} ({}^{(4)}g, {}^{(4)}\!A) \cdot ({}^{(4)}h, {}^{(4)}\!A') = 0
\end{equation}
when the background and the linearized (Maxwell) field equations are satisfied.

Under these circumstances we thus arrive at the homogeneous wave equations
\begin{equation}\label{eq:h29}
\left({}^{(4)}\!{A'}_\nu^{\hphantom{\nu};\nu}\right)_{;\mu}^{\hphantom{;\mu};\mu} = 0,
\end{equation}
and
\begin{equation}\label{eq:h30}
\left({}^{(4)}\bar{h}_{\alpha\mu}^{\hphantom{\alpha\mu};\mu}\right)_{;\beta}^{\hphantom{;\beta};\beta} + \left\lbrack{}^{(4)}Ric ({}^{(4)}g)\right\rbrack_\alpha^{\hphantom{\alpha}\beta}\; {}^{(4)}\bar{h}_{\beta\mu}^{\hphantom{\beta\mu};\mu} = 0
\end{equation}
satisfied by the gauge fixing quantities \({}^{(4)}\!{A'}_\nu^{\hphantom{\nu};\nu}\) and \({}^{(4)}\bar{h}_{\alpha\mu}^{\hphantom{\alpha\mu};\mu}\). By standard results \cite{Ringstrom_09} one concludes that both \({}^{(4)}\!{A'}_\nu^{\hphantom{\nu};\nu}\) and \({}^{(4)}\bar{h}_{\alpha\mu}^{\hphantom{\alpha\mu};\mu}\) vanish throughout the globally hyperbolic, background spacetime with Cauchy surface \(\Sigma\) if and only if conditions (\ref{eq:h22}--\ref{eq:h23}) are satisfied on \(\Sigma\).

While it may seem that we have thus reduced the linearized field equations to a purely hyperbolic problem this conclusion is slightly misleading for the following reason. By combining the gauge fixed, linearized Maxwell equation (\ref{eq:h21}) with the constraint upon the gauge fixing initial conditions (\ref{eq:h22}) one arrives at
\begin{equation}\label{eq:h31}
\left.\left({}^{(4)}\!{A'}_\nu^{\hphantom{\nu};\nu}\right)_{;\mu}\; n^\mu\right|_\Sigma = \left. n^\mu\; K'_\mu\right|_\Sigma = 0
\end{equation}
where the latter equality is precisely the usual, \textit{elliptic} constraint upon linearized Maxwell initial data expressed in 4-dimensional notation (with \(K'_\mu\) the first variation of \(K_\mu\) given explicitly by Eq.~(\ref{eq:h09})). In a similar way, by combining the gauge fixed, linearized Einstein equation (\ref{eq:h20}) with the gauge fixing initial condition (\ref{eq:h23}) one arrives at
\begin{equation}\label{eq:h32}
\begin{split}
&\left.\left\lbrace\left(D\; {}^{(4)}Ein ({}^{(4)}g) \cdot {}^{(4)}h\right)_{\mu\nu} - 8\pi\; \left(D\; {}^{(4)}T ({}^{(4)}g, {}^{(4)}\!A) \cdot ({}^{(4)}h, {}^{(4)}\!A')\right)_{\mu\nu}\right\rbrace\; n^\mu\right|_\Sigma\\
&\quad = \frac{1}{2} \left.\left\lbrace{}^{(4)}C_{\nu;\mu}\; n^\mu + {}^{(4)}C_{\mu;\nu}\; n^\mu - n_\nu\; {}^{(4)}C_\mu^{\hphantom{\mu};\mu}\right\rbrace\right|_\Sigma\\
&\quad = 0
\end{split}
\end{equation}
where
\begin{equation}\label{eq:h33}
{}^{(4)}C_\alpha := {}^{(4)}\bar{h}_{\alpha\beta}^{\hphantom{\alpha\beta};\beta}
\end{equation}
and where the final equality follows from the imposition of the gauge fixing initial data constraints (\ref{eq:h23}). But the resulting equation is precisely the usual, \textit{elliptic} constraint upon the linearized Einstein initial data expressed in 4-dimensional form.

Since we have already shown that the gauge conditions are preserved in time by the gauge fixed field equations it follows that the (linearized) Einstein-Maxwell constraint equations
\begin{equation}\label{eq:h34}
\begin{split}
&\left\lbrace\left(D\; {}^{(4)}Ein ({}^{(4)}g) \cdot {}^{(4)}h\right)_{\mu\nu} - 8\pi\; \left(D\; {}^{(4)}T ({}^{(4)}g, {}^{(4)}\!A) \cdot ({}^{(4)}h, {}^{(4)}\!A')\right)_{\mu\nu}\right\rbrace\; n^\mu\\
&\quad = 0
\end{split}
\end{equation}
and
\begin{equation}\label{eq:h35}
K'_\mu\; n^\mu = 0
\end{equation}
hold on an arbitrary Cauchy surface (with unit normal field \(n^\mu \partial_\mu\)) and not merely on the `initial' one.

The results given in this Appendix are, of course, simply a linearized version of the local existence and uniqueness theorem for the fully nonlinear Einstein-Maxwell equations proven by Choquet-Bruhat in Ref.~\cite{C-B_09}. But in view of the \textit{linear} character of our field equations one can adapt arguments of the type presented in \cite{Ringstrom_09} to establish the \textit{global} extendibility of solutions to the full, maximal Cauchy development of a chosen initial data surface. Thus, in particular, solutions generated from appropriate initial data will automatically extend to the full \textit{domain of outer communications} of a background black hole solution that we choose to perturb.

A well known, important feature of the hyperbolic form of the perturbation equations is that it guarantees the strictly causal propagation of the corresponding solutions. For a Kerr-Newman background, for example, this ensures that Cauchy data having `initially' compact support, bounded away from the horizon and from spatial infinity, will retain this property for all finite Boyer-Lindquist time, \textit{t}. This reflects the fact that Boyer-Lindquist  time slices are `locked down' at \(i_0\) (spacelike infinity) and at the bifurcation 2-sphere lying in the horizon. For the (causally propagating)  purely Maxwellian perturbations of the Kerr spacetime analyzed in Section \ref{sec:pure-electromagnetic-kerr-spacetimes} it follows that, for such compactly supported initial data, the potential energy flux contributions at spatial infinity and the horizon, arising from the `continuity' equation (\ref{eq:145}), will \textit{vanish identically}. This leaves only the possibility of a non-vanishing energy flux at the `artificial' boundary provided by the axes of symmetry at \(\theta = 0, \pi\). To verify that these also vanish for regular perturbations one needs to evaluate \(\mathcal{J}_{\mathrm{reg}}^\theta\), (c.f., Eq.~(\ref{eq:143})) at these axes.

From the discussion in Appendix \ref{app:electric-charge} we know that the perturbative quantities, \(\lambda'\) and \(\eta'\) both vanish along the axes of symmetry, a fact that results from our demand that the electric and magnetic charges of the `background' spacetime remain unperturbed. It then follows from the smoothness criteria developed in Ref.~\cite{Rinne_05} that each of these functions vanishes \(\sim \sin^2{\theta}\) at these axes. From Eqs.~(\ref{eq:121}), (\ref{eq:122}) and (\ref{eq:137}) it then follows that each of \(\underline{\tilde{u}'}\) and \(\underline{\tilde{v}'}\) also vanishes \(\sim \sin^2{\theta}\) at the axes and thus, after a straightforward calculation, that \(\mathcal{J}_{\mathrm{reg}}^\theta\) vanishes \(\sim \sin^2{\theta}\) as well. Consequently the energy \(H^{\mathrm{Reg}}\) defined  via Eqs.~(\ref{eq:135})--(\ref{eq:138}) is strictly conserved for this class of (spatially compactly supported) perturbations. A more comprehensive treatment would allow the Maxwellian perturbations to lie in suitable (weighted) Sobolev spaces and appeal to their (presumed) \textit{dense filling} by the compactly supported solutions to establish the corresponding energy conservation result. While this would seemingly be straightforward to carry out, we shall not pursue it further here.  %B; originally appendix H
\section{The Reduced Hamiltonian Formalism for Axi-Symmetric Spacetimes}
\label{app:reduced-hamiltonian}
This article deals primarily with the linearized Einstein-Maxwell equations restricted to the domain of the outer communications, \textit{V}, of a charged (if \(Q \neq 0\)), rotating (if \(a \neq 0\)) Kerr-Newman black hole. The coordinate systems discussed in Appendix~\ref{app:kerr-newman-spacetimes} cover such domains and are adapted to the stationarity and axial symmetry of the Kerr-Newman solutions. Each such domain is a product of the form \(V = \mathbb{R} \times (\mathbb{R}^3 \backslash B_b)\) where \(B_b\) is a closed ball (or exceptionally a point) of coordinate radius \(b \geq 0\). In the spatially cylindrical (Weyl-Papapetrou) coordinates \(\lbrace t,\rho,z,\varphi\rbrace\) introduced in that appendix, for example,
\begin{equation}
B_b = \left\lbrace (\rho, z, \varphi)\left|\rho^2 + z^2 \leq b^2 = \frac{M^2 - a^2 - Q^2}{4} \geq 0\right.\right\rbrace
\label{eq:a01}
\end{equation}
and the corresponding Kerr-Newman spacetime (restricted to \textit{V}) admits \(\psi = \frac{\partial}{\partial\varphi}\) as a spacelike Killing field.

Since we shall only consider perturbations that preserve this axial Killing symmetry it will be natural to pass to the corresponding quotient space \(V/ U(1)\) (where \(U(1)\) is the circle action generated by \(\psi\)) and to formulate the linearized equations thereon. Since points on the symmetry axis are invariant under this group action (since \(\psi\) vanishes there) the resulting quotient space is a manifold with boundary of the form
\begin{equation}
V / U(1) = \mathbb{R} \times M_b
\label{eq:a02}
\end{equation}
where \(M_b\) is the half-plane \(\lbrace (\rho,z)|\: \rho \geq 0, z \in \mathbb{R}\rbrace\) with the half-disk \(D_b = \left\lbrace (\rho,z)| \rho \geq 0, \rho^2 + z^2 \leq b^2 > 0\right\rbrace\) or point \((\rho = z = 0)\) removed. The boundary points of \(M_b\) (i.e., those on the \textit{z}-axis with \(z^2 > b^2 \geq 0\)) correspond to those on the spacetime's axis of symmetry. In this appendix we shall focus on deriving the requisite linearized field equations at \textit{interior points} of the quotient space \(\mathbb{R} \times M_b\) (i.e. points in the complement of the boundary), keeping in mind that certain geometrically natural \textit{regularity conditions} will need to be imposed on the linearized fields, not only at the boundary axis but also at the background black hole's event horizon (corresponding to \(\rho^2 + z^2 \searrow b^2 = \frac{1}{4} (M^2 - a^2 - Q^2) \geq 0\)) and, asymptotically, at `infinity'. Such regularity conditions will be necessary to ensure that linearized solutions on \(\mathbb{R} \times M_b\) can be `lifted' to yield sufficiently smooth and asympotically acceptable perturbations on \textit{V}.

In coordinates \(\lbrace x^\mu\rbrace = \lbrace t,x^a,x^3\rbrace\) of the aforementioned type for the 4-manifold \(V = \mathbb{R} \times (\mathbb{R}^3 \backslash B_b)\), where \(x^0 = t, \lbrace x^a\rbrace = \lbrace x^1,x^2\rbrace = \lbrace\rho,z\rbrace\) and \(x^3 = \varphi\), we begin by expressing the spacetime line element in Arnowitt, Deser and Misner (ADM) form \cite{MTW_73}:
\begin{equation}
\begin{split}
ds^2 &= {}^{(4)}\!g_{\mu\nu} dx^\mu dx^\nu\\
 &= -N^2 dt^2 + g_{ij} (dx^i + N^i dt)(dx^j + N^jdt)
\label{eq:a03}
\end{split}
\end{equation}
where \(\mu, \nu, \ldots \normalfont\text{range over } \lbrace 0,1,2,3\rbrace\) while \(i,j,\ldots \normalfont\text{range over } \lbrace 1,2,3\rbrace\). For the metric \({}^{(4)}\!g = {}^{(4)}\!g_{\mu\nu} dx^\mu \otimes dx^\nu\) to be properly Lorentzian it is essential that the `lapse function' \textit{N} be nowhere vanishing and that the induced metric, \({}^{(3)}\!g = g_{ij} dx^i \otimes dx^j\), and \(t = \normalfont\text{constant}\) hypersurfaces be Riemannian. To avoid confusion with the lapse function \textit{N} we shall designate the `shift vector field', \(N^i \frac{\partial}{\partial x^i}\), in coordinate free notation, by \textit{X}. When the spacetime \(\left(V,{}^{(4)}\!g\right)\) admits an electomagnetic field \({}^{(4)}\!F = {}^{(4)}\!F_{\mu\nu} dx^\mu \wedge dx^\nu\) that is globally derivable from a `vector potential' \({}^{(4)}\!A = {}^{(4)}\!A_\mu dx^\mu\) then we have
\begin{equation}
{}^{(4)}\!F_{\mu\nu} = {}^{(4)}\!A_{\nu,\mu} - {}^{(4)}\!A_{\mu,\nu}
\label{eq:a04}
\end{equation}
and introduce an ADM parameterization for \({}^{(4)}\!A\) by setting
\begin{equation}
{}^{(4)}\!A = A_0 dt + A_i dx^i.
\label{eq:a05}
\end{equation}

Let \(\Omega \subset V\) be an arbitrary compact domain in \textit{V} with (at least piecewise) smooth boundary \(\partial\Omega\). The Einstein-Maxwell equations (at interior points of \(\Omega\)) follow from the ADM variational principle
\begin{equation}
\begin{split}
\delta_{g_{ij}} I_\Omega &= \delta_{\pi^{ij}} I_\Omega = \delta_{A_i} I_\Omega = \delta_{\mathcal{E}^i} I_\Omega\\
 &= \delta_N I_\Omega = \delta_{N^i} I_\Omega = \delta_{A_0} I_\Omega = 0
\label{eq:a06}
\end{split}
\end{equation}
(subject to suitable boundary conditions on the variations of the fields on \(\partial\Omega\)) with
\begin{equation}
I_\Omega := \int_\Omega d^4 x \left\lbrace\pi^{ij} g_{ij,t} + A_i\mathcal{E}^i_{,t} - N \mathcal{H} - N^i\mathcal{H}_i - A_0\mathcal{E}^i_{,i}\right\rbrace
\label{eq:a07}
\end{equation}
where
\begin{align}
\mathcal{H} &= \frac{1}{\mu_{{}^{(3)}\!g}} \left(\pi^{ij}\pi_{ij} - \frac{1}{2} (\pi_i^i)^2\right) - \mu_{{}^{(3)}\!g} {}^{(3)}\!R + \frac{1}{2} \frac{g_{ij}}{\mu_{{}^{(3)}\!g}} (\mathcal{E}^i\mathcal{E}^j + \mathcal{B}^i\mathcal{B}^j),\label{eq:a08}\\
\mathcal{H}_i &= -2\pi_{i\hphantom{j}|j}^j - \epsilon_{ijk}\mathcal{E}^j\mathcal{B}^k\label{eq:a09}
\end{align}
with
\begin{equation}
\mathcal{B}^i = \frac{1}{2} \epsilon^{ijk} (A_{k,j} - A_{j,k}).
\label{eq:a10}
\end{equation}
Here \(\mu_{{}^{(3)}\!g}\) and \({}^{(3)}\!R\) are the volume element \(\left(\mu_{{}^{(3)}\!g} = \sqrt{\det{|g_{ij}|}\!}\right)\) and scalar curvature of the Riemannian metric \({}^{(3)}\!g = g_{ij} dx^i \otimes dx^j,\) \(|\) designates covariant differentiation with respect to this metric and spatial indices \(i,j,\ldots\) are raised and lowered using \({}^{(3)}\!g\) and its inverse, \({}^{(3)}g^{-1} = g^{ij} \frac{\partial}{\partial x^i} \otimes \frac{\partial}{\partial x^j}\).
The (contravariant) symmetric tensor density \({}^{(3)}\!\pi = \pi^{ij} \frac{\partial}{\partial x^i} \otimes \frac{\partial}{\partial x^j}\) is the momentum canonically conjugate to \({}^{(3)}\!g\) whereas the vector density \({}^{(3)}\!\mathcal{E} = \mathcal{E}^i \frac{\partial}{\partial x^i}\) is (up to sign) that conjugate to \({}^{(3)}\!A = A_i dx^i\). The Levi-Civita symbols \(\epsilon_{ijk}\) and \(\epsilon^{ijk}\) are covariant and contravariant, completely antisymmetric tensor densities (such that \(\mu_{{}^{(3)}\!g} \epsilon_{ijk}\) and \(\frac{1}{\mu_{{}^{(3)}\!g}} \epsilon^{ijk}\) are tensor fields) satisfying \(\epsilon_{123} = \epsilon^{123} = 1\).

A derivation of this action principle from its (perhaps more familiar) Lagrangian form is presented in Chapter 21 of the text ``Gravitation'' by Misner, Thorne and Wheeler (MTW) \cite{MTW_73}. Our notation differs somewhat from theirs in that we have absorbed a factor of 2 into the symbols \({}^{(4)}\!A, {}^{(4)}\!F, A_0, {}^{(3)}\!A = A_i dx^i, {}^{(3)}\!\mathcal{E} = \mathcal{E}^i \frac{\partial}{\partial x^i}\) and \({}^{(3)}\!\mathcal{B} = \mathcal{B}^i \frac{\partial}{\partial x^i}\) in order to simplify the forms of the electromagnetic Hamilton equations. In addition we write \(\mu_{{}^{(3)}\!g}\) for their \(\sqrt{g}\) and use \(\epsilon^{ijk}\) and \(\epsilon_{ijk}\) instead of \(\lbrack ijk\rbrack\) to designate the Levi-Civita tensor densities. To recover the expressions of MTW one should replace our \(A_0, A_i dx^i, \mathcal{E}^i \frac{\partial}{\partial x^i}\) and \(\mathcal{B}^i \frac{\partial}{\partial x^i}\) by \(2A_0, 2A_i dx^i, 2\mathcal{E}^i \frac{\partial}{\partial x^i}\) and \(2\mathcal{B}^i \frac{\partial}{\partial x^i}\) respectively, write \(\sqrt{g}\) in place of our \(\mu_{{}^{(3)}\!g}\) and substitute \(\lbrack ijk\rbrack\) for our \(\epsilon^{ijk}\) and \(\epsilon_{ijk}\).

Now restrict attention to those Lorentizan metrics on \textit{V} which have the circle action generated by \(\psi = \frac{\partial}{\partial\varphi}\) as a (spacelike) isometry group and impose the corresponding \(\left(U(1)\right)\) invariance on \({}^{(4)}\!A\) by demanding that
\begin{equation}
\left(\mathcal{L}_{\frac{\partial}{\partial\varphi}} {}^{(4)}\!g\right)_{\mu\nu} = \frac{\partial}{\partial\varphi} {}^{(4)}\!g_{\mu\nu} = 0
\label{eq:a11}
\end{equation}
and
\begin{equation}
\left(\mathcal{L}_{\frac{\partial}{\partial\varphi}} {}^{(4)}\!A\right)_\mu = \frac{\partial}{\partial\varphi} {}^{(4)}\!A_\mu = 0.
\label{eq:a12}
\end{equation}
One can now express the field equations alluded to above entirely in terms of fields induced on the quotient space \(\mathbb{R} \times M_b\). To this end it is convenient to reparametrize the (\(U(1)\)-invariant) Lorentzian metric \({}^{(4)}\!g\) on \textit{V} by setting
\begin{equation}
\begin{split}
ds^2 &= {}^{(4)}\!g_{\mu\nu} dx^\mu dx^\nu = e^{-2\gamma}\left\lbrace - \tilde{N}^2 dt^2 + \tilde{g}_{ab} (dx^a + \tilde{N}^a dt) (dx^b + \tilde{N}^b dt)\right\rbrace\\
& + e^{2\gamma} \left\lbrace d\varphi + \beta_0 dt + \beta_a dx^a\right\rbrace^2
\label{eq:a13}
\end{split}
\end{equation}
and, correspondingly, to write
\begin{equation}
{}^{(4)}\!A_\mu dx^\mu = A_0 dt + A_a dx^a + A_3 d\varphi
\label{eq:a14}
\end{equation}
for the (\(U(1)\)-invariant) vector potential. Here, \(a, b, \ldots\) range only over \(\lbrace 1,2\rbrace\) the indices for coordinates for \(M_b\). Abusing notation slightly we shall employ the same symbols to designate the fields induced, in Kaluza-Klein fashion, on the quotient space.

At interior points of the quotient space (i.e., on the complement of the symmetry axis) we may regard
\begin{equation}
d\sigma^2 := \left\lbrace -\tilde{N}^2 dt^2 + \tilde{g}_{ab} (dx^a + \tilde{N}^a dt)(dx^b + \tilde{N}^b dt)\right\rbrace
\label{eq:a15}
\end{equation}
as the ADM form of the line element for an induced, 2+1-dimensional, Lorentz metric and view \(e^{2\gamma} = {}^{(4)}\!g \left(\frac{\partial}{\partial\varphi},\frac{\partial}{\partial\varphi}\right)\) and \(A_3 = \left\langle {}^{(4)}\!A,\frac{\partial}{\partial\varphi}\right\rangle\) as induced functions and \(\beta_0 dt + \beta_a dx^a\) and \(A_0 dt + A_a dx^a\) as induced one-forms on (interior points of) the quotient space \(\mathbb{R} \times M_b\). Note however that since \(e^{2\gamma}\) must vanish at boundary points of this quotient (which corresponds to points on the symmetry axis in \textit{V}), the function \(\gamma\) must entail a logarithmic singularity in this limit and, accordingly, \(\tilde{N}\) and \(\tilde{g}_{ab}\) must incorporate a singular (vanishing at the boundary) conformal factor to cancel the singularity coming from \(e^{-2\gamma}\). While one could explicitly remove these singularities from the base fields by a change of parametrization the elegant form of the projected field equations (at interior points of \(\mathbb{R} \times M\)) would thereby be disturbed. To avoid this we shall retain the notation introduced above, keeping in mind that certain fields induced on the quotient must exhibit well-defined singular behaviors at the boundary in order to `lift' naturally to yield smooth fields on \textit{V}. The background Kerr-Newman solutions of course automatically exhibit this (geometrically natural) singular behavior when parametrized as above (c.f., Eqs.~(\ref{eq:b11}--\ref{eq:b13}) of Appendix~\ref{app:kerr-newman-spacetimes}) and we shall need to impose suitable regularity conditions on their perturbations in order that such perturbations lift smoothly back to \textit{V}. For the moment however we shall focus on transforming the projected field equations at interior points of the quotient and postpone the discussion of the regularity conditions needed at the boundary until later.

Letting \(\mathcal{D}\) represent an arbitrary compact domain in \(M_b\), disjoint from the boundary, define momenta \(\left\lbrace\tilde{p}, \tilde{e}^a, \tilde{\pi}^{ab}\right\rbrace\) conjugate to \(\left\lbrace\gamma, \beta_a, \tilde{g}_{ab}\right\rbrace\) by setting
\begin{equation}
\begin{split}
\int_{\mathcal{D}\times S^1} d^3x\> \left\lbrace\pi^{ij} g_{ij,t}\right\rbrace &= \int_{\mathcal{D}\times S^1} d^3x\> \left\lbrace\tilde{\pi}^{ab} \tilde{g}_{ab,t} + \tilde{e}^a \beta_{a,t} + \tilde{p}\gamma_{,t}\right\rbrace\\
 &= 2\pi \int_{\mathcal{D}} d^2x\> \left\lbrace\tilde{\pi}^{ab} \tilde{g}_{ab,t} + \tilde{e}^a \beta_{a,t} + \tilde{p}{\gamma}_{,t}\right\rbrace.
\label{eq:a16}
\end{split}
\end{equation}
This leads, together with (\ref{eq:a03}) and (\ref{eq:a13}) to relations such as
\begin{equation}
\tilde{\pi}^{ab} = e^{-2\gamma} \pi^{ab},\: g_{ab} = e^{-2\gamma} \tilde{g}_{ab} + e^{2\gamma} \beta_a\beta_b,\qquad \text{etc.}
\label{eq:a17}
\end{equation}
which can be read off from the above defining expression. To incorporate the electromagnetic terms introduce also the definitions
\begin{align}
\tilde{f}^a &= (\tilde{e}^a - \mathcal{E}^a A_3),\> F^3 = -(\mathcal{E}^3 + \beta_a\mathcal{E}^a)\label{eq:a18}\\
C_a &= -(A_a - \beta_a A_3),\> C_0 = -(A_0 - \beta_0 A_3)\label{eq:a19}
\end{align}
and reexpress the ADM action in terms of the new variables. The result (modulo an inessential boundary term) is expressible (on domains of the form \(\Omega = [t_0,t_1] \times \mathcal{D}\times S^1\)) as
\begin{equation}
\begin{split}
\tilde{I}_\Omega &:= \int_{t_0}^{t_1} dt\> \int_{\mathcal{D}} d^2x\> \left\lbrace\tilde{\pi}^{ab} \tilde{g}_{ab,t} + \tilde{p}\gamma_{,t} + \tilde{f}^a \beta_{a,t}\right.\\
 &\left. + F^3 A_{3,t} + \mathcal{E}^a C_{a,t} + \beta_0 \tilde{f}^a_{,a} + C_0 \mathcal{E}^a_{,a} - \tilde{N}\tilde{\mathcal{H}} - \tilde{N}^a\tilde{\mathcal{H}}_a\right\rbrace\\
 &= (I_\Omega /2\pi) - (\text{boundary term})\label{eq:a20}
\end{split}
\end{equation}
where
\begin{align}
\tilde{\mathcal{H}} &= \left(\frac{1}{\mu_{{}^{(2)}\!\tilde{g}}}\right) \left\lbrack\tilde{\pi}^{ab} \tilde{\pi}_{ab} - (\tilde{\pi}^a_{~a})^2 + \frac{1}{8} \tilde{p}^2 + \frac{1}{2} e^{-4\gamma} \tilde{g}_{ab} (\tilde{f}^a + \mathcal{E}^a A_3) (\tilde{f}^b + \mathcal{E}^b A_3)\right\rbrack\nonumber\\
 &+ \mu_{{}^{(2)}\!\tilde{g}} \left\lbrack -{}^{(2)}\!\tilde{R} + 2\tilde{g}^{ab} \gamma_{,a}\gamma_{,b} + \frac{1}{4} e^{4\gamma} \tilde{g}^{ac} \tilde{g}^{bd} (\beta_{a,b} - \beta_{b,a}) (\beta_{c,d} - \beta_{d,c})\right\rbrack\nonumber\\
 &+ \left(\frac{1}{2\mu_{{}^{(2)}\!\tilde{g}}}\right) \left\lbrack e^{2\gamma} (F^3)^2 + e^{2\gamma} \left( \epsilon^{ab} (C_{a,b} - A_3\beta_{a,b})\right)^2\right.\nonumber\\
  &\hphantom{\left(\frac{1}{2\mu_{{}^{(2)}\!\tilde{g}}}\right)} \left.+ e^{-2\gamma} \tilde{g}_{ab} (\mathcal{E}^a\mathcal{E}^b + \epsilon^{ac} A_{3,c} \epsilon^{bd} A_{3,d})\vphantom{\left((x)^2\right)^2}\right\rbrack,\label{eq:a21}\\
\tilde{\mathcal{H}}_a &= -2\> {}^{(2)}\!\tilde{\nabla}_b\tilde{\pi}^b_{~a} + \tilde{p}\gamma_{,a} + F^3 A_{3,a} + \tilde{f}^b (\beta_{b,a} - \beta_{a,b}) + \mathcal{E}^b (C_{b,a} - C_{a,b}).\label{eq:a22}
\end{align}
In these formulas indices \(a,b,\ldots\) are raised and lowered using the Riemanian 2-metric \({}^{(2)}\!\tilde{g} = \tilde{g}_{ab}\; dx^2 \otimes dx^b,\> {}^{(2)}\!\tilde{R}\) is the scalar curvature of this metric, \({}^{(2)}\!\tilde{\nabla}_a\) its covariant derivative operator and \(\mu_{{}^{(2)}\!\tilde{g}}\) its volume element \((\mu_{{}^{(2)}\!\tilde{g}} := \sqrt{\det{|\tilde{g}_{ab}|}})\). In addition \(\epsilon^{ab}\) is the antisymmetric tensor density (such that \(\epsilon^{ab}/\mu_{{}^{(2)}\!\tilde{g}}\) is a tensor) satisfying \(\epsilon^{12} = 1\).

The constraint equations are obtained by varying \(\tilde{I}_\Omega\) with respect to \(\tilde{N}, \tilde{N}^a, \beta_0\) and \(C_0\) and are thus given by
\begin{equation}
\tilde{\mathcal{H}} = \tilde{\mathcal{H}}_a = \tilde{f}^a_{,a} = \mathcal{E}^a_{,a} = 0.
\label{eq:a23}
\end{equation}
The evolution equations are obtained by varying \(\tilde{I}_\Omega\) with respect to the canonical variables \(\lbrace\tilde{g}_{ab}, \tilde{\pi}^{ab}, \gamma, \tilde{p}, \beta_a, \tilde{f}^a, A_3, F^3, C_a, \mathcal{E}^a\rbrace\). There are neither constraints nor evolution equations for the quantities \(\tilde{N}, \tilde{N}^a, \beta_0\) and \(C_0\) which must be fixed (either explicitly or implicitly) by a choice of gauge.

At fixed \textit{t} the constraint equations \(\tilde{f}^a_{,a} = 0\) and \(\mathcal{E}^a_{,a} = 0\) may, on the topologically trivial space \(M_b\), be solved in generality by setting
\begin{align}
\tilde{f}^a &= \epsilon^{ab} \omega_{,b}\label{eq:a24}\\
\mathcal{E}^a &= \epsilon^{ab} \eta_{,b}\label{eq:a25}
\end{align}
where \(\omega\) and \(\eta\) are uniquely determined up to additive constants (that can vary with \textit{t}). The Hamilton equations for \(\mathcal{E}^a_{,t}\) and \(\tilde{f}^a_{,t}\) may be manipulated to yield
\begin{align}
\eta_{,t} &= \frac{\tilde{N} e^{2\gamma}}{\mu_{{}^{(2)}\!\tilde{g}}} \epsilon^{ab} (C_{a,b} - A_3 \beta_{a,b}) + \epsilon_{ab}\tilde{N}^b\mathcal{E}^a + f(t)\label{eq:a26}\\
\omega_{,t} &= \frac{\tilde{N} e^{2\gamma}}{\mu_{{}^{(2)}\!\tilde{g}}} A_3\epsilon^{ab} (A_3\beta_{a,b} - C_{a,b}) + \frac{\tilde{N} e^{4\gamma}}{\mu_{{}^{(2)}\!\tilde{g}}} \epsilon^{ab}\beta_{a,b} + \epsilon_{ab}\tilde{N}^b\tilde{f}^a + k(t)\label{eq:a27}
\end{align}
where \(f(t)\) and \(k(t)\) are certain undetermined functions of \textit{t} which arise from passing from the equations for \(\mathcal{E}_{,t}^a = (\epsilon^{ab}\eta_{,t})_{,b}\) and \(\tilde{f}^a_{,t} = (\epsilon^{ab} \omega_{,t})_{,b}\) to those for \(\eta_{,t}\) and \(\omega_{,t}\). Since, however, \(\omega\) and \(\eta\) are only determined by (\ref{eq:a24}) and (\ref{eq:a25}) up to arbitrary additive functions of \textit{t} we may smoothly resolve the ambiguity in their definitions (up to additive, true constants) by demanding that \(f(t) = k(t) = 0\).

Defining
\begin{equation}
\tilde{r} = \epsilon^{ab}\beta_{a,b},\qquad \tilde{u} = \epsilon^{ab} C_{a,b}
\label{eq:a28}
\end{equation}
we therefore fix the equations of motion for the `twist potentials' \(\eta\) and \(\omega\) to be
\begin{align}
\eta_{,t} &= \frac{\tilde{N} e^{2\gamma}}{\mu_{{}^{(2)}\!\tilde{g}}} (\tilde{u} - A_3\tilde{r}) + \tilde{N}^b\eta_{,b}\label{eq:a29}\\
\omega_{,t} &= \frac{\tilde{N} e^{2\gamma}}{\mu_{{}^{(2)}\!\tilde{g}}}  A_3 (A_3\tilde{r} - \tilde{u}) + \frac{\tilde{N} e^{4\gamma}}{\mu_{{}^{(2)}\!\tilde{g}}}\tilde{r} + \tilde{N}^a\omega_{,a}\label{eq:a30}
\end{align}
These equations, together with all the remaining evolution and constraint equations, may be derived from the reduced action \(\tilde{J}_\Omega\) obtained from \(\tilde{I}_\Omega\) by substituting the expressions (\ref{eq:a24}), (\ref{eq:a25}) and (\ref{eq:a28}) and discarding an inessential boundary term. Upon defining
\begin{equation}
\lambda = A_3,\qquad \tilde{v} = F^3
\label{eq:a31}
\end{equation}
we get, for the reduced action,
\begin{equation}
\tilde{J}_\Omega = \int_{t_0}^{t_1} dt\> \int_{\mathcal{D}} d^2x\> \left\lbrace\tilde{\pi}^{ab}\tilde{g}_{ab,t} + \tilde{p} \gamma_{,t} + \tilde{r}\omega_{,t} + \tilde{u}\eta_{,t} + \tilde{v}\lambda_{,t} - \tilde{N}\tilde{\mathcal{H}} - \tilde{N}^a\tilde{\mathcal{H}}_a\right\rbrace
\label{eq:a32}
\end{equation}
where \(\tilde{\mathcal{H}}\) and \(\tilde{\mathcal{H}}_a\) now take the forms
\begin{align}
\tilde{\mathcal{H}} &= \frac{1}{\mu_{{}^{(2)}\!\tilde{g}}} \left\lbrack\tilde{\pi}^{ab}\tilde{\pi}_{ab} - (\tilde{\pi}_{~a}^a)^2 + \frac{1}{8} (\tilde{p})^2 + \frac{1}{2} e^{4\gamma} (\tilde{r})^2 + \frac{1}{2} e^{2\gamma} \left(\tilde{v}^2 + (\tilde{u} - \lambda\tilde{r})^2\right)\right\rbrack\label{eq:a33}\\
& + \mu_{{}^{(2)}\!\tilde{g}} \left\lbrack - {}^{(2)}\!\tilde{R} + 2\tilde{g}^{ab}\gamma_{,a}\gamma_{,b} + \frac{1}{2} e^{-2\gamma}\tilde{g}^{ab} (\eta_{,a}\eta_{,b} + \lambda_{,a}\lambda_{,b}) + \frac{1}{2} e^{-4\gamma} \tilde{g}^{ab}(\omega_{,a} + \lambda\eta_{,a}) (\omega_{,b} + \lambda\eta_{,b})\right\rbrack,\nonumber\\
\tilde{\mathcal{H}}_a &= -2\> {}^{(2)}\!\tilde{\nabla}_b\tilde{\pi}_{~a}^b + \tilde{p}\gamma_{,a} + \tilde{r}\omega_{,a} + \tilde{v}\lambda_{,a} + \tilde{u}\eta_{,a}\label{eq:a34}
\end{align}
Variation of \(\tilde{J}_\Omega\) with respect to \(\tilde{N}\) and \(\tilde{N}^a\) yields the remaining constraints \(\tilde{\mathcal{H}} = 0\) and \(\tilde{\mathcal{H}}_a = 0\) whereas variation with respect to the canonical pairs \(\left\lbrace (\tilde{g}_{ab},\tilde{\pi}^{ab}), (\gamma,\tilde{p}), (\omega,\tilde{r}), (\eta,\tilde{u}), (\lambda,\tilde{v})\right\rbrace\) yields the Hamiltonian evolution equations for the reduced system. It is well-known, though perhaps less evident in the present Hamiltonian setting, that this set of reduced field equations is (at interior points of \(\mathbb{R} \times M_b\)) equivalent to the 2+1-dimensional Einstein equations (for the Lorentz metric given in (\ref{eq:a15})) minimally coupled to a \textit{wave map} defined by the four scalar fields \(\lbrace\gamma,\omega,\lambda,\eta\rbrace\). The naturally occurring target space for this wave map (whose metric can be read off from the expression (\ref{eq:a33}) for \(\tilde{\mathcal{H}}\)) is the Riemannian 4-manifold \((\mathbb{R}^4, dk^2)\) with line element
\begin{equation}
dk^2 = 4(d\gamma)^2 + e^{-2\gamma} (d\eta^2 + d\lambda^2) + e^{-4\gamma} (d\omega + \lambda d\eta)^2
\label{eq:a35}
\end{equation}
which can be recognized as a (global) coordinate representation of complex hyperbolic space. If the Maxwell field is `turned off' so that only vacuum spacetimes are considered then
\begin{equation}
dk^2 \longrightarrow 4(d\gamma)^2 + e^{-4\gamma}(d\omega)^2
\label{eq:a36}
\end{equation}
which, defined over \(\mathbb{R}^2\), is nothing but a coordinate representation for real hyperbolic space. Some background on this 4-dimensional target space and its 8-dimensional isometry group \(\mathcal{S}U(2,1)\) is given in Ref.~\cite{Mon_90} and in further references cited therein, and will not be included here. In particular though Eq.~(2.60) of this reference lists, explicitly, a set of eight (locally) conserved quantities that one builds appealing to Noether's theorem from the eight independent Killing fields of the target metric.

To reconstruct an Einstein-Maxwell field on \textit{V} from a solution to the reduced field equations on \(\mathbb{R} \times M_b\) one needs to reconstruct the one forms \(\beta_0 dt + \beta_a dx^a\) and \(C_0 dt + C_a dx^a\) of which only the `transverse projections' \(\tilde{r} = \epsilon^{ab}\beta_{a,b}\) and \(\tilde{u} = \epsilon^{ab} C_{a,b}\) directly survive (as momenta conjugate to the wave map variables \(\omega\) and \(\eta\)) in the reduced formulation. The time components, \(\beta_0\) and \(C_0\), of the one-forms are essentially gauge variables and can be chosen arbitrarily together with initial data for \(\beta_a\) and \(C_a\) compatible with (\ref{eq:a28}). To recover \(\beta_a\) and \(C_a\) one integrates the Hamiltonian equations for these quantities, which, expressed in terms of wave map variables, take the form
\begin{align}
\beta_{a,t} &= \beta_{0,a} + \tilde{N}^b \epsilon_{ab}\tilde{r} + \frac{\tilde{N}}{\mu_{{}^{(2)}\!\tilde{g}}} e^{-4\gamma} \tilde{g}_{ab}\epsilon^{bc} (\omega_{,c} + \lambda\eta_{,c})\label{eq:a37}\\
C_{a,t} &= C_{0,a} + \frac{\tilde{N}e^{-2\gamma}}{\mu_{{}^{(2)}\!\tilde{g}}} \tilde{g}_{ab}\epsilon^{bc}\eta_{,c} + \tilde{N}^b \epsilon_{ab}\tilde{u} + \frac{\tilde{N}e^{-4\gamma}}{\mu_{{}^{(2)}\!\tilde{g}}} \tilde{g}_{ab} \lambda \left\lbrack \epsilon^{bc} (\omega_{,c} + \lambda\eta_{,c})\right\rbrack\label{eq:a38}
\end{align}
Upon reverting to the original notation one finds that Eqs.~(\ref{eq:a37}) and (\ref{eq:a38}) are indeed equivalent to the original Hamilton equations for these fields (derivable from the action \(\tilde{I}_\Omega\)) and that they guarantee preservation of the defining equations given in (\ref{eq:a28}). The remaining Hamiltonian evolution and constraint equations also revert to their original forms.

Needless to say all of the above equations are automatically satisfied by the Kerr-Newman fields. Our main aim is to study linear perturbations of these `backgrounds' and, in particular, to do so within the reduced Hamiltonian framework sketched above. To this end however it is first necessary to compute the twist potentials \(\omega\) and \(\eta\) for these Kerr-Newman backgrounds since these potentials cannot be simply read off the explicit formulas for \({}^{(4)}\!g\) and \({}^{(4)}\!A\).

From the formulas given in Appendix~\ref{app:kerr-newman-spacetimes} one sees immediately that, in the chosen coordinate systems, \(\beta_a = A_a = 0\) from which it follows, via the definitions (\ref{eq:a19}) and (\ref{eq:a28}) that \(C_a = 0, \tilde{r} = 0\) and \(\tilde{u} = 0\). Noting also that the (2+1-dimensional) shift vector field \(X = \tilde{N}^a \frac{\partial}{\partial x^a}\) vanishes as well one sees, from Eqs.~(\ref{eq:a29}) and (\ref{eq:a30}) that \(\omega_{,t} = \eta_{,t} = 0\), as one should have expected for a stationary solution. From the Hamilton equations for \(\tilde{g}_{ab},\gamma\) and \(\lambda\) it also follows that \(\tilde{\pi}^{ab} = \tilde{p} = \tilde{v} = 0\) for these (stationary) Kerr-Newman backgrounds.

Reading off the (Boyer-Lindquist) coordinate expressions
\begin{align}
\beta_0 &= \frac{-a (2Mr - Q^2)}{\left\lbrack(r^2 + a^2)^2 - a^2 \Delta \sin^2{\theta}\right\rbrack},\label{eq:a39}\\
\lambda &= \frac{2Qra\sin^2{\theta}}{r^2 + a^2\cos^2{\theta}},\label{eq:a40}\\
C_0 &= \frac{2Qr(r^2 + a^2)}{\left\lbrack(r^2 + a^2)^2 - a^2 \Delta \sin^2{\theta}\right\rbrack},\label{eq:a41}\\
\tilde{N} &= \Delta^{1/2} \sin{\theta},\label{eq:a42}\\
e^{2\gamma} &= \left(\frac{\sin^2{\theta}}{r^2 + a^2 \cos^2{\theta}}\right) \left\lbrack(r^2 + a^2)^2 - a^2 \Delta \sin^2{\theta}\right\rbrack\label{eq:a43}
\end{align}
and
\begin{equation}
\left(\frac{1}{\mu_{{}^{(2)}\!\tilde{g}}}\right) \tilde{g}_{ab} dx^a dx^b = \Delta^{-1/2} dr^2 + \Delta^{1/2} d\theta^2
\label{eq:a44}
\end{equation}
where \(\Delta = r^2 - 2Mr + a^2 + Q^2\) and substituting these expressions into Eqs.~(\ref{eq:a37}) and (\ref{eq:a38}) one arrives at a system of first order linear equations for the unknowns \(\omega\) and \(\eta\). The integrability conditions for this system are readily verified and the system integrated to yield (with a particularly simple choice for the arbitrary additive constants)
\begin{align}
\eta &= \frac{-4Q (a^2 + r^2) \cos{(\theta)}}{a^2 + 2r^2 + a^2 \cos{(2\theta)}}\label{eq:a45}\\
\intertext{and}
\omega &= aM \cos{(\theta)}\left(5 - \cos{(2\theta)}\right)\label{eq:a46}\\
 &\mbox + \left(\frac{4a^3\cos{(\theta)}\sin^4{(\theta)}\left\lbrack a^2M + 2r(Q^2 + Mr) + a^2 M\cos{(2\theta)}\right\rbrack}{\left(a^2 + 2r^2 + a^2\cos{(2\theta)}\right)^2}\right)\nonumber
\end{align}
Note that these yield
\begin{equation}
\eta(r,0) - \eta(r,\pi) = -4Q\label{eq:a47}
\end{equation}
and
\begin{equation}
\omega(r,0) - \omega(r,\pi) = 8aM\label{eq:a48}
\end{equation}
for the (unambiguous) differences of these functions on the upper and lower symmetry axes (which thread through `wormholes' in the analytically extended black hole spacetimes and are actually disjoint).

Though one can readily derive the reduced field equations by variation of the reduced action \(\tilde{J}_\Omega\) (c.f. Eqs.~(\ref{eq:a32})--(\ref{eq:a34})) we present them here explicitly to lay the groundwork for their linearization. The evolution equations for the canonical pairs \(\left\lbrace(\gamma,\tilde{p}), (\omega,\tilde{r}), (\eta,\tilde{u}), (\lambda,\tilde{v}), (\tilde{g}_{ab},\tilde{\pi}^{ab}\right\rbrace\) are given by:
{\allowdisplaybreaks
\begin{align}
\gamma_{,t} &= \frac{\tilde{N}\tilde{p}}{4\mu_{{}^{(2)}\!\tilde{g}}} + \mathcal{L}_X\gamma,\label{eq:a49}\\
&\nonumber\\
\tilde{p}_{,t} &= \left\lbrace\frac{-2\tilde{N}}{\mu_{{}^{(2)}\!\tilde{g}}} e^{4\gamma}(\tilde{r})^2 - \frac{\tilde{N}}{\mu_{{}^{(2)}\!\tilde{g}}} e^{2\gamma}\left(\tilde{v}^2 + (\tilde{u} - \lambda\tilde{r})^2\right)\right.\label{eq:a50}\\
& + 4(\tilde{N}\mu_{{}^{(2)}\!\tilde{g}}\> \tilde{g}^{ab}\gamma_{,a})_{,b} + \tilde{N}\mu_{{}^{(2)}\!\tilde{g}}\> e^{-2\gamma}\tilde{g}^{ab} (\eta_{,a}\eta_{,b} + \lambda_{,a}\lambda_{,b})\nonumber\\
& + \left.\vphantom{\frac{-2\tilde{N}}{\mu_{{}^{(2)}\!\tilde{g}}}} 2\tilde{N}\mu_{{}^{(2)}\!\tilde{g}}\> e^{-4\gamma}\tilde{g}^{ab} (\omega_{,a} + \lambda\eta_{,a})(\omega_{,b} + \lambda\eta_{,b}) + \mathcal{L}_X\tilde{p}\right\rbrace,\nonumber\\
&\nonumber\\
\omega_{,t} &= \frac{\tilde{N}e^{4\gamma}}{\mu_{{}^{(2)}\!\tilde{g}}} \tilde{r} + \frac{\tilde{N}e^{2\gamma}}{\mu_{{}^{(2)}\!\tilde{g}}} \lambda (\lambda\tilde{r} - \tilde{u}) + \mathcal{L}_X\omega,\label{eq:a51}\\
&\nonumber\\
\tilde{r}_{,t} &= \left\lbrace\left(\tilde{N}\mu_{{}^{(2)}\!\tilde{g}}\> e^{-4\gamma}\tilde{g}^{ab}(\omega_{,a} + \lambda\eta_{,a})\right)_{,b} + \mathcal{L}_X\tilde{r}\right\rbrace,\label{eq:a52}\\
&\nonumber\\
\eta_{,t} &= \frac{\tilde{N}e^{2\gamma}}{\mu_{{}^{(2)}\!\tilde{g}}} (\tilde{u} - \lambda\tilde{r}) + \mathcal{L}_X\eta,\label{eq:a53}\\
&\nonumber\\
\tilde{u}_{,t} &= \left\lbrace(\tilde{N}\mu_{{}^{(2)}\!\tilde{g}}\> e^{-2\gamma}\tilde{g}^{ab}\eta_{,a})_{,b} + \left(\tilde{N}\mu_{{}^{(2)}\!\tilde{g}}\> e^{-4\gamma}\tilde{g}^{ab} \lambda(\omega_{,a} + \lambda\eta_{,a})\right)_{,b} + \mathcal{L}_X\tilde{u}\right\rbrace,\label{eq:a54}\\
&\nonumber\\
\lambda_{,t} &= \frac{\tilde{N}e^{2\gamma}\tilde{v}}{\mu_{{}^{(2)}\!\tilde{g}}} + \mathcal{L}_X\lambda,\label{eq:a55}\\
&\nonumber\\
\tilde{v}_{,t} &= \left\lbrace\frac{\tilde{N}}{\mu_{{}^{(2)}\!\tilde{g}}}e^{2\gamma}\tilde{r} (\tilde{u} - \lambda\tilde{r}) + (\tilde{N}\mu_{{}^{(2)}\!\tilde{g}}\> e^{-2\gamma}\tilde{g}^{ab}\lambda_{,a})_{,b}\right.\label{eq:a56}\\
 &\left.\vphantom{\frac{-2\tilde{N}}{\mu_{{}^{(2)}\!\tilde{g}}}} - \tilde{N}\mu_{{}^{(2)}\!\tilde{g}}\> e^{-4\gamma}\tilde{g}^{ab} (\omega_{,a} + \lambda\eta_{,a})\eta_{,b} + \mathcal{L}_X\tilde{v}\right\rbrace,\nonumber\\
 &\nonumber\\
\tilde{g}_{ab,t} &= \frac{2\tilde{N}}{\mu_{{}^{(2)}\!\tilde{g}}} (\tilde{g}_{ac}\tilde{g}_{bd} - \tilde{g}_{ab}\tilde{g}_{cd})\tilde{\pi}^{cd} + (\mathcal{L}_X {}^{(2)}\!\tilde{g})_{ab},\label{eq:a57}\\
&\nonumber\\
\tilde{\pi}^{ab}_{,t} &= \left\lbrace\frac{-2\tilde{N}}{\mu_{{}^{(2)}\!\tilde{g}}} \lbrack\tilde{\pi}^{ac}\tilde{\pi}^{bd}\tilde{g}_{cd} - \tilde{\pi}^{ab}\tilde{\pi}^c_c\rbrack + (\mathcal{L}_X\tilde{\pi})^{ab}\right.\label{eq:a58}\\
&+ \frac{1}{2} \frac{\tilde{N}}{\mu_{{}^{(2)}\!\tilde{g}}} \tilde{g}^{ab}\left\lbrack\tilde{\pi}^{cd}\tilde{\pi}_{cd} - (\tilde{\pi}_c^c)^2\right\rbrack + \mu_{{}^{(2)}\!\tilde{g}} (\tilde{N}^{|ab} - \tilde{g}^{ab}\tilde{N}_{\hphantom{|c}|c}^{|c})\nonumber\\
&+ \frac{1}{2} \frac{\tilde{N}}{\mu_{{}^{(2)}\!\tilde{g}}} \tilde{g}^{ab} \left\lbrack\frac{1}{8} (\tilde{p})^2 + \frac{1}{2} e^{4\gamma} (\tilde{r})^2 + \frac{1}{2} e^{2\gamma} \left(\tilde{v}^2 + (\tilde{u} - \lambda\tilde{r})^2\right)\right\rbrack\nonumber\\
&+ \tilde{N}\mu_{{}^{(2)}\!\tilde{g}} \left(\tilde{g}^{ac}\tilde{g}^{bd} - \frac{1}{2}\tilde{g}^{ab}\tilde{g}^{cd}\right) \left\lbrack 2\gamma_{,c}\gamma_{,d} + \frac{1}{2} e^{-2\gamma} (\eta_{,c}\eta_{,d} + \lambda_{,c}\lambda_{,d})\right.\nonumber\\
&\left.\vphantom{\frac{-2\tilde{N}}{\mu_{{}^{(2)}\!\tilde{g}}}}\left. + \frac{1}{2} e^{-4\gamma} (\omega_{,c} + \lambda\eta_{,c}) (\omega_{,d} + \lambda\eta_{,d})\right\rbrack\right\rbrace\nonumber
\end{align}
}
whereas the constraints are now simply
\begin{equation}
\tilde{\mathcal{H}} = 0\qquad \text{ and }\qquad \tilde{\mathcal{H}}_a = 0
\label{eq:a59}
\end{equation}
with \(\tilde{\mathcal{H}}\) and \(\tilde{\mathcal{H}}_a\) defined by Eqs.~(\ref{eq:a33}) and (\ref{eq:a34}). In the above formulas the Lie derivatives with respect to \(X = \tilde{N}^a \frac{\partial}{\partial x^a}\) of scalars (\(\gamma, \omega, \eta, \lambda\)) are simply their directional derivatives with, for example,
\begin{equation}
\mathcal{L}_X\gamma = \tilde{N}^a\gamma_{,a},
\label{eq:a60}
\end{equation}
whereas those of the scalar \textit{densities} (\(\tilde{p}, \tilde{r}, \tilde{u}, \tilde{v}\)) are
\begin{equation}
\mathcal{L}_X\tilde{p} = (\tilde{N}^a\tilde{p})_{,a},\qquad \text{etc.}
\label{eq:a61}
\end{equation}
while those for the tensor \({}^{(2)}\!\tilde{g} = \tilde{g}_{ab} dx^a \otimes dx^b\) and tensor density \({}^{(2)}\!\tilde{\pi} := \tilde{\pi}^{ab} \frac{\partial}{\partial x^a} \otimes \frac{\partial}{\partial x^b}\) are
\begin{equation}
\begin{split}
(\mathcal{L}_X {}^{(2)}\!\tilde{g})_{ab} &= \tilde{N}^c\tilde{g}_{ab,c} + \tilde{N}^c_{,a}\tilde{g}_{cb} + \tilde{N}^c_{,b}\tilde{g}_{ac}\\
&= \tilde{N}_{a|b} + \tilde{N}_{b|a}
\label{eq:a62}
\end{split}
\end{equation}
and
\begin{equation}
(\mathcal{L}_X {}^{(2)}\!\tilde{\pi})^{ab} = (\tilde{N}^c\tilde{\pi}^{ab})_{,c} - \tilde{N}^a_{,c}\tilde{\pi}^{cb} - \tilde{N}^b_{,c}\tilde{\pi}^{ac}
\label{eq:a63}
\end{equation}
respectively. The two dimensional indices \(a, b, \ldots\) are raised and lowered using \({}^{(2)}\!\tilde{g}\) and \({}^{(2)}\!\tilde{g}^{-1} := \tilde{g}^{ab} \frac{\partial}{\partial x^a} \otimes \frac{\partial}{\partial x^b}\) whereas covariant differentiation with respect to \({}^{(2)}\!\tilde{g}\) is designated by a vertical bar.

The last two of equations (\ref{eq:a49})--(\ref{eq:a58}) together with the constraints (\ref{eq:a59}) comprise the 2+1-dimensional Einstein equations with a wave map source whereas the first eight of these equations are the corresponding (curved space) wave map equations in Hamiltonian form.

The Kerr-Newman solutions given explicitly in Appendix~\ref{app:kerr-newman-spacetimes} are of course stationary and have vanishing (2+1-dimensional) shift, \(X = \tilde{N}^{a} \frac{\partial}{\partial x^a} = 0\). It follows immediately from Eqs.~(\ref{eq:a49})--(\ref{eq:a58}) that all of the canonical momenta vanish, i.e., that
\begin{equation}
\tilde{p} = \tilde{r} = \tilde{u} = \tilde{v} = \tilde{\pi}^{ab} = 0
\label{eq:a64}
\end{equation}
and therefore that the evolution equations reduce to
{\allowdisplaybreaks
\begin{align}
&\left\lbrace 4(\tilde{N}\mu_{{}^{(2)}\!\tilde{g}}\> \tilde{g}^{ab}\gamma_{,a})_{,b} + \tilde{N}\mu_{{}^{(2)}\!\tilde{g}}\> e^{-2\gamma}\tilde{g}^{ab}(\eta_{,a}\eta_{,b} + \lambda_{,a}\lambda_{,b})\right.\label{eq:a65}\\
 &\left.\mbox{} + 2\tilde{N}\mu_{{}^{(2)}\!\tilde{g}}\> e^{-4\gamma}\tilde{g}^{ab} (\omega_{,a} + \lambda\eta_{,a})(\omega_{,b} + \lambda\eta_{,b})\right\rbrace = 0,\nonumber\\
&\nonumber\\
&\left\lbrace\left(\tilde{N}\mu_{{}^{(2)}\!\tilde{g}}\> e^{-4\gamma}\tilde{g}^{ab} (\omega_{,a} + \lambda\eta_{,a})\right)_{,b}\right\rbrace = 0,\label{eq:a66}\\
&\nonumber\\
&\left\lbrace(\tilde{N}\mu_{{}^{(2)}\!\tilde{g}}\> e^{-2\gamma}\tilde{g}^{ab}\eta_{,a})_{,b} + \left(\tilde{N}\mu_{{}^{(2)}\!\tilde{g}}\> e^{-4\gamma}\tilde{g}^{ab} \lambda(\omega_{,a} + \lambda\eta_{,a})\right)_{,b}\right\rbrace = 0,\label{eq:a67}\\
&\nonumber\\
&\left\lbrace(\tilde{N}\mu_{{}^{(2)}\!\tilde{g}}\> e^{-2\gamma}\tilde{g}^{ab} \lambda_{,a})_{,b} - \tilde{N}\mu_{{}^{(2)}\!\tilde{g}}\> e^{-4\gamma}\tilde{g}^{ab} (\omega_{,a} + \lambda\eta_{,a})\eta_{,b}\right\rbrace = 0,\label{eq:a68}\\
&\nonumber\\
&\left\lbrace\mu_{{}^{(2)}\!\tilde{g}} (\tilde{N}^{|ab} - \tilde{g}^{ab}\tilde{N}^{|c}_{~|c}) + \tilde{N}\mu_{{}^{(2)}\!\tilde{g}} \left(\tilde{g}^{ac}\tilde{g}^{bd} - \frac{1}{2} \tilde{g}^{ab}\tilde{g}^{cd}\right)\times\right.\label{eq:a69}\\
 &\left.\left\lbrack 2\gamma_{,c}\gamma_{,d} + \frac{1}{2} e^{-2\gamma} (\eta_{,c}\eta_{,d} + \lambda_{,c}\lambda_{,d}) + \frac{1}{2} e^{-4\gamma} (\omega_{,c} + \lambda\eta_{,c}) (\omega_{,d} + \lambda\eta_{,d})\right\rbrack\right\rbrace = 0,\nonumber
\end{align}
}
whereas the Hamiltonian constraint, \(\tilde{\mathcal{H}} = 0\), takes the form
\begin{equation}
\begin{split}
\mu_{{}^{(2)}\!\tilde{g}} \left\lbrack -{}^{(2)}\!\tilde{R} + 2\tilde{g}^{ab}\gamma_{,a}\gamma_{,b} + \frac{1}{2} e^{-2\gamma}\tilde{g}^{ab} (\eta_{,a}\eta_{,b} + \lambda_{,a}\lambda_{,b})\right.\\
 \left.\vphantom{\frac{1}{2}} \mbox + \frac{1}{2} e^{-4\gamma}\tilde{g}^{ab} (\omega_{,a} + \lambda\eta_{,a})(\omega_{,b} + \lambda\eta_{,b})\right\rbrack = 0
\label{eq:a70}
\end{split}
\end{equation}
while the momentum constraint, \(\tilde{\mathcal{H}}_a = 0\), is satisfied identically.

Note especially that the trace of Eq.~(\ref{eq:a69}) results in the formula
\begin{equation}
\tilde{N}^{|c}_{~|c} = 0.
\label{eq:a71}
\end{equation}
This fact that the (2+1-dimensional) lapse for Kerr-Newman solutions is harmonic will play an important role in our treatment of the linearized equations.
 %C; orginally appendix A
\section{Covariance and regularity of the fundamental wavemap fields}
\label{app:coveriance}
It is clear from their definitions in terms of the axial Killing field, \(\psi = \psi^\mu \frac{\partial}{\partial x^\mu} \rightarrow \frac{\partial}{\partial\varphi}\), that the wavemap variables
\begin{equation}\label{eq:g01}
e^{2\gamma} := {}^{(4)}g_{\mu\nu}\; \psi^\mu \psi^\nu \rightarrow {}^{(4)}g_{\varphi\varphi}
\end{equation}
and
\begin{equation}\label{eq:g02}
\lambda := \psi^\mu\; {}^{(4)}\!A_\mu \rightarrow A_3 = A_\varphi
\end{equation}
both transform as spacetime \textit{scalars}\footnote{\label{note04}Note that \(\lambda\) is in fact also invariant with respect to electromagnetic gauge transformations since we only admit those transformations that preserve explicit axial symmetry.}. On the other hand the covariance properties of the complementary variables, \(\omega\) and \(\eta\), are not immediately evident from our (reduced Hamiltonian framework) introduction of these objects in Appendix~\ref{app:reduced-hamiltonian}. As we shall show herein however, all the wavemap fields do indeed transform as spacetime \textit{scalars}. It will then follow that their corresponding first variations, \(\lbrace\gamma', \lambda', \omega', \eta'\rbrace\), undergo linearized gauge transformations of the familiar form
\begin{align}
\gamma' &\rightarrow \gamma' + \mathcal{L}_{{}^{(4)}Y}\gamma & \lambda' &\rightarrow \lambda' + \mathcal{L}_{{}^{(4)}Y}\lambda, \label{eq:g03}\\
\omega' &\rightarrow \omega' + \mathcal{L}_{{}^{(4)}Y}\omega & \eta' &\rightarrow \eta' + \mathcal{L}_{{}^{(4)}Y}\eta \label{eq:g04}
\end{align}
where \({}^{(4)}Y = {}^{(4)}Y^\mu \frac{\partial}{\partial x^\mu}\) is an arbitrary spacetime vector field that commutes with \(\psi\).

Recall that, in the absence of sources, both the electromagnetic 2-form field \({}^{(4)}F\) and its Hodge dual \(\star {}^{(4)}F\) are closed,
\begin{equation}\label{eq:g05}
d\; {}^{(4)}F = 0,\qquad d\; \star {}^{(4)}F = 0.
\end{equation}
Combined with its invariance under axial rotations,
\begin{equation}\label{eq:g06}
\mathcal{L}_\psi\; \star {}^{(4)}F = 0,
\end{equation}
the closure of \(\star {}^{(4)}F\) implies the closure of the corresponding 1-form field
\begin{equation}\label{eq:g07}
\begin{split}
{}^{(4)}\Omega &= {}^{(4)}\Omega_\mu dx^\mu := \psi^\mu\; \star {}^{(4)}F_{\mu\nu} dx^\nu\\
 &= \frac{1}{2} \left(\frac{1}{\sqrt{-\det {}^{(4)}g}}\right) \psi_\mu\; \epsilon^{\mu\nu\alpha\beta}\; {}^{(4)}F_{\alpha\beta} {}^{(4)}g_{\nu\gamma}\; dx^\gamma
\end{split}
\end{equation}
and thus, on any simply connected domain such as the domain of outer communications (DOC) of a black hole, the \textit{exactness} of \({}^{(4)}\Omega\).

In fact, by direct evaluation of the right hand side of the defining formula (\ref{eq:g07}) in terms of our variables one arrives at
\begin{equation}\label{eq:g08}
{}^{(4)}\Omega = d\eta = \eta_{,\gamma} dx^\gamma
\end{equation}
and thus concludes that our wavemaps field \(\eta\) is indeed a spacetime scalar.

Finally, consider the 1-form field
\begin{equation}\label{eq:g09}
\begin{split}
{}^{(4)}\Delta &= {}^{(4)}\Delta_\gamma dx^\gamma :=\\
& {} \frac{1}{2} \frac{1}{\sqrt{-\det {}^{(4)}g}}\; \epsilon^{\mu\nu\alpha\beta} \psi_\mu (\partial_\alpha\psi_\beta - \partial_\beta\psi_\alpha)\; {}^{(4)}g_{\nu\gamma}\; dx^\gamma
\end{split}
\end{equation}
constructed covariantly from the Killing 1-form \(\psi_\mu dx^\mu\) and its exterior derivative. Evaluating the right hand side of this expression in terms of our variables one arrives at
\begin{equation}\label{eq:g10}
\begin{split}
{}^{(4)}\Delta &= -\left\lbrace d\omega + \lambda d\eta\right\rbrace\\
 &= {} - \left\lbrace\omega_{,\mu} dx^\mu + \lambda\eta_{,\mu} dx^\mu\right\rbrace
\end{split}
\end{equation}
and thus concludes that the remaining wavemap field, \(\omega\), does indeed transform as a spacetime scalar.  %D; orginally appendix G
\section{Electric Charge and Angular Momentum Conservation Laws}
\label{app:electric-charge}
The electric flux of a Maxwell field \({}^{(4)}\!F\) through a closed, connected and orientable 2-surface \({}^{(2)}\!\Sigma\) is defined by the integral of its dual 2-form, \(\star {}^{(4)}\!F\), over \({}^{(2)}\!\Sigma\) where, in coordinates,
\begin{equation}\label{eq:i01}
\star {}^{(4)}\!F = \frac{1}{2} \left(\star {}^{(4)}\!F\right)_{\mu\nu}\; dx^\mu \wedge dx^\nu
\end{equation}
with
\begin{equation}\label{eq:i02}
\left(\star {}^{(4)}\!F\right)_{\mu\nu} = \frac{1}{2} \sqrt{-\det{{}^{(4)}\!g}}\; \epsilon_{\mu\nu\alpha\beta}\; {}^{(4)}\!F^{\alpha\beta}.
\end{equation}
If, for example, \({}^{(2)}\!\Sigma\) bounds a 3-ball \textit{B} lying in a spacelike hypersurface then the electric charge \(Q_B\) contained in that ball would be given, in our slightly non-standard conventions\footnote{\label{note05}Recall that we have absorbed a factor of 2 into \({}^{(4)}\!F\) and its ADM representatives \(\left\lbrace{}^{(4)}\!A, {}^{(3)}\mathcal{E}, {}^{(3)}\!B\right\rbrace\) to `normalize' the form of Hamilton's equations.}, by
\begin{equation}\label{eq:i03}
8\pi Q_B = \int_{{}^{(2)}\!\Sigma = \partial B} \star {}^{(4)}\!F\\
\end{equation}

In the case of a black hole however the presence of non-vanishing flux through a 2-surface surrounding its event horizon may simply be a measure of `field lines trapped in the topology of space' with no actual source current \(\star j\) for the Maxwell field necessarily existing in the spacetime (Wheeler's `charge without charge'). This is indeed the case for the maximally analytically extended Kerr-Newman black hole spacetimes which are global solutions to the pure electrovacuum field equations.

By the same (topological trapping) mechanism a stationary black hold solution can exhibit a non-vanishing \textit{magnetic} flux (the integral of \({}^{(4)}\!F\) itself over a surface surrounding the event horizon) without the necessity of actual magnetic monopoles existing in the (topologically non-trivial) spacetime. But since one expects, on astrophysical grounds, that actual black holes in the Universe are created from the collapse of ordinary material sources preexisting in topologically trivial space (e.g., rotating stars), such objects could certainly be electrically, but presumably not magnetically, charged. For this reason we herein exclude the consideration of a non-vanishing magnetic flux, both for the background black hole spacetime and its perturbations.

In view of the axial symmetry of the Kerr-Newman black holes we can exploit the formalism developed herein to evaluate the (electric) charge integral,
\begin{equation}\label{eq:i04}
8\pi Q = \int_{{}^{(2)}\!\Sigma} \star {}^{(4)}\!F,
\end{equation}
(over a surface \({}^{(2)}\!\Sigma\) surrounding the event horizon) in terms of the values of the wave map potential function \(\eta\) taken on the axes of symmetry. For simplicity let us evaluate this integral over the (topologically spherical) surface \({}^{(2)}\!\Sigma\) defined in the Boyer-Lindquist type coordinates of Appendix~\ref{app:kerr-newman-spacetimes} by \(R = R_0 = \text{constant} > R_+\) and \(t = t_0\). Recalling that, in these coordinates, the axial Killing field \(\psi = \psi^\mu \frac{\partial}{\partial x^\mu} \longrightarrow \frac{\partial}{\partial\varphi}\) we get, by direct calculation
\begin{equation}\label{eq:i05}
\begin{split}
\int_{{}^{(2)}\!\Sigma} \star {}^{(4)}\!F &= \int_{{}^{(2)}\!\Sigma} \frac{1}{2} \sqrt{-\det{{}^{(4)}\!g}}\; \epsilon_{\theta\varphi\alpha\beta}\; {}^{(4)}\!F^{\alpha\beta}\; d\theta \wedge d\varphi \\
 &= -\frac{1}{2} \int_{{}^{(2)}\!\Sigma} \frac{1}{\sqrt{-\det{{}^{(4)}\!g}}}\; {}^{(4)}\!F_{\alpha\beta}\; \epsilon^{\mu\nu\alpha\beta}\; {}^{(4)}\!g_{\theta\mu}\; \psi_\nu\; d\theta d\varphi\\
 &= 2\pi \int_0^\pi \frac{1}{2} \frac{1}{\sqrt{-\det{{}^{(4)}\!g}}}\; \psi_\nu\; \epsilon^{\nu\mu\alpha\beta}\; {}^{(4)}\!F_{\alpha\beta}\; {}^{(4)}\!g_{\mu\theta} d\theta\\
 &= 2\pi \int_0^\pi \eta_{,\theta}\; d\theta\\
 &= 2\pi \left(\eta (R_0,\pi) - \eta (R_0,0)\right)
\end{split}
\end{equation}
where we have, in the final steps, appealed to Eqs.~(\ref{eq:g07}) and (\ref{eq:g08}). This result reproduces the observation made incidentally in (\ref{eq:a47}) while now justifying the identification of the parameter \textit{Q} occurring in the Kerr-Newman solution with electric charge.

From the defining formula (\ref{eq:a25}) and the fact that the electric vector density \(\mathcal{E}^a \frac{\partial}{\partial x^a}\) must, for reasons of regularity, have a vanishing \(\theta\)-component along the axes of symmetry it follows that \(\eta\) must be constant along each of these axes so that both \(\eta (R,\pi)\) and \(\eta (R,0)\) are independent of \textit{R}.

A straightforward linearization of the above argument leads to the corresponding perturbative formula
\begin{equation}\label{eq:i06}
4Q' = \eta' (t,\theta = \pi) - \eta' (t, \theta = 0)
\end{equation}
which, at first glance, would seem to allow for a time dependent perturbed charge. However, by combining the linearizations of Eqs.~(\ref{eq:a28}) and (\ref{eq:a29}) with the axis regularity results of Ref.~\cite{Rinne_05} one finds that both \(\tilde{r}'\) and \(\tilde{u}'\) vanish to order \(O(\sin{\theta})\) at the axes of symmetry and, combined with a decay result for \(\mathcal{L}_{X'}\eta\), that \(\eta'_{,t}\) actually vanishes to order \(O(\sin^2{\theta})\) at these axes. It follows that \(\eta'(t, \pi)\) and \(\eta' (t, 0)\) are both independent of \textit{t}.

While one could thus allow the perturbation of \(\eta\) to incorporate a corresponding perturbation of the conserved electric charge there is little or no reason for doing so. One can simply insist that the given `unperturbed' black hole have the full charge desired for the final, perturbed object and thus demand, without serious loss of generality, that \(\eta'\) actually vanish on both symmetry axes. Only this choice is compatible with the natural perturbative boundary condition that \(\eta'\) should vanish at infinity --- an assumption that we shall impose herein.

A similar argument can be given for the evaluation of the total angular momentum of a Kerr-Newman black hole and for that of its axisymmetric perturbations by appealing to Komar's famous flux formula for such cases \cite{Wald_84}. Komar's formula states the total angular momentum \textit{J} is given by the flux integral
\begin{equation}\label{eq:i07}
16\pi J =  \int_{{}^{(2)}\!\Sigma} \star d\psi,
\end{equation}
where \(\psi = \psi_\mu dx^\mu\), the covariant form of the axial Killing field, with the proviso that now, in order to include contributions from material sources such as the electromagnetic field, the integral should be evaluated in the limit that the `radius' of the integration surface \({}^{(2)}\!\Sigma\) tends to infinity. Note that the Killing 1-form \(\psi_\mu dx^\mu\) plays here a role analogous to that of the `vector potential' \(A_\mu dx^\mu\) in the case of electric charge.

A direct evaluation of this flux integral over the (topologically spherical) surface \({}^{(2)}\!\Sigma_{R_0}\) of Boyer-Lindquist `radius' \(R_0\) gives
\begin{equation}\label{eq:i08}
\begin{split}
\int_{{}^{(2)}\!\Sigma_{R_0}} \star d\psi &= \frac{1}{2} \int_{{}^{(2)}\!\Sigma_{R_0}} \frac{1}{\sqrt{-\det{{}^{(4)}\!g}}}\; (\partial_\alpha\psi_\beta - \partial_\beta\psi_\alpha)\; \epsilon^{\nu\mu\alpha\beta}\; \psi_\nu\; {}^{(4)}\!g_{\theta\mu}\; d\theta d\varphi\\
 &= 2\pi \int_0^\pi \frac{1}{2} \left\lbrace\frac{1}{\sqrt{-\det{{}^{(4)}\!g}}}\; \psi_\nu\; \epsilon^{\nu\mu\alpha\beta} (\partial_\alpha\psi_\beta - \partial_\beta\psi_\alpha)\; {}^{(4)}\!g_{\mu\theta}\right\rbrace_{R=R_0}\; d\theta\\
 &= 2\pi \int_0^\pi \left.\left\lbrace\vphantom{\frac{1}{2}} -(\omega_{,\theta} + \lambda\eta_{,\theta})\right\rbrace\right|_{R=R_0}\; d\theta
\end{split}
\end{equation}
where we have, in the final step, appealed to Eqs.~(\ref{eq:g09}) and (\ref{eq:g10}).

In the limit that \(R_0 \rightarrow \infty\) the contribution proportional to \(\lambda\) (c.f. Eq.~(\ref{eq:a40})) drops out leaving
\begin{equation}\label{eq:i09}
8J = \lim_{R_0 \rightarrow \infty} \left\lbrace\omega (R_0,0) - \omega (R_0,\pi)\right\rbrace
\end{equation}
By an argument completely analogous to that given above for \(\eta\) though one finds that \(\omega (R_0, \theta = \pi)\) and \(\omega (R_0, \theta = 0)\) are both independent of \(R_0\) so that one recovers Eq.~(\ref{eq:a48}) together with the identification that \(J = aM\) in terms of the Kerr-Newman parameter \textit{a}.

A straightforward linearization of the above argument leads to the corresponding perturbation formula
\begin{equation}\label{eq:i10}
8J' = \left\lbrace\omega' (t,\theta = 0) - \omega' (t,\theta = \pi)\right\rbrace
\end{equation}
which would seem to allow for a time dependent perturbed angular momentum. But a straightforward linearization of Eq.~(\ref{eq:a30}), combined with the aforementioned results for \(\tilde{r}'\) and \(\tilde{u}'\) and an appeal to Ref.~\cite{Rinne_05} for the evaluation of \(\mathcal{L}_{X'}\omega\), shows that \(\omega'_{,t}\) vanishes to order \(O(\sin^4{\theta})\) at the axes of symmetry. It follows that \(\omega' (t, 0)\) and \(\omega' (t, \pi)\) are both independent of \textit{t}.

While one could thus allow the perturbation of \(\omega\) to reflect a corresponding perturbation in the \textit{conserved} angular momentum there is, as was already noted for the case of electric charge, no reason for doing so. Again one can simply demand that the given `unperturbed' Kerr-Newman black hole have the total angular momentum desired for the final, perturbed object and thus take \(\omega'\) to actually vanish on both symmetry axes. We thus assume herein, without any essential loss of generality, that the perturbations are taken to satisfy \(J' = 0\) and \(Q' = 0\).

The formulas, corresponding to Eqs.~(\ref{eq:i04}) and (\ref{eq:i05}) above, for the \textit{magnetic} flux threading through a 2-surface \({}^{(2)}\Sigma\) surrounding (at \(t = t_0\) and \( R = R_0 > R_+\)) the black hole's event horizon are given (again in our slightly non-standard conventions) by
\begin{equation}\label{eq:i11}
\begin{split}
8\pi Q^{\mathrm{mag}} &= \int_{{}^{(2)}\Sigma} {}^{(4)}F = \int_{{}^{(2)}\Sigma} {}^{(4)}F_{\theta\varphi}\; d\theta d\varphi\\
 &= \int_{{}^{(2)}\Sigma} (\partial_\theta\lambda)\; d\theta d\varphi = 2\pi \int_0^\pi \partial_\theta\lambda\; d\theta\\
 &= 2\pi \left(\lambda (R_0,\pi) - \lambda (R_0,0)\right).
\end{split}
\end{equation}
This expression of course vanishes for our (non magnetically charged) background solution since \(\lambda\) vanishes on the axes of symmetry .

Linearizing the ADM formula for the magnetic field,
\begin{equation}\label{eq:i12}
\mathcal{B}^i = \frac{1}{2} \epsilon^{ijk} (\partial_j A_k - \partial_k A_j),
\end{equation}
one arrives at the 2-dimensional vector density
\begin{equation}\label{eq:i13}
\mathcal{B}^{a'} = \epsilon^{ab} \partial_b A'_\varphi = \epsilon^{ab} \partial_b \lambda'
\end{equation}
which, for reasons of regularity, must have a vanishing \(\theta\)-component along the axes of symmetry. It follows that \(\lambda'\) must be independent of \textit{R} along each of these axes and thus that the linearization for (\ref{eq:i11}) yields
\begin{equation}\label{eq:i14}
4Q^{\mathrm{mag}'} = \lambda' (t, \theta = \pi) - \lambda' (t, \theta = 0)
\end{equation}
which, at first glance, would seem to allow for a time dependent perturbation of the magnetic charge. However a straightforward linearization of Eq.~(\ref{eq:a55}), combined with the axis regularity results of Ref.~\cite{Rinne_05}, shows that \(\lambda'_{,t}\) vanishes to order \(O (\sin^2{\theta})\) at the axes of symmetry and hence that both \(\lambda' (t, \pi)\) and \(\lambda' (t, 0)\) are independent of \textit{t}. As mentioned above we shall demand that these constants of motion both vanish so that even our perturbed black hole is \textit{not} magnetically charged. Thus we demand that \(\lambda'\) vanish on both the axes of symmetry.  %E; orginally appendix I
\section{Gauge Conditions for the Linearized Equations}
\label{app:gauge-conditions}
A fundamental result of Refs.~\cite{CH_08} and \cite{CHN_11} is that one can always express the induced metric, \(g_{ij}\; dx^i \otimes dx^j\), on a Cauchy hypersurface for the DOC of an axisymmetric, non-degenerate, asymptotically flat black hole in coordinates \(\lbrace x^i\rbrace = \lbrace x^a, \varphi\rbrace = \lbrace\rho, z, \varphi\rbrace\) such that (reexpressed in our notation)
\begin{equation}\label{eq:l01}
g_{ij}\; dx^i \otimes dx^j = e^{-2\gamma} \tilde{g}_{ab}\; dx^a \otimes dx^b + e^{2\gamma} (d\varphi + \beta_a dx^a) \otimes (d\varphi + \beta_b dx^b)
\end{equation}
where
\begin{equation}\label{eq:l02}
    \begin{split}
    \tilde{g}_{ab}\; dx^a \otimes dx^b &= e^{2\nu} h_{ab}\; dx^a \otimes dx^b\\
        &= {} e^{2\nu} (d\rho \otimes d\rho + dz \otimes dx),\\
    \end{split}
\end{equation}
\(\psi = \frac{\partial}{\partial\varphi}\) is the generator of (axial) rotations under which \(g_{ij}\; dx^i \otimes dx^j\) is invariant and where \(R := \sqrt{\rho^2 + z^2}\) takes a constant value, \(R \rightarrow R_+ > 0\), on the (topological) sphere corresponding to the black hole's (non-degenerate) horizon (intersected with the chosen Cauchy surface). The coordinates introduced (via Eqs.~(\ref{eq:b08})--(\ref{eq:b12})) for the Kerr-Newman `background' solutions are clearly of this (Weyl-Papapetrou) type.

The flexibility to arrange that the coordinate sphere \(R := \sqrt{\rho^2 + z^2} \rightarrow R_+ = \hbox{ constant } > 0\) coincide with a particular (topological) sphere of geometrical significance (e.g., the black hole's horizon) results from the fact that the (manifestly conformally flat) form (\ref{eq:l02}) for the Riemannian 2-metric \(\tilde{g}_{ab}\; dx^a \otimes dx^b\) is preserved under arbitrary \textit{conformal transformations} whereby the coordinates \(\rho\) and \textit{z} can be replaced by arbitrary, conjugate harmonic functions thereof: \(\rho \rightarrow u(\rho, z), z \rightarrow v(\rho, z)\).

To preserve this metric form under Einsteinian evolution, however, one would need to impose the condition
\begin{equation}\label{eq:l03}
(\mu_{{}^{(2)}\tilde{g}} \tilde{g}^{ab})_{,t} = -2\tilde{N} \left(\tilde{\pi}^{ab} - \frac{1}{2} \tilde{g}^{ab} \tilde{g}_{ef} \tilde{\pi}^{ef}\right) + \mathcal{L}_X (\mu_{{}^{(2)}\tilde{g}} \tilde{g}^{-1})^{ab} = 0
\end{equation}
as a restriction on the (2-dimensional) shift field \(X = \tilde{N}^a \frac{\partial}{\partial x^a}\). Reexpressed in terms of the flat metric \(h_{ab}\; dx^a \otimes dx^b\), Eq.~(\ref{eq:l03}) becomes
\begin{equation}\label{eq:l04}
\left(\sqrt{{}^{(2)}h}\; h^{ab}\right)_{,t} = -2\tilde{N} \left(\tilde{\pi}^{ab} - \frac{1}{2} h^{ab} h_{ef} \tilde{\pi}^{ef}\right) + \mathcal{L}_X \left(\sqrt{{}^{(2)}h}\; h^{-1}\right)^{ab} = 0,
\end{equation}
where \(h^{-1} = h^{ab} \frac{\partial}{\partial x^a} \otimes \frac{\partial}{\partial x^b}\) and \({}^{(2)}h := \det{(h_{ab})}\). Equation (\ref{eq:l04}) ensures, of course, that the manifestly conformally flat form of this metric is preserved under the evolution but, even though we also demand that \(h_{ab}\; dx^a \otimes dx^b\) \textit{remain flat}, it is not uniquely fixed by Eq.~(\ref{eq:l04}) since (as was previously noted in Section~\ref{subsec:evaluation-conformal}) any metric of the form \(h^\lambda_{ab} = e^{2\lambda} h_{ab}\) is also flat whenever the function \(\lambda\) is harmonic (with respect to \(h_{ab}\; dx^a \otimes dx^b\) or any metric conformal thereto).

In other words the requirement that \(h_{ab}\; dx^a \otimes dx^b\) be flat does not uniquely fix the decomposition of \(\tilde{g}_{ab} = e^{2\nu} h_{ab}\) into a flat metric and a conformal factor but we can impose such uniqueness by fiat by absorbing the (harmonic logarithm) \(\lambda\) of any such deformation into the function \(\nu\), letting \(\nu \rightarrow \nu + \lambda\) and holding \(h_{ab}\) fixed.

In this paper, of course, we shall not need to deal with this issue at the fully nonlinear level but the linearized form of Eq.~(\ref{eq:l03}), about a Kerr-Newman background (for which \(\tilde{\pi}^{ab} = 0\) and \(X^a = \tilde{N}^a = 0\)) is:
\begin{equation}\label{eq:l05}
\mathcal{L}_{X'} (\mu_{{}^{(2)}\tilde{g}}\; \tilde{g}^{-1})^{ab} = 2\tilde{N} \left(\tilde{\pi}'^{ab} - \frac{1}{2} \tilde{g}^{ab} \tilde{g}_{ef} \tilde{\pi}'^{ef}\right)
\end{equation}
or, equivalently,
\begin{equation}\label{eq:l06}
\mathcal{L}_{X'} (\sqrt{{}^{(2)}h}\; h^{-1})^{ab} = 2\tilde{N} \left(\tilde{\pi}'^{ab} - \frac{1}{2} h^{ab} h_{ef} \tilde{\pi}'^{ef}\right)
\end{equation}
where \(X'^a = \tilde{N}^{a'}\).

In this article, however, rather than attempt to solve Eq.~(\ref{eq:l05}) or (\ref{eq:l06}) directly for the linearized shift \(X'\) we shall, in Appendix~\ref{app:transforming}, construct the gauge transformation that carries one from an arbitrary gauge to the desired Weyl-Papapetrou gauge at the linearized level. From the vector field \({}^{(4)}Y = {}^{(4)}Y^\mu \frac{\partial}{\partial x^\mu}\) that generates this gauge transformation (c.f., Eqs.~(\ref{eq:f01})--(\ref{eq:f06})) one can then simply compute, among other quantities, the transformed, linearized shift field, \(X' = \tilde{N}^{c'} \frac{\partial}{\partial x^c}\), via Eq.~(\ref{eq:f24}).

Thus we may assume, without essential loss of generality, that the flat, `conformal' metric, \(h_{ab}\; dx^a \otimes dx^b\), preserves its (manifestly flat) form,
\begin{equation}\label{eq:l07}
\begin{split}
h_{ab}\; dx^a \otimes dx^b &= d\rho \otimes d\rho + dz \otimes dz\\
 &= dR \otimes dR + R^2\; d\theta \otimes d\theta
\end{split}
\end{equation}
 under the perturbation and thus take \(h'_{ab} = 0\). Since, in principle, this (Weyl-Papaetrou) gauge condition can be imposed at the fully nonlinear level we may assume, a fortiori, that it holds to higher order at the perturbative level and thus, in particular, set \(h_{ab}^{\prime\prime} = 0\).  %F; equations starting l01
\section{Analysis of the Linearized Constraint Equations}
\label{app:analysis-linearized-constraint-equations}
Upon introducing the `twist' potentials \(\eta\) and \(\omega\) we have solved the electromagnetic (Gauss law) constraint and the azimuthal projection of the (3+1 dimensional) momentum constraint leaving only
\begin{equation}
\tilde{\mathcal{H}} = 0\> \hbox{ and }\> \tilde{\mathcal{H}}_a = 0
\label{eq:c01}
\end{equation}
as constraints for the reduced field equations. A straightforward calculation using the reduced evolution equations (\ref{eq:a49})--(\ref{eq:a58}) with arbitrary lapse \(\tilde{N}\) and shift \(X = \tilde{N}^a \frac{\partial}{\partial x^a}\) shows that these quantities, if not already vanishing, satisfy the evolution equations
\begin{align}
\frac{\partial}{\partial t} \tilde{\mathcal{H}} &= (\tilde{N}^a\tilde{\mathcal{H}})_{,a} + \tilde{N}_{,b}\; \tilde{g}^{ab}\; \tilde{\mathcal{H}}_a + (\tilde{N}\; \tilde{g}^{ab}\; \tilde{\mathcal{H}}_a)_{,b}, \label{eq:c02}\\
\frac{\partial}{\partial t} \tilde{\mathcal{H}}_a &= (\tilde{N}^b \tilde{\mathcal{H}}_a)_{,b} + \tilde{N}^b_{,a} \tilde{\mathcal{H}}_b + \tilde{N}_{,a} \tilde{\mathcal{H}} \label{eq:c03}
\end{align}
which are clearly at least consistent with the preservation of the constraints (\ref{eq:c01}) in time. Linearizing Eqs.~(\ref{eq:c02}) and (\ref{eq:c03}) about a background solution for which (as in the Kerr-Newman cases of interest here) \(X = 0\) yields the corresponding propagation equations for the first variations \((\tilde{\mathcal{H}}^\prime, \tilde{\mathcal{H}}^\prime_a)\) and \((\tilde{\mathcal{H}}, \tilde{\mathcal{H}}_a)\):
\begin{align}
\frac{\partial}{\partial t} \tilde{\mathcal{H}}^\prime &= \tilde{N}_{,b}\; \tilde{g}^{ab}\; \tilde{\mathcal{H}}^\prime_a + (\tilde{N}\; \tilde{g}^{ab}\; \tilde{\mathcal{H}}^\prime_a)_{,b} \label{eq:c04}\\
\frac{\partial}{\partial t} \tilde{\mathcal{H}}^\prime_a &= \tilde{N}_{,a} \tilde{\mathcal{H}}^\prime. \label{eq:c05}
\end{align}
These can also be derived by directly computing the time derivatives of \((\tilde{\mathcal{H}}^\prime, \tilde{\mathcal{H}}^\prime_a)\) by means of the linearized evolution equations.

As a subset of the linearized Einstein-Maxwell field equations  the linearized constraints
\begin{equation}
\tilde{\mathcal{H}}^\prime = 0\> \hbox{ and }\> \tilde{\mathcal{H}}^\prime_a = 0
\label{eq:c06}
\end{equation}
are gauge invariant (provided always that the background, exact field equations are satisfied) and this is reflected in the fact that neither \(\tilde{N}^\prime\) nor \(\tilde{N}^{a^\prime}\) appear in Eqs.(\ref{eq:c04}) and (\ref{eq:c05}). (c.f., the discussion in \cite{Moncrief_74_3}).

In a \textit{free evolution} framework one would impose the linearized constraints \(\tilde{\mathcal{H}}^\prime = \tilde{\mathcal{H}}^\prime_a = 0\) on an initial Cauchy hypersurface and appeal to the propagation equations (\ref{eq:c04})--(\ref{eq:c05}) to establish their preservation in time. Since these propagation equations however are apparently not of a standard type we prefer to adopt the strategy of \textit{constrained evolution} whereby one enforces the linearized constraints on every time slice by solving them for certain `dependent' variables in terms of the unconstrained, `dynamical' variables, namely the first variations \((\gamma^\prime, \omega^\prime, \eta^\prime, \lambda^\prime)\) of the wave map functions and their conjugate momenta \((\tilde{p}^\prime, \tilde{r}^\prime, \tilde{u}^\prime, \tilde{v}^\prime)\).

In the class of gauges that we shall consider and recalling that the background, Kerr-Newman solutions of interest have vanishing canonical momenta, the linearized constraints reduce to:
\begin{align}
\tilde{\mathcal{H}}^\prime &= \sqrt{{}^{(2)}\!h}\; h^{ab}\; \left\lbrack 4\gamma_{,a} \gamma^\prime_{,b} - e^{-2\gamma} \gamma^\prime (\eta_{,a} \eta_{,b} + \lambda_{,a} \lambda_{,b})\right.\label{eq:c07}\\
&\mbox{} + e^{-2\gamma} (\eta_{,a} \eta^\prime_{,b} + \lambda_{,a} \lambda^\prime_{,b}) - 2e^{-4\gamma} \gamma^\prime (\omega_{,a} + \lambda\eta_{,a}) (\omega_{,b} + \lambda\eta_{,b})\nonumber\\
&\mbox{} + \left.e^{-4\gamma} (\omega_{,a} + \lambda\eta_{,a}) (\omega^\prime_{,b} + \lambda\eta^\prime_{,b} + \lambda^\prime\eta_{,b})\right\rbrack\nonumber\\
&\mbox{} + 2 \partial_a \left(\sqrt{{}^{(2)}\!h}\; h^{ab}\; \nu^\prime_{,b}\right) = 0,\nonumber\\
\tilde{\mathcal{H}}^\prime_a &= -2\; {}^{(2)}\!\nabla_b(h)\; {\tilde{r}^\prime_a}^{b} - e^{2\nu} \sqrt{{}^{(2)}\!h}\; \tau^\prime_{,a}\label{eq:c08}\\
& \mbox{} + (\tilde{p}^\prime \gamma_{,a} + \tilde{r}^\prime \omega_{,a} + \tilde{v}^\prime \lambda_{,a} + \tilde{u}^\prime \eta_{,a}) = 0\nonumber
\end{align}
where
\begin{equation}
\begin{split}
{\tilde{r}^\prime_a}^b &:= \tilde{g}_{ad} \left({\tilde{\pi}^\prime}^{bd} - \frac{1}{2}\; \tilde{g}^{bd} \tilde{g}_{ef} {\tilde{\pi}^\prime}^{ef}\right)\\
&= e^{2\nu} h_{ad} \left({\tilde{\pi}^\prime}^{bd} - \frac{1}{2}\; h^{bd} h_{ef} {\tilde{\pi}^\prime}^{ef}\right)\\
&= e^{2\nu} h_{ad} \left({\tilde{\pi}^\prime}^{bd} - \frac{1}{2} \sqrt{{}^{(2)}\!h}\; h^{bd} \tau^\prime\right)\label{eq:c09}
\end{split}
\end{equation}
denotes the traceless part of \({\tilde{\pi}^\prime}^{ab}\) and
\begin{equation}
\tau^\prime := \frac{\tilde{g}_{ab}}{\mu_{{}^{(2)}\!\tilde{g}}} {\tilde{\pi}^\prime}^{ab} = \frac{h_{ab}}{\sqrt{{}^{(2)}\!h}} {\tilde{\pi}^\prime}^{ab}
\label{eq:c10}
\end{equation}
its (scalarized) trace. Here \(h_{ab} = e^{-2\nu}\; \tilde{g}_{ab}\) designates the flat metric on \(M_b\) introduced in Appendix~\ref{app:gauge-conditions} whereas \({}^{(2)}\!\nabla_a(h)\) and \(\sqrt{{}^{(2)}\!h}\) denote covariant differentiation and `volume' element for this metric. Recall that in the Weyl-Papapetrou coordinates \(\lbrace\bar{\rho},\bar{z}\rbrace\) first introduced in Appendix~\ref{app:kerr-newman-spacetimes}, \(M_b\) corresponds to the half plane \(\lbrace(\bar{\rho},\bar{z})\; |\bar{\rho} \geq 0, \bar{z} \in \mathbb{R}\rbrace\) with the `cut' (\ref{eq:b20}) removed and \(\frac{1}{\sqrt{{}^{(2)}\!h}}\; h_{ab} dx^a \otimes dx^b = d\bar{\rho} \otimes d\bar{\rho} + d\bar{z} \otimes d\bar{z}\).

The Hamiltonian constraint (\ref{eq:c07}) is an elementary (flat space) Poisson equation on \(\lbrace M_b, h_{ab}\; dx^a \otimes dx^b\rbrace\) for the first variation, \(\nu'\), of the logarithm of the conformal factor \(e^{2\nu}\). As discussed in Section \ref{subsec:axis-regularity}, however, regularity at the axes of symmetry requires that we impose the Dirictlet boundary condition (c.f., Eq.~(\ref{eq:496})):
\begin{equation}\label{eq:c25}
\left.\nu'\right|_{\substack{\theta = 0,\pi\\
    R \geq R_+}} = \left.2\gamma'\right|_{\substack{\theta = 0,\pi\\
        R \geq R_+}}.
\end{equation}
Additional considerations, such as those discussed in Section \ref{subsec:evaluation-conformal}, can lead to the imposition of a Neumann boundary condition such as the (minimal surface preserving) condition
\begin{equation}\label{eq:c26}
\left.\nu'_{,R}\right|_{R = R_+} = 0
\end{equation}
at the event horizon. Thus one can be naturally led to a \textit{mixed}, elliptic boundary value problem for \(\nu'\) with Dirichlet data required along the axes of symmetry and complementary Neumann data needed along the horizon boundary. Though such problems can be notoriously difficult to solve in general we shall be able to exploit the special features of our particular problem to solve it by elementary means. In this way we simultaneously remove the ambiguity in the construction of \(\nu'\) (which would otherwise be undetermined up to the addition of a harmonic function) and cancel the flux contributions that could otherwise lead to a violation of the conservation of energy.

A standard (Green's theorem) argument shows that if indeed a solution vanishing at infinity exists for this (mixed, elliptic) problem then it will necessarily be unique. Our strategy for constructing this hypothetical solution will be to seek to express it as
\begin{equation}\label{eq:c27}
\nu' = \nu'_D + \nu'_N
\end{equation}
where \(\nu'_D\) is the solution to an associated, inhomogeneous Dirichlet problem chosen to solve Eq.~(\ref{eq:c07}) with the boundary condition (\ref{eq:c25}) imposed, whereas \(\nu'_N\) will be the \textit{harmonic} solution to a complementary, homogeneous Neumann problem chosen to impose the boundary condition (\ref{eq:c26}) and constructed in such a way as to leave the Dirichlet condition on the axes of symmetry undisturbed. The special (2-dimensional, conformally covariant) nature of our problem is what allows this last step to be carried out.

We begin by imposing suitable Dirichlet conditions for \(\nu'_D\) on the boundary of the closure \(\bar{M}_b\) of \(M_b\) (i.e., on the full \(\bar{z}\)-axis of the half-plane \(\lbrace (\bar{\rho}, \bar{z}) | \bar{\rho} \geq 0, \bar{z} \in \mathbb{R}\rbrace\)) and with suitable `regularity' assumed for the free data \(\lbrace\gamma', \omega', \eta', \lambda'\rbrace\) appearing in Eq.~(\ref{eq:c07}). More precisely we choose Dirichlet data for \(\nu'_D\) along the upper and lower axis components to cancel the unwanted flux contributions identified previously (i.e., so as to impose (\ref{eq:c25})) and, as an intermediate step, interpolate along the `strut' separating these disjoint axes with smooth but arbitrarily chosen, complementary Dirichlet data. One could, for example, choose \(\nu'_D = 2\gamma'\) along this strut. Using the explicitly known fundamental solution (Green's function) for this problem (see, for example \cite{GBF_95}), we solve the corresponding Dirichlet problem (i.e., solve Eq.~(\ref{eq:c07}) for \(\nu'_D\) in place of \(\nu'\) with the boundary data so chosen).

The solution for \(\nu'_D\) will of course fail in general to satisfy the Neumann condition (\ref{eq:c26}) along the horizon but if, as in the asymptotically pure gauge problem discussed in Section \ref{subsec:evaluating_dynamical}, \(\gamma'\) has the property that
\begin{equation}\label{eq:c28}
\left.\gamma'_{,R}\right|_{\substack{R = R_+\\
    \theta = 0,\pi}} = 0
\end{equation}
then, from the chosen condition,
\begin{equation}\label{eq:c29}
\nu'_D = 2\gamma'
\end{equation}
along the \(\bar{z}\)-axis we shall automatically have
\begin{equation}\label{eq:c30}
\left.\nu'_{D,R}\right|_{\substack{R = R_+\\
    \theta = 0,\pi}} = 0.
\end{equation}
We now revert to the `half-plane with half disk removed' picture for \(M_b\) discussed in Appendix \ref{app:kerr-newman-spacetimes} and extend this to a `full plane with full disk removed' by reflection across the \(\bar{z}\)-axis. We now choose Neumann data for \(\nu'_N\) on the circle at \(R = R_+\) by setting
\begin{equation}\label{eq:c31}
\left.\nu'_{N,R}\right|_{\substack{R = R_+\\
    \theta \in [0,\pi]}} = \left. -\nu_{D,R}\right|_{\substack{R = R_+\\
        \theta \in [0,\pi]}}
\end{equation}
and then \textit{anti-reflecting} this data across the \(\bar{z}\)-axis to complete the specification on the full circle (i.e., choosing the value of \(\nu'_{N,R}\) to be the negative of that at its mirror image on the circle). While it is not strictly needed for our construction Eq.~(\ref{eq:c30}) will ensure that this extension of the Neumann data will be continuous at those points where the horizon meets the axes of symmetry.

The fundamental solution for the Neumann problem on the plane with a disk removed is explicitly known (c.f., \cite{EDB_10}). Using it together with the chosen Neumann data on the circle \(R = R_+\) we construct the harmonic function \(\nu'_N\). From the uniqueness of this solution and the reflection (anti-) symmetry of its boundary data we see that \(\nu'_N\) will in fact vanish on the axes of symmetry. Expressing \(\nu'_D\) and \(\nu'_N\) in a common coordinate system, adding them and restricting the result to \(\bar{M}_b\) we see that \(\nu' := \nu'_D + \nu'_N\) satisfies (\ref{eq:c07}) together with the mixed boundary conditions (\ref{eq:c25}) and (\ref{eq:c26}). By construction it is the \textit{unique} function vanishing at infinity that has these properties. We thus conclude that\footnote{\label{note08}By construction our solution satisfies the \textit{necessary} condition, \ref{eq:c25}, for regularity at the axes. That it is, moreover, fully regular at the axes follows indirectly from its uniqueness as a solution to the relevant Hamiltonian constraint and an independent, purely 3+1-dimensional treatment of the corresponding Lichnerowicz equation with axisymmetric boundary data. The well-known existence of a unique, globally smooth solution to the latter ensures that our solution, with which it must agree, has the required regularity.}
\begin{theorem}\label{thm:03}
Equation (\ref{eq:c07}) has, for each choice of regular data \(\lbrace\gamma', \omega', \eta', \lambda'\rbrace\), a unique solution \(\nu'\) that vanishes at infinity and satisfies the mixed (Dirichlet/Newmann) boundary conditions (\ref{eq:c25}) and (\ref{eq:c26}).
\end{theorem}

To solve the momentum constraint (\ref{eq:c08}) we exploit the fact that, under suitable boundary and asymptotic conditions (discussed in detail in Appendix~\ref{app:complactly-supported-solutions}), symmetric \textit{transverse traceless} tensors on \(\bar{M}_b\) vanish identically and thus that (the mixed form of) a symmetric traceless density, \({\tilde{r}^\prime_a}^{b} \frac{\partial}{\partial x^b} \otimes dx^a\), can be expressed as
\begin{equation}
{\tilde{r}^\prime_a}^{b} = \sqrt{{}^{(2)}\!h}\> \left\lbrack{}^{(2)}\!\nabla_a (h){Y^\prime}^b + {}^{(2)}\!\nabla^b (h) (h_{ac} {Y^\prime}^c) - \delta_a^b {}^{(2)}\!\nabla_c (h) {Y^\prime}^c\right\rbrack
\label{eq:c11}
\end{equation}
for a suitably chosen vector field \(Y^\prime = {Y^\prime}^c \frac{\partial}{\partial x^a}\). Recalling that \(h_{ab} dx^a \otimes dx^b\) is flat one finds easily that Eq.~(\ref{eq:c08}) reduces to
\begin{equation}
2\sqrt{{}^{(2)}\!h} \left({}^{(2)}\!\nabla_b (h) {}^{(2)}\!\nabla^b (h) (h_{ac} {Y^\prime}^c)\right) = -e^{2\nu} \sqrt{{}^{(2)}\!h} \tau^\prime_{,a} + (\tilde{p}^\prime \gamma_{,a} + \tilde{r}^\prime\omega_{,a} + \tilde{v}^\prime\lambda_{,a} + \tilde{u}^\prime\eta_{,a})
\label{eq:c12}
\end{equation}
which, in Weyl-Papapetrou coordinates, takes the form of elementary, decoupled Poisson equations for the components of \(Y^\prime\). With suitable boundary and regularity conditions imposed upon the relevant data these can be solved explicitly for \(Y^\prime\). Again the relevant elliptic theory is presented in detail in Appendix~\ref{app:transforming} below.

By exploiting the background field equations (\ref{eq:a64})--(\ref{eq:a71}), satisfied by an arbitrary, Kerr-Newman black hole, it is straightforward to show that
\begin{equation}\label{eq:c13}
\begin{split}
\tilde{N} \tilde{\mathcal{H}}^\prime &= \frac{\partial}{\partial x^b} \left\lbrace\tilde{N} \sqrt{{}^{(2)}\!h}\; h^{ab} \left\lbrack 4\gamma_{,a} \gamma^\prime + 2\nu^\prime_{,a}\right.\right.\\
& \left.\left.\mbox{} + e^{-2\gamma} (\eta_{,a} \eta^\prime + \lambda_{,a} \lambda^\prime) + e^{-4\gamma} (\omega_{,a} + \lambda \eta_{,a}) (\omega^\prime + \lambda \eta^\prime)\right\rbrack - 2\sqrt{{}^{(2)}\!h}\; h^{ab} \tilde{N}_{,a} \nu^\prime\right\rbrace
\end{split}
\end{equation}
for arbitrary \((\gamma^\prime, \omega^\prime, \eta^\prime, \lambda^\prime, \nu^\prime)\). That this expression is a spatial divergence reflects the fact, discussed briefly in Section~\ref{subsec:review-linearization-stability}, that \((C,Z) = (\tilde{N},0)\) is an element of the kernel of the adjoint of the linearized constraint operator, corresponding to the occurrence of \(\zeta = \frac{\partial}{\partial t}\) as a Killing field on the quotient manifold \(V/U(1) = \mathbb{R} \times M_b\) (c.f., Appendix~\ref{app:reduced-hamiltonian}). To fully appreciate its implications for the perturbative analysis it is essential to consider the 2nd variation of the Hamiltonian constraint.

Let us abbreviate by \(\lbrace\mathbf{q}\rbrace := \lbrace\gamma, \omega, \eta, \lambda\rbrace\) the wave map variables and by \(\lbrace\mathbf{p}\rbrace := \lbrace\tilde{p}, \tilde{r}, \tilde{u}, \tilde{v}\rbrace\) their canonically conjugate momenta. These are the unconstrained, dynamical `degrees of freedom' for the reduced, axisymmetric Einstein-Maxwell system. The flat, conformal metric \(h_{ab} dx^a \otimes dx^b\) and (2+1-dimensional) mean curvature function, \(\tau\), are restricted, in our reduced Hamiltonian framework, through the imposition of suitable gauge conditions by setting, for example, \(h_{ab} dx^a \otimes dx^b = d\bar{\rho} \otimes d\bar{\rho} + d\bar{z} \otimes d\bar{z}\) and (in the simplest case) taking \(\tau = 0\) (2+1-dimensional maximal slicing). The canonically conjugate partners of these gauge variables, namely the tracefree component of the gravitational momentum,
\begin{equation}
\tilde{r}_{\hphantom{a}b}^a \frac{\partial}{\partial x^a} \otimes dx^b := \tilde{\mathbf{r}}, \label{eq:c24}
\end{equation}
and conformal factor, \(e^{2\nu}\), are to be determined (subject to suitable boundary conditions) through the solution of the elliptic momentum and Hamiltonian constraints on each time slice. Preservation of the gauge conditions throughout the evolution necessitates a corresponding fixation of the lapse and shift fields \((N, X)\) via the solution of an auxiliary set of (linear) elliptic equations (c.f., Appendices~\ref{app:gauge-conditions}, \ref{app:transforming} and \ref{app:maximal-slicing} for details).

Treating black hole stability problems at this fully nonlinear level is currently out of reach but closely related methods have been successfully used to prove the fully nonlinear stability (in the direction of cosmological expansions) of a family of \(U(1)\)-symmetric, spatially compact, vacuum cosmological models \cite{Ycb-Mon_01}.

To derive the linearized and higher order perturbation equations (for axisymmetric perturbations of Kerr-Newman backgrounds, in particular) one can imagine having a smooth one-parameter family of exact solutions, containing the desired background, and differentiating the field equations one or more times with respect to this curve parameter, \textit{e}, and then fixing it to the background value, say \(e = 0\). Thus we now write \(\lbrace\mathbf{q}^\prime, \mathbf{p}^\prime\rbrace\) for \(\left\lbrace(\gamma^\prime, \omega^\prime, \eta^\prime, \lambda^\prime), (\tilde{p}^\prime, \tilde{r}^\prime, \tilde{u}^\prime, \tilde{v}^\prime)\right\rbrace\) where
\begin{equation}
\lbrace\mathbf{q}^\prime, \mathbf{p}^\prime\rbrace := \left.\left\lbrace\frac{\partial}{\partial e}\; \mathbf{q} (e,\cdot), \frac{\partial}{\partial e}\; \mathbf{p} (e,\cdot)\right\rbrace\right|_{e=0}
\label{eq:c14}
\end{equation}
and denote by \(\lbrace\mathbf{q}^{\prime\prime}, \mathbf{p}^{\prime\prime}\rbrace\) the corresponding 2nd variations
\begin{equation}
\lbrace\mathbf{q}^{\prime\prime}, \mathbf{p}^{\prime\prime}\rbrace := \left.\left\lbrace\frac{\partial^2}{\partial e^2}\; \mathbf{q} (e,\cdot), \frac{\partial^2}{\partial e^2}\; \mathbf{p} (e,\cdot)\right\rbrace\right|_{e=0}
\label{eq:c15}
\end{equation}
etc.

The gauge choice made for the flat metric \({}^{(2)}\!h\) implies that \({}^{(2)}\!h^\prime = {}^{(2)}\!h^{\prime\prime} = 0\), etc. (c.f., Appendix~\ref{app:gauge-conditions}), whereas that for \(\tau\) (in the simplest, 2+1-dimensional, \textit{maximal} case) yields \(\tau^\prime = \tau^{\prime\prime} = 0\) as well. To allow however for more general time gauge conditions (3+1-dimensional maximal slicing, for example) we shall retain \(\tau^\prime\) and \(\tau^{\prime\prime}\) in the formulas to follow. The corresponding perturbations \(\lbrace \nu^\prime, \tilde{r}_b^{a^\prime}, \tilde{N}^\prime, X^\prime\rbrace\) and \(\lbrace \nu^{\prime\prime}, \tilde{r}_b^{a\prime\prime}, \tilde{N}^{\prime\prime}, X^{\prime\prime}\rbrace\) of the remaining, dependent variables are determined (with suitable boundary conditions) by solving the corresponding perturbed elliptic equations and are thus, in effect, known functionals of \(\lbrace\mathbf{q}^\prime, \mathbf{p}^\prime, \mathbf{q}^{\prime\prime}, \mathbf{p}^{\prime\prime}\rbrace\).

Let us now denote the 1st variations (\ref{eq:c07}) and (\ref{eq:c08}), of the constraints more explicitly as first order linear operators acting on the relevant linearized variables, via
\begin{equation}
D\tilde{\mathcal{H}} (\mathbf{q}, {}^{(2)}\!h) \cdot (\mathbf{q}^\prime, \nu^\prime) := \tilde{\mathcal{H}}^{\prime}
\label{eq:c16}
\end{equation}
and
\begin{equation}
D\tilde{\mathcal{H}}_a (\mathbf{q}, {}^{(2)}\!h, \nu) \cdot (\mathbf{p}^\prime, \tilde{\mathbf{r}}^\prime, \tau^\prime) := \tilde{\mathcal{H}}^\prime_a
\label{eq:c17}
\end{equation}
so that \(\tilde{N}D\tilde{\mathcal{H}} (\mathbf{q}, {}^{(2)}\!h) \cdot (\mathbf{q}^\prime, \nu^\prime)\) is the total divergence given explicitly by Eq.~(\ref{eq:c13}).

A straightforward calculation, utilizing Eqs.~(\ref{eq:a33},\ref{eq:a34},\ref{eq:c13}--\ref{eq:c17}) now yields
\begin{equation}
\tilde{N}\tilde{\mathcal{H}}^{\prime\prime} = \tilde{N}D\tilde{\mathcal{H}} (\mathbf{q}, {}^{(2)}\!h) \cdot (\mathbf{q}^{\prime\prime}, \nu^{\prime\prime}) + \tilde{N}D^2\tilde{\mathcal{H}} (\mathbf{q}, {}^{(2)}\!h, \nu) \cdot \left((\mathbf{q}^\prime, \mathbf{p}^\prime, \tilde{\mathbf{r}}^\prime, \tau^\prime), (\mathbf{q}^\prime, \mathbf{p}^\prime, \tilde{\mathbf{r}}^\prime, \tau^\prime)\right),
\label{eq:c18}
\end{equation}
where
\begin{equation}\label{eq:c19}
\begin{split}
D^2\tilde{\mathcal{H}} (\mathbf{q}, {}^{(2)}\!h, \nu) &\cdot \left((\mathbf{q}^\prime, \mathbf{p}^\prime, \tilde{\mathbf{r}}^\prime, \tau^\prime), (\mathbf{q}^\prime, \mathbf{p}^\prime, \tilde{\mathbf{r}}^\prime, \tau^\prime)\right)\\
 &:= \left\lbrace\frac{2e^{-2\nu}}{\sqrt{{}^{(2)}\!h}} \left\lbrack{\tilde{r}^\prime}_a^b {\tilde{r}^\prime}_b^a + \frac{1}{8}(\tilde{p}^\prime)^2 +
\frac{1}{2} e^{4\gamma} (\tilde{r}^\prime)^2\right.\right.\\
& \mbox{} + \left.\vphantom{\frac{1}{2}}\frac{1}{2} e^{2\gamma} \left((\tilde{v}^\prime)^2 + (\tilde{u}^\prime - \lambda\tilde{r}^\prime)^2\right)\right\rbrack - e^{2\nu} \sqrt{{}^{(2)}\!h} (\tau^\prime)^2\\
& \mbox{} + \sqrt{{}^{(2)}\!h}\; h^{ab}\; \left\lbrack\vphantom{\frac{1}{2}} 4\gamma^\prime_{,a} \gamma^\prime_{,b} + 2e^{-2\gamma} (\gamma^\prime)^2 (\eta_{,a}\eta_{,b} + \lambda_{,a}\lambda_{,b})\right.\\
& \mbox{} - 4e^{-2\gamma}\; \gamma^\prime (\eta_{,a}\eta^\prime_{,b} + \lambda_{,a}\lambda^\prime_{,b}) + e^{-2\gamma} (\eta^\prime_{,a}\eta^\prime_{,b} + \lambda^\prime_{,a}\lambda^\prime_{,b})\\
& \mbox{} + 8e^{-4\gamma} (\gamma^\prime)^2 (\omega_{,a} + \lambda\eta_{,a})(\omega_{,b} + \lambda\eta_{,b})\\
& \mbox{} - 8e^{-4\gamma}\; \gamma^\prime (\omega_{,a} + \lambda\eta_{,a})(\omega^\prime_{,b} + \lambda\eta^\prime_{,b} + \lambda^\prime\eta_{,b})\\
& \mbox{} + e^{-4\gamma} (\omega^\prime_{,a} + \lambda\eta^\prime_{,a} + \lambda^\prime\eta_{,a})(\omega^\prime_{,b} + \lambda\eta^\prime_{,b} + \lambda^\prime\eta_{,b})\\
& \mbox{} + \left.\left.\vphantom{\frac{1}{2}}e^{-4\gamma} (\omega_{,a} + \lambda\eta_{,a})(2\lambda^\prime\eta^\prime_{,b})\right\rbrack\right\rbrace ,
\end{split}
\end{equation}
and
\begin{equation}\label{eq:c20}
\begin{split}
\tilde{\mathcal{H}}^{\prime\prime}_a &= D\tilde{\mathcal{H}}_a (\mathbf{q}, {}^{(2)}\!h, \nu) \cdot (\mathbf{p}^{\prime\prime}, \tilde{\mathbf{r}}^{\prime\prime}, \tau^{\prime\prime})\\
& \mbox{} - 2e^{2\nu} \sqrt{{}^{(2)}\!h}\; \nu^\prime\tau^\prime_{,a} + (\tilde{p}^\prime \gamma^\prime_{,a} + \tilde{r}^\prime\omega^\prime_{,a} + \tilde{v}^\prime\lambda^\prime_{,a} + \tilde{u}^\prime\eta^\prime_{,a}).
\end{split}
\end{equation}

Combining (\ref{eq:c13}), (\ref{eq:c18}),  and (\ref{eq:c19}) we see that the constraint equations of second order, namely
\begin{equation}
\tilde{\mathcal{H}}^{\prime\prime} = 0\> \hbox{ and }\> \tilde{\mathcal{H}}^{\prime\prime}_a = 0,
\label{eq:c21}
\end{equation}
imply that the density \(\mathcal{E}^{\mathrm{Alt}}\) defined by
\begin{equation}
\mathcal{E}^{\mathrm{Alt}} := \frac{\tilde{N}}{2} D^2\tilde{\mathcal{H}} (\mathbf{q}, {}^{(2)}\!h, \nu) \cdot \left((\mathbf{q}^\prime, \mathbf{p}^\prime, \tilde{\mathbf{r}}^\prime, \tau^\prime), (\mathbf{q}^\prime, \mathbf{p}^\prime, \tilde{\mathbf{r}}^\prime, \tau^\prime)\right)
\label{eq:c22}
\end{equation}
is equal to a spatial divergence and thus that the integral
\begin{equation}
E^{\mathrm{Alt}} := \int_{M_b} d^2x\; \lbrace\mathcal{E}^{\mathrm{Alt}}\rbrace
\label{eq:c23}
\end{equation}
is equal to a boundary integral when the field equations are satisfied. In the limiting case of purely electromagnetic perturbations of a \textit{Kerr} back-ground \(E^{\mathrm{Alt}}\) reduces to the functional \(H^{\mathrm{Alt}}\) defined via Eqs.~(\ref{eq:130}) and (\ref{eq:131}).

It is clear from the divergence form for \(\tilde{N}\tilde{\mathcal{H}}'\) given by Eq.~(\ref{eq:c13}) that, when the linearized Hamiltonian constraint, \(\tilde{\mathcal{H}}' = 0\), is imposed, the integral of \(\tilde{N}\tilde{\mathcal{H}}'\) over \(M_b\) will imply the vanishing of a sum of potential `boundary flux' terms arising at the boundary components corresponding to \(R \nearrow\infty\), \(R\searrow R_+\) and \(\theta \rightarrow 0,\pi\). By exploiting the asymptotic behaviors of the perturbations \(\lbrace\eta', \lambda', \omega', \gamma', \nu'\rbrace\) given via Eqs.~(\ref{eq:409}), (\ref{eq:410}), (\ref{eq:503}), (\ref{eq:436}), (\ref{eq:438}), (\ref{eq:440}) and (\ref{eq:442}) it is straightforward to show that the flux integrand vanishes \textit{pointwise} as \(R \nearrow\infty\) yielding a separately vanishing contribution to the net boundary flux. By exploiting the regularity of the various perturbations at the axes of symmetry, including especially the condition (\ref{eq:496}) on \(\nu' - 2\gamma'\) and the fact discussed in Section~\ref{subsec:axis-regularity} that each of \(\lbrace\omega', \lambda', \eta'\rbrace\) vanishes to order \(O (\sin^2{\theta})\) as \(\theta \rightarrow 0,\pi\), it follows that the boundary flux integrand also vanishes pointwise along the (artificial) boundary components corresponding to \(\theta \rightarrow 0,\pi\).

Finally, by exploiting the regularity of the perturbations \(\lbrace\gamma', \lambda', \eta', \omega'\rbrace\) at the horizon discussed in Section \ref{subsec:evaluating_dynamical} together with the (minimal surface preserving) condition (\ref{eq:c26}) and the pointwise vanishing of the factors \(\lbrace\tilde{N}, \gamma_{,R}, e^{-2\gamma}\eta_{,R}, e^{-2\gamma}\lambda_{,R}, e^{-4\gamma}(\omega_{,R} + \lambda\eta_{,R})\rbrace\) as \(R \searrow R_+\), it is straightforward to show that the only potential flux contribution at this `inner' boundary component must come from the only remaining term in the flux integrand:
\begin{equation}\label{eq:c32}
-\left.\left( 2\sqrt{{}^{(2)}h} h^{Ra} \tilde{N}_{,a} \nu'\right)\right|_{R_+} = -4R_+ \left.\left(\frac{\partial}{\partial\theta} \left(\sin{\theta} {}^{(2)}Y^\theta\right)\right)\right|_{R_+}.
\end{equation}
While not pointwise vanishing as the other terms were this clearly has vanishing integral with respect to \(\theta\) when integrated over the interval \(\theta \in\; [0,\pi]\) corresponding to the horizon component at \(R_+\).

Proceeding now to the second variation of the Hamiltonian constraint, \(\tilde{\mathcal{H}}^{\prime\prime} = 0\), it is now clear from Eqs.~(\ref{eq:c13}), (\ref{eq:c16})--(\ref{eq:c19}) that the sum of boundary flux contributions resulting from the integral of the divergence expression
\begin{equation}\label{eq:c33}
\begin{split}
\tilde{N}D\tilde{\mathcal{H}} (\mathbf{q}, {}^{(2)}h) (\mathbf{q}^{\prime\prime}, \nu^{\prime\prime}) &= \frac{\partial}{\partial x^b} \left\lbrace\tilde{N} \sqrt{{}^{(2)}h}\; h^{ab} \left\lbrack 4\gamma_{,a} \gamma^{\prime\prime} + 2\nu^{\prime\prime}_{,a}\right.\right. \\
 &\left. + e^{-2\gamma} (\eta_{,a} \eta^{\prime\prime} + \lambda_{,a} \lambda^{\prime\prime}) + e^{-4\gamma} (\omega_{,a} + \lambda\eta_{,a}) (\omega^{\prime\prime} + \lambda\eta^{\prime\prime})\right\rbrack\\
 &\left. - 2 \sqrt{{}^{(2)}h}\; h^{ab} \tilde{N}_{,a} \nu^{\prime\prime}\right\rbrace
\end{split}
\end{equation}
over \(M_b\) must equate to the volume integral over this same domain of
\begin{equation}\label{eq:c34}
-2\mathcal{E}^{\mathrm{Alt}} = -\tilde{N} D^2\tilde{\mathcal{H}} (\mathbf{q}, {}^{(2)}h, \nu) \cdot \left( (\mathbf{q}', \mathbf{p}', \tilde{\mathbf{r}}', \tau'), (\mathbf{q}', \mathbf{p}', \tilde{\mathbf{r}}', \tau')\right)
\end{equation}
which, by Eqs.~(\ref{eq:c22}) and (\ref{eq:c23}) is equal to \(-2E^{\mathrm{Alt}}\).

It is straightforward to verify that mere boundedness (or even mild blowup) of the perturbations \(\lbrace\lambda^{\prime\prime}, \eta^{\prime\prime}, \omega^{\prime\prime}\rbrace\) as \(R \nearrow\infty\) is sufficient to ensure their pointwise vanishing flux contributions at the `outer' boundary. Furthermore their regularity as smooth scalar fields at the axes of symmetry and at the horizon guarantees their (pointwise) vanishing contributions to the flux integrands at these boundary components as well. This leaves only possible contributions of \(\gamma^{\prime\prime}\) and \(\nu^{\prime\prime}\) to be considered.

On the other hand the demand for regularity at the axes of symmetry leads, upon appealing again to the Rinne/Stewart results \cite{Rinne_05}, to the restriction
\begin{equation}\label{eq:c35}
\nu^{\prime\prime}|_{\theta = 0,\pi} = 2\gamma^{\prime\prime}|_{\theta = 0,\pi}
\end{equation}
upon the second order perturbations and this suffices to ensure their (pointwise) vanishing contributions to the flux integrands along these axes.

Following up on the seminar work of D. Brill \cite{BD_59}, Sergio Dain derived an elegant integral expression for the ADM mass of an asymptotically flat, axisymmetric Einstein spacetime \cite{DA_08}. Its first variation (about a Kerr-Newman background) vanishes, for the class of perturbations considered herein, in view of the flux integral results described above but its second variation, expressed in our notation, yields the formula
\begin{equation}\label{eq:c36}
M^{\prime\prime}_{\mathrm{ADM}} = -\frac{1}{4} \lim_{R\nearrow\infty} \int_0^\pi R^2\sin{\theta} \left\lbrace\nu^{\prime\prime}_{,R} - \frac{1}{R}\nu^{\prime\prime} + \frac{2}{R} \gamma^{\prime\prime}\right\rbrace d\theta
\end{equation}
and thus allows us to express the `volume' integral of \(\mathcal{E}^{\mathrm{Alt}}\) over \(M_b\) (c.f., Eqs.~(\ref{eq:c22})--(\ref{eq:c23})) as follows:
\begin{equation}\label{eq:c37}
\begin{split}
    M^{\prime\prime}_{\mathrm{ADM}} &= \left.\frac{1}{4} E^{\mathrm{Alt}} + \frac{1}{2} \int_0^\pi d\theta (R_+ \sin{\theta}\; \nu^{\prime\prime})\right|_{R\searrow R_+}\\
        &= \int_{M_b} \left\lbrace\frac{1}{8} \tilde{N}D^2 \tilde{\mathcal{H}} (\mathbf{q}, {}^{(2)}h, \nu) \cdot \left( (\mathbf{q}', \mathbf{p}', \tilde{\mathbf{r}}', \tau'), (\mathbf{q}', \mathbf{p}', \tilde{\mathbf{r}}', \tau')\right)\right\rbrace dRd\theta\\
        &+ \left.\frac{1}{2} \int_0^\pi d\theta (R_+ \sin{\theta}\; \nu^{\prime\prime})\right|_{R\searrow R_+}.
\end{split}
\end{equation}

Recalling the discussion at the beginning of Section \ref{sec:conservation} we see that, at least for the class of (asymptotically-pure-gauge) perturbations considered herein, \(E^{\mathrm{Alt}}\) can be replaced by the manifestly positive definite expression \(E^{\mathrm{Reg}}\). Whether this perturbative contribution to the ADM mass is further `shifted' by the boundary integral over the horizon hinges, of course, upon the boundary condition chosen for \(\left.\nu^{\prime\prime}\right|_{R\searrow R_+}\) in the second variation of the Hamiltonian constraint. If, for example, the perturbations considered are chosen to be symmetric under the mapping (`inversion in the sphere') \(R \rightarrow \frac{R_+^2}{R'}\) (for which \(r = R + M + \frac{R_+^2}{R} \rightarrow r' = R' + M + \frac{R_+^2}{R'}\)) which maps one `end' of the Kerr-Newman solution isometrically to the other, then the resulting boundary integral over the horizon could not distinguish one end from the other and would have to vanish.  %G; F; orginally appendix C
\section{Transforming Compactly Supported Perturbations to Weyl-Papapetrou Gauge}
\label{app:transforming}
As discussed previously (c.f., the discussion near the end of Appendix \ref{app:global}) one can evolve a large class of compactly supported solutions to the linearized constraint equations in a hyperbolic gauge and appeal to finite propagation speed to show that such perturbations remain bounded away from the horizon and spatial infinity for all finite (Boyer-Lindquist) time \textit{t}. On the other hand our energy flux derivation has assumed that the perturbations be expressed in a Weyl-Papapetrou gauge in order to make them amenable to an application of Robinson's identity in its traditional form. Thus we need to consider the transformation of perturbations expressed in say a hyperbolic gauge of Lorenz type to a Weyl-Papapetrou gauge of the `elliptic type' needed for our analysis.

Let \(\lbrace{}^{(4)}g = {}^{(4)}g_{\mu\nu} dx^\mu \otimes dx^\nu,\quad {}^{(4)}A = {}^{(4)}A_\mu dx^\mu\rbrace\) be a Kerr-Newman black hole solution expressed in coordinates \(\lbrace x^0 = t, x^1, x^2, x^3 = \varphi\rbrace\) of the (Boyer-Lindquist) type introduced in Appendix~\ref{app:kerr-newman-spacetimes} (wherein \(\zeta = \frac{\partial}{\partial t}\) and \(\psi = \frac{\partial}{\partial\varphi}\) are the Killing fields corresponding to the given black hole's stationarity and axial symmetry). Relative to this background let \(\lbrace{}^{(4)}g' := {}^{(4)}k = {}^{(4)}k_{\mu\nu} dx^\mu \otimes dx^\nu,\quad {}^{(4)}A' := {}^{(4)}\ell = {}^{(4)}\ell_\mu dx^\mu\rbrace\) designate an \textit{axisymmetric}, spatially compactly supported solution to the corresponding linearized equations. If \({}^{(4)}Y = {}^{(4)}Y^\mu \frac{\partial}{\partial x^\mu}\) is a (sufficiently smooth) vector field invariant with respect to the rotations generated by \(\psi\), i.e., such that
\begin{equation}\label{eq:f01}
\left(\mathcal{L}_\psi\; {}^{(4)}Y\right)^\mu = {}^{(4)}Y^\mu_{\hphantom{\mu},\varphi} = 0,
\end{equation}
then the gauge transformed perturbations
\begin{equation}\label{eq:f02}
\lbrace{}^{(4)}\tilde{k} = {}^{(4)}\tilde{k}_{\mu\nu} dx^\mu \otimes dx^\mu,\quad {}^{(4)}\tilde{\ell} = {}^{(4)}\tilde{\ell}_\mu dx^\mu\rbrace
\end{equation}
defined by
\begin{align}
{}^{(4)}\tilde{k}_{\mu\nu} &= {}^{(4)}k_{\mu\nu} + \left(\mathcal{L}_{{}^{(4)}Y} {}^{(4)}g\right)_{\mu\nu}\label{eq:f03}\\
{}^{(4)}\tilde{\ell}_\mu &= {}^{(4)}\ell_\mu + \left(\mathcal{L}_{{}^{(4)}Y} {}^{(4)}A\right)_\mu\label{eq:f04}
\end{align}
will also satisfy the linearized field equations and preserve \textit{explicit} axi\-symmetry, i.e., obey
\begin{equation}\label{eq:f05}
\left(\mathcal{L}_\psi {}^{(4)}\tilde{k}\right)_{\mu\nu} = {}^{(4)}\tilde{k}_{\mu\nu,\varphi} = 0,
\end{equation}
and
\begin{equation}\label{eq:f06}
\left(\mathcal{L}_\psi {}^{(4)}\tilde{\ell}\right)_\mu = {}^{(4)}\tilde{\ell}_{\mu,\varphi} = 0.
\end{equation}
Recalling that, in our notation,
\begin{equation}\label{eq:f07}
\begin{split}
{}^{(4)}g_{ab} &= e^{-2\gamma} \tilde{g}_{ab} + e^{2\gamma} \beta_a \beta_b\\
 &= e^{-2\gamma + 2\nu} h_{ab} + e^{2\gamma} \beta_a \beta_b
\end{split}
\end{equation}
where \(h_{ab} dx^a \otimes dx^b\) is a \textit{flat} 2-metric which, in Weyl-Papapetrou spatial coordinates \(\lbrace x^a\rbrace = \lbrace\rho, z\rbrace\), satisfies the (conformally invariant) condition
\begin{equation}\label{eq:f08}
\frac{1}{\sqrt{{}^{(2)}h}}\; h_{ab} dx^a \otimes dx^b = d\rho \otimes d\rho + dz \otimes dz
\end{equation}
and recalling as well that \(\beta_a = 0\) on a Kerr-Newman background, we see that a gauge transformed perturbation \({}^{(4)}\tilde{k}_{\mu\nu}\) of \({}^{(4)}g_{\mu\nu}\) will preserve this Weyl-Papapetrou form to linearized order if and only if it satisfies the gauge conditions
\begin{equation}\label{eq:f09}
{}^{(4)}\tilde{k}_{\rho\rho} - {}^{(4)}\tilde{k}_{zz} = 0,\quad {}^{(4)}\tilde{k}_{\rho z} = 0.
\end{equation}

Appealing to Eq.~(\ref{eq:f03}) one can reexpress these conditions in the form
\begin{equation}\label{eq:f10}
\left(\mathcal{L}_{{}^{(4)}Y} {}^{(4)}g\right)_{ab} - \frac{1}{2} h_{ab}\; h^{cd} \left(\mathcal{L}_{{}^{(4)}Y} {}^{(4)}g\right)_{cd} = -\left({}^{(4)}k_{ab} - \frac{1}{2} h_{ab}\; h^{cd}\; {}^{(4)}k_{cd}\right)
\end{equation}
But utilizing the fact that the background metric also satisfies
\begin{equation}\label{eq:f11}
{}^{(4)}g_{ta} = {}^{(4)}g_{\varphi a} = 0,\quad {}^{(4)}g_{\mu\nu,\varphi} = {}^{(4)}g_{\mu\nu,t} = 0
\end{equation}
and that \({}^{(4)}Y^\mu_{\hphantom{\mu},\varphi} = 0\) by assumption one can rewrite Eq.~(\ref{eq:f10}) in the 2-dimensionally covariant form
\begin{equation}\label{eq:f12}
\left(\mathcal{L}_{{}^{(2)}Y} {}^{(2)}h\right)_{ab} - \frac{1}{2} h_{ab}\; h^{cd} \left(\mathcal{L}_{{}^{(2)}Y} {}^{(2)}h\right)_{cd} = -e^{2\gamma - 2\nu} \left({}^{(4)}k_{ab} - \frac{1}{2} h_{ab}\; h^{cd}\; {}^{(4)}k_{cd}\right)
\end{equation}
where \({}^{(2)}Y := {}^{(2)}Y^a \frac{\partial}{\partial x^a} = {}^{(4)}Y^a \frac{\partial}{\partial x^a}\). Note that in the complement of the support of \({}^{(4)}k\) (i.e., in the `asymptotic regions' near the horizon and near spatial infinity) Eq.~(\ref{eq:f12}) reduces to the conformal Killing equation for the flat 2-metric \({}^{(2)}h\). As we shall show below this equation (with its inhomogeniety included) can be solved explicitly for \({}^{(2)}Y\) thus determining the (2-dimensional) `spatial components' of \({}^{(4)}Y\).

These 2-dimensional `spatial' components of \({}^{(4)}Y\) will play a distinctive role in that the induced gauge transformations of the linearized wave map \textit{scalars} \(\left\lbrace\gamma', \omega', \lambda', \eta'\right\rbrace\) generated by \({}^{(4)}Y\), namely
\begin{align}
\tilde{\gamma}' &:= \gamma' + \mathcal{L}_{{}^{(4)}Y} \gamma, \quad \tilde{\omega}' := \omega' + \mathcal{L}_{{}^{(4)}Y} \omega,\label{eq:f13}\\
\tilde{\lambda}' &:= \lambda' + \mathcal{L}_{{}^{(4)}Y} \lambda, \quad \tilde{\eta}' := \eta' + \mathcal{L}_{{}^{(4)}Y} \eta,\label{eq:f14}
\end{align}
simplify to
\begin{align}
\tilde{\gamma}' &= \gamma' + {}^{(2)}Y^a \gamma_{,a},\quad \tilde{\omega}' = \omega' + {}^{(2)}Y^a \omega_{,a},\label{eq:f15}\\
\tilde{\lambda}' &= \lambda' + {}^{(2)}Y^a \lambda_{,a},\quad \tilde{\eta}' = \eta' + {}^{(2)}Y^a \eta_{,a}\label{eq:f16}
\end{align}
in view of the invariance of the background fields \(\lbrace\gamma, \omega, \lambda, \eta\rbrace\) with respect to \textit{t} and \(\varphi\) translations. In particular, in the complement of the support of the (compactly supported) perturbations \(\lbrace\gamma', \omega', \lambda', \eta'\rbrace\) their gauge transformed counterparts \(\lbrace\tilde{\gamma}', \tilde{\omega}', \tilde{\lambda}', \tilde{\eta}'\rbrace\), though no longer in general having compact support, will nevertheless simplify to their \textit{pure gauge} forms
\begin{align}
\tilde{\gamma}' &\longrightarrow {}^{(2)}Y^a \gamma_{,a},\quad \tilde{\omega}' \longrightarrow {}^{(2)}Y^a \omega_{,a}\label{eq:f17}\\
\tilde{\lambda}' &\longrightarrow {}^{(2)}Y^a \lambda_{,a},\quad \tilde{\eta}' \longrightarrow {}^{(2)}Y^a \eta_{,a}.\label{eq:f18}
\end{align}

On the other hand, to compute the gauge transformations of the linearized canonical momenta \(\lbrace\tilde{p}', \tilde{r}', \tilde{v}', \tilde{u}'\rbrace\) we shall need the time component \({}^{(4)}Y^0\) of \({}^{(4)}Y\). To see how this is determined recall that the (3+1-dimensional) lapse function \textit{N} and shift field \(N^m \frac{\partial}{\partial x^m}\) are defined by
\begin{equation}\label{eq:f19}
-\frac{1}{N^2} = {}^{(4)}g^{00}\; \hbox{ and }\; \frac{N^m}{N^2} = {}^{(4)}g^{0m}.
\end{equation}
Thus a first variation \(\delta N\) of \textit{N} induced by \({}^{(4)}\tilde{k}_{\alpha\beta}\) is given by
\begin{equation}\label{eq:f20}
\begin{split}
\frac{2}{N^3} \delta N &= -{}^{(4)}g^{0\alpha}\; {}^{(4)}g^{0\beta}\; {}^{(4)}\tilde{k}_{\alpha\beta}\\
&= -{}^{(4)}g^{0\alpha}\; {}^{(4)}g^{0\beta} \left({}^{(4)}k_{\alpha\beta} + \left(\mathcal{L}_{{}^{(4)}Y} {}^{(4)}g\right)_{\alpha\beta}\right).
\end{split}
\end{equation}
Evaluating the Lie derivative and recalling that, in our notation, \(N = e^{-\gamma}\tilde{N}\) so that
\begin{equation}\label{eq:f21}
\delta N = e^{-\gamma} \delta\tilde{N} - \tilde{N} e^{-\gamma} \delta\gamma
\end{equation}
whereas, since \({}^{(4)}g_{\varphi\varphi} = e^{2\gamma}\),
\begin{equation}\label{eq:f22}
\begin{split}
\delta\gamma &= \frac{1}{2} e^{-2\gamma}\; {}^{(4)}\tilde{k}_{\varphi\varphi}\\
&= \frac{1}{2} e^{-2\gamma}\; {}^{(4)}k_{\varphi\varphi} + {}^{(2)}Y^a \gamma_{,a}
\end{split}
\end{equation}
we arrive at
\begin{equation}\label{eq:f23}
\frac{\tilde{N}'}{\tilde{N}} = \frac{1}{2} e^{-2\gamma} \left({}^{(4)}k_{\varphi\varphi} - \tilde{N}^2\; {}^{(4)}g^{0\alpha}\; {}^{(4)}g^{0\beta}\; {}^{(4)}k_{\alpha\beta}\right) + {}^{(4)}Y^0_{\hphantom{0},0} + {}^{(2)}Y^a \frac{\tilde{N}_{,a}}{\tilde{N}}
\end{equation}
where we now write \(\tilde{N}'\) for \(\delta\tilde{N}\) in accordance with our previously established notation. Thus given a choice for the linearized lapse function \(\tilde{N}'\) in the desired (elliptic) gauge, Eq.~(\ref{eq:f23}) determines \({}^{(4)}Y^0\) by direct time integration.

In a completely analogous way one finds that the components of the linearized shift are given by
\begin{align}
\begin{split}\label{eq:f24}
\tilde{N}^{c'} &= N^{c'} = \tilde{g}^{ac} \left\lbrack e^{2\gamma}\left({}^{(4)}k_{0a} - \beta_0 {}^{(4)}k_{a\varphi}\right)\right\rbrack\\
 & {} + {}^{(4)}Y^c_{\hphantom{c},0} - \tilde{N}^2\; \tilde{g}^{ac}\; {}^{(4)}Y^0_{\hphantom{0},a}
 \end{split}\\
 \begin{split}\label{eq:f25}
N^{\varphi'} &= (\beta_0 - \tilde{N}^a \beta_a)'\\
&= e^{-2\gamma} \left({}^{(4)}k_{0\varphi} - \beta_0 {}^{(4)}k_{\varphi\varphi}\right)\\
& {} + {}^{(4)}Y^0_{\hphantom{0},0} \beta_0 + {}^{(4)}Y^\varphi_{\hphantom{\varphi},0} + {}^{(2)}Y^c \beta_{0,c}\\
&\longrightarrow \beta'_0\qquad \hbox{ (since $\tilde{N}^a = \beta_a = 0$ in the background)}
\end{split}
\end{align}
Note that this last equation provides a means of computing the `last' component, \({}^{(4)}Y^\varphi\), of \({}^{(4)}Y\) provided that a gauge condition for \(N^{\varphi'}\) is specified. However a different way of computing \({}^{(4)}Y^\varphi\) (that would then fix the corresponding choice for \(N^{\varphi'}\)) arises from noting that
\begin{equation}\label{eq:f26}
\begin{split}
(e^{2\gamma} \beta_a)' &\longrightarrow e^{2\gamma} \beta'_a = {}^{(4)}\tilde{k}_{a\varphi}\\
&= {}^{(4)}k_{a\varphi} + \left(\mathcal{L}_{{}^{(4)}Y} {}^{(4)}g\right)_{a\varphi}\\
&= {}^{(4)}k_{a\varphi} + e^{2\gamma} \left({}^{(4)}Y^0_{\hphantom{0},a} \beta_0 + {}^{(4)}Y^\varphi_{\hphantom{\varphi},a}\right)
\end{split}
\end{equation}
so that
\begin{equation}\label{eq:f27}
\beta'_a = {}^{(4)}Y^\varphi_{\hphantom{\varphi},a} + \beta_0 {}^{(4)}Y^0_{\hphantom{0},a} + e^{-2\gamma} {}^{(4)}k_{a\varphi}.
\end{equation}
Thus a choice for \({}^{(4)}Y^\varphi\) allows one to control the `longitudinal part' of \(\beta'_a\) whereas its `transversal part' is governed independently by the linearized wave map momentum variable \(\tilde{r}'\) (c.f., Eq.~(\ref{eq:a28}))

To actually solve Eq.~(\ref{eq:f12}) let us first reexpress it in the more convenient form
\begin{equation}\label{eq:f28}
\begin{split}
\left(\mathcal{L}_{{}^{(2)}Y} \left(\sqrt{{}^{(2)}h}\; {}^{(2)}h\right)^{cd}\right) &= \sqrt{{}^{(2)}\tilde{g}}\; \tilde{g}^{ac}\; \tilde{g}^{bd}\; e^{2\gamma} \left\lbrace{}^{(4)}k_{ab} - \frac{1}{2} \tilde{g}_{ab}\; \tilde{g}^{ef}\; {}^{(4)}k_{ef}\right\rbrace\\
&:= \mathcal{M}^{cd}
\end{split}
\end{equation}
Evaluating this (traceless, symmetric) equation in the \(\lbrace R,\theta\rbrace\) coordinates of Appendix~\ref{app:kerr-newman-spacetimes}, for which
\begin{equation}\label{eq:f29}
\sqrt{{}^{(2)}h}\; h^{cd} \frac{\partial}{\partial x^c} \otimes \frac{\partial}{\partial x^d} = R \frac{\partial}{\partial R} \otimes \frac{\partial}{\partial R} + \frac{1}{R} \frac{\partial}{\partial\theta} \otimes \frac{\partial}{\partial\theta},
\end{equation}
one gets the two independent components
\begin{align}
{}^{(2)}Y_{\hphantom{\theta},\theta}^\theta &= R\left(\frac{{}^{(2)}Y^R}{R}\right)_{,R} + \frac{1}{R} \mathcal{M}^{RR}\label{eq:f30}\\
{}^{(2)}Y_{\hphantom{\theta},R}^\theta &= -\frac{1}{R} \left(\frac{{}^{(2)}Y^R}{R}\right)_{,\theta} - \frac{1}{R} \mathcal{M}^{R\theta} \label{eq:f31}
\end{align}
The integrability condition for this first order system for \({}^{(2)}Y^\theta\) is the Poisson-type equation for \(\frac{{}^{(2)}Y^R}{R}\) given by
\begin{equation}\label{eq:f32}
\frac{1}{R} \left(R \frac{\partial}{\partial R} \left(\frac{{}^{(2)}Y^R}{R}\right)\right)_{,R} + \frac{1}{R^2} \frac{\partial^2}{\partial\theta^2} \left(\frac{{}^{(2)}Y^R}{R}\right) = -\frac{1}{R^2} \mathcal{M}^{R\theta}_{\hphantom{R\theta},\theta} - \frac{1}{R} \left(\frac{1}{R} \mathcal{M}^{RR}\right)_{,R}
\end{equation}
Note that the operator acting on \(\frac{{}^{(2)}Y^R}{R}\) in this equation is identical to the \textit{scalar} Laplacian for the flat metric \({}^{(2)}f\) (conformal to \({}^{(2)}h\)) given by
\begin{equation}\label{eq:f33}
{}^{(2)}f = dR \otimes dR + R^2 d\theta \otimes d\theta
\end{equation}

For reasons of regularity the radial component, \({}^{(2)}Y^R\), of \({}^{(2)}Y\) must admit a Fourier expansion of the form
\begin{equation}\label{eq:f34}
{}^{(2)}Y^R = a_0 (R,t) + \sum_{n=1}^\infty a_n (R,t) \cos{(n\theta)}
\end{equation}
so that, in particular, its \(\theta\)-derivative vanishes on the axes of symmetry corresponding to \(\theta = 0\) and \(\theta = \pi\). For the same reasons \({}^{(2)}Y^\theta\) must itself vanish at these axes and thus admit an expansion of the form
\begin{equation}\label{eq:f35}
{}^{(2)}Y^\theta = \sum_{n=1}^\infty b_n (R,t) \sin{(n\theta)}
\end{equation}
Similar considerations for the vector density resulting from pairing the one-form \(dR\) with the tensor density \(\left\lbrace\mathcal{M}^{ab} \frac{\partial}{\partial x^a} \otimes \frac{\partial}{\partial x^b}\right\rbrace\) lead to Fourier explansions of the latter's components given by
\begin{align}
-\frac{1}{R} \mathcal{M}^{RR} &= c_0 (R,t) + \sum_{n=1}^\infty c_n (R,t) \cos{(n\theta)}\label{eq:f36}\\
-R\mathcal{M}^{R\theta} &= \sum_{n=1}^\infty d_n (R,t) \sin{(n\theta)}\label{eq:f37}
\end{align}

Substituting these expansions into Eqs.~(\ref{eq:f30}--\ref{eq:f31}) leads to the following system for the Fourier coefficients
\begin{align}
a_{0,R} - \frac{1}{R} a_0 &= c_0, \label{eq:f38}\\
a_{n,R} - \frac{1}{R} a_n - nb_n &= c_n, \label{eq:f39}\\
R^2 b_{n,R} - na_n &= d_n\label{eq:f40}
\end{align}
for \(n = 1, 2, \dots\).

While we shall show below how to solve this system explicitly using the method of `variation of parameters' this will not, by itself, deal with the convergence issues presented by the resultant (formal) Fourier series. To prove that global, bounded solutions to Eq.~(\ref{eq:f28}) for \({}^{(2)}Y\) do indeed exist we shall instead first solve the Poisson equation for \(\left(\frac{{}^{(2)}Y^R}{R}\right)_{,\theta}\) which, together with the solution of Eq.~(\ref{eq:f38}), will serve to determine \({}^{(2)}Y^R\) and, at the same time, provide the needed integrability condition for the complementary component \({}^{(2)}Y^\theta\).

For convenience extend the domains of definitions (at fixed \textit{t} which, for simplicity, we suppress in the following) of the source components \(\lbrace\mathcal{M}^{ab}\rbrace\) to the full plane \(\mathbb{R}^2\) with the open disk of radius \(R_+ = \frac{1}{2} \sqrt{M^2 - (a^2 + Q^2)}\) removed. This corresponds to `reflecting' \(\mathcal{M}^{RR}\) and `anti-reflecting' \(\mathcal{M}^{R\theta}\) through the z-axis or, equivalently, through taking the range of \(\theta\) in expansions (\ref{eq:f36}--\ref{eq:f37}) to now be \(\lbrack 0, 2\pi)\). Note accordingly that the source term on the right-hand side of Eq.~(\ref{eq:f32}) will automatically be reflection symmetric whereas its \(\theta\)-derivative, which provides the source in the Poisson equation for \(\left(\frac{{}^{(2)}Y^R}{R}\right)_{,\theta}\), namely
\begin{equation}\label{eq:f41}
\frac{1}{R} \frac{\partial}{\partial R} \left( R \frac{\partial}{\partial R} \left(\left(\frac{{}^{(2)}Y^R}{R}\right)_{,\theta}\right)\right) + \frac{1}{R^2} \frac{\partial^2}{\partial\theta^2} \left(\left(\frac{{}^{(2)}Y^R}{R}\right)_{,\theta}\right) = -\frac{1}{R^2} \mathcal{M}^{R\theta}_{\hphantom{R\theta},\theta\theta} - \frac{1}{R} \left(\frac{1}{R} \mathcal{M}^{RR}_{\hphantom{RR},\theta}\right)_{,R},
\end{equation}
will be reflection \textit{anti}-symetric and thus have a vanishing net `charge' as well as having compact support.

The fundamental solution (Green's function) for this Dirichlet problem is explicitly known and, for arbitrary sufficiently smooth, reflection anti-symmetric Dirichlet data specified on the circle \(R = R_+\), provides a unique, globally bounded, reflection anti-symmetric solution, \(\left(\frac{{}^{(2)}Y^R}{R}\right)_{,\theta}\), that decays asymptotically like \(\sim \frac{1}{R}\) \cite{EDB_10,RHB_06}. Note that terms of the form \(\alpha + \beta \ln{(R/R_+)}\) that might otherwise be expected to occur are excluded by the reflection anti-symmetry of the source and boundary conditions.

To complete the determination of \({}^{(2)}Y^R\) we must solve Eq.~(\ref{eq:f38}) for the Fourier component \(a_0\) which, in view of Eq.~(\ref{eq:f34}), is defined by
\begin{equation}\label{eq:f42}
a_0(R) = \frac{1}{2\pi} \int_0^{2\pi} d\theta\; {}^{(2)}Y^R (R,\theta).
\end{equation}
From Eq.~(\ref{eq:f36}) we see that the source, \(c_0(R)\), for this quantity is in turn given by
\begin{equation}\label{eq:f43}
c_0(R) = \frac{1}{2\pi} \int_0^{2\pi} d\theta\; \left(-\frac{1}{R} \mathcal{M}^{RR}(R,\theta)\right).
\end{equation}
This solution to Eq.~(\ref{eq:f38}) is simply
\begin{equation}\label{eq:f44}
a_0(R) = R\left\lbrack\frac{a_0(R_+)}{R_+} + \int_{R_+}^R dR'\; \left(\frac{1}{R'} c_0(R')\right)\right\rbrack
\end{equation}
but only the unique choice
\begin{equation}\label{eq:f45}
a_0(R_+) = -R_+ \int_{R_+}^\infty dR'\; \left\lbrack\frac{1}{R'} c_0(R')\right\rbrack
\end{equation}
yields a globally bounded solution for \(a_0(R)\) which in fact \textit{vanishes} outside the support of \(c_0\).

After fixing (for reasons of regularity at the axes) \({}^{(2)}Y^\theta (R_+,0) = 0\) one could now integrate the first order system (\ref{eq:f30})--(\ref{eq:f31}) to determine \({}^{(2)}Y^\theta (R,\theta)\). A more elegant approach however is to combine this regularity condition with the integral of Eq.~({\ref{eq:f30}) with respect to \(\theta\) at \(R = R_+\) to determine (reflection anti-symmetric) Dirichlet data, \({}^{(2)}Y^\theta (R_+,\theta)\) for the solution to the Poisson equation
\begin{equation}\label{eq:f46}
\frac{1}{R} \left( R\; {}^{(2)}Y^\theta_{\hphantom{\theta},R}\right)_{,R} + \frac{1}{R^2}\; {}^{(2)}Y^\theta_{\hphantom{\theta},\theta\theta} = -\frac{1}{R} \mathcal{M}^{R\theta}_{\hphantom{R\theta},R} + \frac{1}{R^3} \mathcal{M}^{RR}_{\hphantom{RR},\theta}.
\end{equation}
which, in turn, results from Eqs.~(\ref{eq:f30})--(\ref{eq:f31}). Since the source term in Eq.~(\ref{eq:f46}) and its associated Dirichlet data are both reflection anti-symmetric this Poisson equation has a unique, globally bounded, reflection anti-symmetric solution. From the explicit form of Green's function combined with the source's compact support it further follows that \({}^{(2)}Y^\theta(R,\theta)\) decays asymptotically as \(\sim \frac{1}{R}\).

The above argument has shown that a unique, globally bounded, regular solution for \({}^{(2)}Y\) is determined from specifying Dirichlet data for \({}^{(2)}Y^R\) at the horizon. On the other hand it is still of interest to see more explicitly how the Fourier coefficients \(\lbrace a_0, a_n, b_n\rbrace\) for this solution behave, especially in the asymptotic regions. We have already solved Eq.~(\ref{eq:f38}) and found that
\begin{equation}\label{eq:f47}
\begin{split}
\frac{1}{2\pi} \int_0^{2\pi} d\theta\; {}^{(2)}Y^R(R,\theta) &= a_0(R)\\
&= -R \left\lbrack\int_{R_+}^\infty dR'\; \left(\frac{1}{R'} c_0(R')\right)\right\rbrack
\end{split}
\end{equation}
near \(R = R_+\) and that
\begin{equation}\label{eq:f48}
a_0(R) = 0
\end{equation}
for all \textit{R} outside the support of \(c_0\).

We shall prove below in Appendix~\ref{app:vanishing} that, for the perturbations of interest herein, the `integral invariant' \(a_0 (R_+)\) defined by Eq.~(\ref{eq:f45}) \textit{actually vanishes}. It follows then from Eqs.~(\ref{eq:f44}) and (\ref{eq:f48}) that \(a_0 (R)\) will vanish both inside and outside the support of \(c_0\) (i.e., throughout both asymptotic regions).

To solve Eqs.~(\ref{eq:f39}) and (\ref{eq:f40}) first note that they imply
\begin{equation}\label{eq:f49}
\frac{1}{R}(R\; b_{n,R})_{,R} - \frac{n^2 b_n}{R^2} = \frac{nc_n}{R^2} + \frac{1}{R} \left(\frac{d_n}{R}\right)_{,R}
\end{equation}
and that independent solutions to the corresponding homogenous equations (for \(n = 1, 2, \cdots\)) are given by \(R^n\) and \(R^{-n}\). It is therefore straightforward to apply the method of \textit{variation of parameters} to show that, for each \textit{n}, there is a unique, globally bounded solution \(b_n(R)\) determined by boundary data \(b_n(R_+)\) specified at the horizon. These are of course nothing but the Fourier coefficients for the corresponding solutions \({}^{(2)}Y^\theta\) to Eq.~(\ref{eq:f46}) found previously. For large \textit{R}, outside the source's support, these solutions take the form (at fixed time \textit{t})
\begin{equation}\label{eq:f50}
b_n(R) = \beta_n^{(-)} R^{-n}
\end{equation}
for suitable constants \(\lbrace\beta_n^{(-)}\rbrace\) whereas for \textit{R} sufficiently near \(R_+\) (inside the source's support) they have the form
\begin{equation}\label{eq:f51}
b_n(R) = \alpha_n^{(+)} R^n + \alpha_n^{(-)} R^{-n}
\end{equation}
for suitable constants \(\lbrace\alpha_n^{(+)}, \alpha_n^{(-)}\rbrace\). The constants \(\lbrace\alpha_n^{(+)}, \alpha_n^{(-)}, \beta_n^{(-)}\rbrace\) are all determined explicitly in terms of the chosen Dirichlet data specified at \(R_+\) and by the source functions \(\lbrace c_n(R), d_n(R)\rbrace\).

By now simply setting \(\forall\; n \geq 1\)
\begin{equation}\label{eq:f52}
a_n(R) = \frac{R^2 b_{n,R}}{n}  - \frac{d_n(R)}{n}
\end{equation}
one readily verifies that all of Eqs.~(\ref{eq:f39}) and (\ref{eq:f40}) are satisfied and that the \(\lbrace a_n(R)\rbrace\) take the asymptotic forms
\begin{equation}\label{eq:f53}
a_n(R) = -\beta_n^{(-)} R^{-n+1}
\end{equation}
for \textit{R} sufficiently large and
\begin{equation}\label{eq:f54}
a_n(R) = \alpha_n^{(+)} R^{n+1} - \alpha_n^{(-)} R^{-n+1}
\end{equation}
for \textit{R} sufficiently near \(R_+\). Note in particular that \(a_1(R) \longrightarrow -\beta_1^{(-)}\) for large \textit{R} whereas the higher order coefficients \(\lbrace a_n(R); n = 2, 3, \dots\rbrace\) decay as increasingly negative powers of \textit{R}.

While it may not be specifically needed for our analysis to go through we shall focus henceforth on those particular gauge transformations generated by vector fields \({}^{(2)}Y\) satisfying the `homogeneous' Dirichlet condition
\begin{equation}\label{eq:f55}
{}^{(2)}Y^R_{\hphantom{R},\theta}(R_+,\theta) = 0.
\end{equation}
From Eqs.~(\ref{eq:f34}) and (\ref{eq:f54}) this boundary condition clearly corresponds to setting \(a_n(R_+) = 0\quad \forall\; n \geq 1\) or, equivalently
\begin{equation}\label{eq:f56}
\alpha^{(-)}_n = R_+^{2n}\; \alpha_n^{(+)}.
\end{equation}
Designating the `source' term for Eq.~(\ref{eq:f49}) by \(\sigma_n (R)\) so that
\begin{equation}\label{eq:f57}
\sigma_n (R) := \frac{nc_n (R)}{R^2} + \frac{1}{R}\; \frac{\partial}{\partial R} \left(\frac{d_n (R)}{R}\right)
\end{equation}
one readily finds the unique, globally bounded solution to this equation to be
\begin{equation}\label{eq:f58}
\begin{split}
b_n &= -R^n \left\lbrace\int_R^\infty \frac{\sigma_n (R')}{2n} (R')^{1-n}\; dR'\right\rbrace\\
 &\hphantom{=} {} + R^{-n} \left\lbrace - R_+^{2n} \int_{R_+}^\infty \frac{\sigma_n (R')}{2n} (R')^{1-n}\; dR'\right.\\
 &\hphantom{= {} +}\left. {} - \int_{R_+}^R \frac{\sigma_n (R')}{2n} (R')^{n+1}\; dR'\right\rbrace .
\end{split}
\end{equation}
Specializing this formula to the asymptotic regions corresponding to \(R \searrow R_+\) and \(R \nearrow \infty\) one easily discovers that the coefficients \(\lbrace\alpha_n^{(+)}, \alpha_n^{(-)}, \beta_n^{(-)}\rbrace\) are given by
\begin{align}
\alpha_n^{(+)} &= -\int_{R_+}^\infty \frac{\sigma_n (R')}{2n} (R')^{1-n}\; dR'\nonumber\\
&= \frac{\alpha_n^{(-)}}{R_+^{2n}}\label{eq:f59}\\
\intertext{and}
\beta_n^{(-)} &= \left\lbrace -  R_+^{2n} \int_{R_+}^\infty \frac{\sigma_n (R')}{2n} (R')^{1-n}\; dR'\right.\nonumber\\
&\hphantom{=} \left. {} - \int_{R_+}^\infty \frac{\sigma_n (R')}{2n} (R')^{n+1}\; dR'\right\rbrace \label{eq:f60}
\end{align}
wherein, as above, we have suppressed their time dependence to simplify the notation.

Since we have already argued (c.f., the proof given in Appendix~\ref{app:vanishing}) that \(a_0 (R_+) = 0\) for the perturbations of interest herein, it follows from Eqs.~(\ref{eq:f55}) and (\ref{eq:f56}) that \({}^{(2)}Y^R\) satisfies the Dirichlet condition
\begin{equation}\label{eq:f80}
{}^{(2)}Y^R (R_+,\theta) = 0
\end{equation}
at the horizon boundary.

At several points in our discussion we have encountered occasions wherein the leading order term in an expansion of the form
\begin{equation}\label{eq:f61}
\Psi_1 (R,\theta) := \sum_{k=1}^\infty \beta_k^{(-)}\; \sin{(k\theta)} \frac{1}{R^k}
\end{equation}
cancels out in the expression of interest leaving what appears to have a faster rate of decay as \(R \nearrow \infty\). While this higher rate of decay would be self-evident for a \textit{finite} series it is not obvious in the case of an infinite series that the `remainder' does indeed decay faster than the leading order term. For the functions considered herein however we shall see that this is indeed the case.

In the asymptotic region near \(\infty\) the functions of interest in this context are harmonic, hence analytic and have convergent expansions of the type indicated above. If for some reason the first \(N - 1\) terms (for \(N \geq 2\)) cancelled from a quantity being computed we'd be left with a (convergent) expansion of the form
\begin{equation}\label{eq:f62}
\Psi_N = \sum_{k=N}^\infty \beta_k^{(-)}\; \sin{(k\theta)} \frac{1}{R^k}.
\end{equation}
We wish to consider this in the asymptotic region \(R > R_0 > R_+\). For this purpose define, for convenience, the coordinate \textit{x} by
\begin{equation}\label{eq:f63}
R = \frac{R_0}{1 - R_0x},\qquad x \in \left(-\frac{1}{R_0}, \frac{1}{R_0}\right)
\end{equation}
so that \(x \searrow 0 \Leftrightarrow R \searrow R_0\) and \(x \nearrow \frac{1}{R_0} \Leftrightarrow R \nearrow \infty\).

Writing
\begin{equation}\label{eq:f64}
\Psi_N = \sum_{k=N}^\infty \beta_k^{(-)}\; \sin{(k\theta)} \left(\frac{1 - R_0x}{R_0}\right)^k
\end{equation}
and recalling that the analyticity of \(\Psi_N\) implies the \textit{absolute} convergence of its series expansion we get
\begin{equation}\label{eq:f65}
\begin{split}
|\Psi_N| &= \left|\sum_{k=N}^\infty \beta_k^{(-)}\; \sin{(k\theta)}\; \left(\frac{1 - R_0x}{R_0}\right)^k\right|\\
 &\hphantom{=} {} \leq \sum_{k=N}^\infty \left|\beta_k^{(-)}\; \sin{(k\theta)}\right|\; \left|\frac{1 - R_0x}{R_0}\right|^k\\
 &= \left|\frac{1 - R_0x}{R_0}\right|^N \sum_{k=N}^\infty \left|\beta_k^{(-)}\; \sin{(k\theta)}\right| \left|\frac{1 - R_0x}{R_0}\right|^{k-N}\\
 &= \left|\frac{1 - R_0x}{R_0}\right|^N \sum_{\ell = 0}^\infty \left|\beta_{\ell + N}^{(-)}\; \sin{\left( (\ell + N)\theta\right)}\right| \left|\frac{1 - R_0x}{R_0}\right|^\ell\\
 &\hphantom{=} {} < \infty.
\end{split}
\end{equation}
Every term in the final summation has positive sign and either remains constant or decays monotonically in \textit{R} as \(R \rightarrow \infty\) (i.e., \(x \nearrow \frac{1}{R_0}\)) for any fixed \(\theta\). It follows that the resultant expression for \(|\Psi_N|\) decays at least of order \(O \left(\left(\frac{1 - R_0x}{R_0}\right)^N\right) = O \left(\frac{1}{R^N}\right)\) for \(N \geq 2\) as \(R \nearrow \infty\).

Recalling that
\begin{equation}\label{eq:f66}
\frac{\partial r}{\partial R} = \left( 1 - \frac{R_+^2}{R^2}\right)
\end{equation}
and that the partial derivatives \(\lbrace\gamma_{,r}, \gamma_{,\theta}, \omega_{,r}, \omega_{,\theta}, \lambda_{,r}, \lambda_{,\theta}, \eta_{,r}, \eta_{,\theta}\rbrace\) are all bounded at \(R = R_+\) we see that the corresponding gauge transformed perturbations \(\lbrace\tilde{\gamma}', \tilde{\omega}', \tilde{\lambda}', \tilde{\eta}'\rbrace\) are all \textit{regular at the horizon}. (c.f., Eqs.~(\ref{eq:f15})--(\ref{eq:f18})).

To evaluate the relevant `flux' integrals resulting from, for example, the integrated form of Eq.~(\ref{eq:401}) we shall need the asymptotic forms of the linearized canonical momenta. These are given by the linearized field equations (c.f., Eqs.~(\ref{eq:a49}), (\ref{eq:a51}), (\ref{eq:a53}), (\ref{eq:a55}) and (\ref{eq:a57}))
\begin{align}
\frac{\tilde{N}\tilde{p}'}{\sqrt{{}^{(2)}\tilde{g}}} &= 4(\gamma'_{,t} - \tilde{N}^{a'} \gamma_{,a}),\label{eq:f67}\\
\frac{\tilde{N} e^{2\gamma}\tilde{v}'}{\sqrt{{}^{(2)}\tilde{g}}} &= \lambda'_{,t} - \tilde{N}^{a'} \lambda_{,a},\label{eq:f68}\\
\frac{\tilde{N} e^{2\gamma}(\tilde{u}' - \lambda\tilde{r}')}{\sqrt{{}^{(2)}\tilde{g}}} &= \eta'_{,t} - \tilde{N}^{a'}\eta_{,a},\label{eq:f69}\\
\begin{split}\label{eq:f70}
\frac{\tilde{N} e^{4\gamma}\tilde{r}'}{\sqrt{{}^{(2)}\tilde{g}}} &= \omega'_{,t} - \tilde{N}^{a'}\omega_{,a} + \frac{\tilde{N} e^{2\gamma}}{\sqrt{{}^{(2)}\tilde{g}}} \lambda (\tilde{u}' - \lambda\tilde{r}'),\\
&= \omega'_{,t} - \tilde{N}^{a'} \omega_{,a} + \lambda (\eta'_{,t} - \tilde{N}^{a'} \eta_{,a}),
\end{split}\\
\frac{2\tilde{N}}{\sqrt{{}^{(2)}\tilde{g}}} (\tilde{g}_{ac} \tilde{g}_{bd} - \tilde{g}_{ab} \tilde{g}_{cd})\tilde{\pi}^{'cd} &= \tilde{g}'_{ab,t} - \left(\mathcal{L}_{\tilde{N}^{c'} \partial_c} {}^{(2)}\tilde{g}\right)_{ab}\label{eq:f71}
\end{align}
Using Eqs.~(\ref{eq:f17}), (\ref{eq:f18}) and (\ref{eq:f24}) to evaluate these in the \textit{asymptotic regions} (where \({}^{(4)}k_{\alpha\beta} = 0\)) we obtain, thanks to a fortuitous cancellation of the terms involving \({}^{(2)}Y^a_{\hphantom{a},t}\),
\begin{align}
\frac{\tilde{N}\tilde{p}'}{\sqrt{{}^{(2)}\tilde{g}}} &\longrightarrow 4\tilde{N}^2 \tilde{g}^{ac} \gamma_{,a} {}^{(4)}Y^0_{\hphantom{0},c},\label{eq:f72}\\
\frac{\tilde{N} e^{2\gamma} \tilde{v}'}{\sqrt{{}^{(2)}\tilde{g}}} &\longrightarrow \tilde{N}^2 \tilde{g}^{ac} \lambda_{,a} {}^{(4)}Y^0_{\hphantom{0},c},\label{eq:f73}\\
\frac{\tilde{N} e^{2\gamma}}{\sqrt{{}^{(2)}\tilde{g}}} (\tilde{u}' - \lambda\tilde{r}') &\longrightarrow \tilde{N}^2 \tilde{g}^{ac} \eta_{,a} {}^{(4)}Y^0_{\hphantom{0},c},\label{eq:f74}\\
\frac{\tilde{N} e^{4\gamma} \tilde{r}'}{\sqrt{{}^{(2)}\tilde{g}}} &\longrightarrow \tilde{N}^2 \tilde{g}^{ac} {}^{(4)}Y^0_{\hphantom{0},c} (\omega_{,a} + \lambda\eta_{,a}),\label{eq:f75}\\
\intertext{and}
\frac{2\tilde{N}}{\sqrt{{}^{(2)}\tilde{g}}} (\tilde{g}_{ac} \tilde{g}_{bd} - \tilde{g}_{ab} \tilde{g}_{cd}) \tilde{\pi}^{'cd} &\longrightarrow \left(\mathcal{L}_{{}^{(2)}\mathcal{D}} \tilde{g}\right)_{ab}\label{eq:f76}
\end{align}
where
\begin{equation}\label{eq:f77}
{}^{(2)}\mathcal{D} := \left(\tilde{N}^2 \tilde{g}^{cd} {}^{(4)}Y^0_{\hphantom{0},d}\right) \frac{\partial}{\partial x^c}.
\end{equation}
Note that each of the above expresses the desired (linearized) momentum asymptotically in terms of  Lie derivatives with respect to \({}^{(2)}\mathcal{D}\).

In deriving the above we have made use of the formula
\begin{equation}\label{eq:f78}
\tilde{g}'_{ab} = \left(\tilde{g}_{ab}\; e^{-2\gamma}\; {}^{(4)}k_{\varphi\varphi} + e^{2\gamma}\; {}^{(4)}k_{ab}\right) + \left(\mathcal{L}_{{}^{(2)}Y} \tilde{g}\right)_{ab}
\end{equation}
which results from linearizing the defining equation
\begin{equation}\label{eq:f79}
\begin{split}
\tilde{g}_{ab} &= e^{2\gamma} g_{ab} - e^{4\gamma} \beta_a\beta_b\\
&= {}^{(4)}g_{\varphi\varphi} {}^{(4)}g_{ab} - {}^{(4)}g_{a\varphi} {}^{(4)}g_{b\varphi}
\end{split}
\end{equation}
about the chosen background (c.f. Eqs.~(\ref{eq:111}--\ref{eq:114}).  %H; G; orginally appendix F
\section{Compactly Supported Solutions of the Linearized Constraint Equations}
\label{app:complactly-supported-solutions}
As we have already discussed near the end of Appendix \ref{app:global}, the use of hyperbolic gauge conditions for the linearized field equations allows one to exploit the corresponding, causal propagation of the perturbations to conclude that compactly supported initial data on a Cauchy hypersurface of constant Boyer-Lindquist time, \textit{t}, evolves so as to preserve this property for all finite \textit{t}. Thus data initially bounded away from the horizon and from spacelike infinity evolves to remain so throughout the evolution --- a feature which reflects the fact that Boyer-Lindquist time  slices for Kerr-Newman spacetimes are `locked down' at \(i_0\) (spacelike infinity) and at the bifurcation 2-sphere lying in the horizon. While this property of compactly supported evolution will ultimately be lost upon transformation to an elliptic gauge of the type adopted herein, it will be noteworthy to recognize that the transformed perturbations, though no longer in general having compact support, will necessarily be of `pure gauge type' near the horizon and near infinity.

The utilization of hyperbolic gauge conditions to secure causal evolution for the perturbations does not, however, preclude the need to solve the linearized constraint equations, at least on the initial Cauchy hypersurface. Since these latter are normally treated as an \textit{elliptic} system for certain dependent or constrained variables, it is not immediately clear how to ensure the desired compact support of their resulting solutions. While one could presumably guarantee this outcome by imposing suitable restrictions upon the otherwise `free data' occurring in these equations we shall herein adopt a different strategy whereby one solves the constraints \textit{algebraically} for a subset of this normally regarded free data, reversing somewhat the usual roles of free and constrained variables. This will allow us to ensure the compact support of the solutions so obtained without otherwise unduly restricting their generality.

Consider first the reduced momentum constraints, Eqs.~(\ref{eq:c08}), and assume for definiteness that the background has charge \(Q \neq 0\). Assume also that \(a \neq 0\) since otherwise the background spacetime would be a Reissner-Nordstrom solution which is treatable by much more elementary methods \cite{Moncrief_74_3,Moncrief_74_1,Moncrief_74_2}. Under these assumptions the functions \(\lambda\) and \(\eta\) (c.f. Eqs.~(\ref{eq:a40}) and (\ref{eq:a45})) are both non-vanishing and one can reexpress the momentum constraints as an algebraic system of the form
\begin{equation}\label{eq:e01}
\left(\begin{matrix}
\eta_{,R} & \lambda_{,R}\\
\eta_{,\theta} & \lambda_{,\theta}\end{matrix}\right) \left(\begin{matrix}
\tilde{u}'\\
\tilde{v}'\end{matrix}\right) = \left(\begin{matrix}
\mathcal{S}_R\\
\mathcal{S}_\theta\end{matrix}\right)
\end{equation}
where
\begin{align}
\mathcal{S}_R &:= 2\; {}^{(2)}\nabla_b(h)\; \tilde{r}^{'b}_{\hphantom{'b}R} + e^{2\nu} \sqrt{{}^{(2)}h}\; \tau'_{,R} - \tilde{p}' \gamma_{,R} - \tilde{r}' \omega_{,R}\label{eq:e02}\\
\mathcal{S}_\theta &:= 2\; {}^{(2)}\nabla_b(h)\; \tilde{r}^{'b}_{\hphantom{'b}\theta} + e^{2\nu} \sqrt{{}^{(2)}h}\; \tau'_{,\theta} - \tilde{p}' \gamma_{,\theta} - \tilde{r}' \omega_{,\theta}.\label{eq:e03}
\end{align}
The idea is choose the data \(\lbrace\tilde{p}', \tilde{r}', \tau', \tilde{r}^{'b}_{\hphantom{'b}a}\rbrace\) to have compact support on \(M_b\) and to solve equations (\ref{eq:e01}) for the electromagnetic momenta \(\lbrace\tilde{u}', \tilde{v}'\rbrace\). Clearly the feasibility of this approach hinges upon the invertibility of the matrix function
\begin{equation}\label{eq:e04}
\mathcal{D} := \left(\begin{matrix}
\eta_{,R} & \lambda_{,R}\\
\eta_{,\theta} & \lambda_{,\theta}\end{matrix}\right).
\end{equation}
By a straightforward computation one finds that its determinant
\begin{equation}\label{eq:e05}
\begin{split}
\det{\mathcal{D}} &= \eta_{,R}\lambda_{,\theta} - \lambda_{,R}\eta_{,\theta}\\
&= \frac{\left(1 - \frac{R^2_+}{R^2}\right) (r^2 + a^2) 4Q^2 a\; \sin^3{(\theta)}}{(r^2 + a^2 \cos^2{\theta})^2},
\end{split}
\end{equation}
which is thus non-vanishing except on the horizon (where \(\left(1 - \frac{R^2_+}{R^2}\right) = 0\)) and on the axes (having \(\sin{(\theta)} = 0\)). The formal solution to Eq.~(\ref{eq:e01}) is given explicitly by
\begin{equation}\label{eq:e06}
\left(\begin{matrix}
\tilde{u}'\\
\tilde{v}'\end{matrix}\right) = \mathcal{D}^{-1}\left(\begin{matrix}
\mathcal{S}_R\\
\mathcal{S}_\theta\end{matrix}\right)
\end{equation}
where
\begin{equation}\label{eq:e07}
\mathcal{D}^{-1} = \left(\frac{1}{4Qa}\right)\left(\begin{matrix}
\frac{4ra \cos{\theta}}{\left(1 - \frac{R^2_+}{R^2}\right) \sin^2{\theta}} & \frac{2a (r^2 - a^2 \cos^2{\theta})}{(r^2 + a^2) \sin{\theta}}\\
\\
\frac{-2(r^2 - a^2 \cos^2{\theta})}{\left(1 - \frac{R_+^2}{R^2}\right) \sin^2\theta} & \frac{4r a^2 \cos{\theta}}{(r^2 + a^2) \sin{\theta}}\end{matrix}\right)
\end{equation}
By choosing the free data occurring in \(\mathcal{S}_R\) on \(\mathcal{S}_\theta\) to have not only compact support on \(M_b\) but also to vanish at the axes as suitable powers of \(\sin{\theta}\) one ensures both the compact support of the resulting solution and its regularity at the axes. Note by contrast that one normally thinks of Eqs.~(\ref{eq:e01})--(\ref{eq:e03}) as an elliptic system to be solved for \(\tilde{r}^{'b}_{\hphantom{'b}a}\) instead of an algebraic one for \(\lbrace\tilde{u}', \tilde{v}'\rbrace\).

Now, however, suppose that \(Q = 0\) (but \(a \neq 0\) since otherwise the background would simply be Schwarzchild). The functions \(\gamma\) and \(\omega\) (given by Eqs.~(\ref{eq:a43}) and (\ref{eq:a46}) in the limiting case \(Q \rightarrow 0\)) are still non-vanishing and one can now express the momentum constraints in the alternative form:
\begin{equation}\label{eq:e08}
\tilde{\mathcal{D}} \left(\begin{matrix}
\tilde{p}'\\
\tilde{r}'\end{matrix}\right) = \left(\begin{matrix}
\tilde{\mathcal{S}}_R\\
\tilde{\mathcal{S}}_\theta\end{matrix}\right)
\end{equation}
where
\begin{equation}\label{eq:e09}
\tilde{\mathcal{D}} := \left.\left(\begin{matrix}
\gamma_{,R} & \omega_{,R}\\
\gamma_{,\theta} & \omega_{,\theta}\end{matrix}\right)\right|_{Q=0}
\end{equation}
and where
\begin{equation}\label{eq:e10}
\tilde{\mathcal{S}}_R := \left.\left\lbrace 2\; {}^{(2)}\nabla_b(h)\; \tilde{r}^{'b}_{\hphantom{'b}R} + e^{2\nu} \sqrt{{}^{(2)}h}\; \tau'_{,R}\right\rbrace\right|_{Q=0}
\end{equation}
and
\begin{equation}\label{eq:e11}
\tilde{\mathcal{S}}_\theta := \left.\left\lbrace 2\; {}^{(2)}\nabla_b(h)\; \tilde{r}^{'b}_{\hphantom{'b}\theta} + e^{2\nu} \sqrt{{}^{(2)}h}\; \tau'_{,\theta}\right\rbrace\right|_{Q=0}
\end{equation}
Algebraic solvability now hinges on the invertibility of the matrix function \(\tilde{\mathcal{D}}\). A straightforward computation of the determinant,
\begin{equation}\label{eq:e12}
\det{\tilde{\mathcal{D}}} := \left.\left\lbrace\gamma_{,R} \omega_{,\theta} - \omega_{,R} \gamma_{,\theta}\right)\right|_{Q=0},
\end{equation}
of \(\tilde{\mathcal{D}}\) yields
\begin{equation}\label{eq:e13}
\begin{split}
\det{\tilde{\mathcal{D}}} &= \left(\frac{-Ma \sin^3{\theta} \left( 1 - \frac{R_+^2}{R^2}\right)}{(r^2 + a^2 \cos^2{\theta})^3 \left\lbrack(r^2 + a^2)^2 - a^2 \Delta \sin^{2}{\theta}\right\rbrack}\right) \\
& {} \times \left\lbrace(r^3 - Ma^2) \left\lbrack 6r^6 + 2a^2 \left( r^4 - a^4 \cos^6{(\theta)}\right)\right\rbrack\right.\\
& {} + 6a^6 r (Mr - a^2) \cos^6{(\theta)}\\
& {} + \left\lbrack 10r^3 (Mr - a^2) + 2M (r^4 - a^4) + 2r^2 (r^3 - Ma^2)\right\rbrack a^4 \cos^4{(\theta)}\\
& {} + \left.\left\lbrack 10a^2 r^4 (r^3 - a^2M) + 4r^2 Ma^2 (r^4 - a^4) + 2r^5 a^2 (rM - a^2)\right\rbrack \cos^2{(\theta)}\right\rbrace
\end{split}
\end{equation}
This is easily seen to be non-vanishing except on the axes (where it vanishes as \(\sin^3{(\theta)}\)) and at the horizon where, in the subextremal cases, it vanishes like \(\left( 1 - \frac{R_+^2}{R^2}\right)\) as \(R \searrow R_+\). Curiously, in the extremal cases (\(|a| = M\)), every term in the \(\lbrace~\rbrace\) brackets also vanishes at the horizon \(r \rightarrow r_+ = M = |a|\).

Thus one can now solve the momentum constraints for the gravitational momenta, \(\lbrace\tilde{p}', \tilde{r}'\rbrace\), taking the `free data' \(\lbrace\tilde{r}^{'b}_a, \tilde{\tau}'\rbrace\) to have compact support and to vanish at the axes as suitable powers of \(\sin{(\theta)}\) to ensure regularity of the solution.

Turning now to the (linearized) Hamiltonian constraint, \(\tilde{\mathcal{H}}' = 0\), one sees from Eq.~(\ref{eq:c13}) that this can be expressed in \textit{divergence form} as
\begin{equation}\label{eq:e14}
\begin{split}
\tilde{N}\tilde{\mathcal{H}}' &= \frac{\partial}{\partial x^b} \left\lbrace\tilde{N} \sqrt{{}^{(2)}h}\; h^{ab} \left\lbrack 4\gamma_{,a} \gamma' + 2\nu'_{,a}\right.\right.\\
& {} + \left. e^{-2\gamma} (\eta_{,a}\eta' + \lambda_{,a}\lambda') + e^{-4\gamma} (\omega_{,a} + \lambda\eta_{,a}) (\omega' + \lambda\eta')\right\rbrack\\
& {} - \left. 2 \sqrt{{}^{(2)}h}\; h^{ab} \tilde{N}_{,a}\nu'\right\rbrace = 0.
\end{split}
\end{equation}
Since \(M_b\) is simply connected the vector density appearing in the \(\lbrace~\rbrace\) brackets must take the form \(\lbrace~\rbrace^b = \epsilon^{bc} \sigma'_{,c}\) for some function \(\sigma'\). Thus any solution to (\ref{eq:e14}) must satisfy
\begin{equation}\label{eq:e15}
\begin{split}
4\gamma_{,a}\gamma' &+ e^{-4\gamma} (\omega_{,a} + \lambda\eta_{,a}) (\omega' + \lambda\eta') + e^{-2\gamma} (\eta_{,a}\eta' + \lambda_{,a}\lambda')\\
& = \frac{h_{ab} \epsilon^{bc}}{\tilde{N} \sqrt{{}^{(2)}h}} \sigma'_{,c} + \frac{2\tilde{N}_{,a}}{\tilde{N}} \nu' - 2\nu'_{,a}
\end{split}
\end{equation}
Now if \(Q \neq 0\) (and, as always \(a \neq 0\)) we define \(\Omega' := \omega' + \lambda\eta'\) and regard (\ref{eq:e15}) as an algebraic system for \(\lbrace\eta', \lambda'\rbrace\), taking the `free data' \(\lbrace\sigma', \nu', \gamma', \Omega'\rbrace\) in this case to have compact support and to vanish sufficiently rapidly at the axes.

Since the matrix of coefficients for this algebraic problem is nothing other than the \(\mathcal{D}\) defined previously one solves for \(\lbrace\eta', \lambda'\rbrace\) and then sets \(\omega' = \Omega' - \lambda\eta'\) to complete the solution.

If on the other hand \(Q = 0\) then (\ref{eq:e15}) reduces to the form
\begin{equation}\label{eq:e16}
\left.\left\lbrace\gamma_{,a} (4\gamma') + \omega_{,a} (e^{-4\gamma}\omega')\right\rbrace\right|_{Q=0} = \left.\left\lbrace\frac{h_{ab}\epsilon^{bc}}{\tilde{N} \sqrt{{}^{(2)}h}} \sigma'_{,c} + \frac{2\tilde{N}_{,a}}{\tilde{N}} \nu' - 2\nu'_{,a}\right\rbrace\right|_{Q=0}
\end{equation}
and one can exploit the fact that \(\tilde{\mathcal{D}}\) is invertible (assuming as always that \(a \neq 0\)) to solve this system algebraically for \(\lbrace 4\gamma', e^{-4\gamma}\omega'\rbrace\). Thus all of the reduced constraints can be solved algebraically for compactly supported data that is regular at the axes of symmetry for the background black hole.

There is however a remaining subtlety that must be dealt with. We need to `lift' the Cauchy data defined on the quotient manifold \(M_b\) back up to the actual, 3-dimensional Cauchy surface for the black hole's DOC and ensure that it all has compact support there as well. The potential obstructions to this are the first variations,
\begin{equation}\label{eq:e17}
\tilde{r}' = \epsilon^{ab} \beta'_{a,b},\quad \tilde{u}' = \epsilon^{ab} C'_{a,b}
\end{equation}
of the defining equations (\ref{eq:a28}) for the one-forms \(\beta_a dx^a\) and \(C_a dx^a\). Even if \(\tilde{r}'\) and \(\tilde{u}'\) have compact support the one-forms \(\beta'_a dx^a\) and \(C'_a dx^a\) need not inherent this property without further restrictions upon the `sources' \(\tilde{r}'\) and \(\tilde{u}'\). By contrast note that the first variations of Eqs.~(\ref{eq:a24}) and (\ref{eq:a25}),
\begin{equation}\label{eq:e18}
\tilde{f}^{a'} = \epsilon^{ab} \omega'_{,b},\quad \mathcal{E}^{a'} = \epsilon^{ab} \eta'_{,b}
\end{equation}
automatically yield lifted vector densities \(\tilde{f}^{a'} \frac{\partial}{\partial x^a}\) and \(\mathcal{E}^{a'} \frac{\partial}{\partial x^a}\) of compact support provided only that the base space potentials \(\omega'\) and \(\eta'\) have this property.

Since both equations (\ref{eq:e17}) are identical in form it suffices to show what further restrictions upon \(\tilde{r}'\) are needed to solve for a compactly supported \(\beta'_a dx^a\) since the argument for the pair \(\lbrace\tilde{u}', C'_a dx^a\rbrace\) will follow the same pattern.

Guided by the Hodge decomposition of one-forms on simply connected 2-manifolds we seek a solution to \(\tilde{r}' = \epsilon^{ab} \beta'_{a,b}\) of the form
\begin{equation}\label{eq:e19}
\beta'_a = \zeta_{,a} + \frac{h_{ab}}{\sqrt{{}^{(2)}h}} \epsilon^{bc} \psi_{,c}
\end{equation}
for some undetermined functions \(\lbrace\zeta, \psi\rbrace\). The equation to be solved now takes the \textit{u} form of Poisson's equation for the unknown function \(\psi\),
\begin{equation}\label{eq:e20}
\nabla_b(h) \nabla^b(h) \psi := \frac{1}{\sqrt{{}^{(2)}h}} \partial_b (\sqrt{{}^{(2)}h} h^{bc} \psi_{,c}) = \frac{1}{\sqrt{{}^{(2)}h}} \epsilon^{bc} \beta'_{b,c}.
\end{equation}

In terms of the coordinates \textit{R} and \(\theta\) introduced for \(M_b\) in Appendix~\ref{app:kerr-newman-spacetimes}, and for which the flat metric \(h_{ab} dx^a \otimes dx^b\) takes the form
\begin{equation}\label{eq:e21}
h_{ab} dx^a \otimes dx^b = dR \otimes dR + R^2 d\theta \otimes d\theta,
\end{equation}
any smooth source function
\begin{equation}\label{eq:e22}
s := \frac{1}{\sqrt{{}^{(2)}h}} \epsilon^{bc} \beta'_{b,c}
\end{equation}
that is regular at the axes of \(M_b\) and that has compact support on this space will admit a Fourier expansion of the form
\begin{equation}\label{eq:e23}
s(R,\theta) = \sum_{m=0}^\infty \sigma_m(R) \cos{(m\theta)}
\end{equation}
where each of the Fourier coefficient functions \(\lbrace\sigma_m\rbrace\) will vanish for all \textit{R} such that
\begin{equation}\label{eq:e24}
R \geq R_2 \geq R_1 > R_+
\end{equation}
and that
\begin{equation}\label{eq:e25}
R_+ < R \leq R_1 \leq R_2
\end{equation}
for suitably chosen \(R_1\) and \(R_2\). Any smooth solution \(\psi\) to (\ref{eq:e20}) must admit a corresponding Fourier expansion,
\begin{equation}\label{eq:e26}
\psi (R,\theta) = \sum_{m=0}^\infty \psi_m(R) \cos{(m\theta)}
\end{equation}
with coefficients satisfying the associated ordinary differential system
\begin{equation}\label{eq:e27}
\frac{d^2\psi_m}{dR^2} + \frac{1}{R} \frac{d\psi_m}{dR} - \frac{m^2}{R^2} \psi_m = \sigma_m;\qquad m = 0, 1, 2, \ldots
\end{equation}
Each of these equations can be readily solved using the method of variation of parameters. It is straightforward to show that the resulting solution \(\psi\) will take constant values in the two asymptotic regions
\begin{equation}\label{eq:e28}
R_+ < R \leq R_1\quad \hbox{ and }\quad R \geq R_2,
\end{equation}
and thus have compactly supported gradient on \(M_b\) if and only if the source functions \(\lbrace\sigma_m\rbrace\) satisfy the following (definite) integral conditions:
\begin{equation}\label{eq:e29}
\int_{R_1}^{R_2} R^{1-m} \sigma_m(R) dR = 0\qquad \hbox{(no sum on \textit{m})}
\end{equation}
and
\begin{equation}\label{eq:e30}
\int_{R_1}^{R_2} R^{1+m} \sigma_m(R) dR = 0.\qquad \hbox{(no sum on \textit{m})}
\end{equation}
The fact that these two conditions coincide for \(m = 0\) corresponds to the flexibility of allowing \(\psi\) to have two distinct, constant values in the two asymptotic regions.

The remaining function \(\zeta\) arising in the decomposition (\ref{eq:e19}) is unrestricted by Eq.~(\ref{eq:e17}) and thus can be chosen arbitrarily to have compactly supported gradient. The freedom to add an arbitrary gradient to the one-form \(\beta'_a dx^a\) corresponds to that of making a coordinate transformation of the form
\begin{equation}\label{eq:e31}
x^3 = \varphi \rightarrow \varphi + \zeta
\end{equation}
in the \(U(1)\) bundle over \(M_b\).

While we chose above to solve the reduced (Hamiltonian and momentum) constraints algebraically it is straightforward to see from the preceding example that we could, alternatively, have treated them as Poisson type equations for the `usual' unknowns \(\lbrace\nu', \tilde{r}^{'b}_{\hphantom{'b}a} \rbrace\) and still ensured compact support for the solutions by imposing suitable integral constraints (as well as compactness of support) upon the `free data' \(\lbrace(\gamma',\tilde{p}'), (\omega',\tilde{r}'), (\eta',\tilde{u}'), (\lambda',\tilde{v}'), \tau'\rbrace\). This follows from the fact that, when expressed in terms of the `Cartesian' coordinates \(\lbrace\bar{\rho}, \bar{z}\rbrace\) for the flat metric \({}^{(2)}h\) (wherein \({}^{(2)}h = d\bar{p} \otimes d\bar{p} + d\bar{z} \otimes d\bar{z}\)), the linearized constraints reduce to \textit{decoupled equations} of precisely the (flat space) Poisson type that we have just dealt with for the unknowns \(\lbrace\nu', \tilde{r}^{'b}_{\hphantom{'b}a}\rbrace\).

At various stages in our analysis (e.g., solving the momentum constraint in Appendix \ref{app:analysis-linearized-constraint-equations}, preserving Weyl-Papapetrou gauge conditions with a suitably chosen perturbed shift in Appendix~\ref{app:gauge-conditions} and in the proof of vanishing of the `integral invariant' \(a_0(R_+)\) presented in the Appendix below) we have (implicitly or explicitly) exploited the claim that transverse-traceless symmetric 2-tensors, subject to suitable asymptotic and boundary conditions on \(M_b\), vanish identically. To establish this claim let us first work in `isothermal' coordinates \(\lbrace \rho, z\rbrace\) for which the flat metric \({}^{(2)}h = h_{ab}dx^a \otimes dx^b\) takes the form
\begin{equation}\label{eq:e32}
{}^{(2)}h = h_{ab}dx^a \otimes dx^b = d\rho \otimes d\rho + dz \otimes dz.
\end{equation}
An arbitrary \textit{traceless} symmetric 2-tensor, \({}^{(2)}k^{\mathrm{tr}} = k_{ab}^{\mathrm{tr}} dx^a \otimes dx^b\) can be expressed in these coordinates as
\begin{equation}\label{eq:e33}
\begin{split}
    {}^{(2)}k^{\mathrm{tr}} &= k_{ab}^{\mathrm{tr}} dx^a \otimes dx^b\\
    &= u (d\rho \otimes d\rho - dz \otimes dz)\\
    &- v (d\rho \otimes dz + dz \otimes d\rho)
\end{split}
\end{equation}
and is obviously traceless with respect to any metric conformal to \({}^{(2)}h\) as well.

Imposing the independent (and equally conformally invariant) condition that the covariant divergence of \({}^{(2)}k^{\mathrm{tr}}\) vanish is well-known (and straightforwardly seen) to be equivalent to requiring that the component functions \(\lbrace u, v\rbrace\) satisfy the Cauchy-Riemann equations
\begin{equation}\label{eq:e34}
\begin{split}
u_{,p} &= v_{,z}\\
u_{,z} &= -v_{,p}
\end{split}
\end{equation}
which in turn of course imply that each of \(u, v\) is harmonic with respect to the metric \({}^{(2)}h\) (or to any metric conformal thereto):
\begin{equation}\label{eq:e35}
\Delta_{{}^{(2)}h} u = \Delta_{{}^{(2)}h} v = 0.
\end{equation}

Reverting to polar coordinates \(\lbrace R, \theta\rbrace\) on \(M_b\) for which \({}^{(2)}h\) takes the form
\begin{equation}\label{eq:e36}
{}^{(2)}h = dR \otimes dR + R^2 d\theta \otimes d\theta
\end{equation}
with \(R > R_+\) and \(\theta \in\; [0, 2\pi)\) one easily finds that globally harmonic functions that vanish on the horizon as \(R \searrow R_+\) and are bounded on \(M_b\) must in fact vanish identically. Hence we have that
\begin{theorem}\label{thm:04}
Globally defined transverse traceless symmetric 2-tensors, \({}^{(2)}k^{\mathrm{TT}}\), which are bounded on \(M_b\) and which vanish at the horizon corresponding to \(R \searrow R_+ > 0\) vanish identically.
\end{theorem}  %I; H; orginally appendix E
\section{The Vanishing of $a_0(R_+)$}
\label{app:vanishing}
As discussed in Appendix~\ref{app:transforming}, the successful implementation of our chosen (Weyl-Papapetrou) gauge condition hinges upon proving that a certain `integral invariant', \(a_0 (R_+)\), actually vanishes for the class of perturbations considered. In the course of carrying out such a proof we shall see that this quantity is in fact gauge-invariant (with respect to the relevant class of such transformations) and thus justify its characterization as such.

In terms of the (spatially compactly supported) 4-metric perturbation, \({}^{(4)}k = {}^{(4)}k_{\mu\nu}\; dx^\mu \otimes dx^\nu\) of \({}^{(4)}g\) introduced in Appendix~\ref{app:transforming} (and assumed therein to be expressed in a `hyperbolic' gauge), \(a_0 (R_+)\) was defined by the integral formula (c.f., Eq.~(\ref{eq:f45}))
\begin{equation}\label{eq:k28}
a_0 (R_+) = -R_+ \int_{R_+}^\infty dR' \left\lbrack\frac{1}{R'}\; c_0 (R')\right\rbrack
\end{equation}
wherein \(c_0 (R)\) was in turn given by (c.f., Eq.~(\ref{eq:f43}))
\begin{equation}\label{eq:k29}
c_0 (R) = \frac{1}{2\pi} \int_0^{2\pi} d\theta \left(-\frac{1}{R}\; \mathcal{M}^{RR} (R,\theta)\right)
\end{equation}
with \(\mathcal{M}^{cd}\) defined via (c.f., Eq.~(\ref{eq:f28}))
\begin{equation}\label{eq:k30}
\mathcal{M}^{cd} = \sqrt{{}^{(2)}\tilde{g}}\; \tilde{g}^{ac} \tilde{g}^{bd}\; e^{2\gamma} \left\lbrace {}^{(4)}k_{ab} - \frac{1}{2}\; \tilde{g}_{ab} \tilde{g}^{ef}\; {}^{(4)}k_{ef}\right\rbrace.
\end{equation}
Evaluating \(\mathcal{M}^{RR}\) on a Kerr-Newman background and exploiting Eq.~(\ref{eq:112}) to express the relevant components of \({}^{(4)}k\) in terms of (first variations of) our 2+1 dimensional quantities \(\lbrace\gamma, \nu, \tilde{g}_{ab}, h_{ab}\rbrace\) we arrive at
\begin{equation}\label{eq:k31}
\begin{split}
    \mathcal{M}^{RR} &\rightarrow \frac{R}{2}\; e^{2\gamma -2\nu} \left\lbrace{}^{(4)}k_{RR} - \frac{1}{R^2} {}^{(4)}k_{\theta\theta}\right\rbrace\\
        &= \frac{R}{2} \left\lbrace (\delta h_{RR}) - \frac{1}{R^2} (\delta h_{\theta\theta})\right\rbrace
\end{split}
\end{equation}
where \(\delta h_{ab}\) designates the first variation (also signified by a \(\prime\)) of the flat `conformal metric' introduced in Appendix~\ref{app:gauge-conditions}. In view of the axis regularity requirements discussed in Appendix~\ref{app:transforming} this perturbation has an expansion (with its \textit{t}-dependence suppressed, as before, to simplify the notation) of the form, setting \(\ell_{ab} := \delta h_{ab}\)
\begin{align}
\ell_{RR} &= \gamma_0 (R) + \sum_{n=1}^\infty \gamma_n (R) \cos{(n\theta)},\label{eq:k32}\\
\ell_{R\theta} &= \ell_{\theta R} = \sum_{n=1}^\infty \delta_n (R) \sin{(n\theta)},\label{eq:k33}\\
\ell_{\theta\theta} &= \sigma_0 (R) + \sum_{n=1}^\infty \sigma_n (R) \cos{(n\theta)}.\label{eq:k34}
\end{align}
To preserve its flatness this perturbation of the \({}^{(2)}h_{ab}\) metric must satisfy the (necessary and sufficient) condition
\begin{equation}\label{eq:k35}
\begin{split}
    D \left(\sqrt{{}^{(2)}h}\; {}^{(2)}R ({}^{(2)}h)\right) \cdot \ell &= \sqrt{{}^{(2)}h}\; \left\lbrace{}^{(2)}\nabla^a ({}^{(2)}h) {}^{(2)}\nabla^b ({}^{(2)}h) \ell_{ab}\right.\\
    &\qquad\left. -{}^{(2)}\nabla_a ({}^{(2)}h) {}^{(2)}\nabla^a ({}^{(2)}h)\; (h^{cd} \ell_{cd})\right\rbrace\\
    &= -\frac{1}{R} \ell_{RR,\theta\theta} + \ell_{RR,R} + \frac{2}{R} \ell_{R\theta,\theta R}\\
    &\qquad -\frac{2}{R^3} \ell_{\theta\theta} + \frac{2}{R^2} \ell_{\theta\theta, R} - \frac{1}{R} \ell_{\theta\theta, RR}\\
    &= 0.
\end{split}
\end{equation}
This condition is, of course, automatically satisfied by the \textit{pure gauge} perturbations
\begin{equation}\label{eq:k36}
\begin{split}
    v_{ab} &:= \left(\mathcal{L}_{{}^{(2)}Y}\; {}^{(2)}h\right)_{ab}\\
     &= {}^{(2)}Y^c h_{ab,c} + {}^{(2)}Y^c_{\hphantom{c},a} h_{cb} + {}^{(2)}Y^c_{\hphantom{c},b} h_{ac}
\end{split}
\end{equation}
where
\begin{equation}\label{eq:k37}
h_{ab}\; dx^a \otimes dx^b = dR \otimes dR + R^2\; d\theta \otimes d\theta
\end{equation}
and wherein the vector field \({}^{(2)}Y\) admits an expansion of the form given by Eqs.~(\ref{eq:f34}) and (\ref{eq:f35}), namely
\begin{align}
{}^{(2)}Y^R &= a_0 (R) + \sum_{n=1}^\infty a_n (R) \cos{(n\theta)},\label{eq:k38}\\
{}^{(2)}Y^\theta &= \sum_{n=1}^\infty b_n (R) \sin{(n\theta)}.\label{eq:k39}
\end{align}

In view of the simple formula for \(\mathcal{M}^{RR}\) (c.f., Eq.~(\ref{eq:k31})) which now gives
\begin{equation}\label{eq:k40}
\frac{\mathcal{M}^{RR}}{R} = \frac{1}{2} \left\lbrace\ell_{RR} - \frac{1}{R^2} \ell_{\theta\theta}\right\rbrace
\end{equation}
and the angular integral in Eq.~(\ref{eq:k29}) for \(c_0 (R)\) we see that only the (rotationally-invariant) \(n = 0\) terms in the expansions (\ref{eq:k32})--(\ref{eq:k34}) contribute to \(c_0 (R)\) and hence to \(a_0 (R_+)\). For these quantities it is convenient to define a new set of variables
\begin{align}
k_0^{\mathrm{I}} &:= \gamma_0 - \left(\frac{\sigma_0}{R}\right)_{,R}\label{eq:k41}\\
k_0^{\mathrm{II}} &:= \sigma_0\label{eq:k42}
\end{align}
for which the inverse transformation is clearly
\begin{align}
\sigma_0 &= k_0^{\mathrm{II}},\label{eq:k43}\\
\gamma_0 &= k_0^{\mathrm{I}} + \left(\frac{k_0^{\mathrm{II}}}{R}\right)_{,R}.\label{eq:k44}
\end{align}
It is easily verified that \(k_0^{\mathrm{I}}\) is gauge invariant whereas \(k_0^{\mathrm{II}}\) is, in effect, pure gauge. Furthermore, the rotationally invariant component of Eq.~(\ref{eq:k35}) (i.e., its integral with respect to \(\theta\) over the circle) yields the (gauge invariant) constraint
\begin{equation}\label{eq:k45}
k_{0,R}^{\mathrm{I}} = 0
\end{equation}
so that \(k_{0,R}^{\mathrm{I}}\) is, at most, a (possibly \textit{t}-dependent) constant.

Substituting the above results into the formula for \(a_0 (R_+)\) we now arrive at:
\begin{equation}\label{eq:k46}
a_0 (R_+) = \frac{R_+}{2} \int_{R_+}^\infty dR\; \left\lbrack\frac{k_0^{\mathrm{I}}}{R} + \left(\frac{k_0^{\mathrm{II}}}{R^2}\right)_{,R}\right\rbrack.
\end{equation}
But the term in \(k_0^{\mathrm{I}}\) can only give a finite contribution if this constant vanishes whereas the (boundary) contributions of \(k_0^{\mathrm{II}}\) will vanish for any compactly supported perturbation. We conclude that
\begin{theorem}\label{thm:05}
The integral invariant \(a_0 (R_+)\) vanishes when evaluated upon compactly supported perturbations (that vanish on the asymptotic regions near the horizon on infinity).
\end{theorem}

A simpler, more explicit proof of the above result can be given in the non-rotating (\(a = 0\)) case by exploiting the utility of expanding the perturbations of the (spherically symmetric) background Reissner-Nordstr\"{o}m solution in (Regge-Wheeler) tensor harmonics. It is clear from the structure of \(c_0(R)\) (c.f., Eq.~(\ref{eq:f43})) and \(a_0 (R)\) (c.f., Eqs.~(\ref{eq:f44})--(\ref{eq:f45})) that only the spherically symmetric `mode' of the perturbations contributes in this case and, as is well-known, this non-dynamical mode decouples from all of the `higher harmonic' modes. Because of the dynamical triviality of this (spherically symmetric) perturbative mode, as guaranteed by the (generalized) Birkhoff theorem, it was not treated in detail in the earlier, Hamiltonian stability analyses of the Reissner-Nordstr\"{o}m spacetime (c.f. Refs.~\cite{Moncrief_74_3,Moncrief_74_1,Moncrief_74_2}). We therefore provide those `missing' details in the following.

In the \(\lbrace t, R, \theta, \varphi\rbrace\) coordinates of Appendix~\ref{app:kerr-newman-spacetimes} the Reissner-Nordstr\"{o}m line element takes the form
\begin{align}
\begin{split}\label{eq:k01}
ds^2 &= \frac{-\left( 1 - \frac{R_+^2}{R^2}\right)^2\; dt^2}{\left( 1 + \frac{M}{R} + \frac{R_+^2}{R^2}\right)^2}\\
 &\hphantom{=}{} + \left(1 + \frac{M}{R} + \frac{R_+^2}{R^2}\right)^2 \left( dR^2 + R^2 d\theta^2 + R^2\sin^2{\theta}\; d\varphi^2\right)
\end{split}
\intertext{where}
R_+ &= \frac{1}{2} \sqrt{M^2 - Q^2}\label{eq:k02}
\end{align}
with the remaining ADM variable (c.f., Ref.~\cite{Moncrief_74_3}) given by\footnote{\label{note03} We assume throughout that the magnetic field \(B^i\) is derivable from a vector potential and thus vanishes identically (together with its first variation) in the \textit{spherically} symmetric case of interest here. Recall also the slightly non-standard conventions for the designation of the electromagnetic field introduced in Appendix~\ref{app:reduced-hamiltonian}.}
\begin{equation}\label{eq:k03}
\mathcal{E}^R =2Q \sin{\theta}, \mathcal{E}^\theta = \mathcal{E}^\varphi = 0, B^i = 0, \pi^{ij} = 0.
\end{equation}
The axisymmetric perturbations of such a background may be conveniently expanded in the usual way (c.f., \cite{Moncrief_74_3,Moncrief_74_1,Moncrief_74_2}) in terms of Regge-Wheeler tensor harmonics, which in turn, are constructed explicitly in terms of the standard (scalar) spherical harmonics \( \lbrace Y_{\mathrm{L}0}\rbrace\). Since we shall only here be concerned with the (spherically symmetric) case corresponding to \(\mathrm{L} = 0\) and since \(Y_{00} = \sqrt{\frac{1}{4\pi}}\) we shall absorb this ubiquitous constant multiplicative factor into the perburbative functions that it multiplies (i.e., into the quantities \(H_2, K, P_{\mathrm{H}}, P_K, Y^R\), etc. defined below) to simplify the notation.

Defining
\begin{equation}\label{eq:k04}
e^{2\lambda} = \left( 1 + \frac{M}{R} + \frac{R_+^2}{R^2}\right)^2
\end{equation}
we expand the ADM spatial metric perturbation \((h_{ij}) := (\delta g_{ij})\) as
\begin{equation}\label{eq:k05}
(h_{ij}) = \left(\begin{array}{ccc}
 e^{2\lambda}H_2(R,t) & 0 & 0\\
 0 & e^{2\lambda} R^2 K (R,t) & 0\\
 0 & 0 & e^{2\lambda} R^2 \sin^2{\theta} K(R,t)
 \end{array}\right)
\end{equation}
The gauge transformations of \((h_{ij})\) in this case are generated entirely by spatial vector fields, \(Y = Y^i \frac{\partial}{\partial x^i}\), of the form
\begin{equation}\label{eq:k06}
(Y^i) = \left( Y^R (R,t), 0, 0\right)
\end{equation}
and induce the (pure gauge) first variations (c.f., \cite{Moncrief_74_3})
\begin{align}
\delta H_2 &= 2\lambda_{,R} Y^R + 2\; Y_{\hphantom{R},R}^R\label{eq:k07}\\
\delta K &= \frac{2 \left( 1 - \frac{R_+^2}{R^2}\right)}{\left( R + M + \frac{R_+^2}{R}\right)} Y^R.\label{eq:k08}
\end{align}
It is therefore natural to introduce the new variables \(\lbrace k_1, k_2\rbrace\) defined by
\begin{align}
k_1 &:= H_2 + \frac{\left( \frac{M}{R} + \frac{2R_+^2}{R^2}\right) K}{\left( 1 - \frac{R_+^2}{R^2}\right)} - \left\lbrack \frac{\left( R + M + \frac{R_+^2}{R}\right) K}{\left( 1 - \frac{R_+^2}{R^2}\right)}\right\rbrack_{,R}\label{eq:k09}\\
k_2 &:= \frac{1}{2}\; \frac{K \left( R + M + \frac{R_+^2}{R}\right)}{\left( 1 - \frac{R_+^2}{R^2}\right)} \label{eq:k10}
\end{align}
for which the inverse transformation is easily found to be
\begin{align}
H_2 &= k_1 + 2k_{2,R} - \frac{2k_2 \left( \frac{M}{R} + \frac{2R_+^2}{R^2}\right)}{\left( R + M + \frac{R_+^2}{R}\right)}\label{eq:k11}\\
K &= \frac{2k_2 \left( 1 - \frac{R_+^2}{R^2}\right)}{\left( R + M + \frac{R_+^2}{R}\right)}\label{eq:k12}
\end{align}
and for which the pure gauge variations take the form
\begin{equation}\label{eq:k13}
\delta k_1 = 0,\; \delta k_2 = Y^R
\end{equation}
showing that \(k_1\) is gauge invariant.

In view of the Gauss law constraint, \(\mathcal{E}_{\hphantom{i},i}^i = 0\), and its `linearization' about the chosen background, the only allowed spherically syummetric perturbation of \(\mathcal{E}^i \frac{\partial}{\partial x^i}\) must take the form
\begin{align}
\delta\mathcal{E}^R &:= \mathcal{E}^{R'} = 2Q' \sin{\theta},\label{eq:k14}\\
\delta\mathcal{E}^\theta &:= \mathcal{E}^{\theta'} = 0,\label{eq:k15}\\
\delta\mathcal{E}^\varphi &:= \mathcal{E}^{\varphi'} = 0.\label{eq:k16}
\end{align}
Since we shall eventually require that the perturbations of interest have compact support this will necessitate taking the charge perturbation \(Q' = 0\) but we shall retain this for now.

In terms of these new variables the linearized Hamiltonian constraint, \(\mathcal{H}' = 0\), now takes the form
\begin{equation}\label{eq:k17}
\begin{split}
\mathcal{H}' &= \sin{\theta} \frac{\partial}{\partial R} \left\lbrace\frac{-2k_1}{R} \left( R^2 - R_+^2\right)\right\rbrace\\
 &\hphantom{=} {} + \frac{\sin{\theta}}{\left( 1 + \frac{M}{R} + \frac{R_+^2}{R^2}\right)} \left\lbrace\frac{4QQ'}{R^2} - \frac{k_1}{R} \left\lbrack 2M + \frac{8R_+^2}{R} + \frac{2MR_+^2}{R^2}\right\rbrack\right\rbrace\\
 &= 0
\end{split}
\end{equation}
which, of course, is gauge invariant (c.f., the discussion in Section IV of \cite{Moncrief_74_3}). Given a choice for \(Q'\), this constraint is clearly a first order linear equation for the invariant perturbation \( k_1\) whose general solution is given by
\begin{equation}\label{eq:k18}
k_1 = -R \frac{\partial}{\partial R} \left\lbrack\frac{\left(\frac{2M'}{R} + \frac{\left( MM' - QQ'\right)}{R^2} \right)}{\left( 1 - \frac{R_+^2}{R^2}\right)}\right\rbrack
\end{equation}
where \(M'\) is the corresponding `constant' of integration which, at this point, could conceivably be a function of time (as could \(Q'\)).

As we shall see however, the linearized evolution equations can be exploited to show that both \(M'\) and \(Q'\) are both necessarily \textit{true constants} which, not surprisingly, designate 1st order variations to the mass and charge parameters of the (Reissner-Nordstr\"{o}m) `background' solution. Indeed the most straightforward way of solving Eq.~(\ref{eq:k17}) is simply to evaluate \(k_1\) for this `trivial' perturbation which, by the generalized Birkhoff theorem, is the most general, spherically symmetric perturbation that could induce a variation of this gauge invariant quantity.

It is now clear however that the only such \textit{compactly supported} perturbations must have \(Q' = M' = 0\) with \(k_1 = 0\) and \(\mathcal{E}^{i'} \frac{\partial}{\partial x^i} = 0\) and \(\mathcal{B}^{i'} \frac{\partial}{\partial x^i} = 0\) accordingly. Note furthermore that these quantities must vanish \(\forall\; t\) since their otherwise non-compact support at any finite value of \textit{t} would contradict the causal propagation of perturbations in `hyperbolic' gauge (c.f., the discussion in Appendix~\ref{app:global}). Below however we shall give an independent proof of the `conservation' of \(M'\) and \(Q'\).

Evaluating \(\mathcal{M}^{RR}\) (c.f., Eq.~(\ref{eq:f28}) on these spherically symmetric perturbations (prior to imposing their compact support) one arrives at
\begin{equation}\label{eq:k19}
\begin{split}
\frac{-\mathcal{M}^{RR}}{R} &= -\frac{1}{2}\; k_1 - R \left(\frac{k_2}{R}\right)_{,R}\\
&= R \frac{\partial}{\partial R} \left\lbrace\frac{\left(\frac{M'}{R} + \frac{\left( MM' - QQ'\right)}{2R^2}\right)}{\left( 1 - \frac{R_+^2}{R^2}\right)} - \frac{k_2}{R}\right\rbrace
\end{split}
\end{equation}
so that (c.f., Eqs.~(\ref{eq:f43})--(\ref{eq:f45}))
\begin{equation}\label{eq:k20}
\begin{split}
c_0(R) &= \frac{1}{2\pi} \int_0^{2\pi} d\theta \left( -\frac{1}{R} \mathcal{M}^{RR}\right)\\
&= R \frac{\partial}{\partial R} \left\lbrace\frac{\left(\frac{M'}{R} + \frac{\left( MM' - QQ'\right)}{2R^2}\right)}{\left( 1 - \frac{R_+^2}{R^2}\right)} - \frac{k_2}{R}\right\rbrace
\end{split}
\end{equation}
and, consequently,
\begin{equation}\label{eq:k21}
a_0(R_+) = -R_+ \left.\left\lbrace\frac{\left(\frac{M'}{R} + \frac{\left( MM' - QQ'\right)}{2R^2}\right)}{\left( 1 - \frac{R_+^2}{R^2}\right)} - \frac{k_2}{R}\right\rbrace\right|_{R_+}^\infty
\end{equation}
which clearly thus vanishes for any allowed perturbation of compact support.

The linearized Maxwell equations for \(\partial_t \mathcal{E}^{R'}\) gives immediately the expected result that \(\frac{d}{dt} Q' = 0\) (i.e., conservation of charge). To derive directly the corresponding result for \(M'\) we introduce the linearized, spherically symmetric, gravitational momenta \((\delta\pi^{ij}) := (p^{ij})\) with
\begin{equation}\label{eq:k22}
(p^{ij}) = \left(\begin{array}{ccc}
 e^\lambda R^2 \sin{\theta} P_{\mathrm{H}} (R,t) & 0 & 0\\
 0 & e^\lambda \sin{\theta} P_K (R,t) & 0\\
 0 & 0 & \frac{e^\lambda}{\sin{\theta}} P_K (R,t)\end{array}\right)
\end{equation}
and define the `new variables'
\begin{align}
p_1 &:= R^2 e^{3\lambda} P_{\mathrm{H}},\label{eq:k23}\\
\begin{split}
p_2 &:= 4Re^{2\lambda} P_K \left( 1 - \frac{R_+^2}{R^2}\right) - 2Re^{2\lambda} P_{\mathrm{H}} \left(\frac{M}{R} + \frac{2R_+^2}{R^2}\right)\\
 &\hphantom{:=} {} - \frac{\partial}{\partial R} \left\lbrack 2R^2 e^{3\lambda} P_{\mathrm{H}}\right\rbrack\label{eq:k24}
\end{split}
\end{align}
so that \(\lbrace p_1, p_2\rbrace\) are (after absorbing the normalization factor of \(Y_{00} = \sqrt{\frac{1}{4\pi}}\)) precisely the canonical momenta conjugate (respectively) to \(\lbrace k_1, k_2\rbrace\).

In terms of these quantities the linearized momentum constraint becomes
\begin{equation}\label{eq:k25}
(\mathcal{H}'_i) = \left( p_2 \sin{\theta}, 0, 0\right)
\end{equation}
which, as expected (c.f. section IV of \cite{Moncrief_74_3}) reveals this constraint as the generator of the gauge transformations (\ref{eq:k13}). The linearized evolution equations for \(\lbrace k_1, k_2\rbrace\) yield
\begin{align}
k_{1,t} &= \frac{-p_2}{2Re^{3\lambda}} \approx 0,\label{eq:k26}\\
k_{2,t} &= \frac{-p_1}{2Re^{3\lambda}} + X^{R'}\label{eq:k27}
\end{align}
where \((X^{i'}) = (X^{R'}, 0, 0)\) is the linearized shift field. Note that the first of these gives the independent proof that \(\frac{d M'}{dt} = 0\).
 %J; I
\section{Maximal Slicing Gauge Conditions}
\label{app:maximal-slicing}
For the `background' Kerr-Newman metric, expressed in Boyer-Lindquist coordinates via Eq.~(\ref{eq:b01}), both the 2+1-dimensional \textit{mean curvature} of the constant time hypersurfaces,
\begin{equation}\label{eq:j01}
\tau := \left(\frac{\tilde{g}_{ab} \tilde{\pi}^{ab}}{\mu_{{}^{(2)}\tilde{g}}}\right),
\end{equation}
and its 3+1-dimensional analogue,
\begin{equation}\label{eq:j02}
\mathrm{tr}_{{}^{(3)}g} {}^{(3)}K := \frac{\frac{1}{2} g_{ij} \pi^{ij}}{\mu_{{}^{(3)}g}} = e^\gamma \left(\tau + \frac{\tilde{p}}{4\mu_{{}^{(2)}\tilde{g}}}\right)
\end{equation}
vanish so that these slices are `maximal' in both senses of the term.

To impose, on the other hand, a (linearized) maximal slicing gauge condition on the perturbations one must choose between setting \(\tau' = 0\) (maximal slicing in the 2+1-dimensional sense) or \(\tau' + \frac{\tilde{p}'}{4\mu_{{}^{(2)}\tilde{g}}} = 0\) (its 3+1-dimensional analogue) since, in general, these are inequivalent.

The linearized field equations yield
\begin{equation}\label{eq:j03}
\tau'_{,t} = -\frac{1}{\mu_{{}^{(2)}\tilde{g}}} \partial_c (\mu_{{}^{(2)}\tilde{g}} \tilde{g}^{cd} \tilde{N}'_{,d})
\end{equation}
so that, to enforce 2+1-dimensional maximal slicing, one needs to require that the linearized lapse function \(N'\) satisfy the `harmonic' condition
\begin{equation}\label{eq:j04}
\partial_c \left(\sqrt{{}^{(2)}h} h^{cd} \tilde{N}'_{,d}\right) = 0.
\end{equation}
Taken together with the simplest (homogeneous) boundary conditions this equation has the unique, trivial solution \(\tilde{N}' = 0\). This is the gauge condition we have exploited above in our discussion of energy conservation (c.f., Section~\ref{sec:conservation}) since it automatically `kills off' several of the terms in the energy flux formula (c.f., Eq.~(\ref{eq:401})) that would, otherwise, need to be evaluated and dealt with.

Consider however the alternative condition needed to preserve 3+1-dimensional maximal slicing, namely \(\tau'_{,t} + \left(\frac{\tilde{p}'}{4\mu_{{}^{(2)}\tilde{g}}}\right)_{,t} = 0\). In this case the linearized field equations yield the more intricate elliptic equation for \(\tilde{N}'\) given by:
\begin{equation}\label{eq:j05}
\begin{split}
-\partial_a \left\lbrack\sqrt{{}^{(2)}h} h^{ab} e^\gamma (e^{-\gamma} \tilde{N}')_{,b}\right\rbrack &+ \tilde{N}' \sqrt{{}^{(2)}h} h^{ab} \left\lbrack\frac{e^{-2\gamma}}{4} (\eta_{,a} \eta_{,b} + \lambda_{,a} \lambda_{,b})\right.\\
 & \hphantom{=} {} \left. + \frac{e^{-4\gamma}}{2} (\omega_{,a} + \lambda\eta_{,a}) (\omega_{,b} + \lambda\eta_{,b})\right\rbrack\\
 & \hphantom{=} {} + \tilde{N} \sqrt{{}^{(2)}h} h^{ab} \left\lbrack\frac{e^{-2\gamma}}{4} (\eta_{,a} \eta_{,b} + \lambda_{,a} \lambda_{,b})\right\rbrack'\\
 & \hphantom{=} {} + \tilde{N} \sqrt{{}^{(2)}h} h^{ab} \left\lbrack\frac{e^{-4\gamma}}{2} (\omega_{,a} + \lambda\eta_{,a}) (\omega_{,b} + \lambda\eta_{,b})\right\rbrack'\\
 & \hphantom{=} {} + \left(\tilde{N} \sqrt{{}^{(2)}h} h^{ab} \gamma'_{,b}\right)_{,a}\\
 & = 0.
\end{split}
\end{equation}
One would want to solve this equation, if possible, with boundary conditions chosen so that no non-vanishing energy flux contributions result from the terms involving \(\tilde{N}'\) in Eq.~(\ref{eq:401}).

Equation (\ref{eq:j05}) will be more recognizable and tractible to analyze if we first `lift' it back to 3-dimensions and reexpress it as an equation for the first variation, \(N'\), of the \textit{3+1-dimensional} lapse function \(N = e^{-\gamma} \tilde{N}\), namely
\begin{equation}\label{eq:j06}
N' = e^{-\gamma} \tilde{N}' - \gamma' e^{-\gamma} \tilde{N} = e^{-\gamma} \tilde{N}' - \gamma' N.
\end{equation}
At this point of course \(\gamma'\) and \textit{N} will be known quantities that can be `shifted' into the `source terms' for the single unknown \(N'\). The lifted equation, expressed in terms of the ADM spatial metric \({}^{(3)}g = g_{ij}\; dx^i \otimes dx^j\) (c.f., Appendix~\ref{app:reduced-hamiltonian}) takes the form
\begin{equation}\label{eq:j07}
\begin{split}
-\partial_i \left(\sqrt{{}^{(3)}g} g^{ij} N'_{,j}\right) &+ N' \sqrt{{}^{(3)}g} g^{ij} \left\lbrack\frac{e^{-2\gamma}}{4} (\eta_{,i} \eta_{,j} + \lambda_{,i} \lambda_{,j})\right.\\
 &\hphantom{=} {} + \left.\frac{e^{-4\gamma}}{2} (\omega_{,i} + \lambda\eta_{,i}) (\omega_{,j} + \lambda\eta_{,j})\right\rbrack\\
 &= -Ne^{-\gamma} \sqrt{{}^{(3)}g} g^{ij} \left\lbrack\frac{e^{-\gamma}}{4} (\eta_{,i} \eta_{,j} + \lambda_{,i} \lambda_{,j})\right.\\
 &\hphantom{=} {} + \left.\frac{1}{2} e^{-3\gamma} (\omega_{,i} + \lambda\eta_{,i}) (\omega_{,j} + \lambda\eta_{,j})\right\rbrack'\\
 &\hphantom{=} {} \partial_i \left(\sqrt{{}^{(3)}g} g^{ij} \gamma' N_{,j}\right)
\end{split}
\end{equation}
wherein, for the sake of uniform notation, we have included terms that actually vanish by virtue of axisymmetry \(\lbrace \hbox{e.g.~} \eta_{,3} = \eta_{,\varphi}, \eta'_{,\varphi}, \hbox{~etc.}\rbrace\). By the same token we are only interested in axisymmetric solutions for which of course \(\partial_i \left(\sqrt{{}^{(3)}g} g^{ij} N'_{,j}\right) \rightarrow \partial_a \left(\sqrt{{}^{(3)}g} g^{ab} N'_{,b}\right)\). Equation~(\ref{eq:j07}) is of course nothing but the linearized version of the usual 3+1-dimensional lapse equation for maximal slicing reexpressed in terms of our variables and restricted to a Kerr-Newman background solution. We anticipate that well-known arguments (c.f., \cite{Bartnik_84}) can be modified to establish the existence and uniqueness of smooth, axisymmetric solutions to this equation that vanish at infinity with homogeneous Dirichlet data specified on the horizon boundary (i.e., \(N'|_{R_+} = 0\)). In fact a standard uniqueness argument would suffice to guarantee that any such (i.e., smooth, bounded with vanishing Dirichlet data) solution would automatically be axisymmetric and hence project naturally to the original quotient space whereon Eq.~(\ref{eq:j05}) was formulated.

But would such a solution contribute unwanted flux terms to Eq.~(\ref{eq:401}) and disrupt the argument for conservation of energy?

The terms in Eq.~(\ref{eq:401}) involving \(\tilde{N}'\) can be expressed as the divergence of the vector density
\begin{equation}\label{eq:j08}
\begin{split}
\varXi^b &:= \left(\tilde{N}\tilde{N}'_{,a} - \tilde{N}'\tilde{N}_{,a}\right)\; 2\tilde{\pi}^{\prime ab}\\
 &= \left( e^\gamma \tilde{N}N'_{,a} + \tilde{N}^2\gamma'_{,a} - N' e^{2\gamma} N_{,a}\right) 2\tilde{\pi}^{\prime ab}
\end{split}
\end{equation}
where
\begin{equation}\label{eq:j09}
\tilde{\pi}^{\prime ab} = \tilde{g}^{bc} \left\lbrack (\tilde{r}_c^{\hphantom{c}a})' + \frac{1}{2}\; \delta_c^a \tau' \mu_{{}^{(2)}\tilde{g}}\right\rbrack
\end{equation}
and \(N'\) is given by (\ref{eq:j06}).

From Eqs.~(\ref{eq:f76}--\ref{eq:f77}) one sees that, in the asymptotic regions near \(R \searrow R_+\) and \(R \nearrow \infty\), one has
\begin{equation}\label{eq:j10}
\tilde{\pi}^{\prime cd} = \frac{\tilde{g}^{ac}\tilde{g}^{bd}}{2\tilde{N}}\; \mu_{{}^{(2)}\tilde{g}} \left\lbrack (\mathcal{L}_{{}^{(2)}\mathcal{D}}\tilde{g})_{ab} - \tilde{g}_{ab}\tilde{g}^{ef} (\mathcal{L}_{{}^{(2)}\mathcal{D}}\tilde{g})_{ef}\right\rbrack
\end{equation}
where
\begin{equation}\label{eq:j11}
{}^{(2)}\mathcal{D} := \left(\tilde{N}^2\tilde{g}^{cd}\; {}^{(4)}Y_{,d}^0\right) \frac{\partial}{\partial x^c}.
\end{equation}
In terms of the `conformal data' \(\nu\) and \(h_{ab}\) (for which, as before, \(\tilde{g}_{ab} = e^{2\nu}\; h_{ab}\)) this becomes
\begin{equation}\label{eq:j12}
\tilde{\pi}^{\prime cd} = \mu_{{}^{(2)}h} \frac{h^{ac} h^{bd}}{2\tilde{N}} \left\lbrace -2 {}^{(2)}\mathcal{D}^e\; \nu_{,e}\; h_{ab} + (\mathcal{L}_{{}^{(2)}\mathcal{D}} h)_{ab} - h_{ab} h^{ef} (\mathcal{L}_{{}^{(2)}\mathcal{D}} h)_{ef}\right\rbrace .
\end{equation}
Utilizing the asymptotic properties of \({}^{(2)}\mathcal{D}\) derived in Section~\ref{subsec:evaluating_dynamical} and imposing the (homogeneous) Dirichlet boundary condition \(N' |_{R_+} = 0\) upon the desired solution of Eq.~(\ref{eq:j07}) one can show that, if a regular such solution exists, then one has
\begin{equation}\label{eq:j13}
\varXi^R |_{R_+} = 0
\end{equation}
i.e., pointwise vanishing of the energy flux integrand at the horizon boundary. Furthermore the corresponding flux integrand vanishes as \(R \nearrow \infty\) for any solution \(N'\) that grows sufficiently slowly. In particular any solution that is bounded with bounded first derivatives would yield a (pointwise) vanishing energy flux integrand as \(R \nearrow \infty\).

Finally, by exploiting the regularity results for axi-symmetric fields and their perturbations derived in \cite{Rinne_05}, it is straightforward to verify that potential flux contributions at the (artificial) boundaries provided by the axes of symmetry at \(\theta = 0,\pi\) vanish (pointwise) as \(O(\sin^2{\theta})\). Thus, modulo the aforementioned need for an existence proof for Eq.~\ref{eq:j07}, it follows that conservation of our energy functional holds as well in the 3+1-dimensional maximal slicing gauge.  %K; J; origninally appendix I
\section{The Weyl Tensor for Vacuum Axisymmetric Spacetimes}
\label{app:weyl-tensor}
In section~\ref{sec:pure-electromagnetic-kerr-spacetimes} we analyzed the (axisymmetric) purely electromagnetic perturbations of a Kerr black hole spacetime by introducing a complete set of (electromagnetic) gauge and infinitesimal diffeomorphism-invariant canonical variables for the (linearized) Maxwell field and deriving a conserved, positive definite energy functional expressible in terms of these quantities. An advantage of the use of such variables is their insensitivity to the non-local features of any elliptic gauge condition that one might choose to employ. By contrast the variables we introduced later for the full, linearized Kerr-Newman problem  were gauge dependent---a feature directly reflected in the dependence of their evolution equations on the elliptically determined (hence non-local) linearized lapse and shift fields \(\lbrace\tilde{N}^{\,\prime},\tilde{N}^{\,{a'}}\rbrace\).

It is therefore natural to ask whether, at least for the purely gravitational perturbations of a Kerr background, a corresponding set of fully gauge-invariant canonical variables might also be available for the (linearized) metric component of the problem. Since the (complex) field satisfying Teukolsky's equation is gauge-invariant one might well expect that it provides (upon specialization to the axisymmetric setting considered here) a natural answer to this question. If one could affect a canonical transformation to a new set of variables that includes Teukolsky's field and its conjugate momentum as a gauge invariant subset than one would expect that our energy functional, which is itself gauge-invariant, could be reexpressed purely in terms of this (invariant) subset.

Since Teukolsky's field is defined in terms of the linearization of the Weyl tensor about a Kerr background we present here the actual Weyl tensor for \textit{vacuum}, \textit{axisymmetric} metrics expressed in terms of our symmetry reduced canonical variables from Appendix~\ref{app:reduced-hamiltonian}. Since, in the case of a vacuum background, the linearized gravitational and electromagnetic perturbations decouple from one another and since we have already dealt with the Maxwell component in Section~\ref{sec:pure-electromagnetic-kerr-spacetimes}, we focus exclusively here on the pure metric component and specialize the formulas of Appendix~\ref{app:reduced-hamiltonian} accordingly.

As is well-known \cite{AM_04,MR_09} the Weyl tensor for a \textit{vacuum} spacetime can be expressed in terms of ADM Cauchy data \(\lbrace{}^{(3)}g = g_{ij} dx^i \otimes dx^j, {}^{(3)}\pi = \pi^{ij} \frac{\partial}{\partial x^i} \otimes \frac{\partial}{\partial x^j}\rbrace\) on a 3-manifold M as a pair of (traceless, symmetric) tensor densities, an `electric' field \({}^{(3)}\mathcal{E} = \mathcal{E}^{ij} \frac{\partial}{\partial x^i} \otimes \frac{\partial}{\partial x^j}\) given by
\begin{equation}\label{eq:d01}
\mathcal{E}^{ij} := \left\lbrace\mu_{{}^{(3)}g} {}^{(3)}R^{ij} ({}^{(3)}g) - \frac{1}{\mu_{{}^{(3)}g}} \left(\pi^i_{\hphantom{i}m} \pi^{mj} - \frac{1}{2}\pi^{ij} \pi^m_{\hphantom{m}m}\right)\right\rbrace
\end{equation}
and a corresponding `magnetic' field \({}^{(3)}\mathcal{B} = \mathcal{B}^{ij} \frac{\partial}{\partial x^i} \otimes \frac{\partial}{\partial x^j}\) defined by
\begin{equation}\label{eq:d02}
\mathcal{B}^{ij} := \frac{\epsilon^{m\ell j}}{\mu_{{}^{(3)}g}} \left\lbrace\pi_{i\hphantom{m}|\ell}^m - \frac{1}{2} \delta_m^i (\pi^k_{\hphantom{k}k})_{|\ell}\right\rbrace.
\end{equation}
Note that \({}^{(3)}\mathcal{E}\), though manifestly symmetric, is traceless only by virtue of the (vacuum) Hamiltonian constraint
\begin{equation}\label{eq:d03}
\mathcal{E}^m_{\hphantom{m}m} = - \mathcal{H} \rightarrow 0\qquad \hbox{(in vacuum)}
\end{equation}
whereas \({}^{(3)}\mathcal{B}\), though identically traceless, is symmetric only by virture of the (vacuum) momentum constraint
\begin{equation}\label{eq:d04}
\epsilon_{ijk} \mathcal{B}^{ij} = -\frac{1}{2} \frac{\mathcal{H}_k}{\mu_{{}^{(3)}g}} \rightarrow 0\quad \hbox{(in vacuum)}.
\end{equation}

One can now evaluate \({}^{(3)}\mathcal{E}\) and \({}^{(3)}\mathcal{B}\) in terms of the canonical pairs, \(\left\lbrace (\tilde{g}_{ab},\tilde{\pi}^{ab}), (\beta_a,\tilde{e}^a), (\gamma,\tilde{p})\right\rbrace\), for the (axial-) symmetry reduced system defined in Eqs.~(\ref{eq:a13}--\ref{eq:a17}). A final transformation to wave map variables would then result from the substitution
\begin{gather}
\tilde{e}^a \rightarrow \epsilon^{ab}\omega_{,b},\label{eq:d05}\\
\epsilon^{ab}\beta_{a,b} \rightarrow \tilde{r}\label{eq:d06}
\end{gather}
On the other hand the one-form field \(\beta_a dx^a\), which appears in the spatial metric \({}^{(3)}g\), is non-local in terms of the wave map field \(\tilde{r}\) and, moreover, incorporates a (longitudinal) component that varies as
\begin{equation}\label{eq:d07}
\beta_a \rightarrow \beta_a + \lambda_{|a}
\end{equation} under a coordinate transformation of the form
\begin{equation}\label{eq:d08}
x^3 = \varphi \rightarrow \varphi + \lambda
\end{equation}
whereas \(\tilde{r}\) is invariant with respect to such a transformation. For this reason we prefer to express the results in terms of the intermediate canonical pairs listed above.

Only a certain set of `mixed' components of \({}^{(3)}\mathcal{E}\) and \({}^{(3)}\mathcal{B}\) are invariant with respect to the aforementioned `gauge' transformation of \(\beta_a dx^a\), namely,
\begin{equation}\label{eq:d09}
\left\lbrace\mathcal{E}^{ab},\mathcal{E}_{33}, \mathcal{E}_3^{\hphantom{3}a}, \mathcal{B}^{ab}, \mathcal{B}_{33}, \mathcal{B}_3^{\hphantom{3}a}\right\rbrace.
\end{equation}
Using the spatial metric \({}^{(3)}g\) to raise or lower indices one can easily express all of the contravariant or covariant components of these fields in terms of the specified `mixed' components but only at the expense of foregoing the aforementioned invariance.

Without further ado we present here the relevant, `mixed' components of the Weyl tensor expressed in terms of the symmetry reduced canonical variables:
\begin{equation}\label{eq:d10}
\begin{split}
\mathcal{E}^{ab} &= e^{3\gamma} \mu_{{}^{(2)}\tilde{g}} \left\lbrace\vphantom{\frac{1}{2}}{}^{(2)}\tilde{R}^{ab} - {}^{(2)}\tilde{\nabla}^b {}^{(2)}\tilde{\nabla}^a \gamma + \tilde{g}^{ab} \left({}^{(2)}\tilde{\nabla}_c {}^{(2)}\tilde{\nabla}^c \gamma\right)\right.\\
& {}- 3 \left({}^{(2)}\tilde{\nabla}^a \gamma\right) \left({}^{(2)}\tilde{\nabla}^b \gamma\right) + \tilde{g}^{ab} \left({}^{(2)}\tilde{\nabla}_c \gamma\right) \left({}^{(2)}\tilde{\nabla}^c \gamma\right)\\
& \left.{}- \frac{1}{2} e^{4\gamma} \tilde{g}^{ac} \tilde{g}^{df} \tilde{g}^{be} \left(\beta_{d,e} - \beta_{e,d}\right) \left(\beta_{f,c} - \beta_{c,f}\right)\right\rbrace\\
& {}- \frac{e^\gamma}{\mu_{{}^{(2)}\tilde{g}}} \left\lbrace-\frac{1}{2} e^{2\gamma} \tilde{\pi}^{ab} \left(\frac{1}{2} \tilde{p} + 2\tilde{g}_{cd} \tilde{\pi}^{cd}\right)\right.\\
& \quad\left.{}+ e^{2\gamma} \tilde{g}_{cd} \tilde{\pi}^{ad} \tilde{\pi}^{bc} + \frac{1}{4} e^{-2\gamma} \tilde{e}^a \tilde{e}^b\right\rbrace ,
\end{split}
\end{equation}
\begin{equation}\label{eq:d11}
\begin{split}
\mathcal{E}_{33} &= -\frac{e^\gamma}{\mu_{{}^{(2)}\tilde{g}}} \left\lbrace\frac{1}{4} e^{-2\gamma} \tilde{g}_{ab} \tilde{e}^a \tilde{e}^b + \frac{e^{2\gamma}}{4} \tilde{p} \left(\frac{1}{2} \tilde{p} + \tilde{g}_{ab} \tilde{\pi}^{ab}\right)\right\rbrace\\
& {}- e^{3\gamma} \left\lbrace\vphantom{\frac{1}{4}}\partial_a \left(\mu_{{}^{(2)}\tilde{g}} \tilde{g}^{ab} \gamma_{,b}\right) + \mu_{{}^{(2)}\tilde{g}} \tilde{g}^{ab} \gamma_{,a} \gamma_{,b}\right.\\
& \left.{}+ \frac{1}{4} e^{4\gamma} \mu_{{}^{(2)}\tilde{g}} \tilde{g}^{ac} \tilde{g}^{bd} \left(\beta_{c,b} - \beta_{b,c}\right) \left(\beta_{d,a} - \beta_{a,d}\right)\right\rbrace ,
\end{split}
\end{equation}
\begin{equation}\label{eq:d12}
\begin{split}
\mathcal{E}_3^{\hphantom{3}a} &= e^{5\gamma} \mu_{{}^{(2)}\tilde{g}} \left\lbrace\frac{1}{2} {}^{(2)}\tilde{\nabla}_b \left\lbrack\tilde{g}^{bd} \tilde{g}^{ac} \left(\beta_{d,c} - \beta_{c,d}\right)\right\rbrack\right.\\
& \left.{}- \frac{5}{2} \left({}^{(2)}\tilde{\nabla}^b \gamma\right) \tilde{g}^{ac} \left(\beta_{b,c} - \beta_{c,b}\right)\right\rbrace\\
& {}- \frac{e^\gamma}{\mu_{{}^{(2)}\tilde{g}}} \left\lbrack\frac{1}{2} \tilde{e}^c\; \tilde{g}_{bc} \tilde{\pi}^{ab} + \frac{1}{8} \tilde{p}\; \tilde{e}^a\right\rbrack,
\end{split}
\end{equation}
\begin{equation}\label{eq:d13}
\begin{split}
\mathcal{B}^{ab} &= \epsilon^{bc3} \left\lbrace\frac{1}{2} e^{5\gamma} \frac{\tilde{g}_{ce}}{\mu_{{}^{(2)}\tilde{g}}} \tilde{\pi}^{de} \tilde{g}^{af} \epsilon_{fd} \left(\epsilon^{mn} \beta_{m,n}\right)\right.\\
& {}- \frac{\tilde{e}^d}{\mu_{{}^{(2)}\tilde{g}}} e^\gamma \tilde{g}^{af} \gamma_{,f} \tilde{g}_{dc} + \frac{1}{2} \frac{e^\gamma \tilde{e}^d}{\mu_{{}^{(2)}\tilde{g}}} \gamma_{,d} \delta_c^a\\
& {}- \frac{1}{2} \frac{e^{5\gamma}}{\mu_{{}^{(2)}\tilde{g}}} \left(\frac{1}{2} \tilde{p} + \tilde{g}_{mn} \tilde{\pi}^{mn}\right) \tilde{g}^{af} \epsilon_{fc} \left(\epsilon^{rs} \beta_{r,s}\right)\\
& \quad\left.{}- \frac{1}{2} e^\gamma\; {}^{(2)}\tilde{\nabla}_c \left(\frac{\tilde{e}^a}{\mu_{{}^{(2)}\tilde{g}}}\right)\right\rbrace ,
\end{split}
\end{equation}
\begin{equation}\label{eq:d14}
\begin{split}
\mathcal{B}_{33} &= \epsilon^{ac3} \left\lbrace e^{2\gamma} \left(\frac{1}{2} \frac{e^{-\gamma}}{\mu_{{}^{(2)}\tilde{g}}} \tilde{e}_a\right)_{,c} - \frac{1}{2} \frac{e^\gamma \tilde{e}_a}{\mu_{{}^{(2)}\tilde{g}}} \gamma_{,c}\right.\\
& {}+ \left(\frac{e^{5\gamma}}{2 \mu_{{}^{(2)}\tilde{g}}}\right) \left(\beta_{a,c} - \beta_{c,a}\right) \left(\frac{1}{2} \tilde{p} + \tilde{g}_{bd} \tilde{\pi}^{bd}\right)\\
& \left.{} -\left(\frac{1}{2} \frac{e^{5\gamma}}{\mu_{{}^{(2)}\tilde{g}}}\right) \tilde{\pi}_a^f \left(\beta_{f,c} - \beta_{c,f}\right)\right\rbrace,
\end{split}
\end{equation}
\begin{equation}\label{eq:d15}
\begin{split}
\mathcal{B}_3^{\hphantom{3}a} &= \epsilon^{3ca} \left\lbrace -\frac{e^{3\gamma}}{\mu_{{}^{(2)}\tilde{g}}} \gamma_{,b} \left(\tilde{\pi}_c^b - \delta_c^b \tilde{g}_{mn} \tilde{\pi}^{mn}\right)\right.\\
& \left.{}- \frac{1}{4} \frac{e^{3\gamma}}{\mu_{{}^{(2)}\tilde{g}}} \tilde{e}^b \left(\beta_{b,c} - \beta_{c,b}\right) + \frac{1}{4} \left(\frac{e^{3\gamma} \tilde{p}}{\mu_{{}^{(2)}\tilde{g}}}\right)_{,c}\right\rbrace .
\end{split}
\end{equation}
In these formulas indices \(a, b, c, \ldots\) are raised and lowered with the Riemannian 2-metric \({}^{(2)}\tilde{g} = \tilde{g}_{ab} dx^a \otimes dx^b\), \({}^{(2)}\tilde{\nabla}_a\) designates covariant differentiations with respect to this metric whereas \(\mu_{{}^{(2)}\tilde{g}}\) and \({}^{(2)}\tilde{R}^{ab}\) are its `volume' element and Ricci tensor.

Whereas the explicit symmetry of \({}^{(3)}\mathcal{E}\) implies, for example, that
\begin{equation}\label{eq:d16}
\mathcal{E}_3^{\hphantom{3}a} = \mathcal{E}^a_{\hphantom{a}3}\quad \hbox{ and }\quad \epsilon_{ab} \mathcal{E}^{ab} = 0
\end{equation}
the corresponding equations for \({}^{(3)}\mathcal{B}\) only hold `weakly' (i.e., modulo the momentum constraints). More precisely one finds that
\begin{align}
\mathcal{B}_3^{\hphantom{3}a} - \mathcal{B}^a_{\hphantom{a}3} &= -\frac{1}{2} \frac{e^{3\gamma}}{\mu_{{}^{(2)}\tilde{g}}} \epsilon^{ab3} \tilde{\mathcal{H}}_b,\label{eq:d17}\\
\epsilon_{ab} \mathcal{B}^{ab} &= \frac{1}{2} \frac{e^\gamma}{\mu_{{}^{(2)}\tilde{g}}} \tilde{e}^c_{\hphantom{c},c}\label{eq:d18}
\end{align}
and
\begin{equation}\label{eq:d19}
\mathcal{B}^{3a} - \mathcal{B}^{a3} = -\frac{1}{2} \frac{\epsilon^{ac3} e^\gamma}{\mu_{{}^{(2)}\tilde{g}}} \left(\tilde{\mathcal{H}}_c - \beta_c \tilde{e}^d_{\hphantom{d},d}\right).
\end{equation}
where
\begin{equation}\label{eq:d20}
\mathcal{B}^{3a} = e^{-2\gamma} \mathcal{B}_3^{\hphantom{3}a} - \beta_b \mathcal{B}^{ba}
\end{equation}

To see that all of the components of \({}^{(3)}\mathcal{E}\) are indeed determined by the mixed set we have presented above one computes that
\begin{equation}\label{eq:d21}
\mathcal{E}^{a3} = \mathcal{E}^{3a} = e^{-2\gamma} \mathcal{E}^a_{\hphantom{a}3} - \beta_b \mathcal{E}^{ab}
\end{equation}
and that
\begin{equation}\label{eq:d22}
\mathcal{E}^{33} = e^{-4\gamma} \mathcal{E}_{33} - 2e^{-2\gamma} \beta_a \mathcal{E}_3^{\hphantom{3}a} + \beta_a \beta_b \mathcal{E}^{ab}.
\end{equation}
As we have already mentioned, the trace of \({}^{(3)}\mathcal{E}\) only vanishes \textit{weakly} since, in fact
\begin{equation}\label{eq:d23}
g_{ij} \mathcal{E}^{ij} = -\mathcal{H} = -e^\gamma \tilde{\mathcal{H}}.
\end{equation}
Similar formulas hold for the (contravariant) components of \({}^{(3)}\mathcal{B}\), allowing for the fact that it, unlike \({}^{(3)}\mathcal{E}\), is not explicitly symmetric.

A straightforward further calculation gives
\begin{equation}\label{eq:d24}
\begin{split}
\mathcal{E}^{ij} \mathcal{E}_{ij} &= e^{-4\gamma} \left\lbrace\left(\mathcal{E}_{33}\right)^2 + 2\tilde{g}_{ab} \mathcal{E}^a_{\hphantom{a}3} \mathcal{E}^b_{\hphantom{b}3}\right.\\
& \left.{}+ \tilde{g}_{ac} \tilde{g}_{bd} \mathcal{E}^{ab} \mathcal{E}^{cd}\right\rbrace
\end{split}
\end{equation}
which, being independent of \(\beta_a\), is invariant under the `gauge' transformation \(\beta_a \rightarrow \beta_a + \lambda_{|a}\). A similar formula can of course be derived for \(\mathcal{B}^{ij}\mathcal{B}_{ij}\), again allowing for the lack of explicit symmetry of \({}^{(3)}\mathcal{B}\). Taken together these quantities constitute the `Bel Robinson energy density'.  %L; J; orignally appendix D

\newpage

%\bibliographystyle{amsplain}
%\bibliography{References_BH}

\end{document}